\documentclass[phd,tocprelim]{cornell}
\let\ifpdf\relax
\usepackage[font=md,captionskip=8pt]{subfig}
\usepackage{amssymb}
\usepackage{epsfig}
\usepackage{epstopdf}

\usepackage{moreverb}
\usepackage{txfonts}
\usepackage{palatino}
\usepackage[bookmarks=false]{hyperref}

\newcommand{\cD}{{\cal D}}

\newcommand{\cL}{{\cal L}}
\newcommand{\cM}{{\cal M}}
\newcommand{\cN}{{\cal N}}
\newcommand{\cO}{{\cal O}}

\newcommand{\cX}{{\cal X}}
\newcommand{\cZ}{{\cal Z}}

\newcommand{\oD}{{\overline D}}
\newcommand{\opsi}{{\overline\psi}}

\newcommand{\Vp}{{V_{2}}}
\newcommand{\Vm}{{V_{1}}}

\newcommand{\Tr}{\mbox{Tr}}

\renewcommand{\Re}{\text{Re}\ }
\renewcommand{\Im}{\text{Im}\ }

\newcommand{\ra}{\rightarrow}
\newcommand{\be}{\begin{equation}}
\newcommand{\ee}{\end{equation}}
\newcommand{\bea}{\begin{eqnarray}}
\newcommand{\eea}{\end{eqnarray}}

\DeclareMathSymbol{\mg}{\mathrel}{symbols}{"1D}

\def\ba{\begin{array}}
\def\bem{\begin{displaymath}}

\def\ea{\end{array}}
\def\eem{\end{displaymath}}

\def\Im{\mathop{\rm Im}}

\def\ov{\overline}

\def\Re{\mathop{\rm Re}}

\def\s2w{\sin^2 \theta_W}
\def\Tr{\mathop{\rm Tr}}

\relax
\def\dalpha{{\dot\alpha}}
\def\dbeta{{\dot\beta}}

\def\crbig{\\\noalign{\vspace {3mm}}}
\def\bigint{{\displaystyle\int}}

\def\Dint{\bigint d^2\theta d^2\ov\theta\,}
\def\Fint{\bigint d^2\theta\,}
\def\Fbarint{\bigint d^2\ov\theta\,}
\relax

\title {Aspects of Effective Supersymmetric Theories}
\author {Panteleimon Tziveloglou}
\conferraldate {January}{2011}
\degreefield {Ph.D.}
\copyrightholder{Panteleimon Tziveloglou}
\copyrightyear{2011}

\begin{document}

\maketitle
\makecopyright

\begin{abstract}
This work consists of two parts. In the first part we construct the complete extension of the Minimal Supersymmetric Standard Model by higher dimensional effective operators and then study its phenomenology. These operators encapsulate the effects on LHC physics of any kind of new degrees of freedom at the multiTeV scale. The effective analysis includes the case where the multiTeV physics is the supersymmetry breaking sector itself. In that case the appropriate framework is nonlinear supersymmetry. We choose to realize the nonlinear symmetry by the method of constrained superfields. Beyond the new effective couplings, the analysis suggests an interpretation of the `little hierarchy problem' as an indication of new physics at multiTeV scale.

In the second part we explore the power of constrained superfields in extended supersymmetry. It is known that in $\cN = 2$ supersymmetry the gauge kinetic function cannot depend on hypermultiplet scalars. However, it is also known that the low energy effective action of a D-brane in an $\cN = 2$ supersymmetric bulk includes the DBI action, where the gauge kinetic function does depend on the dilaton. We show how the nonlinearization of the second SUSY (imposed by the presence of the D-brane) opens this possibility, by constructing the global $\cN$ = 1 linear + 1 nonlinear invariant coupling of a hypermultiplet with a gauge multiplet. The constructed theory enjoys interesting features, including a novel super-Higgs mechanism without gravity.

\end{abstract}

\begin{biosketch}
\vspace{2cm}
Pantelis was born in Thessaloniki, Greece, on the 19th of December 1983. In 2001, he enrolled in the Physics department of Aristotle University of Thessaloniki. In 2005, after obtaining the physics degree, he was accepted to Cornell University for postgraduate studies in physics. His academic interests brought him in 2007 to CERN in order to conduct his doctoral research under the guidance of Prof. Ignatios Antoniadis. He stayed there for three years until the completion of his Ph.D. thesis. From October 2010 he is situated in Paris and \'Ecole Polytechnique for postdoctoral research.
\end{biosketch}


\begin{acknowledgements}
\vspace{2cm}
I had the great privilege of being guided throughout my Ph.D. by Ignatios Antoniadis. I want to express my gratitude for our excellent collaboration and for his kind support during these three years.

I am particularly grateful to my collaborators for the first part of this work, Emilian Dudas and Dumitru Ghilencea for a very fruitful collaboration from which I gained a lot. For the second part I was fortunate enough to work with Jean - Pierre Derendinger and Nicola Ambrosetti who I wanted to thank for our joyful collaboration.

I would like to thank Henry Tye and Csaba Csaki for giving me the permission to complete my Ph.D. studies at CERN. I am especially grateful to Ritchie Patterson who, as Director of Graduate Studies at the physics department of Cornell, provided the necessary support for my transfer to CERN. I perceive her attention as a genuine example of an academic that truly cares about the ambitions of her students.

My Ph.D. studies would never have been such an amazing experience if it hadn't been for all the people I met, I shared apartments and offices, I experienced unforgettable moments of life. I want to thank them all and I hope that our paths will continue to cross in the future.
\end{acknowledgements}

\contentspage
\figurelistpage

\normalspacing \setcounter{page}{1} \pagenumbering{arabic}
\pagestyle{cornell} \addtolength{\parskip}{0.5\baselineskip}

\chapter{Introduction}

\section{The Importance of Supersymmetry}

Probably the most significant manifestation of the beauty of Supersymmetry is that this simple idea of a symmetry that relates fermions and bosons has proven to be one of the most fruitful proposals in theoretical high energy physics of the last forty years.

At the level of phenomenology, supersymmetry offers a complete or partial solution to almost all shortcomings of the Standard Model (SM). For example, the beautiful properties of SM under renormalization are based on the fact that it is a model of fermions and gauge bosons. However, its cornerstone, the Higgs mechanism, is bound to the existence of a scalar mode. The Higgs scalar seems very unnatural within the framework of the SM. It is the only scalar field and it doesn't share the same renormalization properties with the others. More specifically, the natural value for its mass is at the Planck scale, which would obviously destroy the validity of the model. This puzzle comes with the name ``hierarchy problem" and it's believed to be one of the main reasons for leaving SM behind. The solution by supersymmetry is based on treating scalars on equal footing with all other fields. Not only it contains a variety of scalars, degrading them from the special role they enjoyed in SM, their normalization properties are also no different than all other fields. Their masses scale logarithmically with the cutoff scale which then offers a resolution to the hierarchy problem.

Another source of skepticism towards the SM comes from cosmology. There is a set of cosmological and astrophysical observations that lead to the same conclusion. The stable matter described by the SM, which is the matter that surrounds us, is nothing but a tiny fraction of the full matter content of the universe. `Out there', stable particles exist that we have never observed and that are not described by the SM. The observations can also inform us about the basic properties of these particles. It comes out that they have to be massive and weakly interacting. Once again, supersymmetry has the answer. Supersymmetric models generically come with one stable particle that enjoys the desired properties.

We should also mention that supersymmetry seems to complete the program for unification of gauge interactions. The SM had the striking success of unifying the numerous processes between particles observed in colliders (and seeming extremely complicated in the early years of particle physics) into three fundamental gauge interactions parametrized by three independent coupling constants. The unification would be complete by further unifying into a single gauge group, which would then lead us to a ``Grand Unified Theory". Unfortunately it was calculated that the renormalization group (RG) equations of the SM don't meet at a single point for unification to occur. New degrees of freedom are needed to shift the RG in a way they meet. It has been shown that the degrees of freedom brought by supersymmetry do the job and the predicted unification occurs at around $10^{16}$ GeV.

The above arguments favor supersymmetric models as a candidate for departure from the SM. It seems however that it doesn't merely offer a model for a successful replacement of the SM but it's basic concepts play a fundamental role in quantum field theory. This can be seen as follows. In a paper of 1967 by S. Coleman and J. Mandula \cite{ColemanMandula} it was shown that the most general Lie algebra of symmetries of the S-matrix is the Poincar\'e algebra plus a number of Lorentz scalar generators that form the algebra of a compact Lie group. This was a conclusive no-go theorem about the allowed symmetries of the S-matrix and in particular about the impossibility of a nontrivial combination of a spacetime symmetry with an internal one. The Coleman-Mandula theorem was extremely powerful as it was based on generic assumptions that would apply to any quantum field theory. However, it was later discovered that the assumption that the algebras need to be Lie algebras was too restrictive as one could add fermionic generators forming what is called ``graded Lie algebras". In a paper by Haag, Sohnius and Lopuszanski seven years later, it was shown that the only graded Lie algebras that generate symmetries allowed by the generic assumptions of quantum field theory are the supersymmetric algebras \cite{Haag}. In a few words, the exploration of the largest symmetry allowed by the S-matrix has inevitably led us to supersymmetry.

Last but definitely not least, supersymmetry opens a window for the holy grail of theoretical physics, the unification of gravity with the other three forces. The combination of the principle that gravity is the manifestation of the curvature of spacetime, coming from general relativity, and the fact that supersymmetry is a spacetime symmetry, coming from the Haag-Sohnius-Lopuszanski theorem above, implies that a theory with local supersymmetry is a theory of gravity. Such a theory is called ``supergravity". Supergravities themselves appear as the low energy effective theories of various settings of string theory, the only framework where gravity and the other forces are unified into a single and finite theory. In summary, following the path: Global Supersymmetry $\ra$ Local Supersymmetry $\ra$ String Theory we obtain, for the first time, a complete picture of how the unification of particles and interactions works. Furthermore, the principle of supersymmetry is built in string theory. The very first appearance of a symmetry that exchanges bosons and fermions first appeared in the context of dual models \cite{Ramond, NeveuS}, which is what was later reinterpreted as string theory. Without supersymmetry, string theory would not be a consistent theory. In a few words, the most basic ingredient of the only known path to a theory where matter and forces are unified, is supersymmetry.

This thesis touches upon both model building in supersymmetric theories and more formal aspects, especially related to string theory. It is then naturally devided in two parts which are weakly related to each other and can be read independently. It is based on publications \cite{Paper1,Paper2,Paper3,Paper4,Paper5}.

\section{Effective and Nonlinear Field Theory in the Minimal Supersymmetric Standard Model}

In the first part we apply the techniques of Effective Field Theory (EFT) on the Minimal Supersymmetric Standard Model (MSSM) and study their phenomenological consequences. The MSSM is the minimal extension of the SM and is used as a prototype model for phenomenological studies of supersymmetry \cite{Nilles:1983ge}. Our method involves the addition of higher mass dimension terms in the MSSM Lagrangian. From an EFT point of view, the appearance of such terms is not a sign that the model is sick but rather an indication that it is valid only up to the mass scale that suppresses those terms. Their purpose is to parametrize the effects of any kind of new physics that might exist at a scale that is not approachable by LHC and in the same time not too high, so in the range of a few TeV.

In a few lines, the method of our analysis is as follows. We construct the effective Lagrangian by adding to that of MSSM nonrenormalizable terms of higher mass dimension. These are terms that would appear in a low energy effective model of some UV renormalizable theory by integrating out degrees of freedom above a certain mass scale $M$. However, in a bottom-up point of view we don't focus on the origin of these terms but rather on a generic analysis of their effects. To this purpose, we choose at a first level to add to MSSM all possible mass dimension five operators that are all allowed by the gauge symmetries and by R-parity. In this way, EFT allows us to draw conclusions that are completely model independent. For a more detailed discussion of supersymmetric EFT, see sec.~\ref{EffectiveFieldTheory}.

Generally this constitutes a huge set of extra free parameters, limiting the predictability of the model. Nevertheless, many of these operators are actually redundant as they can be eliminated by proper field redefinitions. In our analysis, we perform such redefinitions reducing to a model with less parameters and thus more distinct phenomenology. We firstly focus on the Higgs sector because of its special importance in view of the little hierarchy problem and because its extension by effective terms is quite restricted, facilitating drawing clear conclusions. After that we pass on to other couplings and processes that may be interesting for LHC physics. Below we summarize the content of the chapters of part I.

In chapter \ref{MSSM5}, we focus on the most general set of R-parity conserving, mass dimension five operators that can exist in the MSSM \cite{Paper1}. We also employ spurion superfields to include any soft supersymmetry breaking effects that these operators parametrize. It turns out that not all of these operators are actually independent. We perform spurion dependent field redefinitions to remove the redundancy thus obtaining the minimal, irreducible set of dimension five operators within MSSM. By incorporating further constraints coming from flavor changing neutral currents (FCNC), we end up with the final model which we call ``MSSM$_5$".

In chapter \ref{PhenomenologyMSSM5}, we go on to study the phenomenological consequences of MSSM$_5$ \cite{Paper1}. One consequence is the generation of new effective interactions of the type quark-quark-squark-squark with potentially large effects in squark production  compared to those generated in the MSSM, especially for the top/stop quarks. This  can be important for LHC supersymmetry searches by direct squark production. Additional ``wrong'' Higgs couplings, familiar in the MSSM at the loop level \cite{Haber,M0,M1}, are also generated with a coefficient that can be larger than the loop-generated MSSM one. Again, these are largest for the top and also bottom sector at large $\tan\beta$. Furthermore, we study the effect of the new terms in the Higgs potential. It turns out that the mass of the Higgs can be shifted in a way that it alleviates the little hierarchy problem. This implies that we can obtain a novel point of view towards this apparent shortcoming of MSSM. Instead of considering it as a weakness of the theory, we can think of it as an indication for new massive particles at the energy range of few TeV. 

Consideration about the stability of the effective potential as well as an observed $\tan\beta$ suppression of the correction to the Higgs mass by five dimensional operators leads to the inevitable inclusion of mass dimension six operators in the Higgs sector \cite{Paper2}. In chapter \ref{MSSMHiggs56}, we perform this analysis insisting on a generic approach, including all possible dimension six operators allowed by the symmetries of the model. In the large $\tan\beta$ region, these two classes of operators can have comparable contributions to the Higgs mass which implies a further alleviation of the little hierarchy.

In chapter \ref{NonlinearMSSM}, we move on to study a different type of EFT, this time realized by nonlinear supersymmetry \cite{Paper3}. In models of low energy SUSY breaking, the gravitino acquires a sub-eV mass and thus it cannot be excluded from the spectrum of the low energy model. If this model is MSSM, we have to study couplings of the gravitino to MSSM. The ``equivalence theorem", which states that in scenaria with very low gravitinos the latter can be effectively replaced by their goldstino component which dominates over the dynamics, greatly simplifies such studies \cite{cddfg}. Nonlinear supersymmetry offers then the most convenient formalism for studying goldstino self interactions and goldstino-matter couplings. We use the method of constrained superfields to realize the nonlinear SUSY algebra and study the most general couplings of the goldstino with MSSM fields.

An important effect of these couplings is the increase in the mass of the Higgs, which can be significant for a SUSY breaking scale at the range of few TeV. This offers one more way for alleviating the little hierarchy. The difference is that in this case we don't even have to assume some kind of new physics at the high scale. The SUSY breaking mechanism itself brings the correction. In addition, we calculated the invisible decay of Higgs to neutralinos and goldstinos and found that it can be comparable with the standard MSSM decay rate of Higgs to photons. Finally, we found that, in the case that the mass of Z is larger than that of the lighest neutralino, there is a bound on the SUSY breaking scale at around 400 - 700 GeV coming from the invisible Z boson decay.

\section{Dilaton - DBI couplings in $\cN=2$ supersymmetry}

In the second part of the thesis we turn towards aspects of supersymmetry closely related to supergravity and string theory. Our target now is to understand how the coupling of a D-brane to the bulk arises in field theory.

The stage that we choose to focus on is type II strings on $\mathcal{R}_{3,1}\times \mathcal{CY}_3$. The geometry of the Calabi Yau manifold breaks SUSY, giving rise to a 4D $\cN=2$ effective supergravity theory. Generically, the presence of a D-brane in such background spontaneously breaks half supersymmetry on its worldvolume giving rise to an $\cN=1+1$ supersymmetric theory where the second supersymmetry is realized nonlinearly. The effective D-brane action is described by a Dirac-Born-Infeld (DBI) theory. It is an effective action for the gauge multiplets of the D-brane as well as for their coupling to the bulk fields. The latter can be described by hypermultiplets, single-tensor multiplets or double-tensor multiplets. All descriptions are Poicar\'e dual to each other.

Reproducing this action from field theory is the main aim of this second part. This task is nontrivial for two reasons. First, it is known that  $\cN=2$ linear supersymmetry, global or local, forbids a dependence of gauge kinetic terms on hypermultiplet scalars. For instance, in $\cN=2$ supergravity, the scalar manifold is the product of a quaternion-K\"ahler manifold for hypermultiplet scalars \cite{BW} and a K\"ahler manifold of a special type for vector multiplet scalars \cite{dWLVP}. In global $\cN=2$ supersymmetry, the quaternion-K\"ahler manifold of hypermultiplet scalars is replaced by a Ricci-flat hyper-K\"ahler space \cite{AGF}. Second, consistency of compactification of type II strings with D-branes requires the presence of orientifolds necessary for tadpole cancellation. These objects break supersymmetry explicitly globally, although is still preserved locally around the D-branes and away from the orientifold plane. It is then not clear if it is possible to construct from field theory the action that couples the bulk and brane multiplets, even those that would be truncated by the orientifold projection.

The DBI action appearing in D-brane dynamics suggests that the restrictions on the coupling between bulk and brane fields in $\cN=2$ supersymmetry are expected to change if (at least) one of the supersymmetries is nonlinearly realized. This is the path that we follow. In chapter \ref{secDBI}, we construct an $\cN=2$ action for the coupling of a single tensor multiplet with a gauge multiplet. This coupling is essentially the supersymmetrization of the Chern-Simons $B\wedge F$ coupling of the antisymmetric NSNS 2-form and the gauge field strength. We then impose nonlinear realization of the second SUSY by applying a supersymmetric constraint on the gauge multiplet. This is the generalization for $\cN=2$ superspace of the constrained superfield method used in the first part of the thesis. The resulting action is invariant under $\cN=1$ linear + 1 nonlinear SUSY and involves the Maxwell goldstino multiplet coupled to a single tensor multiplet \cite{Paper4}. If we remove this multiplet, the action reduces to the standard super-Maxwell DBI theory derived in the past \cite{BG, RT, ADM}.

We have chosen to group the bulk fields in a single tensor multiplet because it is the only one that admits a simple off shell superspace formulation. Hypermultiplets also can be formulated off-shell in the context of harmonic superspace but only in the expense of introducing infinite number of auxiliary fields \cite{harm}. In any case we can always switch between hyper-, single-tensor and double-tensor multiplets by performing Poicar\'e dualities.

By appropriate field redefinitions we obtain another equivalent description of the system, in terms of the Higgs phase of $\cN=1+1$ QED~\cite{APT, Itoyama}. This basis reveals some very interesting features of the system. The goldstino multiplet combines with a chiral superfield to form a $\cN=1$ massive vector multiplet while the other chiral superfield remains massless. This is a novel type of super-Higgs mechanism that does not require a gravitino (which would normally `eat' the goldstino as in the standard super-Higgs mechanism). Also, at one point along the flat direction of the potential, the vector multiplet becomes massless and the $U(1)$ gauge symmetry is restored. This is a known phenomenon from D-brane dynamics, where the $U(1)$ world-volume field becomes generically massive due to the CS coupling.

Having constructed the $\cN=1+1$ DBI action, the next step would be to identify its field content in terms of string fields. As already mentioned, the analog of this construction in string theory is that of type IIB  strings compactified on a Calabi Yau and interacting with a D-brane. The bulk fields under consideration are the dilaton scalar (associated to the string coupling),  the (Neveu-Schwarz) NS--NS antisymmetric tensor and the (Ramond) R--R scalar and two-form. Its natural basis is a double-tensor supermultiplet,\footnote{This representation of $\cN=2$ global supersymmetry has been only recently explicitly constructed \cite{DTmult}. See also ref.~\cite{TV}.} having three perturbative isometries associated to the two axionic shifts of the antisymmetric tensors and an extra shift of the R--R scalar. These isometries form a Heisenberg algebra, which at the string tree-level is enhanced to the quaternion-K\"ahler and K\"ahler space $SU(2,1)/SU(2)\times U(1)$. We can also use an equivalent formulation where the NS--NS and R--R 2-forms are replaced by their Poincar\'e dual scalars. In this formulation, the aforementioned isometries are realized on the scalar manifold of the four scalars which form a hypermutiplet called the ``universal hypermultiplet".

Therefore, we need to determine the proper `global supersymmetry' limit of the universal hypermultiplet and match it with the hyperK\"ahler scalar manifold of the global action. At the level of global $\cN=2$, imposing the Heisenberg algebra of isometries determines a unique hyperk\"ahler manifold of dimension four, depending on a single parameter. This is in close analogy with the local case of a quaternionic space where the corresponding parameter is associated to the one-loop correction~\cite{oneloop}. These similar results suggest a correspondence between the local and global cases which could be studied using a Ricci-flat limit of the quaternion-K\"ahler manifold preserving the Heisenberg algebra.

Obtaining the global SUSY limit of the universal hypermultiplet is not a trivial task. In $\cN=2$, the scalar curvature comes out to be proportional to the gravitational coupling $k$ so in the global SUSY limit we unavoidably obtain a Ricci-flat manifold. However, if we naively send $k$ to zero we reduce to the trivial case of a flat scalar manifold with canonical kinetic terms. To obtain a non-trivial space, an appropriate limit must be defined, involving a new mass scale that should remain finite as Planck mass goes to infinity. This mechanism has only been explicitly displayed for some particular cases, mostly using the quaternionic quotient method \cite{G1, G2}. In chapter \ref{UniversalHypermultipletLocalGlobalSupersymmetry}, we use this procedure to obtain the one-loop effective supergravity of the dilaton hypermultiplet and to then describe the appropriate zero-curvature limit, using the perturbative Heisenberg symmetry as a guideline \cite{Paper5}.

\part{\textsc{Beyond the Minimal Supersymmetric Standard Model}}

\chapter{Effective Field Theory}
\label{EffectiveFieldTheory}

\section{Physics is Effective}
\label{PhysicsEffective}

The ultimate goal of physics is believed to be the formulation of the theory that will disclose all mysteries of nature. There is a lot of discussion about the kind of truths that will be unveiled to us, however physicists generally agree that this final ``Theory of Everything" will provide an exact description of all physical phenomena that occur at any place and any time of the universe. Of course, we don't have this theory yet. We rather have various theories each one being a good description for some class of physical phenomena while failing for others. ``Good" here is used in the sense of being precise enough for our needs. If we want to think in terms of the ``parameter space" of nature, where the parameters can be distance, energy, velocity et c., then we can say that our theories are valid in a certain parameter subspace but not outside. For example, in the study of a system that interacts gravitationally, Newtonian gravity is a good description if interactions are non-relativistic but needs replacement by General Relativity if they are relativistic.

Theories that are valid only in a certain region of the full parameter space are called ``effective". This definition might sound redundant since all physical theories would be effective. Nevertheless, this simple idea has an surprisingly rich structure in quantum field theory (QFT). The most relevant parameter here is distance. After almost a century of experiments in particle physics we have learned that, as we probe smaller distances, nature appears to reveal richer structure. In the context of QFT, this is expressed by the appearance of new degrees of freedom, describing new particles. These are invisible at longer distances either because they are unstable, decaying to known long lived particles, or because they are components of particles that at longer distances seem fundamental. This suggests that a QFT model with a given set of degrees of freedom is valid only at distance scales larger than the threshold for production of new particles, not included in the set. If we agree that the principles of QFT are valid beyond the threshold distance, we will need to exchange the old model with a new one, where the new particles (and the new interactions that they reveal) are included. This process essentially builds a ladder of effective field theory (EFT) models separated by the threshold distances where new particles appear. Various interesting questions arise: How to smoothly switch from one EFT to another, what is their behavior very close to the threshold et c. Another thing that makes EFT nontrivial is the need for regularization. Since regularization involves the behavior of a QFT model at high energies (short distances), it has to be treated with special care \footnote{For a review, see \cite{Georgi1}.}.

Let's attempt a discussion motivated by the questions mentioned above. We focus on two neighbor theories, call them the `UV' and the `IR' theory, seperated by threshold energy $M$ (we prefer to talk in terms of energy than distance). The UV theory contains all the modes of the IR plus those modes with mass of order $M$ that do not appear in the IR. We expect that as we approach $M$ from below, the new physics that the heavy particles bring will become more and more apparent. The way to incorporate these effects in the IR is by integrating out the massive modes. This inevitably introduces a series of higher dimensional, nonrenormalizable operators in the Lagrangian of the IR, suppressed by the threshold scale. From the EFT point of view, the fact that they are nonrenormalizable is not an indication that the model is sick but simply that it is valid up to the threshold scale, as expected \cite{Wilson1}. This point, even if it sounds obvious nowadays, was entirely disregarded in the early days of QFT when nonrenormalizable models were considered pathological.  In the expansion of the operator series, we choose to cut off at some order in $1/M$ depending on the accuracy we want to achieve. The coefficients of the new terms are determined by matching the S-matrix elements of the UV and the IR models. One might ask why should we bother reducing to an effective IR theory when the full UV theory is known. The reason is that in many cases, calculations in the low energy regime are much simpler in the IR theory where the very massive modes do not appear explicitly.

There are many examples of EFT models. For some of them the UV completion is known while for others it isn't. To mention a few, Fermi theory is an effective theory of weak interactions while chiral perturbation theory and nucleon effective theory are low energy effective descriptions of QCD. On the other hand, the Standard Model (SM) itself is an effective theory (it is renormalizable only when gravity is ignored) but its UV completion is still unknown. The same is true for General Relativity.

In order to elucidate the derivation of an effective theory from a known UV completion, we focus on the popular case of the Fermi theory as an effective theory for electroweak interactions. In the SM, consider the tree level exchange of a massive $Z$ gauge boson between charged fermions
\be
{\cal L}\supset
i\,\overline\psi\, \gamma^\mu\,(\partial_\mu+igZ_\mu)\,\psi
-\frac{M^2}{2}\,\,Z_\mu\,Z^{\mu}
\ee
By integrating out $Z_\mu$ we generate the higher dimensional operator
\be
\Delta {\cal L}=
\frac{g^2}{2\,M^2}\,(\overline \psi \gamma_\mu\psi)^2
\ee
which is a nonrenormalizable four-fermion contact term. Similarly, for scalars $H$:
\begin{equation}
{\cal L}\supset \big\vert
(\partial_\mu+ig\,Z_\mu)\,H\big\vert^2-\frac{M^2}{2}\,Z_\mu\,Z^\mu
\end{equation}
and
\begin{equation}
\Delta {\cal L}=
 \frac{g^2}{M^2}\,(H^\dagger \partial_\mu\,H)^2\qquad
\end{equation}

It is also possible that the effective operator is a higher derivative one. Here, we retrieve such operators by the kinetic mixing of light with heavy states, upon integrating out the latter. For example, from
\be
{\cal L}=\frac{1}{2}\,(\partial_\mu\phi)^2
+\frac{1}{2}\,(\partial_\mu\chi)^2
+c\,\partial^\mu\phi\,\,\partial_\mu\chi
-\frac{1}{2} M^2 \chi^2-\frac{1}{2}\lambda^\prime\phi^2\chi^2
\ee
one finds after integrating out the massive field $\chi$:
\begin{eqnarray}
{\cal L}&=&\!\!\frac{1}{2}\,(\partial_\mu\phi)^2
+\frac{c^2}{2}\,\Box\phi\,\frac{1}{M^2+\Box+\lambda^\prime \phi^2}
\,\Box\phi\nonumber\\[5pt]
&=&
\frac{1}{2}\,(\partial_\mu\phi)^2
+\frac{c^2}{2\,M^2}\,(\Box\phi)^2+\cdots\label{truncation}
\end{eqnarray}
which contains higher derivative terms \cite{NH1,NH2}. In both examples above, the UV completion of the effective theory is known. EFT is then a practical reformulation of the relevant degrees of freedom in the low energy regime. However, does EFT have anything to offer when the UV side is unknown?

This answer is definitely positive. EFT has proved to be a very useful tool for exploring new physics in a bottom-up approach \cite{GeorgiPolitzer1,Witten1,Weinberg2,Weinberg1}. Since the effects of inaccessible massive states can be incorporated into nonrenormalizable operators, we can simply add such terms in the IR Lagrangian without referring to a particular UV scheme. In a systematic analysis, we include all possible terms up to a given order in $1/M$ that are allowed by the symmetries of the theory, keeping the coefficients arbitrary. This constitutes a model independent way of exploring new physics beyond the validity of the pure IR model. Any possible UV candidate will essentially reduce to a subset of the nonrenormalizable terms with fixed values for the coefficients.

Even at first order in $1/M$, there is usually a long list of terms allowed by the symmetries of the theory, introducing many new arbitrary parameters. Nevertheless, such set is in general highly reducible. This means that we can write the Lagrangian in a way that a smaller number of new operators appears but physics be the same. There are three different methods to perform such reduction. By setting the higher dimensional operators ``on shell" \cite{Georgi:1991ch,Politzer:1980me,Arzt:1993gz}, by performing field redefinitions \cite{SalamStrathdee,KalloshTyutin} and, if the operator is higher derivative, by applying the ``unfolding" technique \cite{Antoniadis:2007xc,Antoniadis:2006pc}. Since we will be using the first two in the phenomenological analysis of the following chapters, we will briefly present them below in the relevant case of supersymmetric field theories. After restricting to an irreducible set of higher dimensional operators, one can further cut down the parameter space by comparing the model with low energy phenomenology. In the end, the hope is that the effective model will provide concrete testable predictions for the effects that very massive modes can have on low energy observables.

\section{Effective Description of Supersymmetric Theories}
\label{EffectiveDescriptionofSupersymmetricTheories}

EFT has a lot to offer in the yet unexplored territory of TeV physics. By popular belief, the most promising candidate theory for physics around that scale is supersymmetry. It is then reasonable to construct phenomenological supersymmetric models by means of EFT techniques and this is what we do in the following chapters. In order to familiarize with the concept and tools of EFT in the framework of supersymmetric theories, we present here some representative study cases.

\subsection{Integrating out Massive Superfields}
\label{IntegrateMassiveSfields}

Consider the following Lagrangian of dimensionful scales $M$ and $m$ with $M>>m$:
\bea\label{l0}
\cL=\int d^4\theta\, \Big[\Phi^\dagger \Phi +\chi^\dagger \chi\Big]
+\bigg\{
\int d^2\theta \bigg[\frac{M}{2}\,\chi^2+m\,\Phi\,\chi+
\frac{\lambda}{3}\,\Phi^3\bigg]+h.c.\bigg\}
\eea
We want to acquire an effective description by integrating out the heavier mode. We  will follow two different paths; either diagonalize the mass matrix and then integrate or directly integrate. Then we will show that the resulting effective theories are all equivalent by using the ``field redefinitions" method and the ``on shell" method mentioned earlier.

In the first path, we perform the transformation $\Phi=(\cos\theta \,\Phi_1 -\sin\theta\,\Phi_2)$ and
$\chi= (\sin\theta \,\Phi_1 +\cos\theta\,\Phi_2)$. In the diagonal basis of $\Phi_1$ and $\Phi_2$, one finds
\bea
\cL&=&\!\int d^4\theta\, \Big[\Phi_1^\dagger \Phi_1+\Phi_2^\dagger
\Phi_2\Big]\! \nonumber
\\
&+&\bigg\{
\int d^2\theta \bigg[\frac{m_1}{2}\,\Phi_1^2+\frac{m_2}{2}\,\Phi_2^2
+\frac{\lambda}{3}\,(\cos\theta \Phi_1 -\sin\theta\,\Phi_2)^3
\bigg]\!+\! h.c.\!\bigg\}
\eea
where
\bea
m_1 &=& \frac{M}{2}\,\Big(1-(1+4 m^2/M^2)^{1/2}\Big)
=-\frac{m^2}{M}\,\Big(1-\frac{m^2}{M^2}\Big)+\cdots \ ,
\nonumber\\[4pt]
m_2&=&\frac{M}{2}\,\Big(1+(1+4 m^2/M^2)^{1/2}\Big)
=M\,\Big(
1+ \frac{m^2}{M^2}+\cdots\Big) \ ,
\eea
so $\Phi_2$ is the massive field. Then, we integrate out $\Phi_2$ via its equation of motion
\medskip
\bea
-\frac{1}{4} \overline D^2  \Phi_2^\dagger+m_2
\,\Phi_2-\lambda\sin\theta\,
\,\big(\Phi_1\,\cos\theta-\Phi_2\sin\theta\big)^2=0 \ ,
\eea
with solution
\bea
\Phi_2=\frac{\lambda}{m_2} \,\cos^2\theta\,\sin\theta\,\,
 \Phi_1^2
-\frac{\lambda^2}{4 m_2^2}\,\sin^3 2\theta\,\Phi_1^3
+
\frac{\lambda}{4\,m_2^2}\cos^2\theta\,\sin\theta\,
\overline D^2\Phi_1^{\dagger\, 2} +\cO(M^{-3}) .
\eea
The effective Lagrangian that we obtain is:
\bea\label{m1}
\cL^{eff}&=&\int d^4\theta\, \Phi_1^\dagger \Phi_1 \nonumber
\\
&+&\bigg\{
\int d^2\theta \bigg[\frac{-m^2}{2M}Z\Phi_1^2
+\frac{\lambda}{3}Z^{3/2}\Phi_1^3 -\frac{m^2\lambda^2}{2 M^3}
\Phi^4_1 \bigg]+h.c.\bigg\}+\cO(M^{-4})
\eea
where
\bea
Z=1-\frac{m^2}{M^2}+\cO(1/M^4) \ .
\eea
This is an effective description of (\ref{l0}) where only the light mode propagates.

Alternatively, one can choose to directly integrate out $\chi$ from eq.~(\ref{l0}) without firstly diagonalizing. Its e.o.m. is
\be
{\overline D}^2\chi^\dagger -4\,(M\,\chi+m\,\Phi)=0
\ee
with an iterative solution
\bea\label{eq_sol}
\chi=
\frac{1}{M}\Big[-m\,\Phi-\frac{m}{4M}\,{\overline D}^2\Phi^\dagger
+\frac{1}{16}\,\frac{-m}{M^2}\,{\overline D}^2 \,D^2\Phi
-\frac{m}{64\, M^3}\,\overline D^2\, D^2\,\overline D^2\Phi^\dagger
+\cdots\Big]\ .
\eea
Plugging this back in (\ref{l0}), we find
\bigskip
\bea\label{ff1}
\cL^{eff}&=&
\int d^4\theta\,\bigg\{
\bigg[\,1+\frac{m^2}{M^2}\bigg]
\,\Phi^\dagger\Phi+
\frac{m^2}{8\,M^3} \,\Big[\Phi\,D^2\,\Phi+h.c.\Big]
+\frac{m^2}{16\, M^4}\,(\overline D^2\Phi^\dagger)\, (D^2 \Phi)
\bigg\}
\nonumber\\[10pt]
&+&
\bigg\{
\int d^2\theta
\,\,\bigg[\frac{-m^2}{2 M}\,\Phi^2+ \frac{\lambda}{3}\,\Phi^3\bigg]
+h.c.\bigg\}+\cO({1}/{M^5})
\eea
which, after an appropriate rescaling, is written as
\bigskip
\bea\label{ff2}
\cL^{eff}&=&
\int d^4\theta\,\bigg\{
\,\Phi^\dagger\Phi+
\frac{m^2}{8\,M^3}\, \,\Big[\Phi\,D^2\,\Phi+h.c.\Big]
+\frac{m^2}{16\, M^4}\,(\overline D^2\Phi^\dagger)\, (D^2 \Phi)
\bigg\}
\nonumber\\[10pt]
&+&
\bigg\{
\int d^2\theta
\,\,\bigg[\frac{-m^2}{2 M}\,Z\,\Phi^2+
\frac{\lambda}{3}\,Z^{3/2}\,\Phi^3
\bigg]
+h.c.\bigg\}+\cO({1}/{M^5}) \ ,
\eea
where $Z=1/(1+m^2/M^2)$. In this path, we obtained an effective Lagrangian with higher derivative terms. Equations (\ref{m1}) and (\ref{ff2}) look different, however, the physics they describe is the same. We will demonstrate this in two ways.

In the first way, we set ``on shell'' the higher dimensional operator. By use of the e.o.m.
\bea\label{ggg1}
\overline D^2 \Phi^\dagger =
-\frac{4 m^2}{M}\,\Phi+4\,\lambda\,\Phi^2+\cO(1/M^2)
\eea
we can rewrite (\ref{ff2}). The new Lagrangian will contain the term $\Phi\Phi^{\dagger 2}$ which can be removed by a suitable shift
\bea
\Phi=\tilde\Phi-\frac{\lambda \,m^2}{2\,M^3}\,{\tilde\Phi}^{2}
\eea
to find
\bea\label{ggg2}
\cL^{eff}&=&
\!\!\!
\int d^4\theta \,\,
\tilde\Phi^\dagger\tilde\Phi
\\[10pt]
&+&\!\!\!\bigg\{
\int d^2\theta\, \bigg[-\frac{m^2}{2M}\,Z\,\tilde\Phi^2
+
\frac{\lambda}{3}\,\tilde\Phi^3\,\Big(\,1-\frac{3}{2}
\frac{m^2}{M^2}
\,\Big)-
\frac{\lambda^2\,m^2}{2\,M^3}\,\tilde\Phi^4\bigg]\!+\!h.c.\!\bigg\}
+\cO\Big(\frac{1}{M^4}\Big)\nonumber
\eea
where $Z=1/(1+m^2/M^2)$. It is obvious now that this Lagrangian coincides with
that of (\ref{m1}) in the approximation  $\cO(1/M^4)$. This confirms that
 setting the higher derivative operators ``on shell'' via equations of motion is a correct procedure,  within the approximation considered. We obtained again a higher dimensional operator and a scale dependence acquired classically by the couplings of the low energy effective theory.

In the second way, we perform field redefinitions in eq.~(\ref{ff2}) so as to eliminate the $\Phi D^2\Phi$ term. We use
\be\label{hh1}
\Phi = \Phi'+c\,\overline D^2\Phi^{' \dagger}
\ee
where the dimensionful coefficient $c$ is such that the coefficient of $\Phi D^2\Phi$ vanish in the new Lagrangian. This gives $c=-m^2/(8 M^3)$ and, after some calculations, the Lagrangian in (\ref{ff2}) becomes
\bea\label{hh2}
\cL^{eff}&=&\int d^4\theta \,\,\Big[
\Phi^{' \dagger} \,\Phi'+\frac{m^2\,\lambda}{2
  \,M^3}\,\big(\Phi^{' 2}\,\Phi^{' \dagger}+h.c.\big)\Big]\nonumber\\[10pt]
&+&
\bigg\{\!
\int d^2\theta\,\,
\Big[-\frac{m^2}{2 \,M}
  \,Z\,\Phi^{' 2}+
\frac{\lambda}{3}\,Z^{3/2}\,\Phi^{' 3}\Big]+h.c.\bigg\}
+\cO(1/M^4)
\eea
By a final shift $\Phi'= \tilde \Phi-{m^2\,\lambda}/(2\,M^3)\,\tilde\Phi^{2}$
we obtain an effective Lagrangian identical to that in (\ref{m1}) and
 (\ref{ggg2}).
 
We have shown that the three apparently different paths to the reduced Lagrangian, leading to either eq.~(\ref{m1}), (\ref{ff2}) or (\ref{hh2}), are actually different formulations of same physics at the expansion order studied. The correction at $1/M$ is solely a wavefunction renormalization while higher dimensional operators appeared only at higher order.

\subsection{Gauge Interactions and Component Analysis}
\label{GaugeInteractionsComponentAnalysis}

We proceed to study further examples of effective theories, now with gauge interactions present. We will also verify the superfield analysis at the component level. The effective operators that will be generated are the same with those used in the phenomenological model of the subsequent chapters. Therefore, the analysis here provides us intuition about the kind of UV physics that these effective operators encapsulate.

Consider the Lagrangian of an $\cN=1$ supersymmetric non-Abelian gauge theory\footnote{For the link to the MSSM, replace $V\ra V_1\equiv g_2 V_w^i\sigma^i-g_1 V_Y$ with $V_w, (V_Y)$ the $SU(2)$, ($U(1)_Y$) gauge fields respectively;
also $\Phi_2\ra H_2^T\,(i\sigma_2)$, $\Phi_1\ra H_1$ with $\Phi_3$
$(\Phi_4)$ with same quantum numbers to $\Phi_1$ ($\Phi_2$) and
$(i\sigma_2)\exp( -\Lambda)=\exp(\Lambda^T)\,(i\sigma_2)$,
then 
$\Phi_2\,e^{-V}\,\Phi_2^\dagger\ra  H_2^\dagger\,e^{V_2}\,H_2$,
 with $V_2\equiv g_2 V_w^i\sigma^i+g_1 V_Y$.}
\medskip
\bea\label{LO}
\cL&=&\int d^4\theta
\,\,\Big[
\,\Phi_1^{ \dagger}\,\,e^{V}\,\Phi_1
+\,\Phi_3^{ \dagger}\,\,e^{V}\,\Phi_3
+\,\Phi_2\,\,e^{-V}\,\Phi_2^\dagger
+\,\Phi_4\,\,e^{-V}\,\Phi_4^\dagger\,
+S^\dag S\,\Big]
\nonumber\\[6pt]
&+&
\int\,d^4\theta\,\,
\Big[\nu_1\,\Phi_1^{ \dagger}\,\, e^{V}\, \Phi_3+
\nu_2\,\Phi_4\, e^{- V}\, \Phi_2^\dagger+h.c.\Big]
\nonumber\\[6pt]
&+&
\int d^2\theta
\,\,\Big[\,\mu\,\Phi_1\,\Phi_2+M\,\Phi_3\,\Phi_4
+{M\over 2}S^2+\lambda\, S\,\Phi_1\Phi_2\,\Big]+h.c.
\eea
where $M\gg \mu$ and $V$ is the standard vector superfield in the Wess-Zumino gauge. The equations of motion for the massive fields $\Phi_{3,4}$ and $S$ give
\medskip
\bea
-\frac{\nu_1}{4}\,\oD^2\,\Big(\,\Phi_1^\dagger\,e^{V}\Big)-\frac{1}{4}
\,\oD^2\Big(\Phi_3^\dagger \,\,e^{V}\Big)+M\,\Phi_4&=&0
\nonumber
\\[8pt]
-\frac{\nu_2}{4}\,\oD^2\,\Big(\,e^{-
V}\,\Phi_2^\dagger\,\Big)-\frac{1}{4} \oD^2\Big(\,e^{-
V}\,\Phi_4^\dagger\,\Big)+M\,\Phi_3&=&0
\nonumber\\[8pt]
-\frac{1}{4} \oD^2S^\dagger+M\,S+\lambda\, \Phi_1\,\Phi_2&=&0
\eea
As in the previous section, we use these equations to integrate out the massive fields
$\Phi_{3,4}$ to find
\bea\label{lag4}
\cL^{eff}&=&\int d^4\theta
\,\,\Big[
\,\Phi_1^{ \dagger}\,\,e^{V}\,\Phi_1
+\,\Phi_2\,\,e^{- V}\,\Phi_2^\dagger
+\Big(
\,\xi\,\Phi_1^\dagger\,\,e^{V}\,\oD^2\,e^{-V}\,\Phi_2^\dagger
+h.c.\Big)
\,\Big]
\nonumber\\[8pt]
&+&
\int d^2\theta
\,\,\Big[\,\mu\,\Phi_1\,\Phi_2+\xi'(\Phi_1\,\Phi_2)^2\Big]
+h.c.+\cO(M^{-2})
\eea
where $\xi=\frac{\nu_1\,\nu_2}{4M}$, $\xi'=- {\lambda^2 \over 2M}$ and we ignored higher orders in $M^{-1}$. If the superpotential in (\ref{LO}) also contains trilinear  couplings of heavy doublets $\Phi_{3,4}$ to quarks and leptons
\begin{equation}
\Delta {\cal L} \ = \ \int d^2 \theta \Big[ 
Q\, \sigma_u  U^c\,\Phi_4 + 
Q\, \sigma_d  D^c \,\Phi_3 + 
L\, \sigma_e  E^c \,\Phi_3  \Big]+h.c. \ , \label{others}
\end{equation}
then, following the same procedure, we would get the extra effective terms
\begin{eqnarray}\label{B7}
\Delta \cL^{eff}&= & - \frac{1}{M} \int d^4\theta 
\Big[\nu_1\, \Phi_1^\dagger\,e^{V}\,Q\,  \sigma_u U^c\
+ \nu_2\,(Q\,  \sigma_d D^c) \, e^{-V} \,\Phi_2^\dagger 
+ \nu_2\,(L\, \sigma_e E^c) \, e^{-V} \,\Phi_2^\dagger +
h.c.\Big] \nonumber\\[5pt]
&+&\frac{1}{M}\int d^2 \theta \Big[
(Q\sigma_u U^c) ( Q\sigma_d D^c )+(Q\sigma_u U^c) (L\,\sigma_e E^c)\Big]+h.c. \ ,
\end{eqnarray}
where $\sigma_{u,d,e}$ are 3x3 matrices in the families space.

Focusing on (\ref{lag4}), let us set on shell the higher derivative operator by using the equations of motion for $\Phi_{1,2}$:
\bea\label{eqm1}
D^2 \,\big[\,e^{V}\, \Phi_1\,\big]
=4 \,\mu\,\Phi_2^\dagger \ , \qquad\qquad
\overline D^2 \,\big[\, e^{- V}\, \Phi_2^\dagger\,\big] =4\,\mu\,\Phi_1 \ .
\eea
We insert these in (\ref{lag4}) and rescale $\Phi_i\ra \Phi_i' \,(1-2\,\mu\,\xi)$, $i=1,2$, to find:
\bea\label{adasd}
\cL^{eff}&=&\int d^4\theta
\,\,\Big[
\,\Phi_1^{ \dagger}\,\,e^{V}\,\Phi_1
+\,\Phi_2\,\,e^{-V}\,\Phi_2^\dagger
\,\Big]
\nonumber\\[8pt]
&+&\int d^2\theta \,\,\Big[\,\mu\,(1- 4\mu\,\xi)\,\,
\Phi_1\,\Phi_2+ \xi'\,(\Phi_1\,\Phi_2)^2 \Big]+h.c.+\cO(M^{-2})
\eea
It is obvious that the specific operator, when put on shell, brings solely a wavefunction renormalization. We now go on to verify at the component level that both Lagrangians are equivalent. First, we expand (\ref{lag4})\footnote{We use
$-4\,\psi_2\,\cD_\mu\,\cD^\mu\,\psi_1
=-4\,\psi_2\,[\sigma^\nu\,\overline\sigma^\mu-2\,i\,\sigma^{\mu\nu}]
\cD_\nu\,\cD_\mu\,\psi_1=-4\,\psi_2\,\sigma^\nu\,\overline\sigma^\mu\,\cD_\nu\,\cD_\mu
\,\psi_1+4\,\,\psi_2\,\sigma^{\mu\nu}
\,F_{\mu\nu}\,\psi_1$
and the first term in the rhs is that entering the final expression of $\cL$. Here $F_{\mu\nu}=\partial_\mu V_\nu/2- \partial_\nu V_\mu/2+i\,[V_\mu/2,V_\nu/2]$.}:
\bea\label{off-shell}
\cL^{eff}&=&
-\, \phi_1^*\,\cD_\mu\cD^\mu \phi_1
+i\,
\overline\psi_1\, \overline\sigma^\mu\,\cD_\mu\,\psi_1
-\frac{1}{\sqrt 2}\,\Big[
\overline\psi_1\,\overline\lambda\,\phi_1
+h.c.
\Big]
+\phi_1^*\,\frac{D}{2}\,\phi_1
+\vert F_1\vert^2
\nonumber\\[7pt]
&-&
\phi_2\,\cD_\mu\cD^\mu \phi_2^*
+i\,
\psi_2\,\sigma^\mu\,\cD_\mu\,\overline\psi_2
+\frac{1}{\sqrt 2}\,\Big[
\phi_2\,\overline\lambda\,\overline\psi_2
+h.c.
\Big]
-\, \phi_2\,\frac{D}{2}\,\phi_2^*+\vert F_2\vert^2
\nonumber\\[7pt]
&+&
\xi^*\,\bigg\{
4\, \Big[\,F_2 \,\cD_\mu\,\cD^\mu\,\phi_1+
\phi_2\,\cD_\mu\,\cD^\mu\,F_1
\Big]
+
2\sqrt 2\,i\,\,\Big[\psi_2\,\sigma^\mu\,
\overleftarrow\cD_\mu\,\overline\lambda\,\phi_1+
\phi_2\,\overline\lambda\,\,\overline\sigma^\mu\,\cD_\mu\,\psi_1\Big]
\nonumber\\[7pt]
&+&
2 \,(\phi_2\, D\, F_1-F_2\, D \,\phi_1)
-
2\sqrt 2 \,\,\Big[
\psi_2\,\lambda\,F_1-F_2\,(\lambda\,\psi_1)\Big]
-2\,\phi_2\,(\overline\lambda\,\overline\lambda)\,\,\phi_1
\nonumber\\[7pt]
&-&
4\,\psi_2\sigma^\nu\,\overline\sigma^\mu\,\cD_\nu\,\cD_\mu\psi_1
\bigg\}+
\mu\,\Big[\phi_1\,F_2+F_1\,\phi_2-\psi_1\,\psi_2\Big]
\nonumber\\[7pt]
&+&\xi'\,\Big[-
\big(\phi_1\psi_2+\psi_1\phi_2\big)^2+
2\,\big(\phi_1\phi_2\big)\,\big(\phi_1\,F_2
+F_1\phi_2-\psi_1\psi_2\big)\Big]\nonumber\\[7pt]
&+&h.c.+\cO(1/M^2)
\eea
with
\bea
\cD_\mu=\partial_\mu + i\,\,\frac{V_\mu}{2},\qquad
\overleftarrow\cD_\mu=\overleftarrow
\partial_\mu- i\,\,\frac{V_\mu}{2},\qquad
\eea
and the ``h.c." refers to all terms in the last four lines. Notice that in the off shell component form of the Lagrangian we have an interesting tensor coupling $\psi_2\,\sigma^\nu\,\overline\sigma^\mu\,\cD_\nu\cD_\mu\,\psi_1$ in spite of the minimal gauge coupling in (\ref{LO}) (see also \cite{Ferrara}). This coupling could be relevant for tree level calculations of the Feynman diagrams. Next, we eliminate the auxiliary fields
$F_{1,2}$ using their e.o.m.
\medskip
\bea
F_1^*
&=&
-\phi_2\,\Big(\mu+2\,\xi'\,(\phi_1\,\phi_2)\Big)
+
\xi^*
 \,\Big(-4\,\phi_2 \overleftarrow\cD_\mu
\overleftarrow\cD^\mu
-4\,\,\phi_2\,\frac{D}{2}+2\sqrt 2 \,\psi_2\,\lambda\Big)
\nonumber\\[8pt]
F_2^*&=&
-\phi_1\,\Big(\mu+2\,\xi'\,(\phi_1\,\phi_2)\Big)
 +
\xi^*\,
\,\Big(-4\,\cD_\mu\cD^\mu\,\phi_1
+4\,\,\frac{D}{2}\,\phi_1-2\sqrt 2 \,\lambda\,\psi_1\Big)
\eea
In the terms proportional to $\xi$ in $\cL^{eff}$ we can replace the derivatives of the fermions by their equations of motion, since the error would be of higher order. We use
\bea
i\,\overline\sigma^\mu\,\cD_\mu\psi_1&=&
\mu\,\overline\psi_2+\frac{1}{\sqrt
  2}\,\,\overline\lambda\,\phi_1+\cO(\xi) \ ,
\nonumber\\[10pt]
-i\,\psi_2\,\sigma^\mu\,\overleftarrow
\cD_\mu&=&\mu\,\overline\psi_1-\frac{1}{\sqrt
  2}\,\,\phi_2\overline\lambda+\cO(\xi) \ .
\eea
We then rescale the scalars and Weyl fermions and after neglecting terms $\cO(\xi\,\xi')$
we obtain the on shell Lagrangian
\medskip
\bea\label{qqqq}
\cL&=&
-\,\phi_1^\dagger\,\cD^2\,\phi_1+
i\,\overline\psi_1\,\overline\sigma^\mu\,\cD_\mu\,\psi_1
- \frac{1}{\sqrt 2}\,\Big[
\overline\psi_1\,\overline\lambda\,\phi_1+h.c.\Big]
+\,\phi_1^\dagger\,\frac{D}{2}\,\phi_1
\nonumber\\[7pt]
&-&
\phi_2\,\cD^2\,\phi_2^\dagger +
i\,\psi_2\,\sigma^\mu\,\cD_\mu
\overline\psi_2+
\,\frac{1}{\sqrt
  2}\Big[\,\phi_2\,\overline\lambda\,\overline\psi_2+h.c.\Big]
-\,\phi_2\,\frac{D}{2}\,\phi_2^\dagger
\nonumber\\[7pt]
&-&
\mu^2\,\vert 1-4\,\mu\,\xi\,\vert^2\,
\Big[\,\phi_1^\dagger\phi_1+\phi_2\,\phi_2^\dagger
\Big]
-\mu\,\Big[\,(1-4\,\mu\,\xi\,)\,\,\psi_1\,\psi_2 + h.c.\Big]
\nonumber\\[7pt]
&-&
2\,\xi'\,\mu\,\Big[(\phi_1\phi_2)+h.c.\Big]\,
\Big[\,\phi_1^\dagger\phi_1+\phi_2\,\phi_2^\dagger
\,\Big],\qquad \cD^2=\cD^\mu\,\cD_\mu
\eea
This Lagrangian is in agreement with that of (\ref{adasd}), which shows that on shell and in the absence of other interactions, only a wavefunction renormalisation effect is present, giving a  new  $\mu'= \mu\,\,(1-4\,\mu\,\xi)$. To conclude, integrating out the massive superfields $\Phi_{3,4}$ generated  a dimension-five operator $\Phi_2\,e^{-V} D^2\,e^V\,\Phi_1$ which however, brings only a (classical) wavefunction renormalisation, in the absence of additional trilinear interactions. Thus this five dimensional operator doesn't bring new physics in the absence of additional interactions. One could ask if this conclusion remains valid when we include soft supersymmetry breaking terms. Also, if additional trilinear interactions were present, other five dimensional operators of type shown in (\ref{B7})  could also be generated. All these issues are studied in the subsequent chapters.

\section{Nonlinear Realizations and Constrained Goldstino Superfield}
\label{NonlinearRealizationsConstrainedGoldstinoSuperfield}

Consider a field theory invariant under the symmetry group $\mathcal{G}$. The field content of the theory is divided between fields that are invariant and fields that transform under $\mathcal{G}$. The latter can transform either linearly under all generators of $\mathcal{G}$ or linearly under a subgroup $\mathcal{H}$ and nonlinearly under the coset $\mathcal{G}/\mathcal{H}$. In the first case the theory is in its unbroken phase and the classification of all possible transformation laws for the fields is described by representation theory. In the second case the symmetry parametrized by the generators of $\mathcal{G}/\mathcal{H}$ is broken with the breaking scale $M_b$ sent to infinity. In other words, a theory with a nonlinear realization of a symmetry group can be seen as an effective description of the far IR limit of a theory where this group is broken spontaneously \cite{Coleman1,Coleman2}. The Goldstone fields that appear are in 1-1 correspondance with the generators of $\mathcal{G}/\mathcal{H}$. 

If $\mathcal{G}$ is the super-Poincar\'e and $\mathcal{H}$ the Poincar\'e algebra, we have a nonlinear realization of the supersymmetry algebra and this would describe the far IR regime of a spontaneously broken supersymmetric theory. Since the broken generators are fermionic, the corresponding Goldstone mode has to be a fermion, too. To distinguish it from the standard Goldstone fields, we call it ``goldstino". It is quite surprising that Supersymmetry in four dimensions first appeared in its nonlinear version \cite{Volkov:1973ix}. The nonlinear transformation of the goldstino $\lambda_\alpha(x)$ can be written as:
\be\label{nonlin1}
\delta\lambda_\alpha=f\,\eta_\alpha+{i\over f}(\lambda\sigma^\mu\bar{\eta}-\eta\sigma^\mu\bar{\lambda})\partial_\mu\lambda_\alpha \ ,
\ee
where $\eta$ is the supersymmetry transformation parameter and $f$ is a parameter of mass dimension 2 characterizing the susy breaking scale ($\sqrt{f}=M_b$). The commutator of this transformation
\be
[\delta_\eta\,,\,\delta_\xi]\lambda_\alpha=2i(\eta\sigma^\mu\bar{\xi}-\xi\sigma^\mu\bar{\eta})\partial_\mu\lambda_\alpha
\ee
is a spacetime translation and proves that the above transformation closes off shell the super-Poincar\'e algebra.

In order to take advantage of nonlinear realizations we need to know how to construct Lagrangians describing interactions of the goldstino with itself and with other fields. Several strategies have been developed in the past. In the ``geometric" method \cite{Volkov:1973ix,Clark:1996aw,Clark:1998aa} the transformation (\ref{nonlin1}) is interpreted as an extension of the standard superspace transformation
\bea
\theta&\rightarrow&\theta+\eta \ ,\nonumber
\\
x^\mu&\rightarrow&x^\mu+i\theta\sigma^\mu\bar{\eta}-i\eta\sigma^\mu\bar{\theta} \ ,
\eea
to the chiral goldstino field $\lambda(x)$ by identifying $\theta$ with $\lambda/f$. The same analogy between $\theta$ and $\lambda$ can be extended to the superspace differentials $d\theta$ and $d\bar{\theta}$ leading to the construction of a volume element invariant (up to total derivative) under the nonlinear transformations. From this we can extract the Lagrangian density
\be
\cL=-f^2\, \mathrm{det}A\ ,\qquad \mathrm{with} \quad A^\mu_\nu=\delta^\mu_\nu+{i\over f^2}(\lambda\sigma^\mu\partial_\nu\ov{\lambda}-\partial_\nu\lambda\sigma^\mu\ov{\lambda})\ . 
\ee
It is the Volkov-Akulov Lagrangian describing the dynamics of a single goldstino up to higher derivative terms. By nonlinearly realizing the algebra on matter fields $\phi$ as well,
\be
\delta\phi=-{i\over f}(\lambda\sigma^\mu\ov{\eta}-\eta\sigma^\mu\ov{\lambda})\partial_\mu\phi\ ,
\ee
we can construct goldstino-matter couplings ($\phi$ denotes any kind of field). For any operator $\cO(\phi,\partial_\mu\phi)$ we simply have to replace partial derivatives by appropriate covariant derivatives so that $\cO$ transforms in the standard way:
\be
\delta\cO=-{i\over f}(\lambda\sigma^\mu\ov{\eta}-\eta\sigma^\mu\ov{\lambda})\partial_\mu\cO\ .
\ee
Then any action of the type
\be
\mathcal{S}=-f^2\int d^4x\, \mathrm{det}(A)\, \cO
\ee
is invariant under nonlinear transformations. It can be easily shown that, in the geometric method, the lowest order couplings between goldstinos and matter are of the type:
\be
{1\over f^2} T^{\mu\nu}t_{\mu\nu}\ ,
\ee
where $T_{\mu\nu}$ and $t_{\mu\nu}$ are the stress energy tensors of the goldstino and matter field.

Another method for constructing goldstino-matter Lagrangians involves promoting the goldstino to a superfield $\Lambda$ \cite{Ivanov:1977my,Ivanov:1978mx,Samuel:1982uh,Antoniadis:2004uk,Antoniadis:2004se}. This is done in a way compliant with the nonlinear supersymmetry transformations of the goldstino so that in the end, the only physical degree of freedom in $\Lambda$ is simply $\lambda$. Since the basic concepts of goldstino Lagrangians have been presented along with the geometric method, we will skip this method and go directly to the next, which is the one used extensively in chapters \ref{NonlinearMSSM} and \ref{secDBI} (in its $\cN=2$ generalization).

This is the method of constrained superfields \cite{Rocek:1978nb,Lindstrom:1979kq,Casalbuoni:1988xh,SK}. It draws inspiration from a similar technique applied in bosonic symmetries for example in the context of $\sigma$ models. One starts from the full manifold made up from the linear symmetry transformations and then restrict to a certain submanifold by imposing constraints on the coordinates. This breaks the original symmetry down to the subgroup that is left invariant under the constraints. E.g. in an $O(4)$ $\sigma$ model of fields $(\sigma,\overrightarrow{\pi})$ ($\overrightarrow{\pi}$ is a vector of pions), we can break the symmetry down to $O(3)$ by imposing the constraint $\sigma^2+ \overrightarrow{\pi}\cdot\overrightarrow{\pi}=f^2$. It is the same manifold that we would obtain by starting from a vacuum state $(f,0)$ and applying the elements of the coset space $O(4)/O(3)$.

In the context of $\cN=1$ supersymmetry, this technique is realized in the following way. We start from a standard chiral superfield that describes a full supersymmetric multiplet and impose a specific constraint on it. Using the constraint, we eliminate the scalar d.o.f. in favor of the fermion. In particular, the constraint
\be
X_{nl}^2=0\ ,
\ee
delivers
\bea\label{goldstino1}
X_{nl}&=&\phi_X +\sqrt 2\,\, \theta\psi_X
+\theta\theta\,\,F_X,\qquad\rm{with}\qquad
\phi_X=\frac{\psi_X \psi_X}{2\, F_X}\ .
\eea
The simplest possible Lagrangian of $X_{nl}$:
\bea\label{X}
\int\! d^4\theta \,X_{nl}^\dagger X_{nl}
+\bigg{[}\!\int\! d^2\theta fX_{nl}+h.c.\bigg{]}
\!=\!\vert \partial_\mu\phi_X\vert^2\!+\!F_X^\dagger F_X\!+\!
\Big[\frac{i}{2}\overline\psi_{X}\overline\sigma^\mu
\partial_\mu\psi_{X}+f\,F_X+\!h.c.\!\Big]
\eea
reproduces the Volkov-Akulov Lagrangian upon integrating out the auxiliary $F_X$ and identifying $\psi_X$ with the goldstino.

The advantage of this method is the use of superfield formalism. For example, the couplings of goldstinos to matter are easily constructed by treating $X_{nl}$ as any other superfield and following the standard rules of superspace. As a demonstration, consider a supersymmetric theory with chiral multiplets $\Phi_i\equiv(\phi_i,\psi_i, F_i)$ and vector multiplets $V\equiv (A_\mu^a, \lambda^a, D^a)$ coupled in a general way to $X_{nl}$:
\bea
{\cal L}& =&\!\! \int d^4 \theta\,\, \Big[ X_{nl}^{\dagger} X_{nl} 
+ \Phi_i^{\dagger} (e^{V} \Phi)_i - ({m_i^2}/{f^2})\,
 X_{nl}^{\dagger} X_{nl} \Phi_i^{\dagger} (e^{V} \Phi)_i  \Big]
+\bigg\{ \int d^2 \theta \,\Big[ f X_{nl} + W(\Phi_i) 
\nonumber\\[-3pt]
&+&\!
\frac{B_{ij}}{2 f} X_{nl}\, \Phi_i \Phi_j
+ \frac{A_{ijk}}{6\,f} X_{nl} \Phi_i \Phi_j \Phi_k
+ \frac{1}{4} \Big(1+ \frac{2 \,m_\lambda}{f} X_{nl}\Big)\,
\mbox{Tr}\, W^{\alpha} W_{\alpha} 
\Big]
+ {\rm h.c.}\bigg\}, \quad\label{01}
\eea
where $m_i^2,B_{ij}, A_{ijk}$ are soft terms for the scalars and $m_{\lambda}$ is the gaugino mass. From this, one can find the Goldstino ($\psi_X$) couplings to ordinary matter and gauge superfields.

Furthermore, this formalism seems to be more general than the geometric method since it can reproduce couplings that were missed by the latter \cite{Brignole:1997pe,Luty:1998np}. In particular, from the equivalence theorem of spontaneously broken theories \cite{cddfg}, we know that for low energy SUSY breaking, the coupling of the gravitino to matter is dominated by the coupling of its goldstino component and has the form
\be
({1}/{f}) \ \partial^{\mu} \psi_X \ J_{\mu} = \
 - ({1}/{f}) \,\psi_X \ \partial^{\mu}  J_{\mu} +\mbox{(total space-time 
derivative)} , \label{02}
\ee
Here $J_{\mu}$ is the supercurrent of the theory corresponding to that in (\ref{01}) in which the goldstino is  essentially replaced by the spurion, with the corresponding explicit soft breaking terms:
\medskip
\bea
{\cal L}' &=&\!\!\!\! \int\!\! d^4 \theta\,
\Big[1-m_i^2\,\theta^2\overline\theta^2\Big]\, 
\Phi_i^{\dagger} (e^{V} \Phi)_i 
+\!\! \int\! d^2 \theta \,\Big[ 
W(\Phi_i)\! - \,{B_{ij}\over 2}\,\theta^2 \Phi_i \Phi_j
\nonumber\\
&-&\! {A_{ijk}\over 6}\,\theta^2\, \Phi_i \Phi_j \Phi_k 
+\frac{1}{4}\,(1-2 m_\lambda\theta^2) \,
\mbox{Tr}\, W^{\alpha} W_{\alpha} 
\Big] + {\rm h.c.}\ . \label{01prime}
\eea
With this, eq.~(\ref{02}) shows that, on shell, all goldstino couplings are proportional to soft terms. Indeed, the supercurrent of (\ref{01prime}) is given by
(with $\cD_{\mu,ij}=\delta_{ij}\,\partial_\mu +i\,g\,A^a_\mu\,T^a_{ij}$)
\bea
J^\mu_\alpha=
-[\sigma^\nu \overline\sigma^\mu\psi_i]_\alpha
\,[\cD_{\nu,\,ij}\phi_j]^\dagger
+i\,[\sigma^\mu\overline\psi_i]_{\alpha}
F_i
-\frac{1}{2\sqrt 2}[\sigma^\nu\overline\sigma^\rho\,\sigma^\mu
\overline\lambda^a]_\alpha\,F_{\nu\rho}^a
+\frac{i}{\sqrt 2} D^a\,[\sigma^\mu\overline\lambda^a]_\alpha\nonumber
\eea
so
\vspace{-0.2cm}
\bea\label{jjj}
\partial_{\mu} J^{\mu}_\alpha \ = \, \psi_{i,\alpha} 
\,\,(m_{i}^2  {\phi^\dagger}_j
 +  B_{ij} \phi_j + (1/2) A_{ijk} \phi_j \phi_k\, ) +
\frac{m_{\lambda}}{\sqrt{2}} 
\Big[({\sigma^{\mu\nu}})_\alpha^{\,\,\beta}\,
\lambda^a_\beta F_{\mu\nu}^a + D^a\,\lambda^a_\alpha\Big] \ .
\eea
From (\ref{02}), (\ref{jjj}) one then recovers the couplings with one goldstino that are missed in the geometric method.

 Finally, in addition to usual SUSY and goldstino couplings, eq.~(\ref{01}) brings new goldstino-independent couplings induced by eliminating $F_X$.
Indeed, we get
\be
\Big(1-\frac{m_i^2}{f^2}\,\vert \phi_i\vert^2\Big)\,
F_X^\dagger=-\Big(  f +\frac{B_{ij}}{2\,f}\, \phi_i \phi_j + 
 \frac{A_{ijk}}{6\,f}\, \phi_i \phi_j \phi_k +
 \frac{m_{\lambda}}{2\,f} \lambda \lambda+\cdots\Big) \ .
\ee
So $\vert F_X\vert^2$ generates new couplings in $\cL$, such as quartic scalar terms. As we will see in chapter \ref{NonlinearMSSM}, when applied to MSSM, they bring new corrections  to the Higgs scalar potential.

\chapter{MSSM$_5$}
\label{MSSM5}

We apply the methods of EFT on the Minimal Supersymmetric Standard Model (MSSM). Our aim is to study the phenomenological consequences of the complete set of mass dimension five operators that obey the gauge symmetries of MSSM and R-parity \cite{Farrar}. Since not all of them are physically relevant, we will use spurion dependent superfield redefinitions to find the irreducible set of operators. But before getting there, we need to provide the Lagrangian of the model.

\section{The Lagrangian}
\label{Lagrangian}

We denote the Lagrangian as:

\be\label{L1}
\cL=\cL_{MSSM}+\cL^{(5)}
\ee

$\cL_{MSSM}$ is the standard Lagrangian of the MSSM. In particular:
\bea\label{LMSSM}
\cL_{MSSM}
 &=&
\int d^4\theta \,\Big[\,
\cZ_1\,H_1^\dagger \,e^\Vm\,H_1+
\cZ_2\,H_2^\dagger \,e^\Vp\,H_2\Big]
 +\cL_K
\nonumber\\[8pt]
+ && \!\!\!\!  \!\!\!\! \!\!\!\! 
\bigg\{\int d^2\theta\,\Big[-\,H_2 \,Q\,\lambda_U\,U^c
- Q\,\lambda_D\,D^c\,H_1
-  L\,\lambda_E\,E^c\,H_1+
\mu\,H_1\,H_2
\Big]+h.c.\bigg\}
\eea
Here $\cL_{K}$ accounts for the gauge kinetic part and the kinetic terms of the quark and lepton superfields 
$Q,U^c,D^c,L,E^c$ as well as their associated soft
breaking terms obtained using the spurion field formalism. $U^c$, $D^c$ and $E^c$
denote anti-quark/lepton singlet chiral superfields of components  $f_R^c\equiv (f^c)_L$ and $\tilde f_R^*$, $f=u,d,e$, while
$Q$ and $L$ denote the left-handed quark and lepton superfields doublets. Furthermore, since the hypercharge of $H_1$ is $-1$ and that of $H_2$ is $+1$, the vector superfields are $\Vm\equiv \,g_2\, V_W^i\, \sigma^i -g_1 \, V_Y$ and $\Vp\equiv \,g_2\,V_W^i\,\sigma_i+g_1 \, V_Y$.
$V_Y$ and $V_W$ are the vector superfields of $U(1)_Y$ and
$SU(2)_L$ respectively with $g_1$ and
$g_2$ being the corresponding couplings. All SUSY breaking terms are included by allowing spurion dependence in the quantities $\cZ_i$, $\mu$ and the $3\times 3$ flavor matrices $\lambda_{U,D,E}$:
\bea
\cZ_i\equiv \cZ_i(S,S^\dagger),
\qquad \mu\equiv \mu(S),\qquad \lambda_F\equiv
\lambda_F(S),\,\,\,\,\,F:U,D,E\qquad
\eea
where  $S \equiv m_0\,\theta^2$ is the spurion parametrising the soft
supersymmetry breaking and $m_0$ is the supersymmetry breaking scale in the visible sector (e.g. if `$f$' is the v.e.v. of the auxiliary field that breaks SUSY, $m_0$ in gravity mediation is $f/M_{Planck}$ and in gauge mediation $f/M_{messenger}$). Since we  assume a spontaneously broken effective
Lagrangian, consistency of the integration procedure
 implies the restriction
\be
m_0 \ \ll \ M \ .
\ee
$\cL^{5}$ denotes the complete set of mass dimension five operators that preserve
R-parity\footnote{
For a general discussion of D=5 operators with discrete symmetries
 see \cite{Ibanez:1991pr}.}:
\medskip
\bea\label{ll5}
\cL^{(5)}&=&
\frac{1}{M}\bigg\{
\int d^2\theta
\,\,\Big[ Q\,U^c\,T_Q\,Q\,D^c+ Q\,U^c\,T_L\,L\,E^c+ \lambda_H
 (H_1 H_2)^2\,\Big]+h.c.\bigg\}
\nonumber\\[8pt]
&+&
\frac{1}{M}
\int d^4\theta\,\Big[
H_1^\dagger \,e^{\Vm} Q\,Y_U\,U^c\,
+H_2^\dagger\,e^\Vp  Q\,Y_D\,D^c\,
+H_2^\dagger\,e^\Vp L\,Y_E\,E^c\,+h.c.
\Big]\nonumber\\[8pt]
&+&
\frac{1}{M}
\int d^4\theta\,
\Big[
A(S,S^\dagger)\,D^{\alpha}\,\Big(B(S,S^\dagger)\,H_2 \,e^{-\Vm}\Big)
D_{\alpha}\,\Big(\Gamma(S,S^\dagger)\,e^\Vm\,H_1\Big)+h.c.\Big]
\label{dim5}
\eea

\medskip\noindent
The notation is such that
$$
Q\,\, U^c\,T_Q\,Q\,D^c \equiv (Q\,\, U^c)^T\,(i\sigma_2)\,T_Q\,Q\,D^c
$$
Similarly,
$$
D^{\alpha}[B(S,S^\dagger)H_2 e^{-\Vm}] D_{\alpha}[\Gamma(S,S^\dagger)e^{\Vm} H_1] \equiv D^{\alpha}[B(S,S^\dagger)H_2^T(i\sigma_2)e^{-\Vm}]D_{\alpha}[\Gamma(S,S^\dagger)e^{\Vm} H_1 ].
$$
$T_{Q,L}$ are matrices of parameters both in the up and the down sector, thus they carry four indices. In addition, all SUSY breaking terms are parametrized in the usual way, with spurions:
\bea
T_Q\equiv T_Q(S),\,\,\,\, T_L\equiv T_L(S),\,\,\,\,\,
\lambda_H\equiv \lambda_H(S),\qquad
Y_F\equiv Y_F(S,S^\dagger),\,\,\,F:U,D,E\,
\eea
$M$ is the mass scale up to which the effective approach remains valid. It is associated with the mass of the heavy states that have been integrated out in order to obtain the effective operators.

The spurion dependence associated to these operators is
the most general one can have. For the kinetic terms it is:
\medskip
\bea
&& \cZ_1 = 
1 + a_1 S +  a_1^* \,S^{\dagger} + a_2 S S^{\dagger} \ ,
\nonumber \\
&& \cZ_2 = 
1 + b_1 S +  b_1^* \,S^{\dagger} + b_2 S S^{\dagger}
\ .
\eea
and for the higher derivative effective operator:
\bea\label{definitions3}
A(S,S^\dagger)&=& \alpha_0+\alpha_1\,S+\alpha_2\,S^\dagger+
\alpha_3\,S\,S^\dagger\nonumber\\
B(S,S^\dagger)&=&\beta_0+\beta_1\,S+\beta_2\,S^\dagger+
\beta_3\,S\,S^\dagger\nonumber\\
\Gamma(S,S^\dagger)&=&
\gamma_0+\gamma_1\,S+\gamma_2\,S^\dagger+\gamma_3\,S\,S^\dagger
\eea

\section{Keeping the essential: The irreducible Lagrangian}
\label{IrreducibleLagrangian}

The parameter space of Lagrangian (\ref{L1}) is huge. However, big parts of it are redundant since they describe the same physics. We would like to simplify the Lagrangian by removing this redundancy. One way to do this is by performing appropriate field redefinitions. A familiar set of holomorphic superfield redefinitions is
\be
\Phi_i \ \rightarrow \ (1 - k_i \ S ) \ \Phi_i \ , \label{holom}
\ee
which are commonly used in MSSM in order to restrict the so called ``soft'' breaking terms. We shall use this freedom later on. Less familiar  are the
following (super)field transformations\footnote{To avoid a complicated index notation, the transformations in (\ref{tra}) are written in matrix notation for the Higgs $SU(2)$ doublets. For clarity, $(i\sigma_2)$ appears explicitly even if it is implicit in the superpotential.}
\medskip
\bea\label{tra}
H_1 \ \ra \ H_1' &=& H_1-\frac{1}{M}\,
\overline D^2\,\Big[\Delta_1\,H_2^\dagger\,
e^{\Vp}\,(i\,\sigma_2)\Big]^T
+\frac{1}{M} \,Q\,\rho_U\,U^c
\nonumber\\[10pt]
H_2 \ \ra \ H_2' &=& H_2
+
 \frac{1}{M}\, \overline
D^2\,\Big[\Delta_2\,H_1^\dagger\,e^\Vm\,(i\sigma_2)\Big]^T
+\frac{1}{M}\,Q\,\rho_D\,D^c+\frac{1}{M}\,L\,\rho_E\,E^c \eea
where
\bea
\rho_F=\rho_F(S);\,\,\,\,F:U,D,E,\,\,\qquad
\Delta_i=\Delta_i(S,S^\dagger)\qquad\,\,i=1,2 \label{ft}
\eea
are arbitrary functions of the spurion. Also, $\rho_F$, $F=U,D,E$ are $3\times 3$ matrices. The coefficients of their Taylor expansion in $S$ are free parameters. We are free to fix them in a way to eliminate redundant dimension-five operators. These coefficients should have values smaller than $M$. The expansion of $\Delta_i$ is:
\bea\label{delta1and2}
\Delta_1(S,S^\dagger)&=&s_0+s_1\,S+s_2\,S^\dagger+s_3\,S\,S^\dagger
\nonumber\\
\Delta_2(S,S^\dagger)&=&s_0'+s_1'\,S+s_2'\,S^\dagger+s_3'\,S\,S^\dagger
\eea
Notice that R-parity conservation does not allow for a similar set of transformations  (\ref{tra}) on quark and lepton superfields. In addition, these field redefinitions, along with mixing operators from ${\cal L}_{MSSM}$  and ${\cal L}^{(5)}$, generate operators of the type
\medskip
\bea
\frac{1}{M^2} \, \int d^4 \theta \ D^2
\big[ H_2 \,e^{-\Vm} \Delta_1^{\dagger}\big]
\, e^{\Vm} \ {\bar D}^2
\big[ \Delta_1  \, e^{-\Vm}\, H_2^{\dagger} \big]
\eea
plus a similar one for $H_1$. Since these operators are of higher-order in $1/M$, their effects are further suppressed with respect to the dimension-five operators
considered and we shall neglect them for the time being.

One then finds that the  original Lagrangian transforms into:
\medskip
\bea\label{ll6}
\cL &=&\cL_K+
\int d^4\theta\,\,\Big[
  \cZ_1'\,H_1^\dagger \,e^\Vm\,H_1+
  \cZ_2'\,H_2^\dagger \,e^{\Vp}\,H_2\Big]
\nonumber\\[7pt]
&+&
\int d^2\theta\,\,
\Big[
- H_2\,Q\,\lambda'_U\,U^c
- \, Q\,\lambda'_D\,D^c\,H_1
- \, L\,\lambda'_E\,E^c\,H_1 +
\mu\,H_1\, H_2\Big]+h.c.
\nonumber\\[7pt]
&+&
\frac{1}{M}
\int d^2\theta\,\,
\Big[\,
Q\,U^c\,T_Q'\,Q\, D^c+Q\,U^c\,T_L'\,L\,E^c+
\lambda_H \,(H_1\,H_2)^2\,\Big]+h.c.
\nonumber\\[7pt]
&+&
\frac{1}{M}
\,\int d^4\theta\,\Big[
H_1^\dagger\,e^\Vm\,Q\,Y_U'\,U^c+
H_2^\dagger\,e^{\Vp} Q\,Y_D'\,D^c+
H_2^\dagger\,e^{\Vp} L\,Y_E'\,E^c+h.c.\Big]
+\Delta\cL\qquad
\eea
where
\bea\label{deltao}
\Delta\cL&=&\frac{1}{M}
\int d^4\theta \,\Big[
-
\Delta_1^\dagger\,H_2 \,e^{-\Vm}  D^2 (\cZ_1 \ e^{\Vm} H_1)
-
\cZ_2\,H_2\,e^{-\Vm}\, D^2(\Delta_2^\dagger\,e^\Vm\,H_1)
+h.c.\Big]\nonumber\\[8pt]
&+&
\frac{1}{M} \int d^4\theta\,\, \Big[
A(S,S^\dagger)\,D^{\alpha}\,\big(\,B(S,S^\dagger)\,
H_2 \,e^{-\Vm}\big)\,\,D_{\alpha}\,
\big(\Gamma(S,S^\dagger)\,e^\Vm\,H_1\,\big)
+
h.c.\Big]\qquad
\label{redef}\eea

\bigskip
\noindent
The relation between primed and unprimed fields is
\medskip
\bea\label{lambdas}
\lambda'_F(S)=\lambda_F(S)+\frac{\mu(S)}{M}\,\rho_F(S),\qquad F:U,D,E
\eea
also
\bea\label{whys}
Y_U'(S,S^\dagger)&=&Y_U(S,S^\dagger)-4\,\Delta_2(S,S^\dagger)\,\lambda_U(S)
+\cZ_1(S,S^\dagger)\,\rho_U(S)\nonumber\\[8pt]
Y_D'(S,S^\dagger)&=&Y_D(S,S^\dagger)-4\,\Delta_1(S,S^\dagger)\,\lambda_D(S)
+\cZ_2(S,S^\dagger)\,\rho_D(S)\nonumber\\[8pt]
 Y_E'(S,S^\dagger)&=&
 Y_E(S,S^\dagger)-4\,\Delta_1(S,S^\dagger)\,\lambda_E(S)
+\cZ_2(S,S^\dagger)\,\rho_E(S)
\eea
and
\bea\label{ts}
 T'_Q(S)&=&T_Q(S) \ + \ \lambda_U(S)\, \otimes\, \rho_D(S) \ +
\rho_U(S)\, \otimes\, \lambda_D(S)\, \nonumber\\[8pt]
 T_L'(S)&=&T_L(S) \ + \ \lambda_U(S)\, \otimes \rho_E(S) \
 + \ \rho_U(S)\, \otimes  \lambda_E(S) \ .
\eea
Finally,
\bea\label{zs}
\cZ_1'(S,S^\dagger)&=&\cZ_1(S,S^\dagger)-\frac{1}{M}\,
\Big(4 \,\mu(S)\,
\Delta_2(S,S^\dagger)+h.c.\Big)
,\nonumber\\[8pt]
 \cZ_2'(S,S^\dagger)&=&
\cZ_2(S,S^\dagger)-\frac{1}{M}\, \Big(\, 4 \,\mu(S)\,
\Delta_1(S,S^\dagger)+h.c.\Big) \ .
\eea
We perform a second set of field redefinitions to canonically normalize the kinetic terms:
\bea\label{fields}
H_1\ra \frac{1}{\sqrt{a_0'}}\,
 \big[1-k_1\,S\big]\,H_1,\quad
H_2\ra\frac{1}{\sqrt{b_0'}}\,
 \big[1-k_2\,S\big]\,H_2,\quad k_1\equiv \frac{a_1'}{a_0'},\quad
 k_2\equiv \frac{b_1'}{b_0'}
\eea
with
\bea\label{aandb}
a_0'\equiv\cZ_1'\Big\vert_{S,S^\dagger=0},\qquad
a_1'\equiv \cZ_1'\Big\vert_{S},\qquad
b_0'\equiv\cZ_2'\Big\vert_{S,S^\dagger=0},\qquad
b_1'\equiv\cZ_2'\Big\vert_{S}
\eea

\medskip\noindent
which can be directly computed using
the definition of $\cZ_{1,2}'$, $\cZ_{1,2}$ and $\Delta_{1,2}$ given  above.
The Lagrangian then becomes
\bea\label{ll7}
\cL&=& \cL_K+\Delta\cL+
\int d^4\theta\,\,\Big[
  \Big(1- \frac{m_1^2}{m_0^2} \,S\,S^\dagger\Big) \,H_1^\dagger \,e^\Vm\,H_1
+\Big(1- \frac{m_2^2}{m_0^2} \,S\,S^\dagger\Big) \,H_2^\dagger \,e^{\Vp}\,H_2
\Big]\nonumber\\[8pt]
&+&
\int d^2\theta\,\,
\Big[
- H_2\,Q\,\lambda''_U\,U^c
- \, Q\,\lambda''_D\,D^c\,H_1
- \, L\,\lambda''_E\,E^c\,H_1 +
\mu'\,H_1\, H_2\Big]+h.c.
\nonumber\\[8pt]
&+&
\frac{1}{M}
\int d^2\theta\,\,
\Big[\,
Q\,U^c\,T_Q'\,Q\, D^c+Q\,U^c\,T_L'\,L\,E^c+
\lambda_H' \,(H_1\,H_2)^2\,\Big]+h.c.
\nonumber\\[8pt]
&+&
\frac{1}{M}
\,\int d^4\theta\,\Big[
H_1^\dagger\,e^\Vm\,Q\,Y_U''\,U^c+
H_2^\dagger\,e^{\Vp}\,Q\,Y_D''\,D^c+
H_2^\dagger\,e^{\Vp}\,L\,Y_E''\,E^c+
h.c.\Big]\qquad
\eea

\bigskip\noindent
Double primed quantities are given by
\medskip
\bea\label{lambdas2}
\lambda_U''(S)&=&\frac{1}{\sqrt{b_0'}}\,
 \,(1-k_2\,S)\,\,\lambda_U'(S)=(1-b_1\,S)\,\lambda_U(S)+\cO(1/M),
\nonumber\\
\lambda_F''(S)&=& \frac{1}{\sqrt{a_0'}}\,
\,(1-k_1\,S)\,\,\lambda_F'(S)
=(1-a_1\,S)\,\lambda_{F}(S)+\cO(1/M),\qquad F \equiv D, E.
\nonumber\\
\mu'(S) &=&
\frac{1}{\sqrt{a_0'\,b_0'}}\,[1- (k_1+k_2)S]\,\,\mu(S)
=(1-(a_1+b_1)\,S)\,\mu(S)+\cO(1/M).
\eea

\medskip\noindent
Since $a_0', b_0'$ are $M$-dependent, the couplings
$\lambda_{U,D,E}''(S)$ and also $\mu'(S)$
have  acquired, already at the classical level,
a dependence on the scale $M$ of the higher dimensional
operators.
This is denoted above by $\cO(1/M)$ and  can
be easily computed using (\ref{zs}) and (\ref{aandb}). Note that this
 $\cO(1/M)$  correction is relevant for the Lagrangian
(\ref{ll7}). Similar considerations apply to
$m_{1,2}$ that appear in the same Lagrangian. Their exact expressions in terms of initial parameters can be computed in a similar way.
Further
\medskip
\bea\label{yukawas}
\lambda_H'(S)\,=
 \Big(1-2 (a_1+b_1)\, S\Big)\,\,\lambda_H(S),&\,&
Y_U''(S,S^\dagger)\, =
\,\, (1-a_1^* \,S^\dagger\,)\,\,Y_U'(S,S^\dagger)
\nonumber\\[7pt]
Y_D''(S,S^\dagger) =
 (1-b_1^*\, S^\dagger\,)\,\,Y_D'(S,S^\dagger),\qquad &\,&
 Y_E''(S,S^\dagger) =
 (1-b_1^* \, S^\dagger\,)\,\,Y_E'(S,S^\dagger)\qquad\,\,\,
\eea

\medskip\noindent
where we ignored terms which bring $\cO(1/M^2)$
corrections to (\ref{ll7}).
Finally, $\Delta\cL$ in (\ref{ll7})  is that of
(\ref{deltao})  after applying
transformations (\ref{fields}). Its component expansion up to 1/M order is:
\medskip
\bea\label{deltaL1}
\Delta\cL&
=&
-\frac{1}{M}
\int d^4\theta
\,\,t_0\,\,H_2\,e^{-\Vm} \,D^2 \,\Big[e^\Vm \, H_1\Big]
\nonumber\\[7pt]
&+&
\frac{m_0}{M}\,\Big[\,\,
4 \,\big[t_1+t_2+t_0(a_1+b_1)\big]\,h_2 \,
\cD_\mu\cD^\mu\,h_1
-
2 \, \big[t_1-t_2+t_0(b_1-a_1)\big]\,h_2\,D_1\,h_1\nonumber\\[7pt]
&+&
2\sqrt 2 \,(t_1+b_1\,t_0)\,h_2\,\lambda_1\,\psi_{h_1}
-
2\sqrt 2 \,(t_2+a_1\,t_0)\,\psi_{h_2}\,\lambda_1\,h_1
- 4\,t_3\,F_{h_2}\,F_{h_1}
\Big]\nonumber\\[7pt]
&+&
\frac{m_0^2}{M}\,\Big[
-4 \,(t_4-b_1\,t_3)\, h_2\,F_{h_1}
-4 \,(t_5-a_1\,t_3)\,F_{h_2}\,h_1+
2\,t_6\,\psi_{h_2}\psi_{h_1}
\Big]\nonumber\\[7pt]
&+&
\frac{m_0^3}{M}\,\Big[-4\,(t_7-a_1\,t_4-b_1\,t_5+a_1 \,b_1 \,t_3)
\,\,h_2h_1\Big]+h.c.
\eea

\bigskip\noindent
where $D_1$ and $\lambda_1$ are components of the vector superfield $V_1$ and we also used the component notation $H_i=(h_i,\psi_{h_i},F_{h_i})$.
We also replaced  $k_{1}, (k_2)$ by $a_1$, ($b_1$) respectively,
which is correct in the approximation of ignoring $1/M^2$ terms in the
Lagrangian. The coefficients $t_i$ are given by
\bea
t_0&=&\alpha_0\beta_0\gamma_0
+ s_0^*+ s_0^{' *},
\hspace{2.6cm}
t_4= d_4- 
\,s_3^*-a_1^*\,s_2^*- b_2\,s_0^{' *} -b_1\,s_1^{' *},
\nonumber\\[7pt]
t_1&=&d_1-
s_2^* - b_1\,s_0^{' *},
\hspace{2.9cm}
t_5=d_5-a_2\,s_0^*-a_1\,s_1^*-
s_3^{' *}- b_1^*\,s_2^{' *},
\nonumber\\[7pt]
t_2&=&d_2-a_1\,s_0^*-
s_2^{' *},
\hspace{2.8cm}
t_6=d_6,
\nonumber\\[7pt]
t_3&=&d_3-
s_1^*- a_1^*\,s_0^*- 
s_1^{' *}- b_1^*\,s_0^{' *},
\hspace{0.7cm}
t_7=d_7-a_2\,s_2^*-a_1s_3^*-b_1s_3^{' *}-b_2 s_2^{' *}
\eea
\medskip\noindent
with $d_i$ being combinations of input parameters
$\alpha_i, \beta_i,\gamma_i$ of eq.~(\ref{definitions3})
\bea\label{ds}
d_1&\equiv& - \beta_1\, \alpha_0\,\gamma_0\,
-\,\alpha_1\,\beta_0\,\gamma_0/2,
\hspace{2cm}
d_4\equiv  -\beta_3\,\alpha_0\,\gamma_0-\beta_1\,\alpha_2\,\gamma_0
-\alpha_0\beta_1\gamma_2
\nonumber\\[7pt]
d_2&\equiv&  -\gamma_1\,\beta_0\,\alpha_0
-\,\alpha_1\,\beta_0\,\gamma_0/2,
\hspace{2.cm}
d_5 \equiv -\gamma_3\,\beta_0\,\alpha_0-\gamma_1\,\alpha_2\,\beta_0
-\alpha_0\beta_2\gamma_1,
\nonumber\\[7pt]
d_3 &\equiv &-\alpha_2\,\beta_0\,\gamma_0
-\alpha_0\beta_2\gamma_0-\alpha_0\beta_0\gamma_2,
\hspace{0.9cm}
d_6 \equiv \alpha_3\,\gamma_0\,\beta_0
+\alpha_1\beta_2\gamma_0+\alpha_1\beta_0\gamma_2
\nonumber\\[7pt]
&&\hspace{5.9cm}
d_7\equiv -\gamma_3\,\beta_1\,\alpha_0-\gamma_1\,\beta_3\,\alpha_0-
\gamma_1\,\beta_1\,\alpha_2. \qquad
\eea

\medskip\noindent
A suitable choice of coefficients  $s_0, s_0', s_2', s_2$
entering in transformation (\ref{tra}) allows us to set
\bea\label{constraints}
t_i=0,\qquad i=0,1,2, 3.
\eea
This ensures that the nonstandard terms in 
the first, second and third  lines of
$\Delta\cL$ above are not present. The remaining terms
proportional to $m_0^2$ and $m_0^3$
bring solely a renormalisation of
soft terms, which are present anyway in Lagrangian
(\ref{ll7}) and can be ignored. Finally, the term $t_6\, \psi_{h_2}\psi_{h_1}$
brings a renormalisation of the supersymmetric $\mu'$ term ($\mu'
H_1 H_2$) of (\ref{ll7}) and is invariant under the general field transformations (\ref{tra}).
In principle one could set additional coefficients of the last two
lines in $\Delta\cL$
to vanish by a suitable choice of remaining $s_{1,3}, s_{1,3}'$; 
we choose not to do so and instead  save these remaining coefficients
 for additional conditions that can be used to simplify the Lagrangian even further.

We have finally obtained the minimal set of dimension-five operators beyond the MSSM Lagrangian:
\smallskip
\bea
\label{FL}
\cL&=&\cL_K+
\int d^4\theta\,\,\Big[\Big(
1- \frac{m_1^2}{m_0^2} S^{\dagger} S\Big) \,H_1^\dagger \,e^\Vm\,H_1
+ \Big(1- \frac{m_2^2}{m_0^2} S^{\dagger} S\Big)
\,H_2^\dagger \,e^{\Vp}\,H_2 \Big]\nonumber\\[8pt]
&+&
\int d^2\theta
\Big[
- H_2\, Q\,\lambda_U'' (S)U^c
- Q\,\lambda_D'' (S) D^cH_1
- L\,\lambda_E'' (S) E^cH_1
+ \mu'' (S) \,H_1H_2\Big]+h.c.
\nonumber\\[8pt]
&+&
\frac{1}{M}
\int d^2\theta
\,\,
\Big[\,
Q\,U^c\,T_Q'(S) \,Q\, D^c+Q\,U^c\,T_L' (S) \,L\,E^c+
\lambda_H' (S) \,(H_1\,H_2)^2\,\Big]+h.c.
\nonumber\\[8pt]
&\!\!\!\!+&\!\!\!\!
\frac{1}{M}\!
\int\! d^4\theta\Big[
H_1^\dagger e^\Vm  QY_U''(S,S^{\dagger})U^c\!
+
H_2^\dagger e^\Vp QY_D''(S,S^{\dagger})\,D^c\!
+
H_2^\dagger e^\Vp LY_E''(S,S^{\dagger})E^c\!+h.c.\Big]
\nonumber\\
\eea

\noindent
$\cL_K$ stands for gauge kinetic terms and kinetic terms of MSSM fields other than
$H_{1,2}$, together with their spurion dependence. Also,
$\mu''$ here includes the renormalisation due to $t_6$ (not shown). As explained
above, there is still some remaining freedom to further reduce the parameter space and we will use it in the next section. The couplings that appear are given in equations (\ref{lambdas}), (\ref{whys}), (\ref{ts}), (\ref{lambdas2}) and (\ref{yukawas}) in terms of those in the original Lagrangian. The couplings $\lambda''_{U,D,E}(S)$ acquired a threshold correction $\cO(1/M)$, which can be obtained from (\ref{lambdas2}).
The dimension-five operator that was present in the last line
of (\ref{ll5}) is completely ``gauged away'' in the new fields basis,
up to effects which renormalised the soft terms or the supersymmetric $\mu$ term. Since physics is independent of the fields basis we choose, in this new basis it is manifest that the last operator in (\ref{ll5}) cannot affect the relations among physical masses
of the Higgs sector. We discuss this in detail in section~\ref{HiggsMassCorrectionsBeyondMSSM}.


\chapter{Phenomenology of MSSM$_5$}
\label{PhenomenologyMSSM5}

\section{Further Restrictions from Flavor Changing Neutral Currents}
\label{Ansatz}

The couplings in Lagrangian (\ref{FL}) can
have dramatic implications if the scale $M$ is not
too high, in particular due to FCNC effects.
Indeed, if $T'_{Q,L}$ and $Y_{U,D,E}''$ 
have arbitrary family dependent 
couplings, one expects stringent limits from FCNC bounds 
\cite{Gabbiani:1996hi}.  It is possible however,
 under some  mild  assumptions for the original
$\cL$ of (\ref{L1}), to remove the dangerous couplings in (\ref{FL}). For example, assume that the flavor matrices in (\ref{ll5})
and the $\rho_{U,D,E}$ in (\ref{tra}), (\ref{ft}) are proportional to the ordinary Yukawa couplings\footnote{The ansatz is motivated by the discussion in subsection~\ref{GaugeInteractionsComponentAnalysis}, eq.~(\ref{B7}) where a
similar structure of  $T_{Q,L}$ and $\rho_F$ is generated by integrating
out massive $SU(2)$ superfields doublets.}: 
\bea\label{ansatz0}
T_Q(S)& = & c_Q(S)\,\,\lambda_U(0)\otimes \,\lambda_D(0)\nonumber\\
T_L(S)& = & c_L(S)\,\,\lambda_U(0)\otimes \,\lambda_E(0)\nonumber\\
\rho_F(S)& = & c_F(S)\,\,\lambda_F(0),\,\,\,\,\,\,\,\, F: U,D,E
\eea
and, as usual
\bea\label{lambdas3}
\lambda_F(S)&=&\lambda_F(0)\,(1+A_F \,S),\,\,\,\,\, F: U,D,E.
\eea
Here $c_{Q,L}(S)$ are some arbitrary
input functions of $S$; $\lambda_F(S)$ are  $3\times 3$ matrices,
while $A_F$ are trilinear couplings.
In the following
$c_F(S)\equiv c_0^F+S\,\,c^F_1$, $F=U,D,E$ are considered free
parameters which can be adjusted, together with the remaining
$s_{1,3}$, $s_{1,3}'$, to remove some of the couplings in (\ref{FL}).
Indeed, if
\bea
c_U(S)=- c_L(S)-c_E(S), \quad
c_D(S)=- c_Q(S)+c_L(S)+c_E(S)
\eea
while  $c_E(S)$ remains arbitrary, one obtains
\bea \label{TQL}
T_Q'(S)=0,\quad  T_L'(S)=0
\eea
We can therefore remove the associated couplings in (\ref{FL}),
that is the first two terms in the third line. Finally, let us assume that in (\ref{ll5}) we also have
\bea\label{ansatz1}
Y_F(S,S^\dagger)=\,f_F(S,S^\dagger)\,\lambda_F(0),\qquad F:U,D,E
\eea
where $f_F$ are spurion dependent but family independent functions of arbitrary
coefficients:
\bea
f_F(S,S^\dagger)=f_0^F+ S\,f_1^F+ S^\dagger\,f_2^F
+S\,S^\dagger\,f_3^F
\eea
Using (\ref{yukawas}),
we find that the couplings in (\ref{FL}) are
\medskip
\bea
Y_F^{''}(S,S^\dagger) =
\lambda_F(0)\,\Big[ x^F_0 + x^F_1 \,\,S
+ x^F_2\,\,S^\dagger\,
+ x^F_3  \,\,S\,S^\dagger\Big],\quad
F=U,D,E
\eea

\medskip\noindent
One finds
\bea
x^U_0&= &f^U_{0}- 4 s_0' +  \,c_0^U
\nonumber\\
x^U_1&= &f^U_{1}-4\,s_1'+ \,c^U_1+a_1\,c_0^U
\nonumber\\
x^U_2&=&f^U_2-4\,s_2'+a_1^*\,c^U_0- 
a_1^*\, x_0^U
\nonumber\\
x^U_3&=&f^U_3-4\,s_3'+a_1^*\,c^U_1+a_2\,c^U_0 - 
a_1^* \,\,x_1^U
\eea

\medskip\noindent
Similar equations exist for the fields in the $D$ and $E$ multiplets. We just need to replace $U\ra D$ (or $E$), $s_i'\ra s_i$ and $a_i\ra b_i$.

Let us examine  if the form of $Y''_F(S,S^\dagger)$
can  be simplified using the free parameters that we are left
with: these  are $s_{1,3}, s_{1,3}'$  from general transformations
$\Delta_{1,2}$ and $c_E(S)=c^E_0+S\,c^E_1$, a total of 6 free
parameters. We can use $s_{1,3}'$  ($s_{1,3}$)
to eliminate $S$ and $S\,S^\dagger$ parts of
$Y_U''$ \, ($Y_D''$), respectively.
Using $c_0^E$ and $c_1^E$
we can also eliminate the  $S$ and $S\,S^\dagger$ of $Y_E''$.
In conclusion, we used the remaining
 6 free parameters to bring $Y''_F$ to the form
\medskip
\bea\label{newY}
Y''_F(S^\dagger)\equiv Y_F''(0,S^\dagger)
=\lambda_F(0)\,(x_0^F+ x_2^F\,\,S^\dagger),\qquad
F:U,D,E
\eea

\medskip\noindent
The coefficients $x_{0,2}^F$ depend on the arbitrary coefficients  $f_i^F$, $i=0,1,2,3$,  $a_i$, $b_i$, $c_i$ of the original Lagrangian (\ref{L1}). Other simplifications can occur
if we ignore the couplings $Y$ of the first two families.
With these considerations,
 the Lagrangian in (\ref{FL}) takes the form
\medskip
\bea\label{lastL}
\cL&=&\cL_K+
\int d^4\theta\,\,\Big[
\Big(1-  \frac{m_1^2}{m_0^2} S^{\dagger} S\Big) 
\,H_1^\dagger \,e^\Vm\,H_1
+
\Big(1- \frac{m_2^2}{m_0^2} S^{\dagger} S\Big) 
\,H_2^\dagger \,e^{\Vp}\,H_2 \Big]
\nonumber\\[7pt]
&+&
\int d^2\theta\,\,
\Big[
- H_2\, Q\,\lambda_U'' (S)\,U^c -
Q\lambda_D'' (S) D^c\,H_1-
L\,\lambda_E'' (S) E^cH_1+\mu'' (S) H_1\, H_2\Big]+h.c.
\nonumber\\[7pt]
&+&\!\!\!\!
\frac{1}{M}\!
\,\int\! d^4\theta\,\Big[
H_1^\dagger\,e^\Vm\,Q\,Y_U''(S^{\dagger})\,U^c
+
H_2^\dagger\,e^{\Vp} Q\,Y_D''(S^{\dagger})\,D^c
+
H_2^\dagger\,e^{\Vp}  L\,Y_E''(S^{\dagger})\,E^c+h.c.\Big]\qquad
\nonumber\\[7pt]
&+&\frac{1}{M}\int
d^2\theta\,\,\lambda_H'(S)\,(H_1\,H_2)^2+h.c.
\eea

\bigskip\noindent
with  couplings (\ref{newY}) and (\ref{lambdas2})\footnote{
$\lambda_F''(S)$ acquired a threshold correction in $M$:
$\lambda''_U(0)=\lambda_U(0)\,
\big[1+{1}/{M}\,\big(\mu(0)\,c_U(0)+2\,(\mu(0)\,s_0+\mu^*(0)\,s_0^*
)\big)\big]
$
with similar relations for  $D$, $E$
 obtained by
$s_0\ra s_0'$ and  $U\ra D$, ($U\ra E$).
In terms of original parameters, 
$
s_0=-[-4 \alpha_0^*\beta_0^*\gamma_0^*\,b_1-4
\,d_3^*+
(f_1^U+f_1^D+c_1^U+c_1^D+a_1\,c_0^U+b_1\,c_0^D)]/
4\,(a_1-b_1)$
with $d_3$ as in (\ref{ds}); for the $D, E$ sectors we use
$s_0'=-\alpha_0^*\beta_0^*\gamma_0^*-s_0$.
Similar relations exist for
non-supersymmetric counterparts, see (\ref{lambdas2}),
(\ref{yukawas}).}. This Lagrangian defines MSSM$_5$; the extension of MSSM by
mass dimension five operators.

\section{Phenomenological Implications}
\label{PhenomenologicalImplications}

In the following we explore the new couplings that MSSM$_5$ brings with respect to standard MSSM \cite{Buchmuller,Pospelov:2005ks}. We begin with couplings proportional to $m_0$. Part of these are coming from the terms in the second-last line of (\ref{lastL}). These  include nonanalytic  Yukawa couplings \cite{M0}
\medskip
\bea\label{cset3}
&& \frac{m_0}{M}\,x_2^U\,(\lambda^U_0)_{ij}\,\,
(h_1^\dagger\,q_{L\,i})\,\,u_{R\,j}^c+h.c.\nonumber\\
&&  \frac{m_0}{M}\,x_2^D\,(\lambda^D_0 )_{ij}\,\,
(h_2^\dagger\,q_{L\,i})\,\,d_{R\,j}^c+h.c.\nonumber\\
&& \frac{m_0}{M}\,x_2^E\,(\lambda^E_0 )_{ij}\,\,
 (h_2^\dagger\,l_{L\,i})\,\,e_{R\,j}^c+h.c.,\qquad \lambda^F_0\equiv
\lambda_F(0),\,\,\, F:U,D,E.
\eea

\medskip\noindent
These  couplings
are not soft in the sense of \cite{Girardello:1981wz},
but ``hard'' supersymmetry breaking terms in the sense of \cite{M0,M1}. They are  less suppressed than those listed in \cite{M0} where they were generated
at order $m_0^2/M^2$. Such couplings can bring about a $\tan\beta$
enhancement of a prediction for a physical observable,
such as the bottom quark mass relative to bottom quark
Yukawa coupling \cite{Haber, Carena:2001bg}. This effect  is also present  in the electroweak scale effective Lagrangian of the MSSM alone,
after integrating out massive squarks at one loop level,
with a result for bottom quark mass
\cite{Haber,Carena:2001bg,Pierce:1996zz,Hall:1993gn,Carena:1994bv}
\bea
m_b=\frac{v\cos\beta}{\sqrt
  2}\,\Big( \lambda_b +{\delta}{\lambda_b}+
{\Delta}{\lambda_b}\tan\beta\Big)
\eea

\medskip\noindent
where $\lambda_b$ is the ordinary bottom quark
Yukawa coupling, $\delta\lambda_b$ its one loop correction
 and $\Delta\lambda_b$ is a ``wrong'' Higgs
bottom quark  Yukawa coupling, generated by integrating out massive
squarks. In our case, $\Delta\lambda_b$ receives an additional
contribution from the  second line in (\ref{cset3}). The size of
this extra contribution due to  higher dimensional operators, can be
comparable and even substantially larger than the one generated
in the MSSM at one loop level (for a suitable value for
$x_2^D\,m_0/M$ - recall that $x_2^D$ is not fixed). Such
contributions  can bring a $\tan\beta$ enhanced correction of the
Higgs decay rate to bottom quark pairs. Similar considerations apply
to the $U$ and $E$ sectors.

Other similar  couplings derived from (\ref{lastL})
and proportional to $m_0$ are
\medskip
\bea\label{cset4}
&&
\frac{m_0}{M}\,x_2^U\,(\lambda_0^{U\dagger} \lambda_0^U)_{ij}
\,\, (h_1^\dagger\,h_2^\dagger)\,\,\tilde u_{R\,i}\,\tilde u_{R\,j}^*
+h.c.\nonumber\\
&&
\frac{m_0}{M}\,x_2^U\,(\lambda_0^U\, \lambda_0^{U\dagger})_{ij}
\,\, (h_1^\dagger\,\tilde q_{L\,i})\,\,(h_2^\dagger 
\,\tilde q_{L\, j}^\dagger)
+h.c.
\eea

\medskip\noindent
where we used that
$\lambda_0^{F''}$ and $\lambda_0^{F}$ are equal
up to $\cO(1/M)$ corrections,
see (\ref{lambdas}) and (\ref{lambdas2}).
The above terms are strongly suppressed due to the square
of the Yukawa coupling, in addition to $m_0/M\ll 1$,
so their effects are expected to be small, except for the third
generation.
Their counterparts in the down ($D$) sector are
\medskip
\bea\label{cset5}
&&
\frac{m_0}{M}\,x_2^D\,(\lambda_0^{D\dagger}\, \lambda_0^{D})_{ij}
\,\, (h_2^\dagger\,h_1^\dagger)\,\,\tilde d_{R\,i}\,\tilde d_{R\,j}^*
+h.c.\nonumber\\
&&
\frac{m_0}{M}\,x_2^D\,(\lambda_0^D\, \lambda_0^{D\dagger})_{ij}
\,\, (h_2^\dagger\,\tilde q_{L\,i})\,\,(h_1^\dagger
 \,\tilde q_{L\, j}^\dagger)
+h.c.
\eea

\medskip\noindent
In the lepton sector similar couplings
are present, obtained  from eq.~(\ref{cset5})
with $Q\ra L$, $D\ra E$.
All the quartic couplings listed above are renormalisable, 
but naively they would seem to
break supersymmetry in a hard way if inserted into
loops with a cutoff larger than $M$. This, of course, is just an
artifact of using a cutoff larger than the energy scale of the heavy
states that we integrated out.

It is  interesting to note that there is no
``wrong" Higgs-gaugino-higgsino coupling generated \cite{M0}, even
though the original Lagrangian in eq.~(\ref{ll5})  included it, see
eq.~(\ref{deltaL1}) where
\medskip
\bea
\frac{m_0}{M}\,\,\big( \psi_{h_2}\,\lambda_1\,\,h_1+
h_2\,\lambda_1\,\,\psi_{h_1}\big)+h.c.
\eea
was present. Such a coupling can  be generated
at one loop level \cite{Haber} but in our case it was removed by the Higgs fields transformation  (\ref{tra}). This  shows that not all
``wrong'' Higgs couplings are actually independent
(this  may also apply when such couplings are generated at the loop
level). 
 
Note that in the MSSM$_5$  defined by eq.~(\ref{lastL}),
couplings proportional to $m_0$ involving ``wrong'' Higgs A-terms
are not present, given our ansatz (\ref{ansatz0}) and (\ref{ansatz1}) 
leading to (\ref{newY}). If
this ansatz is not imposed on the third generation,
then  one could have such terms from  (\ref{FL})
\medskip
\bea
\frac{m_0^2}{M}\,\Big[
y_{u,3}\,
 h_1^\dagger\,\tilde q_{L,3}\,\,\tilde u_{R,3}^*
+
\,y_{d,3}\, 
\,h_2^\dagger\,\tilde q_{L,3}\,\,\tilde d_{R,3}^* 
+
 y_{e,3}\,\, 
h_2^\dagger\,\tilde l_{L,3}\,\,\tilde e_{R,3}^* \Big]
\eea

\medskip\noindent
where $y_{f,3}$, $f=u,d,e$ are the coefficients of
 component $S\,S^\dagger$ of $Y''(S,S^\dagger)$ of third generation.

There are also new, and perhaps most important,
 supersymmetric couplings that
affect the amplitude of processes like
quark + quark $\ra$ squark + squark or similar with (s)leptons.
These are
\medskip
\bea\label{qqsqsq} &&
\frac{1}{M}\,x_0^U\,(\lambda_0^D)_{ij}\,(\lambda_0^U)_{kl}
\,\,\tilde q_{L\,i}\,\tilde d_{R\,j}^*\,\,
q_{L\,k}\,u_{R\,l}^c+ h.c.
\nonumber\\
 && \frac{1}{M}\,x_0^D\,(\lambda_0^U)_{ij}\,(\lambda_0^D)_{kl}
 \,\,\tilde q_{L\,i}\,\tilde u_{R\,j}^*\,\, q_{L\,k}\,d_{R\,l}^c+h.c.
 \nonumber\\
&&\frac{1}{M}\,x_0^U\,(\lambda_0^E)_{ij}\,(\lambda_0^U)_{kl}
\,\,\tilde l_{L\,i}\,\tilde e_{R\,j}^*\,\, q_{L\,k}\,u_{R\,l}^c
+(L\leftrightarrow Q, E\leftrightarrow U)+h.c.
\label{squarkproduction}
\eea

\medskip\noindent
They can be important particularly for the third generation.
The largest effect would be for squarks pair production
from a pair of quarks; the process could
be comparable to the MSSM tree level
 contribution to the amplitude of the same process
\cite{Dawson:1983fw}. Indeed, let us focus on the $q {\bar q}
\rightarrow {\tilde q} {\tilde q^*}$ in MSSM generated by a tree-level
gluon exchange. The MSSM amplitude behaves as
\medskip
\begin{equation}
A_{q {\bar q} \rightarrow g \rightarrow {\tilde q} {\tilde q^*}} \sim
{\frac{g_3^2}{\sqrt{s}}} \ ,
\end{equation}

\medskip\noindent
where $s$ is the Mandelstam variable. On the other hand, the
operators (\ref{squarkproduction}) generate a contact term
contributing
\medskip
\begin{equation}
A_{q {\bar q} \rightarrow {\tilde q} {\tilde q^*}}^{MSSM_5} \sim
\frac{\lambda_0^U \lambda_0^D}{M} \ .
\end{equation}

\medskip\noindent
The dimension-five operator for the third generation has therefore a
comparable contribution to the MSSM diagrams for energies $E \ge
g_3^2 M$, which can be in the TeV range. In MSSM there are other
diagrams contributing to this process, in particular Higgs exchange.
It can be checked however that at energies above the CP-even
Higgs masses, the MSSM amplitude decreases in energy whereas the
contact term coming from the dimension-five operators gives
a constant contribution which is sizeable
for high energy. Of course, at energies above $M$ we should replace the
contact term by the corresponding tree-level diagram with exchange of
massive $SU(2)$ doublets (or whatever other physics generates
 this effective operator).

 Note that couplings similar to (\ref{qqsqsq}) could
also be generated by the term $\int d^2\theta\, (Q U)\,T_Q (Q D)$ of
(\ref{FL}). This term is not present in MSSM$_5$ of
(\ref{lastL}) due to our FCNC ansatz (\ref{ansatz0}), (\ref{TQL}); however, the ansatz could be relaxed for the third generation. Therefore the above process of squark production can have an even larger amplitude
 from contributions in the third line of (\ref{FL}).

The Lagrangian (\ref{lastL}) also contains other 
 supersymmetric couplings involving
gauge interactions which can be important for
 phenomenology. They arise from any dimension-five D-term in 
 (\ref{lastL}) giving
\bea\label{gg1}
\cL\! &\supset&\!\! \frac{(\lambda_0^U)_{ij} x_0^U}{M}\,
\Big[
-h_1^\dagger \,\cD_\mu\cD^\mu \,(\tilde q_{L\,i} \,\tilde u^*_{R\, j})
-
\frac{1}{\sqrt{2}}\,
h_1^\dagger \lambda_1\,\big(\,\tilde q_{L\,i}\,\,u_{R\,j}^c
+q_{L\, i} \,\,\tilde u_{R\,j}^*\big)
-\frac{1}{\sqrt 2}\,
\overline \psi_{h_1}\,\overline\lambda_1\,\tilde q_{L\,i}\,\,\tilde
u_{R\,j}^*\nonumber\\[4pt]
&+&
\frac{1}{2}\,\,
h_1^\dagger \,D_1\,\tilde q_{L\,i}\, \tilde u_{R\,j}^*
+
i\overline \psi_{h_1}\,\overline\sigma^\mu\,\cD_\mu \,\big(\tilde
q_{L\,i}\,\,u_{R\,j}^c+ q_{L\,i} \,\,\tilde u_{R\,j}^*\big)
\Big]\nonumber\\[7pt]
&+&\!\!\!\!\!(U\rightarrow D, \,H_1\rightarrow H_2,
 \,V_1\rightarrow V_2)+
(Q\rightarrow L, \,H_1\rightarrow H_2, 
\,V_1\rightarrow V_2, U\rightarrow E)+h.c.
\eea

\bigskip\noindent
where $D_1$, $\lambda_1$
are the auxiliary  and gaugino components of $V_1$ vector superfield,
and
\bea
D_1&\equiv& -\frac{g_2^2}{2}\,
\Big[\,h_1^\dagger\,\vec\sigma\,h_1
+
h_2^\dagger\,\vec\sigma\,h_2
+
\tilde q_{L\,i}^\dagger  \vec\sigma\tilde q_{L\,i}
+
\tilde l_{L\,i}^\dagger  \vec\sigma\tilde l_{L\,i}
\Big]\nonumber\\[5pt]
&+&
\!\!\!\!\!\frac{g_1^2}{2}
\Big[-h_1^\dagger h_1 +h_2^\dagger h_2+\frac{1}{3}\tilde
q_{L\,i}^\dagger
\tilde q_{L\,i}-\frac{4}{3}\tilde u_{R\,i}\tilde u_{R\,i}^*
+\frac{2}{3}\tilde d_{R\,i}\tilde d_{R\,i}^*
-\tilde l_{L\,i}^\dagger\,\tilde l_{L\,i}+2 \,\tilde e_{R\,i}\,\tilde
e_{R\,i}^*
\Big]
\eea

\bigskip\noindent
Here $\cD_\mu$ is the covariant derivative, $\cD_\mu=\partial_\mu
+i/2\,V_{1,\mu}$, where $V_{1,\mu}$ is the gauge field of the vector
superfield 
$\Vm\equiv \,g_2\, V_W^i\, \sigma^i -g_1 \, V_Y$, introduced
in eq.~(\ref{LMSSM}).
 Couplings similar to those above are generated by
the substitutions shown in (\ref{gg1}). Some of them can be phenomenologically important, e.g. those involving 2 particles and 2 sparticles such as
Higgs-quark-squark-gaugino or gauge-quark-higgsino-squark,
arising from (\ref{gg1}).  Also, we notice a term with  a ``wrong" Higgs-squark-squark 
derivative coupling.

Yukawa interactions also generate supersymmetric 
couplings of structure similar to some of those in (\ref{gg1}),
involving 4 squarks and a higgs or 2
squarks and 3 higgses or 2 squarks, 2 sleptons and a higgs. However,
these arise at order $\lambda_F^3$, where $\lambda_F$, $F:U,D,E$
are Yukawa couplings entering (\ref{lastL}).
Therefore they are suppressed both by the scale $M$ and,
relative to the above gauge counterparts, by an extra Yukawa 
coupling. This is due to the presence of an extra Yukawa
 coupling in the third line of (\ref{lastL}) relative to ordinary D-terms.
The strength of these interactions
is  also sub-leading to other Yukawa interactions listed so far
which also  involved fewer (s)particles.

Finally, supersymmetric couplings with 3 higgses and 2 squarks or 2 sleptons arise from $(H_1 H_2)^2$ of (\ref{lastL}),
suppressed by two Yukawa couplings and by the scale $M$. Also, there exist
potentially larger couplings of 2 higgses and 2 higgsinos, being
suppressed only by $\lambda_H(0)$ and the scale $M$.
In addition, there are non-supersymmetric  couplings with 4 higgs fields 
whose effects are discussed in section~\ref{HiggsMassCorrectionsBeyondMSSM}.
This concludes our discussion of all the new couplings generated
by dimension-five operators in the MSSM$_5$.

\section{The MSSM Higgs Sector with Mass Dimension Five Operators}
\label{MSSM5HiggsSector}

In the following we  restrict the analysis to the MSSM
Higgs sector extended by mass dimension five operators and analyse their
implications. In this  sector there are in general two dimension-five operators
that affect the Higgs fields masses, shown in eq.~(\ref{lls}) below.
According to our previous discussion
the last operator in (\ref{lls}) is redundant and can be
``gauged away''.   However, in this section
we choose to keep it, in order to show explicitly
that it does not bring new physics of its own.
The relevant part of  MSSM Higgs Lagrangian
with dimension-five operators is
\medskip
\bea\label{lls}
\cL_1&=&\int d^4 \theta \,\,\Big[
\cZ_1(S, S^\dagger)
\,\, H_1^\dagger \,e^{\Vm}\,H_1
+
\,\, \cZ_2(S,S^\dagger)
\,\, H_2^\dagger \,e^{\Vp}\,H_2 \Big]\\[8pt]
&+&
\int d^2\theta \,\,\Big[\,\tilde\mu\,\, (1+c_1\,S)\,\,H_1\,H_2
+ \,\frac{c_3}{M}
\,\,\,(1+c_2\, S)\,(H_1\,H_2)^2 \Big]+h.c.\nonumber\\[8pt]
&+&\!\!\!\!
\frac{1}{M}\!
\int\! d^4\theta\,\,\Big\{
A(S,S^\dagger)\,D^\alpha\,\Big[B(S,S^\dagger)\,H_2\,
e^{-\Vm}\,\Big]
D_\alpha\,\Big[\Gamma(S,S^\dagger)\,
e^{\Vm}\,H_1\,\Big]+h.c.\Big\}\nonumber
\eea

\bigskip\noindent
Additional spurion dependence arises from
the dimension-five  operators considered.
For the definitions of  $A(S,S^\dagger)$,
$B(S,S^\dagger)$, $\Gamma(S,S^\dagger)$ see eq.~(\ref{definitions3}).
After elimination of  the auxiliary fields and
a rescaling of scalar fields,
the scalar part of $\cL_1$ in (\ref{lls}) becomes:
\medskip
\bea\label{finalscalar}
\cL_{1,scalar}&=&
-\frac{1}{8}\,(g_1^2+g_2^2)\,
\big(\vert h_1\vert^2-\vert h_2\vert^2\big)^2
+\frac{m_0}{M}\,(g_1^2+g_2^2)\,\,\big(\vert h_1\vert^2-\vert
  h_2\vert^2\big)\,
\big(\delta_1\,h_1\,h_2\,+h.c.\big)\nonumber\\[8pt]
&+&
\frac{2\,c_3}{M}\,\,\big(\vert h_1\vert^2+\vert h_2\vert^2\big)
\big(\tilde\mu^*\,h_1\,h_2\,+h.c.\big)
-\frac{m_0}{M}\,\,c_3\, \big(\delta_2\, (h_1\,h_2)^2 +h.c.\big)
\\[8pt]
&-&\!
\big(\vert\tilde\mu\vert^2+ m_1^2\big) 
\vert h_1\vert^2\,
-\big(\vert\tilde\mu\vert^2+ m_2^2\big)
\vert h_2 \vert^2
-\big(h_1\,h_2
Bm_0\mu
+h.c.\big)
-h_1^*\cD^2\,h_1 - h_2^*\,\cD^2\,h_2\nonumber
\eea
where
\bea\label{m1m2}
m_1^2
&=& m_0^2\,\Big( \vert\,a_1\,\vert^2-
a_2\Big) +\cO(m_0/M)
\nonumber\\
m_2^2
&=&
m_0^2\,\Big( \vert\,b_1\,\vert^2-
b_2\Big)+\cO(m_0/M)
\nonumber\\
Bm_0\mu
&=&
\tilde\mu\,m_0\,\Big(c_1-a_1-b_1\Big)
+\cO(m_0/M)\label{Bmu}
\eea

\medskip\noindent
The $\cO(m_0/M)$ corrections in (\ref{m1m2})
 are not shown explicitly  since they only renormalise
 $m_{1,2}$ and $Bm_0\mu$ which are anyway unknown parameters
of the MSSM.
We  denoted
\bea\label{delta12only}
\delta_1& = & -\beta_1\,\alpha_0\,\gamma_0
+\gamma_1\,\beta_0\,\alpha_0\,
-\alpha_0\beta_0\gamma_0\,(a_1-b_1),
\quad\,\,
\delta_2= c_2+ 2 (a_1+b_1),
\eea
We notice  the presence
of three  contributions in the scalar potential, introduced by
our dimension-five operators. The
contributions proportional to $c_3$ are due to $(H_1 H_2)^2$ in (\ref{lls})
and were discussed in \cite{Dine} (also \cite{Blum:2008ym,Strumia,Berg:2009mq,Ham:2009gu}; 
 for a review  see \cite{Brignole:2003cm}). The one proportional to $\delta_1$
\bea
\big(\vert h_1\vert^2-\vert h_2\vert^2\big)\,
\big(h_1\,h_2+h.c.\big),
\eea
was introduced by the dimension-five operator in the last line of
(\ref{lls}). This is a new contribution to the scalar potential,
and  is vanishing if $\alpha_0=\beta_0=\gamma_0$.
An interesting feature is that its one loop contribution to $h_{1,2}$
self energy  remains soft (no quadratic divergences) despite
its higher dimensional origin.

\section{Higgs Mass Corrections Beyond MSSM}
\label{HiggsMassCorrectionsBeyondMSSM}

Let us consider the implications of
(\ref{finalscalar}) for the Higgs masses.
The scalar potential is
\medskip
\bea\label{scalarV}
V&=&\tilde m_1^2\,\vert h_1\,\vert^2
+\tilde m_2^2\,\vert h_2\,\vert^2
+\Big( \,Bm_0\mu \,h_1 \,h_2+h.c.\Big)
+\frac{g^2}{8}\,\Big(\vert \,h_1\,\vert^2-\vert\,
h_2\,\vert^2\Big)^2
\nonumber\\[8pt]
&+& \Big(\vert\,h_1\,\vert^2
-\vert\,h_2\,\vert^2\Big)\,\Big(\eta_1 \,h_1\,h_2+h.c.\Big)
+
\Big(\vert\,h_1\,\vert^2+\vert\,h_2\,\vert^2\Big)\,
\Big(\eta_2 \,h_1\,h_2+h.c.\Big)
\nonumber\\[8pt]
&+&
\frac{1}{2}\,\Big(\,\eta_3\,(h_1\,h_2)^2+h.c.\Big)
\eea
where the  definition  of
 $\eta_{1,2,3}\sim 1/M$ can be read
from eq.~(\ref{finalscalar}). We take for simplicity 
 $\eta_{i}$ real,  and  therefore
$\eta_3\geq 0$, $\vert\eta_2\vert\leq \eta_3/4$. 
Also
\bea
\tilde m_1^2&\equiv &m_1^2+\vert\tilde \mu\vert^2,\qquad\qquad
\tilde m_2^2\equiv m_2^2+\vert\,\tilde \mu\,\vert^2,\qquad\qquad
g^2\equiv g_1^2+g^2_2
\eea
Consider quantum fluctuations of $h_i$ around a vacuum expectation value
\bea
h_i=\frac{1}{\sqrt  2}\,(  v_i+\tilde h_i+i\tilde \sigma_i
),\qquad i=1,2
\eea
From the two minimum conditions for the scalar potential $V$ of eq.~(\ref{scalarV})
one can  express $\tilde m_{1,2}$ in terms of $Bm_0\mu$, $  v_1,   v_2$ to find:
\bea\label{tildemi}
\tilde m_1^2 &=& {-Bm_0\mu}\,\,\frac{  v_2}{  v_1}
-\frac{1}{8}\,g^2\,(  v_1^2-  v_2^2)
-\frac{\eta_1}{2}\,\frac{  v_2}{  v_1}\,
(3\,  v_1^2-  v_2^2)-\frac{\eta_2}{2}
\frac{  v_2}{  v_1}\,(3\,  v_1^2+  v_2^2)
-\frac{\eta_3}{2}\,{  v_2}^2 \nonumber\\[10pt]
\tilde m_2^2 &=&\!\!\!
 {-Bm_0\mu}\,\frac{  v_1}{  v_2}+\frac{1}{8}\,g^2({
   v_1}^2-{  v_2}^2)
-\frac{\eta_1}{2}\,\frac{  v_1}{  v_2}\,(
v_1^2-3\,  v_2^2)
-\frac{\eta_2}{2}\,\frac{  v_1}{  v_2}
\,(3\,  v_2^2+  v_1^2)-\frac{\eta_3}{2}\,  v_1^2\qquad
\eea
which shall be used in the following.
The mass matrix is
\bea\label{mij}
\cM_{ij}&=&
\frac{1}{2}\frac{\partial^2 V}{\partial h_i\partial
  h_j}\bigg\vert_{h_i=  v_i/\sqrt{2},\,\tilde\sigma_i=0}
 =X_{ij}+Z_{ij}
\eea
where
\bea
X_{ij}=\frac{1}{2}
\left(
\begin{array}{cc}
2\tilde m_1^2+\frac{1}{4} \,g^2\,(3 v_1^2-v_2^2) &  2 Bm_0\mu-\frac{1}{2}\,g^2
v_1 \,v_2\\[12pt]
2 \,Bm_0\mu-\frac{1}{2}\,g^2\,v_1 \,v_2  &  2\tilde m_2^2
-\frac{1}{4}\,g^2\,(v_1^2-3\,v_2^2)
\end{array}
\right)
\eea
and
\bea
Z_{ij}=\frac{1}{2}
\left(
\begin{array}{cc}
\!\!\!\!\!\!\!\!\!\!\!\!\!\!\!\!\!\!\!\!\!\!\!\!\!\!\!\!\!\!\!\!6\,(\eta_1+\eta_2)\,v_1\,v_2+\eta_3\,v_2^2  &
\!\!\!\!\!\!\!\!\!\!\!\!\!\!\!\!\!\!\!\!\!\!\!\!\!\!\!\!\!\!
3\,(\eta_1+\eta_2)\,v_1^2
+3 (\eta_2-\eta_1)\,v_2^2+2 \eta_3\,v_1\,v_2
\\[12pt]
3\,(\eta_1+\eta_2)\,v_1^2
+3 (\eta_2-\eta_1)\,v_2^2+2 \eta_3\,v_1\,v_2 &
6\,(\eta_2-\eta_1)\,v_1\,v_2+\eta_3\,v_1^2
\end{array}
\right)
\eea
The mass eigenvalues  $m_{h,H}^2$ of
$\cM_{ij}$ are
\bea\label{difference}
m_{h, H}^2&=&M_{h, H}^{2}
\mp
 \frac{6\eta_1}{\sqrt w}
\bigg[Bm_0\mu\,(  v_1^2-  v_2^2)+  v_1   v_2
\Big(\tilde m_1^2-\tilde
  m_2^2+\frac{g^2}{4}(  v_1^2-  v_2^2) \Big)\bigg]
\nonumber\\[10pt]
&+&3\eta_2\,\bigg[  v_1   v_2
\pm\frac{1}{2\sqrt w}(  v_1^2+  v_2^2)(-4Bm_0\mu
+ g^2 \,   v_1   v_2)\bigg]
\nonumber\\[10pt]
&+&
\frac{\eta_3}{4}\bigg[
  v_1^2+  v_2^2
\pm \frac{1}{\sqrt w}
\Big(2 (\tilde m_1^2-\tilde m_2^2)(  v_1^2-
v_2^2)+g^2(  v_1^2+  v_2^2)^2\nonumber
\\
&-&16Bm_0\mu   v_1 v_2\Big)\bigg]
\eea
where upper (lower) signs correspond to the lighter $ m_h^2$
(heavier $ m_H^2$) Higgs field and $M_{h,H}^2$ expresses the pure MSSM part:
\bea
M_{h,H}^{2}\equiv
\frac{1}{2}\bigg[\tilde m_1^2+\tilde
  m_2^2+
\frac{g^2}{4}\,(  v_1^2+  v_2^2)\mp \frac{1}{2}\sqrt w\bigg]
\eea
Also,
\bea\label{rw}
w& \equiv &
(4 Bm_0\mu-g^2   v_1   v_2)^2+4\Big(\tilde m_1^2-\tilde m_2^2+
\frac{g^2}{2}(  v_1^2-  v_2^2)\Big)^2 
\eea
With the values of $\tilde m_{1,2}$ expressed in terms of
$v_{1,2}$ and $Bm_0\mu$ from minimum conditions (\ref{tildemi}),
one can express  $ m_{h,H}^2$ of (\ref{difference})
as follows
\bea\label{mhH11}
 m_{h, H}^2&=&\frac{m_Z^2}{2}
-\frac{Bm_0\mu (u^2+1)}{2\,u} \mp\frac{\sqrt w'}{2}
+{  v}^2\,\Big[\eta_1\,\,q_1^{\pm}
+\,\eta_2\,\,q_2^{\pm}
+\,\eta_3\,\,q_3^{\pm}\Big]
\eea
with
\bea
\! q_1^\pm & = & \frac{u^2-1}{4\,u}\pm
\frac{(u^2-1)}{4 u^2(1+u^2)^2\sqrt w'}\,\Big[
m_Z^2\,u (1-6 u^2+u^4)+ Bm_0\mu\,(1+u^2)(1+18 u^2+u^4)\Big]
\nonumber\\[8pt]
q_2^\pm & = &\!\!\!
- \frac{1-6u^2+u^4}{4\,u\,(1+u^2)}
\mp \frac{m_Z^2 u(1-14 u^2+u^4)+Bm_0\mu (1+u^2) (1+10 u^2 +u^4)}{4 \, u^2\,(1+u^2)\sqrt w'}
\nonumber\\[8pt]
q_3^\pm & = &
\mp \frac{2 u}{(1+u^2)^2\sqrt
  w'}\,\Big[Bm_0\mu(1+u^2)-m_Z^2\,u\Big]
\eea
where
\bea
w'\equiv m_Z^4 + \big[Bm_0\mu (1+u^2)^3+2 m_Z^2 u(1-6 u^2 +u^4)
\big]\frac{Bm_0\mu}{u^2(1+u^2)}
\eea
and where we also used $  v_1=  v\cos\beta,   v_2=  v\,\sin\beta$,
$u=\tan\beta$, $m_Z^2=g^2\,{  v}^2/4$ and $Bm_0\mu<0$.

Similar considerations apply for the pseudoscalar Higgs/Goldstone boson sector. The mass matrix in this case is
\medskip
\bea\label{nij}
N_{ij}&=&\frac{\partial^2 V}{\partial
  \tilde\sigma_i\partial\tilde\sigma_j}\bigg\vert_{
{ h_i=  v_i}/{\sqrt 2},\,\tilde\sigma_i=0}
\eea
with entries
\bea
N_{11}&=&
\tilde m_1^2+\frac{g^2}{8}\,(  v_1^2-  v_2^2)
+(\eta_1+\eta_2)  v_1  v_2 -\frac{\eta_3}{2}  v_2^2
\nonumber\\[8pt]
N_{12}&=&
-\frac{\eta_1}{2}(  v_1^2-  v_2^2)-\frac{\eta_2}{2}
(  v_1^2+  v_2^2)-\eta_3  v_1  v_2-Re(Bm_0\mu)
\nonumber\\[8pt]
N_{22}&=&
\tilde m_2^2-\frac{g^2}{8}\,(  v_1^2-  v_2^2)+
(\eta_2-\eta_1)  v_1  v_2 -\frac{\eta_3}{2}  v_1^2
\eea
The eigenvalues of $N$ are
\medskip
\bea\label{mga}
m_{G, A}^2&=&
\frac{1}{2}\,\big(\tilde m_1^2+\tilde m_2^2)\mp
\frac{1}{8}\sqrt \kappa
\nonumber\\[8pt]
&\mp&
\frac{4 \eta_1}{\sqrt \kappa}\,
\Big[  Bm_0\mu (  v_1^2-  v_2^2)
+
   v_1\,    v_2\,  \Big(\tilde  m_1^2 - \tilde  m_2^2 +
\frac{g^2}{4} (  v_1^2 -   v_2^2)\Big)\Big]\nonumber
\\
&+&
\eta_2\,\Big[  v_1  v_2\mp \frac{4  Bm_0\mu}{\sqrt \kappa}
\,(  v_1^2+  v_2^2)\Big]
\nonumber\\[8pt]
&+&
\eta_3\,\Big[
-\frac{1}{4}\,(  v_1^2+  v_2^2)
\mp\frac{1}{\sqrt \kappa }\Big(
8 Bm_0\mu   v_1  v_2\, + (  v_1^2-  v_2^2)(\tilde
m_1^2- \tilde m_2^2)+\frac{g^2}{4} \, (  v_1^2-  v_2^2)^2\Big)
\Big]\nonumber
\\&&
\eea
with
\bea
\kappa=
16\Big[4 (Bm_0\mu)^2 + \Big(\tilde m_1^2 - \tilde m_2^2 +
  \frac{g^2}{4}\,(  v_1^2-  v_2^2)\Big)^2\Big]
\eea
where the upper sign corresponds to the Goldstone $m_G$  and
the lower sign to $m_A^2$.
One can
use (\ref{tildemi}) to
replace $\tilde m_{1,2}$ in terms of  $  v_{1,2}$ and $m_A$ .
Using (\ref{tildemi}) one shows  that  $m_G=0$ and
\medskip
\bea\label{ma11}
m_A^2&=& -\frac{  v_1^2+  v_2^2}{2   v_1   v_2}
\,\Big[\,2 \,Bm_0\mu+
\eta_1\,(  v_1^2-  v_2^2)
+
\eta_2\,(  v_1^2+  v_2^2)
+
2 \eta_3 \,  v_1\,  v_2\Big]
\nonumber\\[10pt]
&=&
-
\frac{1+u^2}{u}\,Bm_0\mu
+
\, \frac{u^2-1}{2 \,u}\, \eta_1 \,v^2\,
-
\,\frac{1+u^2}{2\,u}\,\eta_2\,v^2\,
-
\eta_3\,v^2
\eea
By eliminating $Bm_0\mu$ between (\ref{mhH11}) and (\ref{ma11}),  one obtains the masses
$m_{h,H}$:
\medskip
\bea\label{finalmh}
m_{h,H}^2&=&
\frac{1}{2}\Big[m_A^2+m_Z^2\mp\sqrt{w''}\Big]
\mp
\frac{4\,m_A^2\,\eta_1\,u\,(u^2-1)\,{  v}^2}{(1+u^2)^2\,\sqrt{w''}}
\nonumber\\[10pt]
&+&
\frac{2\eta_2\,u\,{  v}^2}{1+u^2}
\,\bigg[1\pm\frac{m_A^2+m_Z^2}{\sqrt{w''}}\bigg]+
\frac{\eta_3\,{  v}^2}{2}\,\bigg[1\mp
\frac{(m_A^2-m_Z^2)\,(u^2-1)^2}{\sqrt{w''}\,\,\,(1+u^2)^2\,}\bigg]
\eea
where the upper (lower) signs correspond to $h$ ($H$) respectively
and
\bea\
w'' & \equiv &
 m_A^4+m_Z^4-2\,m_A^2\,m_Z^2\,\frac{1-6u^2+u^4}{(1+u^2)^2}
=
(m_A^2+m_Z^2)^2-4\,m_A^2\,m_Z^2\,\cos^2 2\beta\qquad
\eea
Replacing $u=\tan\beta$ in  $m_{h,H}$ one obtains
an equivalent form of $m_{h,H}$
\medskip
\bea\label{mh2}
 m_{h,H}^2&=&
\frac{1}{2}\Big[m_A^2+m_Z^2\mp\sqrt{w''}\Big]
\pm
\,\eta_1\,{  v}^2\,\sin 4\beta\,\,\frac{m_A^2}{\sqrt{w''}}
\nonumber\\[10pt]
&+&
{\eta_2\,{  v}^2\,\sin 2\beta}
\,\bigg[1\pm\frac{m_A^2+m_Z^2}{\sqrt{w''}}\bigg]+
\frac{\eta_3\,{  v}^2}{2}\,\bigg[1\mp
\frac{(m_A^2-m_Z^2)\,\cos^2 2\beta}{\sqrt{w''}}\bigg]
\eea
For $\eta_2=\eta_3=0$ one finds from (\ref{mh2}):
\bea
m_h^2 +  m_H^2 = m_A^2+m_Z^2
\eea
which is independent of $\eta_1$. Then $\eta_1$ does not
affect the relation among physical masses which  is
consistent with  the result of section~\ref{MSSM5HiggsSector},
where the last term in (\ref{lls}), responsible
 for the $\eta_1$ term in $V$, could be removed
by a suitable field redefinition.

In  the limit of large $\tan\beta$ with $m_A$ fixed at a value $m_A>m_Z$ one finds:
\medskip
\bea\label{rr1}
m_h^2&=&m_Z^2+\frac{4 m_A^2 \,{  v}^2}{m_A^2-m_Z^2}\,(\eta_2-\eta_1)
\,\cot  \beta\nonumber\\[10pt]
&-&\frac{4\,m_A^2\,m_Z^2}{m_A^2-m_Z^2}
\,\bigg[
1-\eta_3 \,{  v}^2\,\frac{m_A^4+m_Z^4}{2\,m_A^2\,m_Z^2 \,(m_A^2-m_Z^2)}
\bigg]\,\cot^2\beta+\cO(\cot^3\beta)
\eea
and
\bea\label{rr2}
m_H^2&=& m_A^2+\eta_3\,{  v}^2 +\frac{4
  \,(m_A^2\,\eta_1-m_Z^2\,\eta_2)\,{  v}^2}{m_A^2-m_Z^2}\,\cot\beta
\nonumber\\[10pt]
&+&
\frac{4\,m_A^2\,m_Z^2}{m_A^2-m_Z^2}\,\bigg[1-\eta_3\,{  v}^2
  \,\frac{m_A^4+m_Z^4}{2 \,m_A^2\,m_Z^2\,(m_A^2-m_Z^2)}
\bigg]\,\cot^2\beta+\cO(\cot^3\beta)
\eea
Therefore
\bea\label{rr3}
\delta
m_h^2&=&\frac{4\,m_A^2\,{  v}^2}{m_A^2-m_Z^2}\,(\eta_2-\eta_1)\,\cot\beta
+\cO(\cot^2\beta)\nonumber\\[10pt]
\delta m_H^2&=&
\eta_3\,{  v}^2 +\frac{4
  \,(m_A^2\,\eta_1-m_Z^2\,\eta_2)\,{  v}^2}{m_A^2-m_Z^2}\,\cot\beta
+\cO(\cot^2\beta)
\eea

\medskip\noindent
in agreement with \cite{Dine} for  $\eta_1=0$.
The above expansions for large $\tan\beta$ should be regarded with due
care since they are the result of a double series expansion
in $\eta_i$ and $1/\tan\beta$.
Assuming $\eta_3=0$ (then $\eta_2=0$, too), 
the  term proportional to $\cot\beta$ in (\ref{rr1})
is larger than the sub-leading one ($\cot^2\beta$),
giving $ m_h^2-m_Z^2>0$ if $\vert\eta_1/g^2\vert\geq
1/(4\tan\beta)$. 
This bound is however outside the validity of the perturbative
 expansion  in $\eta_{1}$ as we shall see shortly and then the
large $\tan\beta$ expansion is not useful.
If $\eta_{1,2}=0$ and  $\eta_3>0$ one could obtain $m_h > m_Z$ if also the square bracket in (\ref{rr1}) is negative, which is more easily satisfied (for a small
$\eta_3$) if $m_A$ is very close to $m_Z$, but then 
 the above large  $\tan\beta$ expansion is not reliable.

Let us therefore analyse  the validity of the corrections to
$m^2_{h,H}$ from eq.~(\ref{mh2}) in the approximation used. 
For our perturbative  expansion in $\eta_i$ to be accurate
we  require that the $\eta_{i}$-dependent
entries in the mass matrix
 $\cM_{ij}$  (\ref{mij}) be much smaller than the
corresponding values of  these matrix elements
 in the  MSSM case. From this condition one finds
\medskip
\bea\label{cond3}
&&\Big\vert
\, 3\, (\eta_1+\eta_2)\,  v_1^2+3 \,(\eta_2-\eta_1)\,  v_2^2
+2\eta_3\,  v_1\,  v_2\,
\Big\vert\ll
\frac{1}{2}\,g^2\,
  v_1\,v_2 
\nonumber\\[3pt]
&&
\Big\vert
\, 6 \,(\eta_2-\eta_1)\,  v_1\,  v_2+\eta_3\,  v_1^2
\,\Big\vert \ll 
\frac{1}{4}\,g^2\,\Big\vert\,
v_1^2-3 \,  v_2^2 \Big\vert
\nonumber\\[3pt]
&&
\Big\vert
\,6 \,(\eta_2+\eta_1)\,  v_1\,  v_2+\eta_3\,  v_2^2
\,\Big\vert \ll 
\frac{1}{4}\,g^2\,\,\Big\vert\,
3\,  v_1^2-  v_2^2 \Big\vert
\eea

\medskip\noindent
Similar conditions are derived
from the pseudoscalar Higgs mass matrix 
elements $N_{ij}$ (\ref{nij}). 
One may find this condition too restrictive;
in principle it may not be necessary to impose the
leading $\eta_i\sim \cO(1/M)$ contribution to the mass matrix entries 
 be suppressed relative to the MSSM  
zeroth order and that one should instead ask that the $\cO(1/M)$
correction  dominate over the
higher order terms $\cO(1/M^2)$ \cite{private}. However, at the quantitative level
this leads, for the present case,
  to results  which are similar or even stronger
(for example for $\eta_3$) than those derived here from comparing
the MSSM zeroth order against the $\cO(1/M)$ terms.
From these one can obtain upper bounds  for each $\eta_i$. Having imposed these bounds, we can examine if the dimension-five operators bring a significant contribution to the higgs mass and in particular if we can surpass the tree level bound $m_h \leq m_Z$.

That would mean to also impose some lower bounds in order to achieve the desired increase. In the approximation considered, these bounds are derived from (\ref{finalmh}) with (\ref{cond3}) and give
\medskip
\bea\label{ssx3}
&&\qquad \frac{(\sqrt\omega+1-\rho)\,(1+u^2)^2\,\sqrt\omega}{
32 \,u\,(u^2-1)}
\leq  -\frac{\eta_1}{g^2}
\ll
\min\bigg\{\frac{u}{6(u^2-1)}, \frac{3\,u^2-1}{24 u},
\frac{\vert u^2-3\vert }{24\,u}
\bigg\}\nonumber\\[8pt]
&&\frac{(\sqrt\omega+1-\rho)\,(1+u^2)^2\,\sqrt\omega}{
4 \,[(1+u^2)^2 \sqrt \omega-(\rho-1)(1-u^2)^2]}
\leq \frac{\eta_3}{g^2}
\ll
\min\bigg\{\frac{1}{4},
\frac{\vert  u^2-3\vert}{4\,u^2},
\frac{u^2-1}{4\,u^2},\frac{u^2-1}{4}
\bigg\}\qquad\qquad
\eea
\medskip
\noindent
with $\omega\equiv (\rho-1)^2+16 u^2\rho/(1+u^2)^2$ and $\rho\equiv m_A^2/m_Z^2$.

Assuming $\eta_2=0$, then $m_h>m_Z$ is possible if one or both
eqs in  (\ref{ssx3}) are  respected.
On the other hand, it has no solution for $\eta_{1}$ within $1\leq   \tan\beta\leq 50$ and $ m_A/m_Z \geq 1$; $\eta_1$ alone  cannot change the MSSM bound $m_h\leq m_Z$ within our approximation.
If  $1\leq m_A^2/m_Z^2 \leq 2.43$ there is
a  somewhat ``marginal''  solution for $\eta_3$,
with $m_A/m_Z$ close to unity and large $\tan\beta$ preferred, to enforce the 
``$\ll$'' inequalities in (\ref{cond3}) and (\ref{ssx3}). For example,
if $m_A=m_Z$  and $\tan\beta=50$, the lower  bound on $\eta_3/g^2$ is
$\eta_3/g^2\geq 0.02$ while  $\eta_3/g^2\ll 0.25$ is also required.
In this case, for
$\tan\beta=50$ the increase of $m_h^2$ relative to $m_Z^2$,
 $\delta_r=(m_h^2-m_Z^2)/m_Z^2$
equals $\delta_r=-100/2501+2\,\eta_3/g^2$. Therefore
 $\delta_r=12\%$ or $m_h\approx 102$ GeV
 if $\eta_3/g^2=0.08$, corresponding to $\eta_3=4.4
\times 10^{-2}$. Larger values for $m_h$ 
should  be regarded  with care, since they would correspond to
cases when  ``$\ll$" of (\ref{ssx3}) is not
comfortably respected; if $\eta_3/g^2\approx 0.04$ then
 $\delta_r\approx 4\%$ or $m_h\approx 95$ GeV.
Further, if we now increase $m_A$ even by a small amount relative to $m_Z$,
 $m_A^2=1.5\, m_Z^2$\,  and $\tan\beta=50$, the lower  bound on
$\eta_3/g^2$ is  $0.118$ which is difficult to 
comply by a good margin 
with an upper bound unchanged at $\eta_3/g^2\ll 0.25$.
Even so, the relative difference would be only $\delta_r=2\times 10^{-3}\%$,
($\eta_3/g^2=0.118$), therefore the increase of $m_h$ is negligible.
So far we took $\eta_2=0$. If we allow a non-zero value for $\eta_2$,
which also requires non-zero $\eta_3$,
their combined effect on increasing $m_h$
is not larger and the above results remain valid.
Note also that for large $\tan\beta$ regions
 $1/M^2$-suppressed operators can be important 
and can affect the results \cite{Dine}.

From this analysis we see that  $\eta_{1}$ alone cannot
change   the MSSM tree level bound $m_h\!\leq\! m_Z$
within the approximation we discuss.
This  is consistent with  section~\ref{MSSM5HiggsSector}, where it was shown that
the operator  which induced the $\eta_1$ term could be removed
by a general field redefinition of suitable 
coefficients\footnote{To see this
 one can also start from (\ref{lls}) and perform
a ``smaller'' version of redefinition (\ref{tra}), with $\rho_F\!=\!0$.}.
However, $\eta_3$ can increase $m_h$ to values 
$\approx 95-100$ GeV  if $m_A\approx m_Z$, with the higher values close
to the limit of our approximation.
Therefore it is  the susy breaking term
associated to $(H_1\,H_2)^2$ that could
relax the MSSM tree level bound.
This increase brings a  small improvement. 
To conclude, adding the quantum corrections
is still needed \cite{Dine} to bring $m_h$ above the 
LEP II bound of 114 GeV \cite{higgsboundLEP}.

These findings show that the MSSM Higgs sector is rather
stable under the addition of dimension-five operators, in the approximation
 we considered (expansion in $1/M$) of integrating out a
massive singlet or a pair of massive $SU(2)$ doublets which generated the $\eta_{1,2,3}$ contributions. If $M$ is low enough, the approximation used by
integrating out these massive fields becomes
unreliable, and one should recompute the full spectrum keeping all fields dynamical. Then the quartic interactions that the initial massive fields brought can be larger or of similar order to their  MSSM counterparts and in principle they can change the above conclusions.

\section{Including Loop Corrections}\label{OneLoopWithMassDimensionFiveOperators}

It is worth mentioning the value of $m_h$ in the presence of one loop corrections from top - stop and dimension five operators \cite{Cassel:2009ps}, mentioned in the text:
\begin{eqnarray} \label{mh1loop5}
m_{h}^{2}\!\!\! &=&\!\!\!\frac{1}{2}\Big[m_{A}^{'\, 2}+m_{Z}^{2}
-\sqrt{{\tilde w}^{'}}+\xi \Big]  
\nonumber \\
&\!\!\!+&\!\!\!\!\!
{(2\zeta_{10}\mu_0) {\ v}^{2}\sin 2\beta }
\bigg[1+\frac{m_{A}^{'\, 2}+m_{Z}^{2}}{
\sqrt{\tilde w^{'}}}\bigg]+\frac{(-2\,\zeta_{11}\,m_0)\,{\ v}^{2}}{2}\,\bigg[1-
\frac{(m_{A}^{'\,2}-m_{Z}^{2})\,\cos ^{2}2\beta }{\sqrt{\tilde w^{'}}}\bigg] 
\qquad
\end{eqnarray} 
where 
\begin{eqnarray} 
{\tilde w}^{'}
 &\equiv &[(m_{A}^{'\,2}-m_{Z}^{2})\,\cos 2\beta +\xi ]^{2}+\sin ^{2}2\beta 
\,(m_{A}^{'\,2}+m_{Z}^{2})^{2}  \nonumber \\[5pt]
m_{A}^{'\,2} &=&\tilde{m}_{1}^{2}+\tilde{m}_{2}^{2}+\xi /2+
(2\,\zeta_{10}\mu_0)\,v^{2}\,\sin 2\beta +\zeta _{11}\,m_0\,v^{2};\quad 
\xi \equiv \delta\, m_{Z}^{2}\,\sin ^{2}\beta
\end{eqnarray}

\medskip\noindent
where $\delta$ is the one-loop correction from top-stop Yukawa
sector to $\lambda_2^0$ of (\ref{ms}) which changes according 
to $\lambda_2^0\ra
\lambda_2^0\,(1+\delta)$ where \cite{Giudice:2006sn,Carena:1995bx}
\medskip  
\begin{eqnarray} 
\delta  &=&\frac{3\,h_{t}^{4}}{g^{2}\,\pi ^{2}\,}\bigg[\ln \frac{M_{\tilde{t}
}}{m_{t}}+\frac{X_{t}}{4}+\frac{1}{32\pi ^{2}}\,\Big(3\,h_{t}^{2}-16
\,g_{3}^{2}\Big)\Big(X_{t}+2\ln \frac{M_{\tilde{t}}}{m_{t}}\Big)\ln \frac{M_{
\tilde{t}}}{m_{t}}\bigg],  \nonumber\\[6pt]
X_{t} &\equiv &\frac{2\,(A_{t}\,m_{0}-\mu \cot \beta )^{2}}{M_{\tilde{t}}^{2}
}\,\,\Big[1-\frac{(A_{t}\,m_{0}-\mu \cot \beta )^{2}}{12\,\,M_{\tilde{t}}^{2} 
}\,\Big]. 
\end{eqnarray}
with $M_{\tilde{t}}^{2}\equiv m_{\tilde{t}_{1}}\,m_{\tilde{t}_{2}}$, 
and $g_{3}$ the QCD coupling. 
The combined effect of $d=5$ operators and top Yukawa coupling $h_t$
is that $m_h$ can reach values of $130$ GeV for $\tan\beta\leq 7$ 
with a small fine-tuning \cite{Barbieri:1998uv,Chankowski:1998xv,Chankowski:1997zh,Kane:1998im} $\Delta\leq 10$ \cite{Cassel:2009ps}
and with the supersymmetric coefficient 
$\zeta_{10}$ giving a larger effect than the non supersymmetric one, 
$\zeta_{11}$. Even for a modest increase of $m_h$ from $d=5$ operators
alone  of order $\cO(10 \mathrm{GeV})$, their impact on the effective
quartic coupling of the Higgs field is  significant (due to the small value of 
the MSSM gauge couplings), and this explains the reduction
of fine-tuning by the effective operators \cite{Batra:2003nj,r3}.


\chapter{MSSM Higgs with Operators of Mass Dimension 5 and 6}
\label{MSSMHiggs56}

We generally expect that corrections to observables from higher order operators will be subdominant to those from the leading, dimension five ones. Nevertheless, we saw in section \ref{HiggsMassCorrectionsBeyondMSSM} that in the limit of large $\tan\beta$, the correction to the mass of the Higgs due to mass dimension five operators is $\tan\beta$ suppressed. In that limit, corrections from dimension six operators can become comparable to dimension five since $1/M^2 \sim 1/ (M\tan\beta)$. Therefore, in order to complete the study of the leading Higgs mass corrections from effective operators, we need to include the contribution from dimension six operators. Since the latter is not $\tan\beta$ suppressed, the sequence ends here, as dimension seven or further will always be subdominant.

\section{The Relevant Operators}
\label{RelevantOperators}

We focus on the Higgs sector of the complete Lagrangian \cite{Carena:2009gx}. This is comprised of the MSSM higgs sector $\cL_0$ and the complete set of mass dimension-five and six operators. For $\cL_0$  we have
\medskip\bea\label{lo}
\cL_0=\int d^4\theta \,\,\sum_{i=1,2}
\cZ_i(S,S^\dagger)\,H_i^\dagger\,e^{V_i}\,H_i
+\bigg\{\int d^2\theta \,\,\mu\,(1+B\,m_0
\,\,\theta\theta)\,H_1\cdot H_2+h.c.\bigg\}
\eea
in standard notation. Here $\cZ_{i}(S,S^\dagger)=1-c_{i}\,m_0^2\,\theta\theta\overline\theta\overline\theta$ with $i=1,2$, $c_i=\cO(1)$ and $m_0$ is the supersymmetry breaking scale as presented in the previous chapter.

We extend this Lagrangian by higher dimensional operators. In dimension-five we have the usual contributions studied in the previous chapter:
\medskip  
\begin{eqnarray} 
\mathcal{L}_{1} &=&\!\!\frac{1}{M}\int d^{2}\theta \,
\,\zeta(S)\,(H_{2}\cdot H_{1})^{2}\!+\!h.c.
\nonumber
\\
&=&2\,\zeta_{10}\,(h_2\cdot h_1)(h_2\cdot F_1+F_2\cdot h_1)+\zeta_{11}\,m_0\,(h_2\cdot h_1)^2+h.c, 
\nonumber\\
\mathcal{L}_{2} &=&\!\!\frac{1}{M}\int d^{4}\theta \,\,\Big\{%
\,A(S,S^{\dagger })D^{\alpha }\Big[B(S,S^{\dagger })\,H_{2}\,e^{-V_{1}}\Big]%
D_{\alpha }\Big[\Gamma(S,S^{\dagger })\,e^{V_{1}}\,H_{1}\Big] 
+h.c.\Big\} \label{dimensionfive}
\end{eqnarray}
where\footnote{We switch to a notation best suited for the analysis here. The dictionary is: $\eta_2\!=\!2\zeta_{10}\mu^*$, $\eta_3\!=\!-2\,m_0 \zeta_{11}$. With respect to the literature: In \cite{Cassel:2009ps}
$\eta_2\!\ra\! \zeta_1$, $\eta_3\!\ra\! \zeta_2$ and in \cite{Dine}
$\eta_2\!\ra\! 2\epsilon_{1r}$, $\eta_3\!\ra\! 2\epsilon_{2r}$ .}
\bea
\frac{1}{M}\,\zeta(S)
= \zeta_{10}+ \zeta_{11}\,m_0\,\theta\theta,\,\,
\qquad \zeta_{10},\, \zeta_{11}\sim 1/M,
\eea
with $S=\theta\theta m_0$ the spurion superfield.
We assume that 
\bea
m_0\ll M
\eea
so that the  effective theory approach is reliable.

$\cL_2$ is eliminated by generalised, spurion-dependent field redefinitions as it was  shown in detail in the previous chapter. For this reason we keep only $\cL_1$ for the discussion below. These redefinitions  bring however a renormalisation
of the usual MSSM soft terms and of the $\mu$ term as well as additional corrections of order $1/M^2$. Since in the following we will write down and study the full set of $d=6$ operators, the latter will be automatically included.

The list of $d=6$ operators is \cite{Piriz:1997id}
\bea
\mathcal{O}_{j} &=&
\frac{1}{M^{2}}\int d^{4}\theta \,\,
Z_{j}(S,S^{\dagger })\,\,
(H_{j}^{\dagger }\,e^{V_{j}}\,H_{j})^{2}, \quad j\equiv 1,2. 
\nonumber\\[-2pt]
\mathcal{O}_{3} &=&
\frac{1}{M^{2}}\int d^{4}\theta \,\,Z _{3}(S,S^{\dagger
})\,\,
(H_{1}^{\dagger 
}\,e^{V_{1}}\,H_{1})\,(H_{2}^{\dagger }\,e^{V_{2}}\,H_{2}),
\nonumber\\
\mathcal{O}_{4} &=&\frac{1}{M^{2}}\int d^{4}\theta \,\,
Z_{4}(S,S^{\dagger })\,\,(H_{2}.\,H_{1})\,(H_{2}.\,H_{1})^{\dagger },
\nonumber\\
\mathcal{O}_{5} &=&\frac{1}{M^{2}}\int d^{4}\theta \,\, Z%
_{5}(S,S^{\dagger })\,\,(H_{1}^{\dagger 
}\,e^{V_{1}}\,H_{1})\,\,H_{2}.\,H_{1}+h.c.
\nonumber\\
\mathcal{O}_{6} &=&
\frac{1}{M^{2}}\int d^{4}\theta \,\,Z_{6}(S,S^{\dagger
})\,\,
(H_{2}^{\dagger }\,e^{V_{2}}\,H_{2})\,\,H_{2}.\,H_{1}+h.c. 
\nonumber\\
\mathcal{O}_{7} &=&\frac{1}{M^{2}}\int d^{2}\theta \,\,
Z_{7}(S,0)\,\frac{1}{16\,g^2\,\kappa}\,{\rm Tr}\,
W^{\alpha }\,W_{\alpha }\,(H_{2}\,H_{1})+h.c.
\nonumber\\
\mathcal{O}_{8} &=&\frac{1}{M^{2}}\int d^{4}\theta
 \,\,\Big[Z_{8}(S,S^{\dagger
   })\,\,(H_{2}\,H_{1})^{2}+h.c.\Big]
\label{operators18}\eea

\bigskip\noindent
where $W^\alpha=(-1/4)\,\overline D^2 e^{-V} D^\alpha\, e^V$
is the chiral field strength
of $SU(2)_L$ or $U(1)_Y$ vector superfields $V_w$ and $V_Y$ respectively.
 Also  $V_{1,2}=V_w^a
(\sigma^a/2)\mp 1/2\,V_Y$ with the upper sign for $V_1$. The remaining $d=6$ operators are:
\bea
\mathcal{O}_{9} &=&\frac{1}{M^{2}}\int d^{4}\theta \,\,
Z_{9}(S,S^{\dagger })\,\,H_{1}^{\dagger }\,
\overline{\nabla}^{2}\,e^{V_{1}}\,\nabla ^{2}\,H_{1}  \nonumber \\[-2pt]
\mathcal{O}_{10} &=&\frac{1}{M^{2}}\int d^{4}\theta \,\,
Z_{10}(S,S^{\dagger })\,\,H_{2}^{\dagger }\,\overline{\nabla } 
^{2}\,e^{V_{2}}\,\nabla ^{2}\,H_{2}  \nonumber \\
\mathcal{O}_{11} &=&\frac{1}{M^{2}}\int d^{4}\theta \,\,Z
_{11}(S,S^{\dagger })\,\,H_{1}^{\dagger }\,e^{V_{1}}\,\nabla ^{\alpha 
}\,W_{\alpha }^{(1)}\,H_{1}  \nonumber \\
\mathcal{O}_{12} &=&\frac{1}{M^{2}}\int d^{4}\theta \,\,Z 
_{12}(S,S^{\dagger })\,\,H_{2}^{\dagger }\,e^{V_{2}}\,\nabla ^{\alpha 
}\,W_{\alpha }^{(2)}\,H_{2}  \label{der0} 
 \nonumber \\ 
\mathcal{O}_{13} &=&\frac{1}{M^{2}}\int d^{4}\theta \,\,Z
 _{13}(S,S^{\dagger })\,\,H_{1}^{\dagger }\,e^{V_{1}}\,
 \,W_{\alpha }^{(1)}\,\nabla^\alpha\,H_{1}  \nonumber \\ 
\mathcal{O}_{14} &=&\frac{1}{M^{2}}\int d^{4}\theta \,\,Z 
 _{14}(S,S^{\dagger })\,\,H_{2}^{\dagger }\,e^{V_{2}}\,
 \,W_{\alpha }^{(2)}\,\nabla^\alpha\,H_{2}  \label{der} 
\eea
Also
$\nabla_{\alpha }\,H_{i}=e^{-V_{i}}\,D_{\alpha }\,e^{V_{i}}
 H_i$ and  $W_\alpha^{(i)}$ is the field strength of $V_i$.
In the most generic case, the above operators should actually include spurion dependence of arbitrary coefficients under any $\nabla_\alpha$, in order to include supersymmetry breaking  effects associated to them. 
The wavefunction coefficients introduced above have the structure
\bea
\frac{1}{M^2}\,Z_i(S,S^\dagger)=\alpha_{i0}
+\alpha_{i1}\,m_0\,\theta\theta
+\alpha_{i1}^*\,m_0\,\overline\theta\overline\theta
+\alpha_{i2}\,m_0^2\,\theta\theta\overline\theta\overline\theta,\qquad
\alpha_{ij}\sim 1/M^2.
\eea
Regarding the origin of these operators: $\cO_{1,2,3}$ can be
 generated in MSSM with an additional,
 massive $U(1)'$ gauge boson or $SU(2)$ triplets integrated out \cite{Dine}.
$\cO_4$ can be generated by a massive gauge singlet or $SU(2)$
 triplet while $\cO_{5,6}$ can be generated by a combination of $SU(2)$ doublets
and massive gauge singlet. $\cO_7$ is essentially a 
threshold correction to the gauge coupling with a 
moduli field replaced by the Higgs.  $\cO_8$ 
exists only in broken supersymmetry but is generated when redefining away
the $d=5$ derivative operator, thus we keep it.

Let us consider for a moment the operators $\cO_{9,...14}$ in
the exact supersymmetry case. We can use the equations of motion to set some of them on shell\footnote{
Superpotential convention: $\int d^2\theta\mu\,H_1.H_2=
\int d^2\theta\,\mu
 \,H_1^T\, (i\sigma_2)\,H_2\equiv \int d^2\theta\, 
 \mu\,\epsilon^{ij}\,H_1^i\,H_2^j$;
 \,\,$\epsilon^{12}=1=-\epsilon^{21}$.}:
\medskip
\bea\label{hdo2}
&& -\frac{1}{4}\,\overline D^2\,(H_2^\dagger \,e^{V_2})
+\mu\,H_1^T\,(i\sigma_2)=0,\qquad
\frac{1}{4}\,\overline D^2\,(H_1^\dagger \,e^{V_1})
+\mu\,H_2^T (i\sigma_2)=0
\eea

\medskip\noindent
We find that in the supersymmetric case\footnote{Also using $(i\sigma_2)\,e^{-\Lambda}=
e^{\Lambda^T}\,(i\sigma_2)$;\, $\Lambda\equiv \Lambda^a\,T^a$;\,
 $(i\sigma_2)^T =-( i\sigma_2)$;\, $(i\sigma_2)^2=-1_2$}:
\medskip
\bea\label{hdo1}
\cO_{9}\sim \int d^4\theta\,\,
H_1^\dagger \,\overline \nabla^2\, e^{V_1}\,\nabla^2\,H_1
=16\,\vert\mu\vert^2\,\int d^4\theta\,H_1^\dagger\,e^{V_1}\,H_1
\eea

\medskip\noindent
and similar for $\cO_{10}$. 
Regarding $\cO_{11,12}$, they vanish in the  supersymmetric
case, following the definition of $\nabla^\alpha$ and
an integration by parts. Furthermore, $\cO_{13,14}$ are similar to $\cO_{9,10}$
which can be seen by using the definition of $W_\alpha^{(i)}$ 
and the relation between  $\nabla^2$, ($\overline\nabla^2$) and
$D^2$, ($\overline D^2$). 

Summarizing, in the exact supersymmetry case the operators $\cO_{9...14}$ give at 
most wavefunction renormalisations of operators already included.
The supersymmetry breaking terms also bring simply soft terms and $\mu$ term renormalization. Since these terms are anyway renormalised by $\cO_{1,...8}$, 
where  spurion dependence is included with {\it arbitrary}
coefficients,  then for what follows there is no loss of generality in ignoring the
supersymmetry breaking effects associated to
$\cO_{9,...14}$. In other words, this discussion shows that
$\cO_{9,...,14}$ are not relevant for the analysis
of the Higgs potential performed below.
Finally, there  can be an additional operator of $d=6$
from the gauge sector, $\cO_{15}=(1/M^2)\int d^2\theta
\,\,W^\alpha\Box W_\alpha$ which could affect 
the Higgs potential\footnote{Its complete gauge invariant form 
is $\int d^4 \theta \ Tr \ e^V W^{\alpha} e^{-V} D^2 (e^V
  W_{\alpha}  e^{-V})$.}.  Using the equations of motion for the
gauge field it can be shown that $\cO_{15}$ gives a renormalisation 
of $\cO_{1,2,3}$, so its effects are ultimately included,
since the coefficients $Z_{1,2,3}$ are arbitrary.

In conclusion, the list of $d=6$ operators that remain for our study of 
the Higgs sector beyond MSSM is that of (\ref{operators18}). 
Let us stress that not all these operators are necessarily  present or generated in a detailed model. Symmetries and details of the ``new physics'' beyond the MSSM that
generated them,  may  forbid or favour the presence of some of them.
Therefore, we regard these remaining
operators as independent of each other, although in specific models
correlations may exist among their coefficients $Z_i$. 
It is  important to keep all these operators in the
analysis, for the purpose of identifying which of
them has the largest individual contribution to the Higgs mass,
one of the main interests of this analysis.
Finally,  some of the $d=6$ operators can in principle
be present even in the absence of the $d=5$ operators, if these classes of
operators are generated by integrating different ``new physics''.
In specific UV completions, one simply keeps the terms generated by the model and sets all the rest to zero.

\section{The Scalar Potential}\label{ScalarPotential}

Following the previous discussion, 
the overall Lagrangian of the model is
\medskip
\bea\label{LL}
\cL_H = \cL_0+\cL_1+\sum_{i=1}^{8}\,\cO_i
\eea

\medskip\noindent
with the MSSM Higgs Lagrangian $\cL_0$ of eq.~(\ref{lo}), $\cL_1$
of eq.~(\ref{dimensionfive}) and $\cO_{1,2,....,8}$ of 
eq.~(\ref{operators18}).

In order to calculate the scalar potential we need the bosonic expansion of the Lagrangian. For the dimension-six operators we have:

\begin{eqnarray} 
\mathcal{O}_{1} &=&
\frac{1}{M^{2}}\int d^{4}\theta \,\,
Z_{1}(S,S^{\dagger })\,\,
(H_{1}^{\dagger }\,e^{V_{1}}\,H_{1})^{2} 
\nonumber \\
&=& 
2 \alpha_{1 0}\,
\Big[
(h_1^\dagger h_1)\,\big[\,(\cD_\mu h_1)^\dagger\,(\cD^\mu h_1)
+h_1^\dagger\,\frac{D_1}{2}\,h_1 + F_1^\dagger F_1\,\big]+
\vert\,h_1^\dagger F_1\vert^2 
+
( h_1^\dagger  \cD^\mu h_1)
(h_1^\dagger \overleftarrow{\cD_\mu} h_1)
\Big]
\nonumber\\
&+&
\Big[2\,\alpha_{1 1}\,m_0 \,(h_1^\dagger h_1)(F_1^\dagger h_1)+h.c.\Big]
+\alpha_{12}\, m_0^2\,(h_1^\dagger h_1)^2+{\rm fermionic \,\,part}
\\[3pt]
\mathcal{O}_{2} 
&=&\frac{1}{M^{2}}\int d^{4}\theta \,\,
Z_{2}(S,S^{\dagger })\,\,
(H_{2}^{\dagger }\,e^{V_{2}}\,H_{2})^{2} 
\nonumber \\
&=& 
2 \alpha_{2 0}\,\Big[(h_2^\dagger h_2)\,\big[\,
(\cD_\mu h_2)^\dagger\,(\cD^\mu h_2)
+h_2^\dagger\,\frac{D_2}{2}\,h_2 + F_2^\dagger F_2\,\big]+
\vert  h_2^\dagger F_2\vert^2 
+
(h_2^\dagger  \cD^\mu h_2)
(h_2^\dagger \overleftarrow{\cD_\mu} h_2)
\Big]
\nonumber\\
&+&
\Big[2\,\alpha_{2 1}\,m_0\, 
(h_2^\dagger h_2)(F_2^\dagger h_2)+h.c.\Big]
+\alpha_{22}\,m_0^2\, (h_2^\dagger h_2)^2+{\rm fermionic \,\,part}
\end{eqnarray}
\begin{eqnarray} 
\mathcal{O}_{3} &=&
\frac{1}{M^{2}}\int d^{4}\theta \,\,Z _{3}(S,S^{\dagger
})\,\,
(H_{1}^{\dagger 
}\,e^{V_{1}}\,H_{1})\,(H_{2}^{\dagger }\,e^{V_{2}}\,H_{2}),
\nonumber\\
&=&
\alpha_{30}\,\Big\{
 (h_1^\dagger h_1)\,
\Big[(\cD_\mu h_2)^\dagger\,(\cD^\mu h_2)
+h_2^\dagger\,\frac{D_2}{2}\,h_2 + F_2^\dagger F_2\Big]+
(h_1^\dagger  F_1)(F_2^\dagger h_2)+(1\leftrightarrow 2)\Big\}
\nonumber\\
&+&
\alpha_{30}\,\Big[
(h_1^\dagger \cD_\mu
h_1)(h_2^\dagger\overleftarrow \cD^\mu h_2)
+h.c.\Big]+\Big\{\alpha_{31}\,m_0\, \Big[
(h_1^\dagger h_1)(F_2^\dagger h_2)
+(h_2^\dagger h_2)(F_1^\dagger h_1)
\Big]+h.c.\Big\}
\nonumber\\
&+&\alpha_{32}\,m_0^2\,
(h_1^\dagger h_1)(h_2^\dagger h_2)+{\rm fermionic \,\,part}
\\[3pt]
\mathcal{O}_{4} &=&\frac{1}{M^{2}}\int d^{4}\theta \,\,
Z_{4}(S,S^{\dagger })
\,\,(H_{2}\,.\,H_{1})\,(H_{2}\,.\,H_{1})^{\dagger },
\nonumber\\
&=&
\alpha_{40}\,\,\partial_\mu (h_2.h_1)\,\partial^\mu(h_2.h_1)^\dagger
+\Big[\alpha_{41}\,m_0\,(h_2.h_1)\,(h_2.F_1 +F_2.h_1)^\dagger +h.c.\Big]
\nonumber\\
&+&\alpha_{42}\,m_0^2\,(h_2.h_1)\,(h_2.h_1)^\dagger
+\alpha_{40}\,\vert h_2\cdot F_1+F_2\cdot h_1\vert^2
+{\rm fermionic \,\,part}
\end{eqnarray}
\begin{eqnarray} 
\mathcal{O}_{5} &=&\frac{1}{M^{2}}\int d^{4}\theta \,\,Z%
_{5}(S,S^{\dagger })\,\,(H_{1}^{\dagger 
}\,e^{V_{1}}\,H_{1})\,H_{2}.\,H_{1}+h.c.  \nonumber \\
&=&
\alpha_{50}
\Big\{\Big[
(\cD_\mu h_1)^\dagger\,(\cD^\mu h_1)
+h_1^\dagger\,\frac{D_1}{2}\,h_1 + F_1^\dagger F_1\Big](h_2.h_1)
+
(h_1^\dagger \overleftarrow\cD_\mu h_1)\,\partial^\mu(h_2.h_1)\Big\}
\nonumber\\
&+&\!\!\!
\Big[\alpha_{50} \,(F_1^\dagger h_1)
+\alpha_{51}^*\,m_0\,(h_1^\dagger\,h_1)\Big]\,(h_2.F_1+F_2.h_1)
+
m_0\,\Big[\alpha_{51}\,(F_1^\dagger h_1)+\alpha_{51}^*\,
(h_1^\dagger F_1)\Big]\,(h_2.h_1)
\nonumber\\
&+&
\alpha_{52}\,m_0^2\,(h_1^\dagger h_1)\,(h_2.h_1)+{\rm h.c.\,of \,all}
+{\rm fermionic \,\,part}
\\[3pt]
\mathcal{O}_{6} &=&
\frac{1}{M^{2}}\int d^{4}\theta \,\,Z_{6}(S,S^{\dagger
})\,\,
(H_{2}^{\dagger }\,e^{V_{2}}\,H_{2})\,\,H_{2}.\,H_{1}+h.c.  \nonumber \\
&=&
\alpha_{60}
\Big\{\Big[
(\cD_\mu h_2)^\dagger\,(\cD^\mu h_2)
+h_2^\dagger\,\frac{D_2}{2}\,h_2 + F_2^\dagger F_2\Big](h_2.h_1)
+
(h_2^\dagger \overleftarrow\cD_\mu h_2)\,\partial^\mu(h_2.h_1)\Big\}
\nonumber\\
&+&\!\!\!
\Big[\alpha_{60} \,(F_2^\dagger h_2)
+\alpha_{61}^*\,m_0\,(h_2^\dagger\,h_2)\Big]\,(h_2.F_1+F_2.h_1)
+
m_0\,\Big[\alpha_{61}\,(F_2^\dagger h_2)+\alpha_{61}^*\,
(h_2^\dagger F_2)\Big]\,(h_2.h_1)
\nonumber\\
&+&
\alpha_{62}\,m_0^2\,(h_2^\dagger h_2)\,(h_2.h_1)+{\rm h.c.\,of \,all}
+{\rm fermionic \,\,part}
\eea
\bea
\mathcal{O}_{7} &=&\frac{1}{M^{2}}\frac{1}{16 g^2 \kappa}
\int d^{2}\theta \,\,
Z_{7}(S,0)\,\,{\rm Tr}\,\, W^{\alpha }\,W_{\alpha }\,(H_{2}\,H_{1})+h.c. 
\nonumber\\[4pt]
&=&
\frac{1}{2}\,(D_w^2+D_Y^2)\,
\Big[\alpha_{70}\,(h_2.h_1)+\alpha_{70}^*\,(h_2.h_1)^\dagger\Big]
+{\rm fermionic \,\,part}\\[3pt]
\mathcal{O}_{8} &=&
\frac{1}{M^{2}}\int d^{4}\theta
 \,\,\Big[Z_{8}(S,S^{\dagger })\,\,
\,[(H_{2}\,H_{1})^{2}+h.c.]\Big] 
\nonumber\\[4pt]
&=&
2\,\alpha_{81}^* \,m_0\,(h_2.h_1)\,(h_2.F_1+F_2.h_1)
+m_0^2\,\alpha_{82}\,(h_2\cdot h_1)^2+h.c.+
{\rm fermionic \,\,part}\qquad
\end{eqnarray}

The notation is as follows: $\cD^\mu h_i=(\partial^\mu+i/2\,V^\mu_i)\,h_i$,\,
$h_i^\dagger \overleftarrow\cD^\mu=(\cD^\mu h_i)^\dagger$.
Further,  $D_1\equiv \vec D_w\,\vec T+(-1/2)\,\,D_Y$ and
  $D_2\equiv \vec D_w\,\vec T+(1/2)\,\,D_Y$, $T^a=\sigma^a/2$.
Finally, one rescales in all $\cO_i$ ($i\not=7$):\,\,\,
$V_{w}\ra 2\,g_2\,V_w$, 
$V_{y}\ra 2\,g_1\,V_y$.
Then $V_{1,2}= 2\,g_2\,\vec V_w\,\vec T +2\,g_1\,(\mp 1/2)\,V_y$ with the
upper sign (minus) for $V_1$, where
 $V_{1,2}$ enter the definition of $\cO_{1,2}$.
Other notations used above:
$H_1\cdot H_2=\epsilon^{ij}\,H_1^i\,H_2^j$. Also
$\vert h_1\cdot h_2\vert^2
=\vert h_1^i\,\epsilon^{ij}\,h_2^j\vert^2
=\vert h_1\vert^2\,\vert h_2\vert^2
-\vert h_1^\dagger \,h_2\vert^2;$
 $\epsilon^{ij}\,\epsilon^{kj}=\delta^{ik}$;\,\,
$\epsilon^{ij}\,\epsilon^{kl}=
\delta^{ik}\,\delta^{jl}-\delta^{il}\,\delta^{jk}$, $\epsilon^{12}=1$,
with
\bea
h_1=\left(\begin{array}{c}
h_1^0 \\[-1pt]
h_1^-\\
\end{array}\right)
\equiv 
\left(\begin{array}{c}
h_1^1 \\[-1pt]
h_1^2\\
\end{array}\right),\,\,Y_{h_1}=-1;
\qquad
h_2=\left(\begin{array}{c}
h_2^+\\[-1pt]
h_2^0
\end{array}\right)
\equiv
\left(\begin{array}{c}
h_2^1\\[-1pt]
h_2^2
\end{array}\right),\,\,\,Y_{h_2}=+1
\eea
With these results we find the following contributions to the scalar potential:
\bea
\label{VFinitial}
V_F=\frac{\partial^2\,K}{\partial\,h_i\,\partial\, h_j^*}
\,F_i\,F_j^*=\vert F_1\vert^2+
\vert F_2\vert^2+
\frac{\partial^2\,K_6}{\partial\,h_i\,\partial \,h_j^*}
\,F_i\,F_j^*\label{VV}
\eea

\smallskip\noindent
where $K_6$ is the contribution of $\cO(1/M^2)$ to the K\"ahler 
potential due to  $\cO_{1,...8}$. Also,
\medskip
 \bea\label{arhos}
 F_1^{* q}&=&
 -\big\{\epsilon^{qp}\,h_2^p\, \big[
 \mu+2\,\zeta_{10}\,(h_1.h_2)
 +\rho_{11}\big]+h_1^{* q}\,\rho_{12}\big\}
 \nonumber\\
 F_2^{* q}&=&
 -\big\{\epsilon^{pq}\,h_1^p \,\big[
 \mu+2\,\zeta_{10}\,(h_1.h_2)
 +\rho_{21}\big]+h_2^{* q}\,\rho_{22}\big\}
 \eea

\medskip\noindent
where $\rho_{ij}$ are functions of $h_{1,2}$:
\medskip
\bea\label{rhos0}
\rho_{11}
&=&
-(2\alpha_{10}\,\mu+\alpha_{40}\mu+\alpha_{51}^*\,m_0)\vert h_1\vert^2
-(\alpha_{30}\,\mu+\alpha_{40}\mu+\alpha_{61}^*\,m_0)\,\vert h_2\vert^2
\nonumber\\
&&-
(\alpha_{41}^*\,m_0+\alpha_{50}^*\,\mu)\,(h_2.h_1)^*+
\big[\,(\alpha_{60} +2\,\alpha_{50})\,\mu+2\alpha_{81}^*\,m_0\big]\,
(h_1.h_2)
\nonumber\\
\rho_{12}
&=&
\,\,\,(2\alpha_{11}^*\,m_0+\alpha_{50}^*\,\mu)\vert h_1\vert^2
+(\alpha_{31}^*\,m_0+\alpha_{50}^*\,\mu)\,\vert h_2\vert^2
\nonumber\\
&&-\big[(2\alpha_{10}+\alpha_{30})\,\mu+\alpha_{51}^*\,m_0\big]\,(h_1.h_2)
+\alpha_{51}^*\,m_0\,(h_2.h_1)^*
\eea
\bea
\rho_{21}
&=&
 -(2\alpha_{20}\,\mu+\alpha_{40}\mu+\alpha_{61}^*\,m_0)\vert h_2\vert^2
 -(\alpha_{30}\,\mu+\alpha_{40}\mu+\alpha_{51}^*\,m_0)\,\vert h_1\vert^2
\nonumber\\
&&-(\alpha_{41}^*\,m_0+\alpha_{60}^*\,\mu)\,(h_2.h_1)^*
+\big[\,(\alpha_{50} +2\,\alpha_{60})\,\mu+2\alpha_{81}^*\,m_0\big]\,
(h_1.h_2)
\nonumber\\
\rho_{22}
&=&\,\,\,
(2\alpha_{21}^*\,m_0+\alpha_{60}^*\,\mu)\vert h_2\vert^2
+(\alpha_{31}^*\,m_0+\alpha_{60}^*\,\mu)\,\vert h_1\vert^2
\nonumber\\
&&-\big[(2\alpha_{20}+\alpha_{30})\,\mu+\alpha_{61}^*\,m_0\big]\,(h_1.h_2)
+\alpha_{61}^*\,m_0\,(h_2.h_1)^*\qquad\qquad\quad
\label{rhos}
\eea

The first two terms in the rhs of (\ref{VFinitial}) give
($h_i$ denote $SU(2)_L$ doublets, $\vert h_i\vert^2\equiv h_i^\dagger \,h_i$):
\medskip
\bea
V_{F,1}&\equiv &\vert F_1\vert^2+
\vert F_2\vert^2
\nonumber\\
&=&\vert \mu+2\,\zeta_{10}\,h_1.h_2\vert^2\,\,
\big(\vert h_1\vert^2+\vert h_2\vert^2\big)
\nonumber\\[3pt]
&+&
\Big[\mu^*\,\Big(
\vert h_1\vert^2\,\rho_{21}+\vert h_2\vert^2\,\rho_{11}
+(h_1.h_2)^\dagger\,(\rho_{22}+\rho_{12})\Big)+h.c.\Big]
\label{VF1}
\eea
The nontrivial field dependent K\"ahler metric gives for the last 
term in $V_F$ of eq.~(\ref{VV}):
\medskip
\bea
V_{F,2}&=&
\vert \mu\vert^2
\Big[
2\,\big(\alpha_{10}+\alpha_{20}+\alpha_{40}\big)
\vert h_1\vert^2\,\vert h_2\vert^2
+(\alpha_{30}+\alpha_{40})\,\big(\vert h_1\vert^4+\vert h_2\vert^4\big)
\nonumber\\[6pt]
&& +\,2\,\big(\alpha_{10}+\alpha_{20}+\alpha_{30}\big)
\,\vert h_1.h_2\vert^2+\,\,
\big(\vert h_1\vert^2+2\,\vert h_2\vert^2\big)
\big(\alpha_{50}\,h_2.h_1+h.c.\big)
\nonumber\\[6pt]
&&+\,\big(2\vert h_1\vert^2+\vert h_2\vert^2\big)
\big(\alpha_{60}\,h_2.h_1+h.c.\big)\Big]
\label{VF2}
\eea
so that
 $V_F=V_{F,1}+V_{F,2}$. Furthermore, from the gauge part we have:
\medskip
 \bea\label{ddd}
 D_w^a&=&-g_2\,\Big[\,\,h_1^\dagger T^a\,h_1\,\,(1+\tilde
 \rho_1)+h_2^\dagger\,T^a\,h_2\,\,(1+\tilde \rho_2)\,\Big],
 \qquad T^a=\sigma^a/2
 \nonumber\\
D_Y&=&-g_1\,\Big[\,\,h_1^\dagger \frac{-1}{2}\,h_1\,\,(1+\tilde
\rho_1)+h_2^\dagger\,\frac{1}{2}\,h_2\,\,(1+\tilde \rho_2)\,\Big]
\eea

\medskip\noindent
with notation:
\bea\label{tilderhoi}
\tilde\rho_1(h_{1,2})\equiv2\alpha_{10}\,\vert h_1\vert^2 +\alpha_{30}\,\vert
h_2\vert^2+
\big[(\alpha_{50}-\alpha_{70})\,\,h_2.h_1+h.c.\big]
\nonumber\\
\tilde\rho_2(h_{1,2})\equiv2\alpha_{20}\,\vert h_2\vert^2 +\alpha_{30}\,\vert
h_1\vert^2+
\big[(\alpha_{60}-\alpha_{70})\,\,h_2.h_1+h.c.\big]
\eea
This gives
\medskip
\bea
 D_w^a\,D_w^a&=&
 \frac{g_2^2}{4}\,\big[\,\,\big(
 (1+\tilde\rho_1)\,\vert h_1\vert^2-
 (1+\tilde\rho_2)\,\vert h_2\vert^2\big)^2
 +4\,(1+\tilde\rho_1)(1+\tilde\rho_2)\,\vert h_1^\dagger\,h_2\vert^2
\big]
\nonumber\\
D_Y^2&=&
\frac{g_1^2}{4}\,\big(
(1+\tilde\rho_1)\,\vert h_1\vert^2-
(1+\tilde\rho_2)\,\vert h_2\vert^2\big)^2\label{dsq}
\eea

\medskip\noindent
So the gauge part of the scalar potential is written as:
\medskip
\bea
V_{gauge}&=&\frac{1}{2}\big( D_w^2+D_Y^2)\,\big[ 
1+ (\alpha_{70}\,h_2.h_1+h.c.)\big]
\nonumber\\[5pt]
&=&
\frac{g_1^2+g_2^2}{8}
\,\big(\vert h_1\vert^2-\vert h_2\vert^2\big)\,
\big[
\big(
1+f_1(h_{1,2}))\,\vert h_1\vert^2-(1+f_2(h_{1,2}))\,
\vert h_2\vert^2\big]
\nonumber\\[3pt]
&+&
\frac{g_2^2}{2}
\,(1+f_3(h_{1,2}))
\vert h_1^\dagger\,h_2\vert^2\qquad\label{VG}
\eea

\medskip\noindent
obtained with (\ref{ddd}) and where $f_{1,2,3}$ are functions of $h_{1,2}$:
\medskip
\bea\label{fs}
f_1(h_{1,2})& \equiv &4\,\alpha_{10}\,\vert h_1\vert^2
+\,\big[\,(2\alpha_{50}-\alpha_{70})\,h_2.h_1+h.c.\big)\big]
\nonumber\\
f_2(h_{1,2})& \equiv &4\,\alpha_{20}\,\vert h_2\vert^2
+\,\big[\,(2\alpha_{60}-\alpha_{70})\,h_2.h_1+h.c.\big)\big]
\nonumber\\
f_3(h_{1,2})& \equiv &\tilde\rho_1+\tilde\rho_2+(\alpha_{70}\,h_2.h_1+h.c.)
\eea

The scalar potential also has corrections $V_{SSB}$ from
supersymmetry breaking, due to spurion dependence in
higher dimensional operators. In addition we also have the usual soft breaking term 
from the MSSM. As a result 
\medskip
\bea
\label{VSSB}
\!\!V_{SSB}\!\!\!&=&- m_0^2\,\Big[
\alpha_{12}\,\,\vert h_1\vert^4
+\,\alpha_{22}\,\,\vert h_2\vert^4
+\,\alpha_{32}\,\,\vert h_1\vert^2\,\vert h_2\vert^2
+\,\alpha_{42}\,\,\vert h_2.h_1\vert^2
\\[3pt]
&&+\,\,\big(\alpha_{52}\,\,\vert h_1\vert^2\,(h_2.h_1)+h.c.\big)
+\big(\alpha_{62}\,\,\vert h_2\vert^2\,(h_2.h_1)+h.c.\big)\Big]
\nonumber\\[3pt]
&&\!\!\!\!-
\Big[\,m_0^2\,\alpha_{82}\,(h_1.h_2)^2+\zeta_{11}\,m_0\,(h_2.h_1)^2
+\mu\,B\,m_0\,(h_1.h_2)\!+\!h.c.\Big]
+ \! m_0^2\,(c_1 \vert h_1\vert^2+\!c_2 \vert h_2\vert^2)\nonumber
\eea

\medskip\noindent
Finally, in $\cO_{1,...8}$
there are  non standard kinetic terms that can contribute to
$V$ when the scalar singlet components (denoted $h_i^0$) of $h_i$ 
acquire a vev. The relevant terms are:
\medskip
\bea
\cL_H\supset(\delta_{ij^*}+ g_{ij^*})\,\,
\partial_{\mu}\,h_i^0\,\partial^\mu h_j^{0
  *},\qquad
i,j=1,2.
\eea
where the field dependent metric is:
\bea
g_{11^*}&=& 4\,\alpha_{10}\,\vert h_1^0\vert^2
+(\alpha_{30}+\alpha_{40})\,\vert h_2^0\vert^2
-2\,(\alpha_{50}\, h_1^0\,h_2^0\,+h.c.)
\nonumber\\
g_{12^*}&=&
(\alpha_{30}+\alpha_{40})\,h_1^{0 *}\,h_2^0
-\alpha_{50}^*\,\,h_1^{0 * 2}
-\alpha_{60}\,\,h_2^{0 \, 2},\qquad g_{21^*}=g_{12^*}^*
\nonumber\\
g_{22^*}&=& 4\,\alpha_{20}\,\vert h_2^0\vert^2
+(\alpha_{30}+\alpha_{40})\,\vert h_1^0\vert^2
-2\,(\alpha_{60}\, h_1^0\,h_2^0\,+h.c.)
\eea

\medskip\noindent
For simplicity we only included the $SU(2)$
higgs singlets contribution, that we actually need in the following,
but the discussion can be extended to the general case.
The metric $g_{ij^*}$
is expanded about a background value $\langle h_i^0\rangle
=v_i/\sqrt 2$, then field redefinitions are performed to
obtain canonical kinetic terms. They are:
\medskip
\bea
h_1^0&\ra& h_1^0\,\,\Big(1-\frac{\tilde g_{11^*}}{2}\Big)\,
-\frac{\tilde g_{21^*}}{2}\,h_2^0
\nonumber\\
h_2^0&\ra& h_2^0\,\,\Big(1-\frac{\tilde g_{22^*}}{2}\Big)\,-
\frac{\tilde g_{12^*}}{2}\,h_1^0,\qquad \tilde g_{ij^*}\equiv 
g_{ij^*}\Big\vert_{h_i^0\rightarrow v_i/\sqrt 2} 
\label{red}
\eea
These bring further corrections to the scalar potential. 

Since the metric has corrections which are $\cO(1/M^2)$,
after (\ref{red}) only the MSSM soft breaking terms and the
MSSM quartic terms are affected.
The other terms in the scalar potential, already suppressed by
one or more powers of the scale $M$  are affected only beyond
the approximation  $\cO(1/M^2)$ considered here.
Following (\ref{red})
the correction terms $\cO(1/M^2)$  induced by the MSSM quartic terms and
by soft breaking terms in $V_{SSB}$ are:
\medskip
\bea
V_{k.t.}&=& 
\tilde m_1^2\, (- \tilde g_{11}^*) \,\vert  \,h_1^0\,\vert^2
+\tilde m_2^2\,(- \tilde g_{22}^*) \,\vert  \,h_2^0\,\vert^2
-\frac{1}{2}\,\big(\tilde m_1^2+\tilde m_2^2\big)\,
\big(\tilde g_{21^*}\,h_1^{0 *}\,h_2^0 + h.c.\big)
\nonumber\\
&+&
\frac{1}{2}\,
\Big[\,B\,m_0\,\mu\,\Big(
\,(\tilde g_{11^*}+\tilde g_{22^*})\,\,h_1^0\,h_2^0
+\,\tilde g_{12^*}\,h_1^{0\,2}+\tilde g_{21^*}\, h_2^{0\,2}
\Big)+h.c.\Big]
\nonumber\\
&-& \frac{g^2}{8}\,\big(\,\vert\, h_1^0\,\vert^2-\vert \,h_2^0\,\vert^2\big)
\,\big(\tilde g_{1 1^*}\,\vert \,h_1^0\,\vert^2-\tilde
g_{22^*}\,\vert\,h_2^0\,\vert^2+h.c.\big)
\label{KT}
\eea

\medskip\noindent
Using equations (\ref{VV}), (\ref{VG}), (\ref{VSSB}) and (\ref{KT}), we find 
the full scalar potential. With notation
$\tilde m_i^2\equiv c_i m_0^2+\vert \mu\vert^2$, 
$i=1,2$ one finally has:
\bea\label{vvv}
V&=&V_{F,1}+V_{F,2}+V_{G}+V_{SSB}+V_{k.t.}
\\[5pt]
&=&V_{k.t.}+ \tilde m_1^2 \vert h_1 \vert^2
+\tilde m_2^2 \vert h_2\vert^2
-\big[\mu\,B\,m_0\,h_1 \cdot h_2
+h.c.\big]\nonumber\\[5pt]
&+&\frac{\lambda_1}{2}\, \,\vert h_1\,\vert^4
+\frac{\lambda_2}{2}\,\,\vert h_2\,\vert^4
+\lambda_3 \,\vert h_1\,\vert^2\,\,\vert h_2\,\vert^2\,
+\lambda_4\,\vert\,h_1\cdot h_2\,\vert^2\nonumber\\[3pt]
&+&
\Big(\,\,
\frac{\lambda_5}{2}\,(h_1\cdot  h_2)^2
+\lambda_6\,\vert\,h_1\,\vert^2\,
(h_1 \cdot h_2)+
\lambda_7\,\vert\,h_2\,\vert^2\,(h_1 \cdot h_2)+h.c.\Big)
\nonumber\\[3pt]
&+&
\frac{g^2}{8}\,\big(\vert h_1\vert^2-\vert h_2\vert^2\big)
\big(f_1(h_{1,2})\,\vert h_1\vert^2-f_2(h_{1,2})\,\vert h_2\vert^2\big)
+4\,\vert \zeta_{10}\vert^2 \vert h_1.h_2\vert^2\,(\vert h_1\vert^2
+\vert h_2\vert^2)
\nonumber\\[3pt]
&+&
\frac{g_2^2}{2}\,f_3(h_{1,2}) \,\vert h_1^\dagger h_2\vert^2
\nonumber
\eea
where $g^2=g_1^2+g_2^2$, and $f_{1,2,3}(h_{1,2})$ are all quadratic in $h_i$, see eq.~(\ref{fs}). Except $V_{k.t.}$, all other fields are in the SU(2) doublets notation. $\lambda_i$ are given by
\medskip
\bea\label{fffV}
{\lambda_1}/{2}&=&
{\lambda_1^0}/{2}
-\vert \mu\vert^2\,(\alpha_{30}+\alpha_{40})
-m_0^2\,\alpha_{12}
-2 m_0\,{\rm Re}\big[\alpha_{51}\,\mu\big]\\
{\lambda_2}/{2}&=&
{\lambda_2^0}/{2}
-\vert \mu\vert^2\,(\alpha_{30}+\alpha_{40})
-m_0^2\,\alpha_{22}
-2 m_0\,{\rm Re}\big[\alpha_{61}\,\mu\big]
\nonumber\\
\lambda_3&=&
\lambda_3^0
-\,2\,\,\vert \mu\vert^2\,(\alpha_{10}+\alpha_{20}+\alpha_{40})
-m_0^2\,\alpha_{32}
-2 m_0\,{\rm Re}
\big[(\alpha_{51}+\alpha_{61})\,\mu\big]
\nonumber\\
\lambda_4&=& \lambda_4^0
-\,2\,\,\vert \mu\vert^2\,
(\alpha_{10}+\alpha_{20}+\alpha_{30})
-m_0^2\,\alpha_{42}
-
2\,m_0\,{\rm Re}\big[(\alpha_{51}+\alpha_{61})\,\mu\big]
\nonumber\\
{\lambda_5}/{2}&=&-\,m_0\,\mu\,(\alpha_{51}+\alpha_{61})-
m_0\,\zeta_{11}-m_0^2\,\alpha_{82}
\nonumber\\
\lambda_6&=&
\,\vert \mu\vert^2\,(\alpha_{50}+2\,\alpha_{60})
+m_0^2\,\alpha_{52}
+m_0\,\mu\,(2\,\alpha_{11}+\alpha_{31}+\alpha_{41})
+2\,m_0\,\mu^*\,\alpha_{81}^*
+\, 2\,\zeta_{10}\,\mu^*
\nonumber\\
\lambda_7&=&
\,\vert \mu\vert^2\,(\alpha_{60}+2\, \alpha_{50})
+m_0^2\,\alpha_{62}
+m_0\,\mu\,(2\,\alpha_{21}+\alpha_{31}+\alpha_{41})
+2\,m_0\,\mu^*\,\alpha_{81}^*
+\,2\,\zeta_{10}\,\mu^*
\nonumber
\eea
where
\medskip\bea\label{ms}
\lambda_1^{0}/2=\frac{1}{8}\,(g_2^2+g_1^{2}),\,\quad
\lambda_2^{0}/2=\frac{1}{8}\,(g_2^2+g_1^{2}),\,\,\quad
\lambda_3^{0}=\frac{1}{4}\,(g_2^2-g_1^{2}),\,\quad
\lambda_4^{0}= -\frac{1}{2}\,g_2^2,\,\quad
\eea

\medskip\noindent
denote the pure MSSM contribution. One can include MSSM loop corrections by replacing $\lambda_i^0$ with radiatively corrected values \cite{Carena:1995bx}.

Equations (\ref{vvv}) and (\ref{fffV}) show the effects of various higher dimensional operators on the scalar potential. As a reminder, note that all $\alpha_{ik}\sim
\cO(1/M^2)$ while $\zeta_{11}, \zeta_{10}\sim \cO(1/M)$. In principle, the dimension-five pieces are the dominant. However, as we will see later, when $\tan\beta$ is large the effect on a physical observable of dimension-five and six terms can be of similar size.
In specific models correlations exist among these coefficients.
The above remarks apply to the case when the 
$d=5$ and $d=6$ operators considered
are generated by the same ``new physics'' beyond the MSSM (i.e. are
suppressed by the same scale).
However, as  mentioned earlier, this may not always be the case;
in various models contributions from some $d=6$ operators
can be independent of those from $d=5$ operators (and present even in
the absence of the latter), if generated by different ``new physics''. 
A case by case study is then needed for a thorough 
analysis of all possible scenarios beyond the MSSM higgs sector.

The overall sign of the $h^6$  terms depends on the relative size of
$\alpha_{j0}$, $j=1,2,5,6,7$, and cannot be fixed even locally, in the absence
of the exact values of these coefficients. $\zeta_{10}$
also contributes to the overall sign, however this alone cannot 
fix it. At large fields' values higher and higher dimensional operators
become relevant and contribute to it. We therefore do not impose that
$V$ be bounded from below at large fields. 
For a discussion of stability with $d=5$ operators only
 see \cite{Blum:2009na}.

Eq.~(\ref{vvv}) is the main result of this section. For simplicity, one can take $\tilde g_{12^*}$ and $\tilde g_{21^*}$ to be real, possible if for example  $\alpha_{50}$ and $\alpha_{60}$ are real and there is no vev for $\mathrm{Im} h_i$. $Bm_0\mu$ can also be taken to be real. In the next section 
we shall adopt these simplifications.

\section{Corrections to the MSSM Higgs Masses}
\label{AnalyticalResults}

Having obtained the general expression for the scalar potential, we proceed with the computation of the mass spectrum. The general expression for the mass of the CP-even Higgs fields $h,H$ is:
\medskip
\bea
m^2_{h,H}\equiv
\frac{1}{2}\frac{\partial^2 V}{\partial h_i^0\partial
  h_j^0}\bigg\vert_{\langle h_i\rangle =v_i/\sqrt 2, 
\langle\,\Im h_i\rangle=0}\eea

\medskip\noindent
In the leading order $\cO(1/M)$ one has (upper signs for $m_h$):
\medskip
\bea\label{mhH}
 m_{h, H}^2\!\!&=&\!\!\!\frac{m_Z^2}{2}
+\frac{B\,m_0\mu (u^2+1)}{2\,u} \mp\frac{\sqrt w}{2}
+{  v}^2\,\Big[
\,(2\,\zeta_{10}\,\mu)\,\,q_1^{\pm}
+\,(-2\, m_0 \,\zeta_{11})\,\,q_2^{\pm}\Big]+\delta m_{h,H}^2
\qquad\eea

\medskip\noindent
with
\medskip
\bea
q_1^\pm\!\! & = &\!\!\!\frac{1}{4 \, u^2\,(1+u^2)\sqrt w}
\nonumber\\[7pt]
&\times &\!\!\!\!\!
\Big[
- (1-6u^2+u^4)\,u\,\sqrt w\mp 
\Big(m_Z^2 u(1-14 u^2+u^4)-B\,m_0\mu (1+u^2) (1+10 u^2
  +u^4)\Big)\Big]
\nonumber\\[5pt]
q_2^\pm\!\! & = &
\mp \frac{2 u}{(1+u^2)^2\sqrt
  w}\,\Big[-B\,m_0\mu(1+u^2)-m_Z^2\,u\Big]
\quad\eea
where
\bea
w\equiv m_Z^4 + \big[-B\,m_0\,\mu (1+u^2)^3+2 m_Z^2 u(1-6 u^2 +u^4)
\big]\frac{(-B\,m_0\mu)}{u^2(1+u^2)},\qquad u\equiv \tan\beta
\eea
In eq.~(\ref{mhH}) 
\bea
\delta m_{h,H}^2= \cO(1/M^2)
\eea
and we also  used that $m_Z=g\,v/2$.
One also shows  that the Goldstone mode has  $m_G=0$ and
the pseudoscalar  A has a mass:
\medskip
\bea\label{ma}
m_A^2=
\frac{1+u^2}{u}\,B\,m_0\,\mu
-\,\frac{1+u^2}{u}\, \zeta_{10}\,\mu\,v^2\,
+2\,m_0\,\zeta_{11}\,v^2+\delta m_A^2,\quad \delta m_A^2=\cO(1/M^2)
\eea
These results agree with the independent calculation up to order $\cO(1/M)$ of the previous chapters.

Ignoring for the moment the corrections $\cO(1/M^2)$,
one eliminates $Bm_0\mu$ between (\ref{mhH}) and (\ref{ma})
to obtain:
\medskip
\bea\label{mhold}
 m_{h,H}^2&=&
\frac{1}{2}\Big[m_A^2+m_Z^2\mp\sqrt{\tilde w}\Big]
\nonumber\\[3pt]
&+&
{(2\,\zeta_{10}\,\mu)\,
{  v}^2\,\sin 2\beta}
\,\Big[1\pm\frac{m_A^2+m_Z^2}{\sqrt{\tilde w}}\Big]+
\frac{(-2\,\zeta_{11}\,m_0)\,{  v}^2}{2}\,\Big[1\mp
\frac{(m_A^2-m_Z^2)\,\cos^2 2\beta}{\sqrt{\tilde w}}\Big]
\nonumber\\[5pt]
&+&
\delta^\prime m_{h,H}^2, \qquad \qquad  \delta^\prime m_{h,H}^2
= \cO(1/M^2)
\eea

\medskip
\noindent
where the upper (lower) signs correspond to $h$ ($H$) respectively
and
\medskip
\bea
\tilde w\equiv
(m_A^2+m_Z^2)^2-4\,m_A^2\,m_Z^2\,\cos^2 2\beta
\eea

\medskip\noindent
This is important if one considers $m_A$ as an input; it is also needed
if one considers the limit of large $\tan\beta$ at fixed $m_A$ (see
later).

The $\cO(1/M^2)$ corrections $\delta m_{h,H}^2$, $\delta
m_A^2$ and $\delta^\prime m_{h,H}^2$ of equations (\ref{mhH}),
 (\ref{ma}) and (\ref{mhold}) in the general case
of including all operators and their associated supersymmetry breaking,
have a rather complicated form. For most purposes, an expansion in $1/\tan\beta$ is accurate enough. The reason for this is that it is only at large $\tan\beta$ that
$d=6$ operators  bring corrections comparable to those of $d=5$. The relative $\tan\beta$ enhancement of $\cO(1/M^2)$ operators compensates for the extra 
suppression factor $1/M$ that these operators have relative 
to $\cO(1/M)$ operators (which involve both $h_1$ and $h_2$ and 
thus are not enhanced in this limit).

If we neglect supersymmetry breaking effects of $d=6$ operators (i.e.
$\alpha_{j1}=\alpha_{j2}=0$, $\alpha_{j0}\not=0$, $j=1,...,8$) 
and with $d=5$ operators contribution, one has\footnote{In the
case of including the supersymmetry breaking effects from effective 
operators, associated with coefficients $\alpha_{j1}$, $\alpha_{j2}$
$j=1,2,..8$, the exact formula is very long and is not
included here.} for the correction $\delta m_{h,H}^2$ in eq.~(\ref{mhH}) 
(upper signs correspond to $\delta m_h^2$)
\smallskip
\bea\label{dmh}
\delta m_{h,H}^2=\sum_{j=1}^{7}\,\,\gamma^\pm_j\,\,\alpha_{j\,0}
+\gamma^\pm_{x}\,\,\zeta_{10}\,\zeta_{11}  
+\gamma^\pm_{z}\,\,\zeta_{10}^2\,   
+\gamma^\pm_{y}\,\,\zeta_{11}^2     
\label{masshiggs}
\eea

\smallskip\noindent
The expressions of the coefficients $\gamma^\pm$
are provided in Appendix~\ref{CoefficientsHiggsMasses}
and can  be used for numerical studies.
While these expressions are exact, they are complicated 
and not very transparent.
It is then instructive to analyse  an approximation of the 
$\cO(1/M^2)$ correction as an expansion in $1/\tan\beta$. 
We present in this limit the correction $\delta m_{h,H}^2$ 
of eq.~(\ref{mhH}), which also includes all  supersymmetry breaking 
effects associated with all $d=5,6$ operators,
(i.e.  $\alpha_{j1}\not=0, \alpha_{j2}\not=0$, 
$\zeta_{11}\not=0$, $j=1,..8$)
in addition to the MSSM soft terms.
This has a simple expression:
\medskip
\bea\label{admh}
\delta m_{h}^2
&=& 
-2 \,v^2\, \Big[
 \alpha_{22} m_0^2+2 \alpha_{61}
\,m_0\mu+(\alpha_{30}+\alpha_{40})\,\mu^2
-  \alpha_{20} \,m_Z^2\Big]
\nonumber\\[-1pt]
&+&\frac{v^2}{\tan\beta}
\Big[ 4\, \alpha_{62}\,m_0^2+4 \mu\,m_0\,( 
2 \alpha_{21}+\alpha_{31}+\alpha_{41}+2 \alpha_{81})
+
4 \mu^2\,\,(2 \alpha_{50}+\alpha_{60})
\nonumber\\[-2pt]
&&\qquad\qquad -\,\, m_Z^2\,(2 \alpha_{60}-3 \alpha_{70})
-\frac{v^2}{(Bm_0\mu)}\,(2\zeta_{10}\,\mu)^2
\Big]+\cO(1/\tan^2\beta)
\qquad
\eea

\medskip\noindent
which is obtained with $Bm_0\mu$ kept fixed. 
The result is dominated by the first line, including both
 SUSY and non-SUSY terms from the effective operators.
 This correction can be comparable
to linear terms in $\zeta_{10}$,\,$\zeta_{11}$
from $d=5$ operators for $(2\,\zeta_{10}\mu) \approx
1/\tan\beta$  (see later). Not all $\cO_{1,2...8}$
are necessarily present, so in some models some 
$\alpha_{ij}$, $\zeta_{10}$, $\zeta_{11}$
could vanish. 
Also:
\medskip
\bea\label{admH}
\delta m_H^2
&=&-
\frac{1}{4} (Bm_0\mu) \,v^2\,\alpha_{60}\,\tan^2\beta
+\frac{v^2\,\tan\beta}{8}
\Big[
-8 Bm_0\mu\,\alpha_{20} -4 \alpha_{62} m_0^2
\nonumber\\[-2pt]
&-&4 \mu\,m_0 ( 2 \alpha_{21}+\alpha_{31}+\alpha_{41}+2 \alpha_{81})
-4 \mu^2\,(2 \alpha_{50}+\alpha_{60}) +(2 \alpha_{60}-\alpha_{70})\,m_Z^2\Big]
\nonumber\\[-1pt]
&+&\frac{3}{4}\,Bm_0\mu\,v^2(\alpha_{50}+\alpha_{60})+
\frac{v^2}{8 \tan\beta}\Big[
-8 B m_0\mu \alpha_{10}+ (12\alpha_{52}-16 \alpha_{62}) m_0^2
\nonumber\\[-1pt]
&-& 4 \mu m_0 (-6 \alpha_{11}+8
 \alpha_{21}+\alpha_{31}+\alpha_{41}+2 \alpha_{81})
-4 \mu^2(5 \alpha_{50}-2 \alpha_{60})
\nonumber\\[-1pt]
&+&(6 \alpha_{50}+20  \alpha_{60}-13 \alpha_{70})\,m_Z^2
+\frac{8\,v^2}{B m_0 \mu}\, (2\,\zeta_{10}\,\mu)^2  
\Big]+\cO(1/\tan^2\beta)
\eea

\medskip\noindent
which is  obtained for $(Bm_0\mu)$  fixed. 
Note the $\cO(1/M^2)$ effects from $d=5$ operators ($\zeta_{10}^2$).

Similar expressions exist for the neutral pseudoscalar $A$.
The results are  simpler in this case
and we present the  exact expression
of $\delta m_A^2$ of (\ref{ma}) in the most general case, that  includes
all supersymmetry breaking effects from the 
 operators of $d=5,6$ and from the MSSM. One finds
\medskip
\bea\label{magen}
\delta m_A^2&=&
\frac{v^2}{8 \tan^2\beta\,(1+\tan^2\beta)}\Big[
-\,2 \,B\,m_0\mu\,\alpha_{50}
+\big[- ( 4 \alpha_{31}+4\alpha_{41}+8 \alpha_{81}+8 \alpha_{11} )
\,m_0\mu 
\nonumber\\
&-& 4 \alpha_{52} m_0^2
 - 8B m_0\mu \alpha_{10}- 4\,(\alpha_{50}+2
  \alpha_{60})\mu^2 
+(2 \alpha_{50}-\alpha_{70})\,m_Z^2\,\big]\,\tan\beta
\nonumber\\
&+&\big[
2 B\,m_0\,\mu\,(10 \alpha_{50}+3\alpha_{60})\,
+16 \alpha_{82}m_0^2
+16 (\alpha_{51}+\alpha_{61})m_0\,\mu\big]\tan^2\beta
\nonumber\\
&+&\!\!\! 2\,\big[\!
-4 B\,m_0\mu (\alpha_{10}+\alpha_{20}+2\,\alpha_{30}+2\,\alpha_{40})
\!-6 (\alpha_{50}+\alpha_{60})\,\mu^2
-(\alpha_{50}+\alpha_{60}-\alpha_{70})\,m_Z^2
\nonumber\\
&-& 2 (\alpha_{62}+\alpha_{52})\,m_0^2
-4 (\alpha_{11}+\alpha_{21}+ \alpha_{31}+ \alpha_{41}+2 \alpha_{81})\,
 m_0\mu
 \big]\,\tan^3\beta
\nonumber\\
&+&
\big[ 2\,B\,m_0\,\mu\,(3 \alpha_{50}+10\,\alpha_{60})\,
+16 \alpha_{82} m_0^2+16 (\alpha_{51} +\alpha_{61})\,m_0\mu
\big]\tan^4\beta
\nonumber\\
&-&
\big[
8 B\,m_0\mu\,\alpha_{20}+
4\,(2 \alpha_{50}+\alpha_{60})\,\mu^2
-(2 \alpha_{60}-\alpha_{70})\,m_Z^2
+4 \alpha_{62}\,m_0^2
\nonumber\\
&+& 4\,\,(2 \alpha_{21}+\alpha_{31}
+\alpha_{41}+2\alpha_{81})\,m_0\,\mu\big]\,\tan^5\beta
-\,\,2 \,B\,m_0\,\mu\,\alpha_{60}\,\tan^6\beta\,\Big]\qquad
\eea

\medskip\noindent
We also showed that
 $\delta m_G=0$ so the Goldstone mode remains massless
in $\cO(1/M^2)$, which is a good consistency check.
A result similar to that in eq.~(\ref{admh}) is found
from an expansion of (\ref{magen}) in the large $\tan\beta$ limit:
\smallskip
\bea
\delta m_A^2&=&
-\frac{1}{4}\,(Bm_0\mu)\,\alpha_{60}\,v^2\,\tan^2\beta
+\frac{\tan\beta}{8}
v^2 \Big[
-8 B m_0\mu \alpha_{20}-4 \alpha_{62} m_0^2\nonumber\\
&-&(8 \alpha_{21} +4 \alpha_{31}+4 \alpha_{41}+8 \alpha_{81})\,m_0\mu
-(8 \alpha_{50}+4 \alpha_{60})\mu^2+2 \alpha_{60}\,m_Z^2-\alpha_{70}\,m_Z^2
\Big]\nonumber\\
&+&\frac{v^2}{4}
\Big[ Bm_0\mu (3\alpha_{50} + 11 \alpha_{60}) + 
        8 m_0^2 \alpha_{82} +
 8 m_0\mu (\alpha_{51} + \alpha_{61})\Big]
\nonumber\\
&+&\!\!\!
\frac{v^2}{8\tan\beta}
\Big[- 8Bm_0\mu\,(\alpha_{10}+2 \alpha_{30}+2 \alpha_{40})
 - 4\,(2 \alpha_{11}+ \alpha_{31}+ \alpha_{41}
+ 2\alpha_{81})\,\,m_0\mu
\nonumber\\
&-&4 \alpha_{52}\,m_0^2
- ( 4\alpha_{50}+8 \alpha_{60})\mu^2
-(2 \alpha_{50}+4 \alpha_{60}-3 \alpha_{70})m_Z^2\Big]
+\cO(1/\tan^2\beta)
\eea

\medskip\noindent
We  emphasise that the large $\tan\beta$ limits presented so far
were done with $(B \,m_0\mu)$ fixed. While this is certainly an 
interesting case, a more natural expression to consider at large $\tan\beta$
 is that in which one keeps $m_A$ fixed and $B m_0\mu$ arbitrary. We  present below the correction $\cO(1/M^2)$ to $m_{h,H}^2$ for 
the case $m_A$ is kept fixed to an appropriate value. 
The result is (assuming $m_A\!>\!m_Z$, otherwise 
$\delta' m_h^2$ and $\delta' m_H^2$ are exchanged):
\bea\label{dd1}
\delta^\prime m_{h}^2
\!\!\!&=&
-2\, v^2\,\Big[ \alpha_{22} \,m_0^2+(\alpha_{30}+\alpha_{40})\mu^2
+2 \alpha_{61} \,m_0\,\mu 
- \alpha_{20}\,m_Z^2\Big]
-\frac{(2\,\zeta_{10}\,\mu)^2\,\,v^4}{m_A^2-m_Z^2}
\nonumber\\
&+&\!\!\!\!\frac{v^2}{\tan\beta}
\bigg[\frac{1}{(m_A^2-m_Z^2)}
\Big( 4 \,m_A^2\,\big(\,
(2 \alpha_{21}\!+\!\alpha_{31}\!+\!\alpha_{41}\! +\!2 \alpha_{81})
\,m_0\,\mu\! +\! (2\alpha_{50}\! +\!\alpha_{60})\,\mu^2
+\alpha_{62}\,m_0^2\big)
\nonumber\\
&-&\, (2 \alpha_{60}-3\alpha_{70})\,m_A^2\,m_Z^2
-(2\alpha_{60}+\alpha_{70})\,m_Z^4\Big)
+\frac{8\,(m_A^2+m_Z^2)\,\,(\mu\,m_0\,\zeta_{10}\,\zeta_{11})
\,v^2}{(m_A^2-m_Z^2)^2}
\bigg]\nonumber\\
&+&\cO(1/\tan^2\beta)
\eea

\medskip\noindent
A similar formula exists for the correction to $m_H$:
\medskip
\bea\label{dd2}
\delta^\prime m_H^2
\!\!\!&=&
\Big[
-2\, \big( m_0\mu\, (\alpha_{51}+\alpha_{61})+\alpha_{82}\,m_0^2
\big)\,v^2 +\frac{(2\,\zeta_{10}\,\mu)^2\,v^4}{m_A^2-m_Z^2}
\Big]\nonumber\\
&\!\!\!+&\!\!\!\!\!
\frac{v^2}{\tan\beta}
\!\Big[ 
\frac{1}{m_A^2\!-\!m_Z^2}
\Big(2 m_A^2\,\big(2\, (\alpha_{11}\!-\!\alpha_{21})\,m_0\mu
+\!(\alpha_{60}\!-\!\alpha_{50})\,\mu^2
+\!(\alpha_{52}\!-\!\alpha_{62})\,m_0^2
-\alpha_{60}\,m_A^2\big)
\nonumber\\
&-&\big[\,
4\, (\alpha_{11}+\alpha_{21}+\alpha_{31}+\alpha_{41}+2
\alpha_{81})\,m_0\mu
+
6(\alpha_{50}+\alpha_{60})\,\mu^2
+
2(\alpha_{52}+\alpha_{62})\,m_0^2
\nonumber\\[3pt]
&\!\!\!\!-&
\!\!\!(\alpha_{50}\! +\! 5\alpha_{60}\! -\! 2\alpha_{70})\,m_A^2 
\big]\,m_Z^2-(\alpha_{50}\!-\alpha_{60})\,m_Z^4\Big) 
\!-\! \,\frac{8\,(m_A^2+m_Z^2)\,
(\mu\,m_0\,\zeta_{10}\,\zeta_{11})
\,v^2}{(m_A^2-m_Z^2)^2}\Big]\nonumber\\
&+&\cO(1/\tan^2\beta)
\eea

\medskip
Corrections (\ref{dd1}) and (\ref{dd2}) must be added to the rhs of eq.~(\ref{mhold})
to obtain the value  of $m_{h,H}^2$ expressed in function of $m_A$. The corrections in equations (\ref{dmh}) to (\ref{dd2}) extend those of the previous chapter to include all $\cO(1/M^2)$ terms.

From equations (\ref{admh}) and (\ref{dd1}) we are able to identify
 the  effective operators of $d=6$ that give the leading contributions  
to $m^2_h$, which is important for model building.
These are $\cO_{2,3,4}$ in the absence of supersymmetry breaking 
and $\cO_{2,6}$ when this is broken. 
It is however preferable  to increase $m_h^2$ by supersymmetric 
rather than supersymmetry-breaking effects
of the effective operators,  because the latter are less under control
in the effective approach and one would favour a supersymmetric
solution to the  fine-tuning problem associated with increasing
the MSSM Higgs mass above the LEPII bound. Therefore $\cO_{2,3,4}$
are the leading operators, with the remark that $\cO_2$ has a 
smaller effect, of order $(m_Z/\mu)^2$ relative to $\cO_{3,4}$
(for similar $\alpha_{j0}$, $j=2,3,4$). At smaller $\tan\beta$,
$\cO_{5,6}$ can also give significant contributions while $\cO_7$ 
has a relative suppression factor $(m_Z/\mu)^2$.

\section{Analysis of the Leading Corrections and Effective Operators}
\label{AnalysisLeadingCorrections}

One expects that when in the Lagrangian appear effective operators of mass dimension five and six, coming from the same UV physics, those of dimension six will be subleading. However, this is not the case when an extra suppression makes the two classes comparable. In our case, some dimension five operators are suppressed by $1/(M\tan\beta)$ but dimension six only have $1/M^2$. Thus, in the limit of large $\tan\beta$ these two classes can be comparable.

In the particular case of the Higgs mass, by comparing $\cO(1/M)$ terms in eq.~(\ref{mhold}) against $\cO(1/M^2)$ terms in equations (\ref{dd1}) and (\ref{dd2}),
one identifies the situation when these two classes of operators
give comparable corrections:
\bigskip
\bea\label{s01}
&&\frac{4 m_A^2}{m_A^2-m_Z^2}\frac{\vert \,\zeta_{10}\,\mu
\,\vert}{\tan\beta}
\approx
\bigg\vert
  \alpha_{22} m_0^2+(\alpha_{30}+\alpha_{40})\mu^2
+2 \alpha_{61} m_0 \mu 
- \alpha_{20}m_Z^2
+\frac{2\,(\zeta_{10}\,\mu)^2\, v^2}{m_A^2-m_Z^2}
\bigg\vert\nonumber\\[8pt]
&&\bigg\vert 
\,\,\zeta_{11}\,m_0 \,\,+\frac{4 m_Z^2}{m_A^2-m_Z^2}
\frac{\zeta_{10}\,\mu}{\tan\beta}\bigg\vert
\approx
\bigg\vert
\,\big( m_0\mu\,(\alpha_{51}+\alpha_{61})+\alpha_{82}\,m_0^2\big)
-\frac{2\,(\zeta_{10}\,\mu)^2 \,v^2}{m_A^2-m_Z^2}\bigg\vert\qquad
\eea

\bigskip\noindent
In this case $\cO(1/(M\,\tan\beta))$ 
and $\cO(1/M^2)$ corrections are approximately equal
(for $M\approx m_0\,\tan\beta$).
Similar relations can be obtained by comparing
(\ref{mhH}) and (\ref{ma}) against $\delta m_{h,H}^2$ of
(\ref{admh}), (\ref{admH}) and (\ref{magen}).

Note that we don't have to consider operators of dimension $>6$ since they do not receive any $\tan\beta$ enhancement in order to become comparable with $d=6$ and will always be subleading.

Let us now examine more closely the corrections to the Higgs masses
due to  $d=6$ operators.  The interest is to maximise the correction
to the MSSM classical value of $m_h$.
From equations (\ref{admh}) and (\ref{dd1}) and ignoring SUSY breaking corrections ($\alpha_{jk}$, $k\not=0$),
we saw that at large $\tan\beta$ $\cO_{3,4}$  bring the largest 
correction and also $\cO_2$ to a lower extent. At smaller $\tan\beta$, $\cO_{5,6,7}$ can have significant
corrections. All this can be seen from the relative variation:
\be\label{opop}
\epsilon_{rel}\equiv
\frac{m_h-m_Z}{m_Z}=\sqrt{\delta_{rel}}-1,
\ee
with
\bea
\delta_{rel}
\!\!\!&\equiv&\!\!\!
1-\frac{4 m_A^2}{m_A^2-m_Z^2}\frac{1}{\tan^2\beta}
+\frac{v^2}{m_Z^2}\,\,\bigg\{
\frac{2\,\zeta_{10}\,\mu}{\tan\beta}\,\frac{4\,m_A^2}{m_A^2-m_Z^2}
+\frac{(-2\,\zeta_{11}\,m_0)}{\tan^2\beta}\,
\frac{2\,(m_A^4+m_Z^4)}{\,\,(m_A^2-m_Z^2)^2}\,
\nonumber\\[3pt]
&-&\!\Big[2 \,\Big( \alpha_{22}\,m_0^2+(\alpha_{30}+\alpha_{40})\,\mu^2
+2\, \alpha_{61}\,m_0\,\mu-\alpha_{20}\,m_Z^2\Big)
+\frac{(2\,\zeta_{10}\,\mu)^2\,v^2}{m_A^2-m_Z^2}\Big]
\nonumber\\
&+&\frac{1}{\tan\beta}\frac{1}{m_A^2-m_Z^2}
\Big[ 4 \,m_A^2\,\mu\,\,
\Big(\,( 2 \alpha_{21}+\alpha_{31}+\alpha_{41}+2\alpha_{81})\,m_0
+(2\,\alpha_{50}+\alpha_{60})\,\mu\Big)
\nonumber\\
&+&\!\!\!\! 4\,\alpha_{62}\,m_0^2\,m_A^2
-\! (2 \alpha_{60}\!- \! 3\,\alpha_{70})\,
m_A^2\,m_Z^2-(2\,\alpha_{60}\!+\!\alpha_{70})\,m_Z^4
\!+\! 8 \,\zeta_{10}\,\zeta_{11}\,\mu\,m_0\,
v^2\frac{m_A^2\!+\!m_Z^2}{m_A^2\!-\!m_Z^2}\Big]
\bigg\}\nonumber\\[4pt]
&+&
\cO(1/\tan^4\beta)+
\!\cO(\tilde m/(M\tan^3\beta))+\cO(\tilde m^2/(M^2\tan^2\beta))
\eea

\bigskip\noindent
where $\tilde m$ is some generic 
mass scale of the theory such as $\mu$, $m_Z$, $m_0$ or $v$.
The arguments of the functions $\cO$ in the last line
show explicitly the origin of these corrections (MSSM, $d=5$ and $d=6$
operators, respectively). Eq.~(\ref{opop})
 gives the overall relative change
of the classical value of $m_h$ in the presence of all possible 
higher dimensional operators of $d=5$ and $d=6$ beyond the MSSM Higgs
sector, for large $\tan\beta$ with $m_A$ fixed.
Depending on the signs of coefficients $\alpha_{jk}$, $\zeta_{10}$ and
$\zeta_{11}$ this relative variation can be positive and increase $m_h$ above the MSSM classical upper bound $m_Z$. The accuracy of the expansion at intermediate $\tan\beta$ depends on $\tilde m/M$; in any case one can use the exact $\delta m_{h,H}^2$ in (\ref{dmh}).

The same expansion in large $\tan\beta$ can also be computed keeping $Bm_0\mu$ fixed, instead of $m_A$. Then
\medskip
\bea
\delta_{rel}\!\!\! & \equiv &
 1-\frac{4}{\tan^2\beta}
+\frac{v^2}{m_Z^2}
\bigg\{
\frac{4\,(2\,\zeta_{10}\,\mu)}{\tan\beta}+
\frac{2}{\tan^2\beta}\,
\Big(\, (-2\,\zeta_{11}\,m_0)   
+\frac{2\,m_Z^2\,(2\,\zeta_{10}\,\mu)}{B\,m_0\,\mu}\Big)
\nonumber\\
&-&2\,\Big[
\alpha_{22}\,m_0^2+2\,\alpha_{61}\,m_0\,\mu+(\alpha_{30}+\alpha_{40})
\,\mu^2-\alpha_{20}\,m_Z^2\Big]
+\frac{1}{\tan\beta}\Big[\frac{(2\,\zeta_{10}\,\mu)^2\,v^2}{-B\,m_0\,\mu}
\nonumber\\
&+\!\!&4\,( 2\,\alpha_{21}\!+\!\alpha_{31}\!+\!\alpha_{41}
\!+\! 2 \alpha_{81})\,m_0\,\mu
+4\,(2\,\alpha_{50}\!+\!\alpha_{60})\,\mu^2
\!+\! 4 \alpha_{62}\,m_0^2
-(2\,\alpha_{60}\!-\! 3\alpha_{70})\,m_Z^2
\Big]\bigg\}\nonumber\\[3pt]
&+&
\cO(1/\tan^4\beta)+\cO(\tilde m/(M\tan^3\beta))+
\cO(\tilde
m^2/(M^2\tan^2\beta))\label{pop}
\eea

In (\ref{opop}) and (\ref{pop}), the $d=6$ operators ($\alpha_{ij}$
dependence) give contributions which are dominated by $\tan\beta$-independent terms.
One particular limit to consider  for $\delta m_h^2$ or $\delta'
m_h^2$ is that in which the effective operators of $d=6$ have coefficients 
such that these contributions add up to maximise $\delta_{rel}$. Since coefficients
$\alpha_{ij}$ are not known, we can choose them equal
in absolute value
\bea
-\alpha_{22}=-\alpha_{61}=-\alpha_{30}=-\alpha_{40}=\alpha_{20}>0
\eea
In this case, at large $\tan\beta$:
\bea\label{maxc}
\delta m_h^2\approx 2\, v^2 \alpha_{20}\big[
m_0^2+2\,m_0\mu+2\,\mu^2+\,m_Z^2\big]
\eea

\medskip\noindent
and similar for $\delta' m_h^2$.
A simple numerical example is illustrative. For $m_0=1$ TeV,
$\mu=350$ GeV and $v\approx 246$ GeV, one has
$\delta m_h^2\approx 2.36\,\alpha_{20}\,\times 10^{11}$ (GeV)$^2$.
Assuming  $\alpha_{20}\sim 1/M^2$ for $M=10$ TeV and the classical MSSM value of $m_h$ to be equal to $m_Z$ (reached for large $\tan\beta$),
we obtain an increase of $m_h$ from $d=6$ operators alone 
of about $\Delta m_h=12.15$ GeV to $m_h\approx 103$ GeV.
An increase of $\alpha_{20}$ by a factor of 2.5 to $\alpha_{20}\sim 2.5/M^2$
would give $\Delta m_h\approx 28$ GeV and $m_h\approx 119.2$ GeV,
which is above the  LEPII bound.

The discussion above indicates that if we persist on using the loop correction to increase the Higgs mass, the effect of these operators will be to relax the strain of the little hierarchy. Indeed, the relative increase  of $\Delta\,m_h$ due to $d=6$ operators alone is mildly reduced, however, the effective quartic coupling of the Higgs is increased. This amounts to a reduction of the fine tuning for the electroweak scale \cite{Cassel:2010px}. The above choice of $M=10$ TeV was partly motivated by the fine-tuning  results of \cite{Cassel:2009ps} and on convergence grounds: The expansion parameter of our effective analysis is $m_q/M$ where $ m_q$ is any scale of the theory, in particular it can be the susy breaking scale $m_0$. For $m_0\simeq 3$ TeV and $c_{1,2}\simeq 2.5$ (of eq.~(\ref{lo})), one finds for $M=10$ TeV that $c_{1,2}\,m_0/M\simeq0.75$ which is already at the limit of validity of the expansion in the effective approach considered. 

These simple estimates demostrate that mass dimension six operators can indeed bring a significant increase of $m_h$ to values compatible with the LEPII bound. However, 
the amount of increase depends on implicit assumptions like
the type and number of operators present and whether their overall sign, as generated by the UV physics, is consistent with an increase of $m_h$.
Take for example the case of the leading contribution to $m_h$
in the large $\tan\beta$ case. One would prefer to generate the leading operators with 
supersymmetric coefficients satisfying
\bea\label{sign}
\alpha_{20}>0,\, \alpha_{30}<0,\, \alpha_{40}<0
\eea
in order to increase $m_h$.
We have already mentioned that $\cO_{1,2,3}$ can be generated
by integrating out a massive gauge boson $U(1)'$ or  $SU(2)$
triplets while $\cO_4$ by a massive gauge singlet
or $SU(2)$ triplets. Let us discuss the signs that these operators are generated with:

\noindent
{\bf (a):} Integrating out a massive vector superfield $U(1)'$
under which Higgs fields have opposite charges (to avoid a
Fayet-Iliopoulos term), 
one finds $\alpha_{20}\!<\!0$ and $\alpha_{30}\!>\!0$ (also $\alpha_{10}\!<\!0$),
which is opposite to  condition  (\ref{sign}). However, this can 
 change if for example there are additional pairs of massive Higgs
doublets also charged under the new $U(1)'$ since then $\cO_3$ could be
generated with  $\alpha_{30}<0$.\,\,
{\bf (b):} Integrating out massive $SU(2)$ triplets that couple to the MSSM
Higgs sector would bring $\alpha_{20}\!>\!0$, $\alpha_{40}\!<\! 0$,
$\alpha_{30} > 0$; the first two of these satisfy (\ref{sign}).
{\bf (c):} Integrating out a massive gauge singlet would bring
$\alpha_{40}>0$ which would actually 
decrease $m_h$. Finally, if we take into account further constraints coming form the $\rho$ parameter \cite{Brignole:2003cm}, it turns out that it is $\alpha_{40}$ and $\alpha_{30}$ that can have the largest correction to $m_h^2$. For generating them, the case of a massive gauge singlet or additional $U(1)'$ 
vector superfield would have the advantage of preserving 
gauge couplings unification at one-loop.

For smaller $\tan\beta$, operators $\cO_{5,6,7}$ could bring
significant corrections to $m_h$ but it is more difficult to generate
these in a renormalisable setup.
For example, $\cO_{5,6}$ can be generated by integrating out a pair
of massive Higgs doublets and a massive gauge singlet but 
the overall sign of $\alpha_{50,60}$ would 
depend on the details of the model. This discussion shows that
while effective operators can in principle increase $m_h$,
deriving a renormalisable model that would generate 
them with appropriate signs for their (supersymmetric) coefficients
is not a simple issue. However this does not exclude the possibility, since the examples given are rather simplistic. Other generating mechanisms for $\cO_i$ could be in place\footnote{For some models with extended 
MSSM Higgs sector see
\cite{Espinosa:1998re,Espinosa:1991gr,Espinosa:1992sk,Espinosa:1992hp}.}
with  appropriate signs to increase $m_h$.

\newpage

\chapter{Nonlinear MSSM}
\label{NonlinearMSSM}

In the previous chapters we used EFT to study in a model independent way the effects of new physics beyond MSSM in the multiTeV scale. Nevertheless, MSSM itself contains new physics at scale $\sqrt{f}$, the SUSY breaking scale. If we take this scale to be around multiTeV, new effects appear by the presence of a goldstino, which is the dominant component of the gravitino. Goldstino couplings are best described in terms of nonlinear supersymmetry, as briefly presented in section \ref{NonlinearRealizationsConstrainedGoldstinoSuperfield}. One way to realize symmetries in a nonlinear fashion is by using appropriate constraints. In supersymmetry, these are constraints on superfields. In the following we apply the method of constrained superfields in order to construct the most general couplings of a goldstino to full MSSM.

\section{The Model}
\label{TheModel}

We couple the constrained superfield $X_{nl}$ of eq.~(\ref{goldstino1}) to the SUSY part of the MSSM, to find the ``nonlinear'' supersymmetry version of MSSM. At energy scales below $m_{soft}$, similar constraints can be applied to the MSSM superfields themselves, corresponding to integrating out the superpartners. Here, the only difference from ordinary MSSM is in the supersymmetry breaking sector.
Supersymmetry is broken spontaneously via a
 vacuum expectation value (v.e.v.) of $F_X$, fixed
by its equation of motion. The  Lagrangian of nonlinear MSSM is:
\bea
\label{LLnonl}
\cL=\cL_0+\cL_X+\cL_{H}+\cL_m+\cL_{AB}+\cL_{g}
\eea
Let us detail these terms. 
$\cL_0$ is the usual MSSM SUSY Lagrangian
\smallskip
\bea\label{mssmsusy}
\cL_0 &=&\!\!\sum_{\Phi, H_{1,2}} \int d^4\theta \,\,
\Phi^\dagger\,e^{V_i}\,\Phi+
\bigg\{\int
d^2\theta\,\Big[\,\mu\,H_1\,H_2+
H_2\,Q\,U^c+Q\,D^c\,H_1+L\,E^c\,H_1\Big]+h.c.\bigg\}
\nonumber\\[-6pt]
&&+\sum_{{\rm SM\,groups}}
\frac{1}{16\, g^2\,\kappa}
\int d^2\theta
\,\mbox{Tr}\,[\,W^\alpha\,W_\alpha] +h.c.,
\qquad\quad \Phi:Q,D^c,U^c,E^c,L\, ,
\eea
where $\kappa$ is a constant canceling the trace factor and the
gauge coupling $g$ is shown explicitly. The family matrices in the superpotential are implicit to lighten the notation.

The SUSY breaking couplings originate from the MSSM fields
couplings to the goldstino superfield; this is done by the
replacement $S\,\ra\, m_{soft} X_{nl}/f$ \cite{SK},
 where $S$ is the usual spurion also used in the previous chapters,
with $S=\theta\theta \,m_{soft}$ and $m_{soft}$ a generic
notation for the soft terms (denoted below $m_{1,2},
m_0$).
One has for the Higgs sector
\medskip
\bea
\cL_{H}\!\! &=&\!\!\sum_{i=1,2} c_i
\int d^4\theta \,\,X_{nl}^\dagger X_{nl}\,\,
H_i^\dagger\,e^{V_i}\,H_i
\nonumber\\[-7pt]
&=&\!\!\! \sum_{i=1,2} c_i \,\Big\{
\vert\phi_X\vert^2\,\Big[
\vert \cD_\mu\, h_i\vert^2
+F_{h_i}^\dagger F_{h_i}+h_i^\dagger\,\frac{D_i}{2}\,h_i+
\Big(\frac{i}{2}\overline\psi_{h_i}\overline\sigma^\mu\,\cD_\mu\psi_{h_i}
-\frac{1}{\sqrt 2} \,h_i^\dagger \lambda_i\,\psi_{h_i}+h.c.\Big)\Big]
\nonumber\\[-6pt]
&+&\!\!\!
\frac{1}{2}\, h_i^\dagger\,(\cD_\mu+\overleftarrow
\cD_\mu)\,h_i\,\,\partial^\mu\vert \phi_X\vert^2
+\overline\psi_X\overline\psi_{h_i}\,\psi_X\psi_{h_i}
-\frac{1}{2}\,[\phi_X^\dagger\,(\partial^\mu-\overleftarrow
\partial^\mu)\,\phi_X]\,[ h_i^\dagger (\cD_\mu - \overleftarrow
\cD_\mu)\,h_i]\nonumber\\[-2pt]
&+&\!\!\!
\Big[
-\frac{i}{2}\phi_X^\dagger \psi_X\, \sigma^\mu\,\overline\psi_{h_i}
(\cD_\mu-\overleftarrow \cD_\mu) {h_i}
-\frac{1}{\sqrt 2}\, \phi_X^\dagger \psi_X \,\,
h_i^\dagger\lambda_i\,{h_i}
-\phi_X^\dagger \psi_X \,F_{h_i}^\dagger\psi_{h_i}
+\phi_X^\dagger F_X\,F_{h_i}^\dagger {h_i}
\nonumber\\[-2pt]
&+&\!\!\!
\frac{i}{2}
\,(\overline\psi_X\,\overline\sigma^\mu\,\psi_X)\,(h_i^\dagger
\,\cD_\mu\,{h_i})
+\frac{i}{2}\,(\phi_X^\dagger \partial_\mu\,\phi_X)
\,(\overline\psi_{h_i}\,\overline\sigma^\mu\,\psi_{h_i})
+\frac{i}{2}
\overline\psi_X\,\overline\sigma^\mu\,
(\partial_\mu-\overleftarrow\partial_\mu)\,\phi_X\,\,
(h_i^\dagger \psi_{h_i})
\nonumber\\[-1pt]
&-&
\overline\psi_X\,F_X\,\,\overline\psi_{h_i}\,{h_i}+h.c.\Big]
+
\Big[
\partial_\mu\phi_X^\dagger \partial^\mu\phi_X+F_X^\dagger F_X
+\Big(\frac{i}{2}
\overline\psi_X\,\overline\sigma^\mu\partial_\mu\psi_X+h.c.\Big)\Big]
\vert\,{h_i}\vert^2\Big\},
\eea

\medskip\noindent
Here  $\cD, \partial$,
($\overleftarrow\cD,\overleftarrow\partial$) act only on the first
field to their right (left) respectively and $h_i$, $\psi_{h_i}$,
$F_{h_i}$ denote SU(2) doublets. Also
\bea
c_{1}=-{m_1^2}/{f^2},\qquad c_2=-{m_2^2}/{f^2}\, .
\eea

\medskip\noindent
Similar terms exist for all matter fields
\bea
\cL_m=\sum_{\Phi} c_\Phi\int d^4\theta \,\,
X_{nl}^\dagger X_{nl}\,\Phi^\dagger
e^V\,\Phi,\qquad c_\Phi
=-\frac{m_\Phi^2}{f^2},\quad \Phi: Q, U^c, D^c, L, E^c,
\eea
One can eventually set $m_\Phi=m_0$ (all $\Phi$).
The\,bi- and trilinear
SUSY breaking couplings are
\medskip
\bea
\cL_{AB}
\!\!&=&\!\!\frac{B'}{f}\,\int d^2\theta \,X_{nl}\,H_1\,H_2
\\[-2pt]
&+&
\frac{A_u}{f}\int d^2\theta\,X_{nl}\,H_2\,Q\,U^c+
\frac{A_d}{f}\,\int d^2\theta\,X_{nl}\,Q\,D^c\,H_1+
\frac{A_e}{f}\,\int d^2\theta\,X_{nl}\,L\,E^c\,H_1+h.c.
\nonumber\\[-3pt]
&=&\frac{B'}{f}\,\Big\{
\phi_X\,\Big[\,h_1\cdot F_{h_2}+F_{h_1}\cdot h_2-\psi_{h_1}\cdot \psi_{h_2}\Big]
-h_1\cdot (\psi_{X}\psi_{h_2})-(\psi_X\psi_{h_1})\cdot h_2+F_X\,h_1\cdot h_2
\Big\}
\nonumber\\[-3pt]
&+&\Big\{
\frac{A_u}{f}\Big[
\phi_X\,h_2\cdot (\phi_Q\,F_U\!-\!\psi_Q\,\psi_U+\! F_Q\,\phi_U)
-\phi_X\,(\psi_{h_2}\cdot \phi_Q \psi_U+\psi_{h_2}\cdot\psi_Q \phi_U
-F_{h_2}\cdot\phi_Q\,\phi_U)
\nonumber\\[-3pt]
&-&
\psi_X\,(h_2\cdot \phi_Q\,\psi_U+h_2\cdot\psi_Q\,\phi_U+\psi_{h_2}\cdot\phi_Q\,\phi_U)
+F_X\,h_2\cdot\phi_Q\,\phi_U\Big]
-\Big[U\ra D, H_2\ra H_1\Big]
\nonumber\\
&-&\Big[U\ra E, H_2\ra H_1,Q\ra L\Big]\Big\}
+h.c.
\eea

\medskip\noindent
where $B'\equiv B\,m_0\mu$. Finally, the supersymmetry breaking couplings in the gauge sector are
\medskip
\bea\label{gb}
\cL_{g}\!\!&=&
\sum_{i=1}^3
\frac{1}{16\, g^2_i\,\kappa}
\frac{2\,m_{\lambda_i}}{f}
\int d^2\theta
\,X_{nl}\,\mbox{Tr}\,[\,W^\alpha\,W_\alpha]_i +h.c.
\nonumber\\[-4pt]
&=&\sum_{i=1}^3\frac{m_{\lambda_i}}{2\,f}\,
\Big\{\phi_X\,\,\Big[
2\,i\,\lambda^a\,\sigma^\mu\,\Delta_\mu\,\overline\lambda^a
-\frac{1}{2}\,F^{a\,\mu\nu}F_{\mu\nu}^a
+D^a D^a-\frac{i}{4}\epsilon^{\mu\nu\rho\sigma}
\,F_{\mu\nu}^a\,F_{\rho\sigma}^a \Big]
\nonumber\\[-6pt]
&&\qquad\qquad\quad -\,\sqrt 2 \psi_X\,\sigma^{\mu\nu}\lambda^a\,F_{\mu\nu}^a
-\sqrt 2\,\psi_X\,\lambda^a\,D^a
+F_X\,\lambda^a \lambda^a\Big\}_i
+h.c.
\eea

\medskip\noindent
with $m_{\lambda_{1,2,3}}$ the masses of the three gauginos
and gauge group index $i$ for $U(1)$, $SU(2)$, $SU(3)$ respectively.
Above we introduced the notation
$\Delta_\mu\overline\lambda^a=\partial_\mu\overline\lambda^a
-g\,t^{abc}\,V_\mu^b\,\overline\lambda^c$.
Equations~(\ref{LLnonl}) to (\ref{gb}), along with (\ref{X}), define the model, with spontaneous
supersymmetry breaking ensured by non-zero $\langle F_X\rangle$.

Since $\phi_X\sim 1/f$,  the Lagrangian contains terms of order
higher than $1/f^2$.  In the calculation of the onshell Lagrangian
we shall restrict the calculations to up to and including $1/f^2$ terms.
This requires solving for $F_\phi$ of matter fields
up to and including  $1/f^2$ terms and for
$F_X$ up to and including $1/f^3$ terms (due to its leading
contribution which is -$f$).
Doing so, in the final Lagrangian no kinetic mixing is present
at this order.
Using the expressions of the auxiliary fields, 
one then computes the $F$-part of the scalar potential 
of the Higgs sector, to find:
\medskip
\bea
V_F=\vert \mu\vert^2\,\Big[\vert h_1\vert^2+\vert
  h_2\vert^2\Big]
+\frac{\vert f+(B'/f)\,h_1\cdot h_2\vert^2}{
1+c_1\,\vert h_1\vert^2+c_2\,\vert  h_2\vert^2}+\cO(1/f^3)
\eea

\medskip\noindent
with $h_1\cdot h_2\equiv h_1^0\,h_2^0-h_1^-\,h_2^+$ and $\vert
h_i\vert^2\equiv h_i^\dagger h_i=
h_i^{0\,*}h_i^0+ h_i^{-\,*} h_i^-$. One can work with this potential, however, for convenience,  if $\vert c_{1,2}\vert \vert h_{1,2}\vert^2\ll~1$, we can approximate $V_F$ by expanding the denominator in a series of powers of these
coefficients. Our analysis below is then valid for
$\vert c_{1,2}\vert \vert h_{1,2}\vert^2\!\ll\!1$.
After adding the gauge contribution, we find the following result 
for the scalar potential of the Higgs sector:
\medskip
\bea
\label{potential0}
V&=&
f^2+
\big(\vert \mu\vert^2+m_1^2\big)\,\,
\vert h_1\vert^2+
\big(\vert \mu\vert^2 +m_2^2\big)
\vert h_2\vert^2
+\big(B'\,h_1\cdot h_2+h.c.\big)
\\[4pt]
&+&\!\!\!
\frac{1}{f^2}\,\Big\vert m_1^2\,\vert h_1\vert^2+m_2^2\,\vert h_2\vert^2+
B'\,h_1\cdot h_2\Big\vert^2
+\frac{g_1^2+g_2^2}{8}\,\Big[\vert h_1\vert^2-\vert h_2\vert^2\Big]^2
+\frac{g_2^2}{2}\,\vert h_1^\dagger\,h_2\vert^2
+\cO(1/f^3)\nonumber
\eea

\medskip\noindent
This is the full Higgs potential. The first term in the last
line  is a new term, absent in MSSM (generated by eliminating $F_X$ of
$X_{nl}$). Its  effects for phenomenology
will be analyzed later. The ignored higher order terms in $1/f$
involve nonrenormalizable  $h_{1,2}^6$ interactions in $V$.

\section{New Couplings in the Lagrangian}
\label{NewcouplingsintheLagrangian}

In this section we compute the new interactions induced by
Lagrangian (\ref{LLnonl}), which are not present
in the MSSM. Many of the new couplings are actually dimension-four in fields,
with a (dimensionless)  $f$-dependent coupling.
The couplings are important in the case of a low SUSY
breaking scale in the  hidden sector and a light gravitino scenario.
Some of the new couplings also involve  the goldstino field
 and  are relevant for phenomenology.

As mentioned earlier, from the SUSY
 breaking part of the Lagrangian only terms up to $1/f^2$  were kept in the total Lagrangian. After eliminating all terms proportional to $F$-auxiliary fields of $X,H_i,Q,D^c,U^c,E^c$ and $L$, one obtains new couplings $\cL^{new}$ beyond those of the usual on shell, supersymmetric part of MSSM,
which are unchanged and not shown. One finds the on shell
Lagrangian
\bea
\cL^{new}\equiv \cL^{aux}_F+\cL^{aux}_D+\cL^{extra}_{m}+\cL^{extra}_{g}
\eea
Let us detail these terms. Firstly,
 \bea
 \cL^{aux}_F=\cL^{aux}_{F\, (1)}+\cL^{aux}_{F\, (2)}
 \eea
with
\medskip
\bea\label{auxLF1}
\cL^{aux}_{F\,(1)}
\!\!&=&\!\!
-\Big[
f^2+\big(m_1^2\vert h_1\vert^2+m_2^2\vert h_2\vert^2
+m^2_{\Phi}\,\vert\phi_\Phi\vert^2\big)\Big]
\nonumber\\
&-&\!\!\!
\Big[ B'\,h_1\cdot h_2
+A_u\,h_2.\phi_Q\,\phi_U
+A_d\,\phi_Q\phi_D .h_1
+A_e\,\phi_L\phi_E .h_1+\frac{1}{2}\,
m_{\lambda_i}\,\lambda_i\lambda_i+h.c.\Big]
\qquad
\eea

\medskip\noindent
recovering the usual MSSM soft terms
and the additional contributions:
\medskip
\bea\label{auxLF2}
\cL^{aux}_{F\,(2)}\!\!\!&=&\!\!\!\!
\Big\{\,
\frac{\opsi_X\opsi_X}{2\,f^2}
\Big[\mu
\big(m_1^2\!+\!m_2^2\big)\,h_1\cdot h_2
\!-\!\big(m_1^2\!+\!m_Q^2\!+\!m_D^2\big) h_1\cdot\phi_Q \phi_D
\!-\!\big(m_1^2\!+\!m_L^2\!+\!m_E^2\big) h_1\cdot\phi_L \phi_E
\nonumber\\
&-&\!\!
\big(m_2^2+m_Q^2+m_U^2\big)\phi_Q\phi_U\cdot h_2
\!+\!
\big(B'\,h_2-A_d\,\phi_Q \phi_D-A_e\,\phi_L\,\phi_E\big)^\dagger
\big(\mu h_2-\phi_Q \phi_D-\phi_L\,\phi_E\big)
\nonumber\\[2.5pt]
&+&
\big(B'\,h_1-A_u\,\phi_Q\,\phi_U\big)^\dagger
\big(\mu\,h_1-\phi_Q \,\phi_U\big)
+
\big(A_d\,\phi_D \,h_1-A_u\,h_2\,\phi_U\big)^\dagger
\big(\phi_D\,h_1-h_2\,\phi_U\big)
\nonumber\\
&+&A_d\,\big( \vert \phi_Q\cdot h_1\vert^2+\vert \phi_E \,h_1\vert^2\big)
+A_u\,\vert h_2\cdot \phi_Q\vert^2+A_e\,\vert \phi_L\cdot h_1\vert^2\Big]
+h.c.\Big\}
-\frac{1}{f^2}\,
\Big\vert B'\,h_1\cdot h_2
\nonumber\\[-3pt]
&+&
\!\!\! A_u h_2\cdot\phi_Q\,\phi_U
\!+\! A_d \phi_Q\,\phi_D \cdot h_1
\!+\! A_e \phi_L\,\phi_E \cdot h_1
\!+\! \frac{m_{\lambda_i}}{2}
\lambda_i\lambda_i
\!+\!
\big( m_1^2\vert h_1\vert^2
\!+\! m_2^2\vert h_2\vert^2
\!+\! m^2_{\Phi}\vert \phi_\Phi\vert^2\!\big)
\Big\vert^2
\nonumber\\[-1pt]
&-&
\frac{1}{f}\,\,\Big[
m_1^2\,\opsi_X\opsi_{h_1}\,h_1+
m_2^2\,\opsi_X\opsi_{h_2}\,h_2+
m^2_{\Phi}\,\opsi_X\opsi_\Phi\,\phi_\Phi+h.c.\Big]
+\cO(1/f^3)
\eea

\bigskip\noindent
A summation is understood over the SM group
indices $i=1,2,3$ in the gaugino term
 and over $\Phi=Q,U^c,D^c,L,E^c$ in the mass terms;
appropriate contractions among $SU(2)_L$
doublets are understood for holomorphic products, when the order
displayed is relevant. The leading interactions $\cO(1/f)$
are those in the last line and are dimension-four in fields. Similar
couplings exist at  $\cO(1/f^2)$ and involve scalar and gaugino fields.
Yukawa matrices are restored in (\ref{auxLF2})
by replacing $\phi_Q\phi_D\ra \phi_Q\gamma_d\phi_D$,
$\phi_Q\phi_U\ra \phi_Q\gamma_u\phi_U$,
$\phi_L\phi_E\ra \phi_L\gamma_e\phi_E$, as already explained.

There are also new couplings
from terms involving the auxiliary components of the vector
superfields of the SM. Integrating them out
one finds:
\medskip
\bea
\cL_D^{aux}\!\!\!\!
&=&
\frac{-1}{2}\,
\Big[
\tilde D_1+ \frac{1}{4\, f^2} \,
\big( \,m_{\lambda_1}\,\psi_X\psi_X+h.c.\big)\,\tilde D_1
+\frac{1}{\sqrt 2 \,f}
\big(\,m_{\lambda_1}\,\psi_X\,\lambda_1+h.c.\big)\Big]^2
\nonumber\\
&+&\!\frac{-1}{2}\,\Big[
\tilde D_2^a +\frac{1}{4 \,f^2}\,
\big(m_{\lambda_2}\,\psi_X\psi_X+h.c.\big)\,\tilde D_2^a
+\frac{1}{\sqrt 2\,f}\big(
m_{\lambda_2}\,\psi_X\,\lambda_2^a+h.c.\big)\Big]^2
\nonumber\\
&+&\!\frac{-1}{2}\,\Big[
\tilde D_3^a +\frac{1}{4 \,f^2}\,
\big(m_{\lambda_3}\,\psi_X\psi_X+ h.c.\big)\,\tilde D_3^a\,
+\frac{1}{\sqrt 2\,f}\,\big(
m_{\lambda_3}\,\psi_X\,\lambda_3^a+h.c.\big)\Big]^2\!\!
+\cO(f^{-3})\qquad
\label{newD}
\eea

\medskip\noindent
with notation:
\medskip
\bea\label{Dmssm}
\tilde D_{1}&=&
-\frac{1}{2}\,g_1\,
\big(- h_1^\dagger h_1+ h_2^\dagger h_2
+1/3\,\,\phi_Q^\dagger\phi_Q
-4/3\,\,\phi_U^\dagger\phi_U
+2/3\,\,\phi_D^\dagger\phi_D
-\phi_L^\dagger\phi_L
+2\,\phi_E^\dagger\phi_E\big)
\nonumber\\[-1pt]
\tilde D_{2}^a&=&
-\frac{1}{2}\, g_2\,
\big( h_1^\dagger\sigma^a h_1+h_2^\dagger\sigma^a
 h_2+\phi_Q^\dagger \sigma^a \phi_Q+\phi_L^\dagger\sigma^a
 \phi_L \big)
\nonumber\\[-1pt]
\tilde D_{3}^a&=&
-\frac{1}{2}\,g_3\,
\big(
\phi_Q^\dagger \,t^a\phi_Q
-
\phi_U^\dagger \,t^a\phi_U
-
\phi_D^\dagger \,t^a\phi_D\big)
\eea

\medskip\noindent
for the corresponding MSSM expressions; here $t^a/2$ are
the SU(3) generators.
From (\ref{newD}) one can easily read the
new, $f-$dependent couplings in the gauge sector,
absent in the MSSM.

The total Lagrangian also contains extra terms, not
proportional to  the auxiliary fields, and {\it not} present in the
MSSM. In the matter sector these are:
\medskip\noindent
\bea
\cL_m^{extra}\!\!\!\! &=&\!\!\frac{1}{4f^2}
\vert \partial_\mu(\psi_X\psi_X)\vert^2+
\Big(\frac{i}{2}\overline\psi_{X}\overline\sigma^\mu\,
\partial_\mu\psi_{X}+h.c.\Big)
\\[-4pt]
&-&
\sum_{i=1}^2
\frac{m_i^2}{f^2}\,\Big\{\,
\overline\psi_X\overline\psi_{h_i}\,\psi_X\psi_{h_i}
\!+\!\Big[\,
\frac{i}{2} \,(\overline\psi_X\,
\overline\sigma^\mu\,\psi_X)\,(h_i^\dagger
\,\cD_\mu\,{h_i})
+\frac{i}{2}\vert h_i\vert^2\,
\overline\psi_X\,\overline\sigma^\mu\partial_\mu\psi_X+h.c.
\Big]\Big\}\nonumber\\[-3pt]
&-&
\Big[m_i^2\ra m_\Phi^2,
H_i\ra \Phi\Big]
+\bigg\{\,\,
\frac{B'}{f}\,\,\Big[\,
\frac{1}{2\,f}\,\,\psi_X\psi_X \,\psi_{h_1}.\psi_{h_2}
-h_1.(\psi_{X}\psi_{h_2})-(\psi_X\psi_{h_1}).h_2\Big]
\nonumber\\[-3pt]
&+&
\frac{A_u}{f}\,\Big[\frac{1}{2\,f}
\psi_X\psi_X\,\big(
\,h_2.\psi_Q\,\psi_U
+\psi_{h_2}.\phi_Q\,\psi_U
+\psi_{h_2}.\psi_Q\,\phi_U\big)
-
\psi_X\,(h_2.\phi_Q\,\psi_U
+h_2.\psi_Q\,\phi_U
\nonumber\\[-2pt]
&+&
\psi_{h_2}.\phi_Q\,\phi_U)\Big]
+
 \Big[\frac{A_d}{f}\,\Big(\frac{1}{2\,f}\,\,
 \psi_X\psi_X\,(\psi_Q\,\psi_D.h_1
 +\,\phi_Q\,\psi_D.\psi_{h_1}+\psi_Q\,\phi_D.\psi_{h_1})\nonumber\\[-2pt]
 &-&\psi_X\,(\phi_Q\,\psi_D.h_1
 +\psi_Q\,\phi_D.h_1\!+\!\phi_Q\,\phi_D.\psi_{h_1})\Big)\!
 +\!(D\!\ra\! E, L\!\ra\! Q)\Big]
\!+\!h.c.\!\bigg\}\!+\!\cO(1/f^3)
\eea
Note the presence of interactions that are dimension-four in
fields ($B'/f \,h_1\psi_X \psi_{h_2}$, etc)
that can be relevant for phenomenology at low $f$.
There are also new couplings in the gauge sector
\bea\label{lgextra}
\cL_g^{extra}&=&
\sum_{i=1}^3
\,\,\frac{m_{\lambda_i}}{2\,f}\,\,\Big[
\frac{\psi_X\psi_X}{-2 \,f}\,\Big(
2\,i\,\lambda^a\sigma^\mu\,\Delta_\mu\,\overline\lambda^a
-\frac{1}{2}\,F^a_{\mu\nu}\,F^{a\,\mu\nu} -\frac{i}{4}
\epsilon^{\mu\nu\rho\sigma}\,F^a_{\mu\nu}\,F^a_{\rho\sigma}
\Big)\qquad\qquad\qquad
\nonumber\\[2pt]
&-&\sqrt 2\,\psi_X\sigma^{\mu\nu}\lambda^a\,F_{\mu\nu}^a
\Big]_i+h.c.+\cO(1/f^3),
\eea

\medskip\noindent
with $i=1,2,3$ is the gauge group index and
$\sigma^{\mu\nu}=i/4\,(\sigma^\mu\overline\sigma^\nu
-\sigma^\nu\overline\sigma^\mu)$.
 The new couplings of $\cL^{new}$ together
with the on shell part of the purely supersymmetric
part of the MSSM Lagrangian (on shell $\cL_0$ of (\ref{mssmsusy}))
gives the final effective Lagrangian
of the model. From this, the full scalar potential is identified.

\section{Implications for the Higgs Masses}
\label{ImplicationsfortheHiggsmasses}

Let us consider the Higgs scalar potential found in (\ref{potential0}) and
analyze its implications for the Higgs masses. From the neutral Higgs
part  of the potential one finds the masses of the
CP even and CP odd Higgs fields. Since eq.\,(\ref{potential0}) is valid up to $1/f^3$ terms, it is sufficient to restrict the expressions up to this order.
Firstly, at the minimum of the scalar potential one has:
\medskip
\bea\label{ttt}
 m_1^2-m_2^2&=&
\cot 2\beta\,\bigg[\,B'+\frac{f^2}{v^2}
\frac{(-1+  \sqrt w_0)(-B'+m_Z^2\,\sin 2\beta)}
{2\mu^2+m_Z^2 \cos^2 2\beta+ B'\sin 2\beta}\,\bigg]
\nonumber\\[10pt]
m_1^2+m_2^2&=&
\frac{1}{\sin 2\beta}\,\bigg[
-B'+\frac{f^2}{v^2}\frac{(-1+\sqrt w_0)(B'+2\,\mu^2\,\sin 2\beta)}{
2\mu^2+m_Z^2 \cos^2 2\beta+ B'\sin 2\beta}\,\bigg]
\eea
where
\bea
w_0\equiv
1-\frac{ v^2}{f^2}\,\big( 4\,\mu^2+2\,m_Z^2\,\cos^2
2\beta+2\,B'\,\sin 2\beta\big)
\eea

\medskip\noindent
One finds  the following results
(upper sign for $m_h^2$):
\medskip
\bea
m_{h,H}^2&=&
\frac{1}{2}\,\,\Big[
m_Z^2+\frac{-2 \,B'}{\sin
  2\beta}\mp \sqrt{w_1}\Big]
+
\frac{v^2}{32 f^2}\,\,
\Big\{
4\,B'\,\Big[\,\,2 B'+(4\mu^2+ 2
  m_Z^2\,\cos^2 2\beta)/\sin
  2\beta\Big]
\nonumber\\[-2pt]
&+&
4\,\,\,\Big[ \,\,2 \,B'^2+ 8 \,\mu^4+ 2\,m_Z^2
(4 \mu^2+m_Z^2)\,\cos^2
  2\beta+8\,B'\,\mu^2\sin
  2\beta\Big]\nonumber\\[-2pt]
&\mp&\!\!
\frac{\csc^2 2\beta}{\sqrt w_1}
\Big[-2 \,(B'^2+4\mu^4)
  m_Z^2+4\mu^2 m_Z^4+m_Z^6
+8\, \big(2  \mu^4 m_Z^2- B'^2\,(4\mu^2+m_Z^2)\big)\cos 4\beta
\nonumber\\[1pt]
&-& m_Z^2
  \,(6\,B'^2+8\mu^4+4\mu^2
  m_Z^2+m_Z^4)\cos 8\beta
- 8\,B'\,(B'^2-8 \mu^4)\sin 2\beta
\nonumber\\[2pt]
&+& B'(-8 B'^2+16 \mu^2 m_Z^2 +m_Z^4)\sin
  6\beta+B' m_Z^4\sin 10\beta\Big]\Big\}
+\cO(1/f^3)
\eea
with
\bea
w_1&=&\Big(m_Z^2+ \frac{-2\,B'}{\sin
    2\beta}\Big)^2
-4\,m_Z^2\, \Big(\frac{-2\,B'}{\sin
  2\beta}\Big)\,\cos^2 2\beta
\eea

\medskip\noindent
Further, the mass $m_A$ of the pseudoscalar Higgs
has a simple form (no expansion):
\medskip
\bea
m_A^2&=&\frac{-2\,B'}{\sin 2\beta}\,\,\bigg\{\,
\frac{3}{4}+\frac{1}{4}\,\,\sqrt w_0
-\,\frac{v^2}{4\,f^2}\,B'\,\sin 2\beta
\bigg\}
\eea

\medskip\noindent
and, as usual, the Goldstone mode has mass $m_G=0$.

\begin{figure}[t!h!]
\def\baselinestretch{1.}
\begin{center}
\begin{tabular}{cc|cr|}
\parbox{6.6cm}{
\subfloat[$m_h$ in function of $\sqrt{f}$, $m_A$ parameter]
{\includegraphics[width=6.6cm]{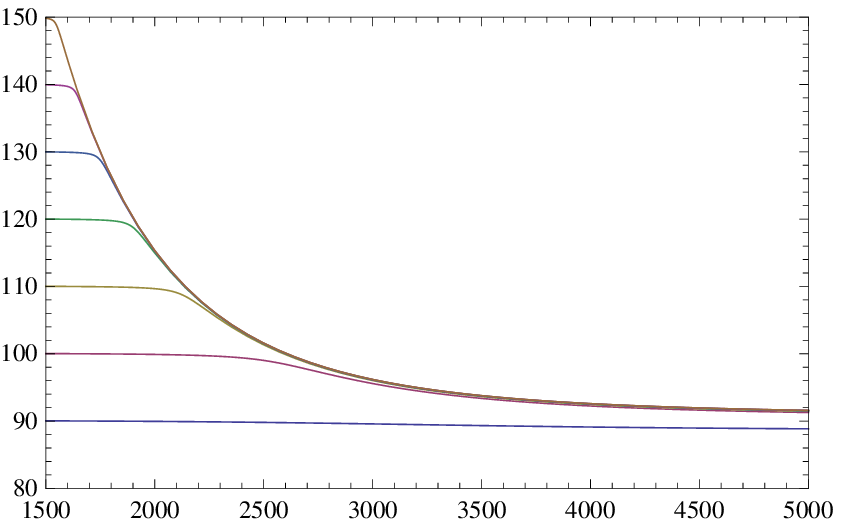}}}
\hspace{4mm}
\parbox{6.6cm}{
\subfloat[{\small $m_H$\,in function of $\sqrt{f}$, $m_A$ parameter}]
{\includegraphics[width=6.6cm]{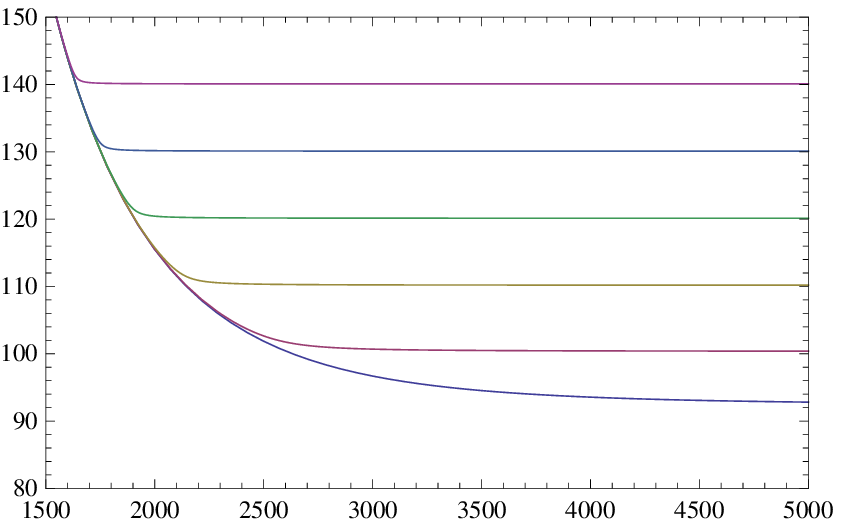}}}
\end{tabular}
\smallskip
\begin{tabular}{cc|cr|}
\parbox{6.6cm}{
\subfloat[{\small $m_h$ in function of $\sqrt{f}$, $\mu$ parameter}]
{\includegraphics[width=6.6cm]{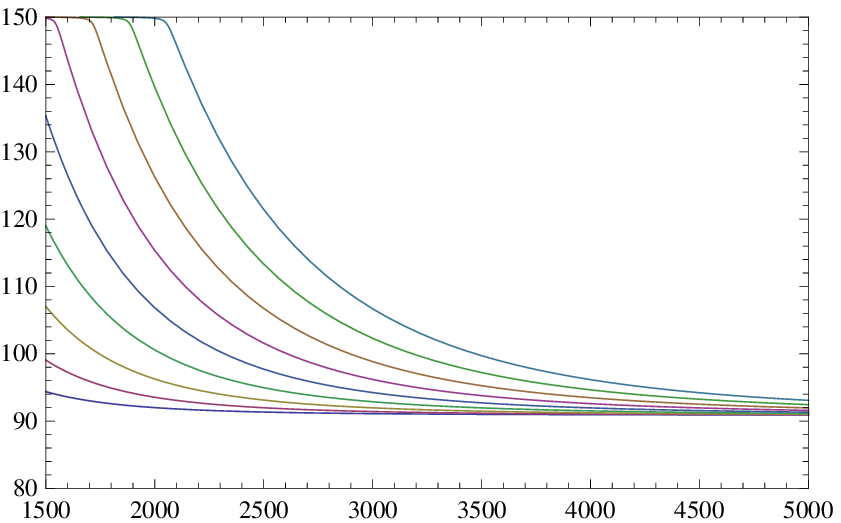}}}
\hspace{4mm}
\parbox{6.6cm}{
\subfloat[{\small $m_h$ in function of $\sqrt{f}$, $\mu$ parameter}]
{\includegraphics[width=6.6cm]{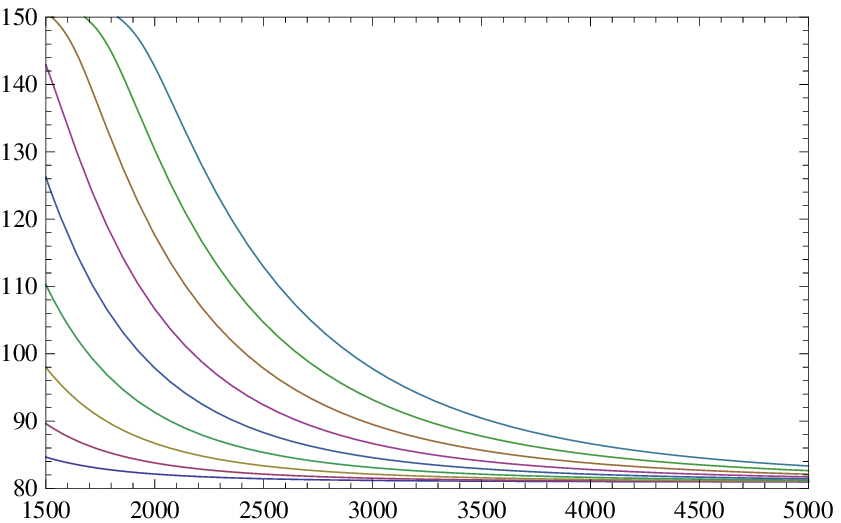}}}
\end{tabular}
\smallskip
\begin{tabular}{cc|cr|}
\parbox{6.6cm}{
\subfloat[{\small $c_1 v^2$ in function of $\sqrt{f}$}]
{\includegraphics[width=6.6cm]{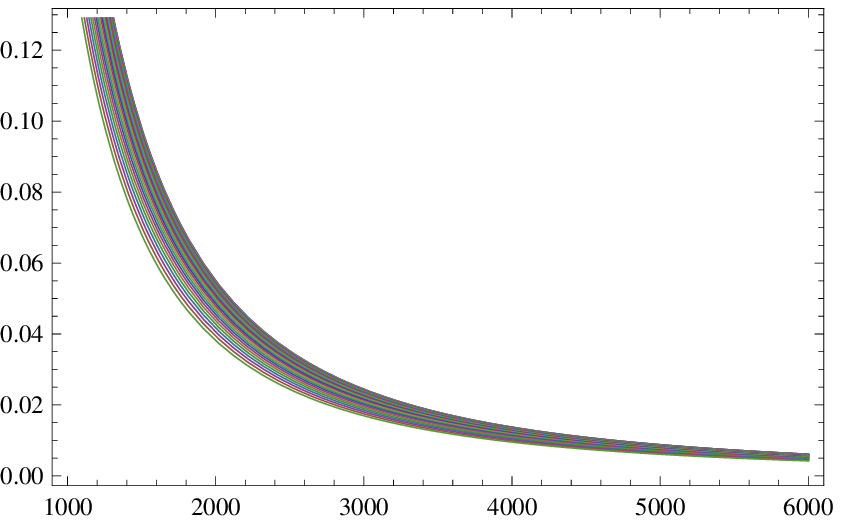}}}
\hspace{4mm}
\parbox{6.6cm}{
\subfloat[{\small $c_2 v^2$ in function of $\sqrt{f}$}]
{\includegraphics[width=6.5cm]{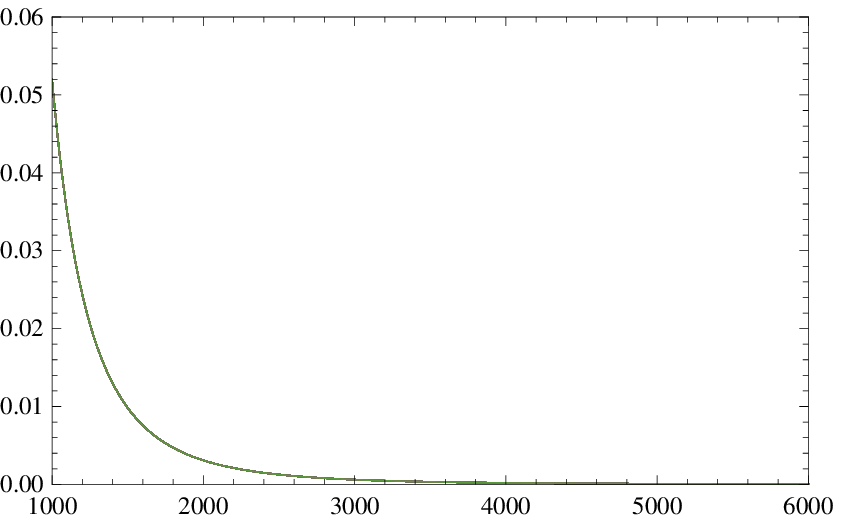}}}
\end{tabular}
\end{center}
\caption{{\protect\small
The tree-level Higgs masses (in GeV) and expansion coefficients as functions of $\sqrt f$ (in GeV). In (a), (b) $\mu=900$ GeV, $\tan\beta=50$, $m_A$ increases upwards from $90$  to $150$ GeV in steps of $10$ GeV. Larger $m_A$ has little impact on $m_h$ for relevant $\sqrt f$. In (c), (d), $m_A=150$ GeV, and $m_h$ increases as $\mu$ varies from 400 to 1200 GeV, in steps of 100 GeV. In (c) $\tan\beta=50$ while in (d) $\tan\beta=5$, showing a milder dependence on $\tan\beta$ than in MSSM. For $\tan\beta\geq 10$ there is little difference from (c). In (e), (f) the expansion coefficients are shown, for $m_A=[90,650]$ GeV with steps of $10$ GeV, $\mu=900$ GeV, $\tan\beta=50$; they are less than unity (even at larger $\mu$), as required for a convergent expansion.}}
\label{higgs1}
\end{figure}

It is instructive to consider the limit of large $u\equiv \tan\beta$, with $B'<0$ fixed, when
\bea
m_h^2\!\!\!&=&\!\!
\Big[m_Z^2+\cO (1/u)\Big]+\frac{v^2}{2\,f^2}\Big[
(2\,\mu^2+m_Z^2)^2+
\frac{4}{u}\,B'\,(2\,\mu^2+m_Z^2)
+\cO(1/u^2)\Big]
+\cO(f^{-3})
\\[3pt]
m_H^2\!\!\!\!&=&\!\!\!
\Big[\frac{-2 B'}{\sin 2\beta}\!+\!\cO(1/u)\Big]\!+\!
\frac{v^2\,B'}{4\,f^2}\,\Big[
(2\,\mu^2+m_Z^2)\,u\!+\! 4\,B'+
\frac{1}{u} (2\,\mu^2\!-\!11 m_Z^2)
\!+\!\cO(1/u^2)\Big]
\!+\!\cO(f^{-3})\nonumber
\eea

\medskip\noindent
which shows that a large $\mu$ can increase $m_h$ (decrease $m_H$).
However, for phenomenology it is customary to use $m_A$ as an input instead of
$B'$, in which case the masses $m_{h,H}$ take the form
\medskip
\bea
m_{h,H}^2& =&
\frac{1}{2}\Big[ m_A^2+m_Z^2\mp \sqrt w\Big]
\pm \frac{v^2}{16 f^2}\frac{1}{\sqrt w}
\Big[
16 m_A^2\mu^4+4 \,m_A^2\,\mu^2\,m_Z^2+(m_A^2-8\,\mu^2)\,m_Z^4
\nonumber\\
&-&
2\,m_Z^6\pm
2\,(-2\,m_A^2\,\mu^2+8\mu^4+4\mu^2\, m_Z^2+m_Z^4)\,
\sqrt{w}+ m_A^2\,m_Z^4\cos 8\beta
\nonumber\\[2pt]
&+&
\!\!
m_A^4\,(m_A^2\!-\! 8\mu^2\!-\! 3 m_Z^2)\sin^2 2\beta
\!+\!
\cos 4\beta\,\big[-2 m_Z^2 \,(8\mu^4\! +\! 4\mu^2\,m_Z^2\! +\!
m_Z^4\!-\! m_A^2 (6\mu^2\!+\! m_Z^2))
\nonumber\\[2pt]
&\pm &
2 \,(2\,m_A^2\mu^2+4\mu^2 m_Z^2 +m_Z^4)\,\sqrt w
-m_A^2 (m_A^2+5 \,m_Z^2)\,\sin^2 2\beta\,
\big]\,\Big]+\cO(1/f^3)
\eea

\medskip\noindent
where the first term (bracket) is just the MSSM contribution.
The upper (lower) signs correspond to $m_h$ ($m_H$) and
$w=(m_A^2+m_Z^2)^2-4\,m_A^2\,m_Z^2\,\cos^2 2\beta$.
At large $\tan\beta$ with $m_A$ fixed one finds\footnote{
In (\ref{lim}) $m_A>m_Z$ is assumed, otherwise just exchange $m^2_{h}$
with $m_H^2$.}  (with $u\equiv \tan\beta$)
\medskip
\bea\label{lim}
m_h^2&=&\Big[m_Z^2+\cO(1/u^2)\Big]
+\frac{v^2}{2\,f^2}\,\Big[(2 \,\mu^2+m_Z^2)^2+\cO(1/u^2)\Big]
+\cO(1/f^3)
\nonumber\\
m_H^2&=&\Big[m_A^2+\cO(1/u^2)\Big]
+\frac{1}{f^2}\,\cO(1/u^2)
+\cO(1/f^3)
\eea
In this limit the increase of $m_h$  is driven by a 
large $\mu$ and is apparently of SUSY origin,
but the quartic Higgs couplings giving this effect
involved combinations of soft masses (see (\ref{potential0})).
 These soft masses combine to give, at the EW minimum, 
the $\mu$-dependent increase in
(\ref{lim}).

Some simple numerical examples are relevant for the size of the corrections
to the Higgs masses, relative to their MSSM values. The largest
correction to $m_h$
for large $\tan\beta$ is  dominated by $\mu$ and $f$.
For example, if $(\mu/\sqrt f)^2=(1/2.25)^2\approx 1/5$, $v=246$ GeV,
with $\mu=900$ GeV then $\sqrt f=2$ TeV, giving $m_h=114.4$ GeV.
Another example is with $\mu=1.2$ TeV, $\sqrt f=2.7$ TeV,  
($(\mu/\sqrt f)^2\approx 1/5$), giving again $m_h=114.4$ GeV.
Smaller  $\mu\approx 600$ GeV can still allow $m_h$ just above the LEP bound
if $\sqrt f=1.35$ TeV, for similar value for $(\mu/\sqrt f)^2=1/5$
and for the rest of the parameters.
This  shows that one can have a {\it classical} value of $m_h$
near or marginally above the LEP bound and larger than the classical
MSSM value ($=m_Z$).
The plots in Figure~\ref{higgs1} illustrate
better this change of  $m_h$ and $m_H$ for various values of $\sqrt f$.
For a low value of $\sqrt f$ near or above $1.35$ TeV, the LEP bound is still
satisfied for $m_h$, while at large $\sqrt f$ the MSSM case is recovered.
By varying  $\sqrt f$ our results can interpolate
between low and high scale (in the hidden sector) SUSY breaking.
Quantum corrections  increase $m_h$ further, just as in the MSSM.

Regarding the usual MSSM tree-level flat direction $\vert
h_1^0\vert=\vert h_2^0\vert$ one can show that the potential in this
direction can have a minimum for the case (not considered in MSSM)
of $m_1^2+m_2^2+2\vert \mu\vert^2< 2\vert B'\vert$, equal to
$V_m=f^2-(1/4)f^2 (m_1^2+m_2^2+2\vert\mu\vert^2+2 B')^2/(m_1^2+m_2^2+B')^2.$
Compared to the usual MSSM minimum, the former can be situated above it
only for values of $f$ which do not comply with the original
assumptions of $m_{1,2}^2, \vert B'\vert<f$.  On the other hand,
the case with $V_m$ situated below the MSSM minimum does not allow
one to recover the MSSM ground state in the decoupling limit of large
$f$, and in conclusion the ``flat'' direction is not of  physical
interest here.

\section{Other Phenomenological Implications}
\label{Otherphenomenologicalimplications}

\subsection{Fine Tuning of the Electroweak Scale}

The increase of $m_h$ beyond the MSSM tree level bound and the
presence of new quartic Higgs couplings have implications in the fine tuning.
In MSSM the smallness of the effective
quartic coupling $\lambda$ (fixed by the gauge sector) is at the
origin of an increased amount of fine tuning of the electroweak scale
for large soft masses.
For soft masses significantly larger than the electroweak (EW) scale,
(also needed  to
increase the MSSM value for $m_h$ above LEP bound via quantum corrections),
fine tuning increases rapidly and may become a potential
problem (sometimes referred to  as the ``little hierarchy'' problem).
Let us see why in the present model
this problem is alleviated. One can write $v^2=-m^2/\lambda$ where
\bea\label{lam}
\lambda &\equiv& \frac{g_1^2+g_2^2}{8}
\Big[\cos^2 2\beta+\delta \sin^4 \beta\Big]
+\frac{1}{f^2}\,\Big\vert
m_1^2\cos^2\beta+m_2^2 \sin^2 \beta +(1/2) \,B'\,\sin  2\beta
\Big\vert^2\nonumber\\[7pt]
m^2&\equiv&  (\vert \mu\vert^2+m_1^2)\cos^2\beta+(\vert \mu\vert^2 +
m_2^2) \sin^2 \beta + \,B'\,\sin 2\beta\
\eea
The first term in $\lambda$ is due to MSSM only, while the
second one, which is positive, is due to the new quartic Higgs terms
in (\ref{potential0}).
Here $\delta$ accounts for
the top/stop quantum effects to $\vert h_2\vert^4$ term in the
potential, which becomes $(1+\delta)\,(g_1^2+g_2^2)/8\,\vert
 h_2\vert^4$; usually
 $\delta\sim \cO(1)$ (ignoring couplings other than top
 Yukawa). This quantum effect is only included for a comparison to
the new quartic Higgs term. The important point to note
is that a larger  $\lambda$ gives a suppression in the
fine tuning measure $\Delta$:
\bea
\Delta=\frac{\partial \ln v^2}{\partial \ln p}
=\frac{\partial \ln (-m^2/\lambda)}{\partial \ln p},
\qquad p=A, B', m_0^2, \mu^2, m_{\lambda_i}^2.
\eea

\medskip\noindent
Here $p$ is an MSSM parameter 
with respect to which fine tuning is evaluated.
In  the large $\tan\beta$ limit, the fine tuning of the electroweak
scale becomes (see the Appendix in \cite{Cassel:2009ps}):
\medskip
\bea
\Delta=-\frac{
(\vert \mu\vert^2+m_2^2)'}{v^2\,{m_2^4}/{f^2}+(1+\delta)\,
{m_Z^2}/{2}}+\cO(1/\tan\beta),\qquad
(\vert \mu\vert^2+m_2)'\equiv \frac{\partial (\vert
  \mu\vert^2+m_2^2)}{\partial \ln p}
 \eea

\medskip\noindent
For small $\tan\beta$ a similar result is obtained in which one
replaces $m_2$ by $m_1$.
The first term in denominator comes from the new correction to
the effective quartic coupling $\lambda$.
Larger soft masses $m_{1,2}$ increase $\lambda$  and
this can  actually reduce fine tuning, see the denominator in $\Delta$.
Therefore, in this case heavier superpartners do not
necessarily bring an increased fine tuning amount (as it usually
happens in the MSSM). The only limitation here is the size of the ratio
$m^2_{1,2}/f\leq 1$ for convergence of the nonlinear formalism. In
the limit this coefficient approaches its upper bound (say $\sim 1/3$),
the two contributions in the denominator have comparable size (for
$\delta\sim 1$ and $v=246$ GeV) and fine tuning is reduced
by a factor $\approx 2$ from that in the absence of the new term in
the denominator ({\it i.e.} the MSSM case).

\subsection{Limiting Cases and Loop Corrections}

Some interesting limits of our ``nonlinear''  MSSM model are worth
considering. Firstly, in the limit of large $f$ 
({\em i.e.} large SUSY breaking scale in the hidden sector) 
and with $m_{1,2}, B'$ fixed, the new quartic term in (\ref{potential0})
vanishes, while the usual explicit soft SUSY breaking terms specific
to the Higgs sector remain. This is just the MSSM case. All other
couplings suppressed by inverse powers of $f$ are negligible in this
limit.
Another limiting case is that of very small $f$. For our analysis to
be valid, one needs to satisfy the condition
$B',\, m_{1,2}^2\leq f.\,\,$
When $f$ reaches this minimal bound, the new quartic couplings in
(\ref{potential0}), not present in the MSSM, increase and eventually
become  closer to unity.
The analysis is then less reliable and additional effective
contributions in the Lagrangian,
suppressed by higher powers like $1/f^4$ and beyond, may  become
relevant for SUSY breaking effects.

Finally, one remark regarding the calculation of
radiative corrections using (\ref{potential0})
and the electroweak symmetry breaking (EWSB).
In our case EWSB was assumed to take place 
by appropriate values of $m_{1,2}^2, B'$. However, 
 the same EWSB mechanism as in the
MSSM is at work here, via quantum corrections to 
these masses, which near the EW
scale turn $m_2^2+\mu^2$ negative and trigger radiative EWSB.
Indeed, if the loops of the MSSM states  are cut off as usual 
at the high GUT scale (well above $\sqrt f$) and with the new Higgs 
quartic  couplings regarded as an effective, classical operator,  
radiative EWSB can take place  as in the MSSM. A similar example is
the case of a MSSM Higgs sector extended  with additional
 effective operators of dimension $d=5$ such as 
$(1/M)\int d^2\theta (H_1 H_2)^2$ giving a dimension-four 
(in fields) contribution to the scalar potential 
$V\supset h_1 h_2\,(\vert h_1\vert^2+\vert h_2\vert^2)$;
this is regarded as an effective operator and radiative EWSB is
 implemented 
 as  in the MSSM, see for example \cite{Dine,Cassel:2009ps}.

It is interesting to remark that that the
loop corrections induced by the (effective) quartic couplings 
proportional to $1/f^2$ in eq.~(\ref{potential0}), can be under control at
large $f$. Indeed, the loop integrals this coupling induces can be 
quadratically divergent and are then cut-off at momentum 
$p^2\leq f$; but the loop effects 
come with  a coupling factor that behaves like $1/f^2$, so overall  they
will be suppressed like $1/f$ and can then be under control 
even at large $f$. It would be interesting to check if for a 
large enough $f$,  radiative EW breaking is still achievable if the usual 
MSSM effects are also cut at this scale (with less an energy range  
to trigger EWSB).

\subsection{Invisible Decays of Higgs and $Z$ Bosons}\label{invisible}

Let us analyze some implications of the interactions involving the
goldstino field, described by the Lagrangian found above.
An interesting possibility, for a light enough neutralino, is the decay
of the neutral higgses into a goldstino and the lightest
neutralino $\chi_1^0$ (this is the NLSP, while the goldstino is the LSP).
The coupling Higgs-goldstino-neutralino is only suppressed by $1/f$. 
It arises from the following terms in $\cL^{new}$ and from the terms
in the on shell, supersymmetric part
of usual MSSM Lagrangian (\ref{mssmsusy}), hereafter denoted
 $\cL_0^{onshell}$:
\medskip
\bea\!\!\!
\cL^{new}+\cL^{onshell}_0\!\!\!
&\supset&\!\!\!\!
-\frac{1}{f}\,\,
\Big[
m_1^2 \,\,\psi_X\psi_{h_1^0}\,h_1^{0\,*}
+m_2^2 \,\,\psi_X\psi_{h_2^0}\,h_2^{0\,*}
\Big]
-\frac{B'}{f}\,\,\Big[\psi_X\psi_{h_2^0}\,h_1^0
+\psi_X\psi_{h_1^0}\,h_2^0\Big]
\nonumber\\
&-&\!\!\!\!\!
\frac{1}{f}\sum_{i=1,2}\,\frac{m_{\lambda_i}}{\sqrt 2}\,
\tilde D_i^a\,\psi_X\lambda_i^a
-
\frac{1}{\sqrt 2}
\Big[g_2\lambda_2^3-g_1\lambda_1\Big]
\Big[h_1^{0\,*}\psi_{h_1^0}
-
h_2^{0\,*}\psi_{h_2^0}\Big]
+h.c.\,\,\,\,\,\,\,\,
\label{tt}
\eea

\medskip\noindent
The last term (present in the MSSM) also brings
a goldstino interaction. This is possible through the goldstino
components of the higgsinos $\psi_{h_{1,2}^0}$
 and EW gauginos $\lambda_{1,2}$.
The goldstino components  are found via the
equations of motion, after EWSB, to give (see also \cite{SK}):
\medskip
\bea\label{Gcom}
\mu\,\psi_{h_1^0}&=&\frac{1}{f\,\sqrt 2}\,\Big(
-m_2^2\,\,v_2-B'\,v_1 -\frac{1}{2}\,\,v_2\,\,
\langle g_2 D_2^3-g_1 D_1\rangle\Big)\,\psi_X+\cdots
\nonumber\\[-2pt]
\mu\,\psi_{h_2^0}&=&\frac{1}{f\,\sqrt 2}\,\Big(
-m_1^2\,\,v_1-B'\,v_2 +\frac{1}{2}\,\,v_1\,\,
\langle g_2 D_2^3-g_1  D_1\rangle\Big)\,\psi_X+\cdots
\nonumber\\
\lambda_1&=&
\frac{-1}{f\,\sqrt 2}\,\langle D_1\rangle\,\,\psi_X+\cdots,\qquad
\lambda_2^3=\frac{-1}{f\,\sqrt 2}\,\langle D_2^3\rangle\,\,\psi_X+\cdots
\eea

\medskip\noindent
which can be further simplified by using the MSSM minimum
conditions in the terms multiplied by $1/f$ (allowed in this
approximation).
As a consistency check we also showed that the determinant of the
neutralino mass matrix  (now a $5\times 5$ matrix, to include the
goldstino)  vanishes up to corrections of order $\cO(f^{-4})$.
This is consistent with our approximation for the Lagrangian,
and verifies the existence of a massless  goldstino
(ultimately ``eaten'' by the gravitino).
Using (\ref{tt}) and (\ref{Gcom}), one finds after
some calculations (for previous calculations of this decay
see \cite{Djouadi:1997gw,hgh,Dimopoulos:1996yq}):

\bea
\cL^{new}+\cL^{onshell}_0
\supset
-\frac{1}{f\sqrt 2}\,\sum_{j,k=1}^4\Big[\,
\psi_X\,\chi_j^0\,H^0 \,\delta_k\,\cX_{jk}^*
+
\psi_X\,\chi_j^0\,h^0 \,\delta_k^\prime\,\cX^*_{jk}\Big]
+h.c.
\eea
where
\bea
\delta_1&=&\,\,\,\,\,m_Z\,\sin\theta_w\,
\big[m_{\lambda_1}\cos(\alpha+\beta)+\mu\sin(\alpha-\beta)\big],
\nonumber\\
\delta_2&=&-
m_Z\cos\theta_w\,
\big[m_{\lambda_2}\cos(\alpha+\beta)+\mu\sin(\alpha-\beta)\big],
\nonumber\\
\delta_3&=&
-m_A^2\sin\beta\,\sin(\alpha-\beta)-\mu^2\cos\alpha
\nonumber\\
\delta_4&=&\,\,\,\,
m_A^2\cos\beta\,\sin(\alpha-\beta)-\mu^2\sin\alpha,
\qquad
\delta_i^\prime=\delta_i\Big\vert_{\alpha\ra \alpha+\pi/2}
\eea

\medskip\noindent
 $\cX$ is the matrix that diagonalizes the MSSM
neutralino mass matrix\footnote{The exact form of $M$ is:
$M_{11}=m_{\lambda_1}$, $M_{12}=0$,
$M_{13}=-m_Z\cos\beta\sin\theta_w$,
$M_{14}=m_Z\sin\beta\sin\theta_w$,
$M_{21}=0$,
$M_{22}=m_{\lambda_2}$,
$M_{23}=m_Z\cos\beta\cos\theta_w$,
$M_{24}=-m_Z\sin\beta\cos\theta_w$,
$M_{33}=0$,
$M_{34}=\mu$, $M_{44}=0$, also $M_{ij}=M_{ji}$. Note
the sign of $\mu$ related to our
definition of the holomorphic product of SU(2) doublets.
With this notation, in the text $\chi_j^0=\cX_{jk}\,\xi_k$,
with $\xi_k^T\equiv (\lambda_1,\lambda^3_2,\psi_{h_1^0},\psi_{h_2^0})$.}:
 $M_d^2=\cX\,M\,M^\dagger\,\cX^\dagger$, and can be easily evaluated
 numerically (see \cite{ElKheishen:1992yv} for its
 analytical expression).
Further $H^0, h^0$ are Higgs mass eigenstates (of mass $m_{h,H}$
 computed earlier) and
$h_i^0=1/\sqrt{2}\,\,(v_i+h_i^{0\, \prime}+i \sigma_i)$ with $\langle
h_i^{0\,\prime}\rangle=0$,  $\langle \sigma_i\rangle=0$; the
 relation of $H^0, h^0$ to $h_{1,2}^{0\,'}$ is a rotation, which in
 this case can be just that of the MSSM (due to extra $1/f$
 suppression in the coupling\footnote{The relation is $h_1^{0\,'}=H^0
\cos\alpha-h^0 \sin \alpha$, and  $h_2^{0\,'}=H^0
\sin\alpha+h^0 \cos \alpha$.}).
The angle $\alpha$ is
\bea
\tan 2\alpha=\tan 2\beta\,\,
\frac{m_A^2+m_Z^2}{m_A^2-m_Z^2},\qquad
-\pi/2 \leq \alpha\leq 0
\eea

\medskip\noindent
If the lightest neutralino is light enough, $m_{\chi_1^0}< m_h$,
 then $h^0, H^0$ can decay into it and a goldstino
 which has a  mass of order $f/M_{Planck}\sim 10^{-3}$ eV; if this is
 not the case, the decay of neutralino into $h^0$ and goldstino
takes place,  examined in \cite{Dimopoulos:1996yq}.
In the former case, the partial decay rate is
\medskip
\bea
\Gamma_{h^0\ra \chi_1^0\,\psi_X}=
\frac{m_h}{16\,\pi\,f^2}\,\,
\Big\vert \sum_{k=1}^4
\delta'_k\,\cX_{1k}\Big\vert^2\,
\,\bigg(1-\frac{m_{\chi_1^0}^2}{m_{h^0}^2}
\,\bigg)^2
\eea

\medskip\noindent
The partial decay rate has corrections coming from both
higgsino ($\cX_{13}$, $\cX_{14}$) and gaugino fields  ($\cX_{11}$,
$\cX_{12}$),  since they both acquire a goldstino
component, see eqs.~(\ref{Gcom}).  The gaugino correction arises after 
gaugino-goldstino mixing, SUSY and EW symmetry breaking,
(as shown by $m_{\lambda_i}$, $m_Z$ dependence in  $\delta_k'$)
and was not included in previous similar studies
\cite{Djouadi:1997gw,hgh,Dimopoulos:1996yq}.

\begin{figure}[t!]
\def\baselinestretch{1.}
\begin{center}
\subfloat[$\Gamma_{h^0\ra\chi\psi_x}$
 in function of $\sqrt{f}$.]
{\includegraphics[width=6.7cm]{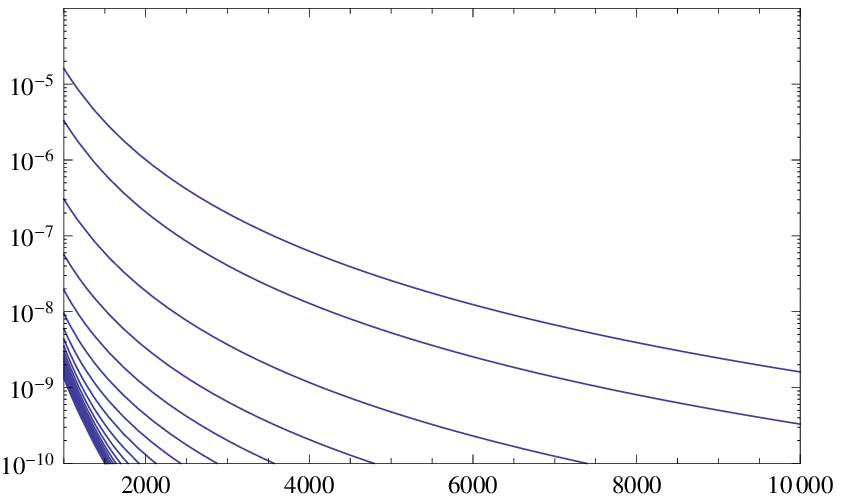}}
\hspace{7mm}
 \subfloat[$\Gamma_{h^0\ra\chi\psi_x}$
 in function of $\sqrt{f}$.]
 {\includegraphics[width=6.7cm]{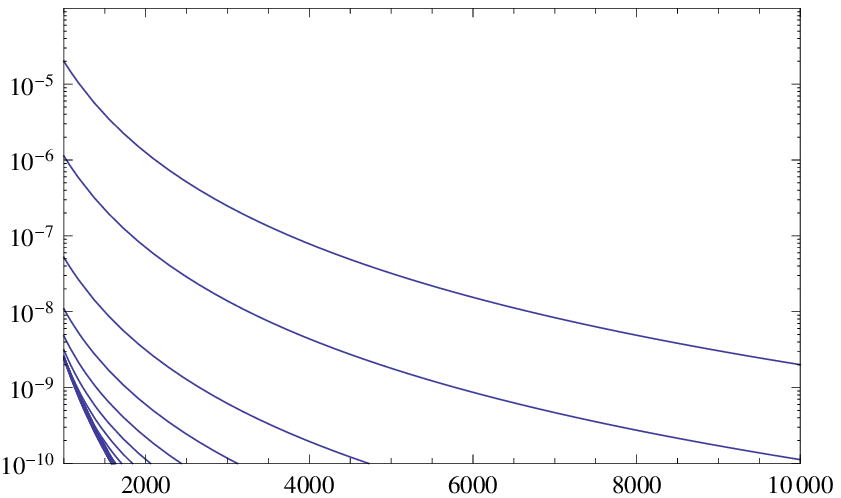}}
\end{center}
\caption{{\protect\small
The partial decay rate of $h^0\ra \psi_X\chi_1^0$ for
(a): $\tan\beta=50$, $m_{\lambda_1}=70$ GeV, $m_{\lambda_2}=150$ GeV,
$\mu$ increases from 50 GeV (top curve) by a step 50 GeV,
$m_A=150$ GeV. Compare against Figure~\ref{higgs1} (c) corresponding
to a similar range for the parameters.
At larger $\mu$, $m_h$ increases, but the partial decay rate decreases.
Similar picture is obtained at low $\tan\beta\sim 5$.
 (b):  As for (a) but with  $\tan\beta=5$. Compare against
 Figure~\ref{higgs1} (d).
 Note that the total SM decay rate, for
$m_h\sim 114$ GeV, is of order $10^{-3}$, thus the branching ratio in
the above cases becomes comparable to that of SM Higgs going into
$\gamma\gamma$  (see Figure 2 in \cite{Djouadi:1997yw}).}}
\label{decayplots}
\end{figure}

The partial decay rate is presented in Figure~\ref{decayplots}
for various values of $\mu$, $m_A$ and $m_{\lambda_{1,2}}$
which are parameters of the model. A larger decay rate requires
a light $\mu\sim \cO(100)$ GeV, when the neutralino $\chi_1^0$ has
a larger higgsino component. At the same time an increase of $m_h$
above the LEP bound requires a larger value for $\mu$,
close to $\mu\approx 700$ GeV if $\sqrt f\approx 1.5 $ TeV, and
$\mu\approx 850$ GeV if $\sqrt f\approx 2$ TeV,
see Figure~\ref{higgs1} (c).  The results in Figure~\ref{decayplots}
show that the partial decay rate can be significant ($\sim 3\times
 10^{-6}$ GeV), if we recall that the total SM Higgs decay rate 
(for $m_h\approx 114$ GeV) is about $3\times 10^{-3}$ GeV,
with a  branching ratio of $h^0\ra \gamma\gamma$ of
 $2\times 10^{-3}$, (Figure 2 in \cite{Djouadi:1997yw}). 
Thus the branching ratio of the process can be
close to that of SM  $h^0\ra \gamma\gamma$.
The decay is not very sensitive to $\tan\beta$ (Figure~\ref{decayplots} (b)),
due to the extra contribution (beyond MSSM) from the quartic Higgs coupling.

An interesting coupling that is also present in the $1/f$ order is
that of goldstino to $Z_\mu$ boson and to a neutralino. Depending on
the relative mass relations, it can bring about a decay of $Z_\mu$
($\chi_j^0$) into $\chi_j^0$ ($Z_\mu$) and a goldstino, respectively. The relevant
terms are
\medskip
\bea\label{sqw}
\cL^{new}+\cL_0^{onshell}&\supset&
-
\frac{1}{4}\,\,\overline \psi_{h_1^0}\overline\sigma^\mu\psi_{h_1^0}
\,\,(g_2 V_2^3-g_1\,V_1)_\mu\,
+
\frac{1}{4}\,\,\overline \psi_{h_2^0}\overline\sigma^\mu\psi_{h_2^0}
\,\,(g_2 V_2^3-g_1\,V_1)_\mu\,\Big\}
\nonumber\\
&-&
\sum_{i=1}^{2}
\frac{m_{\lambda_i}}{\sqrt 2\,\,f}\,\psi_X\,\sigma^{\mu\nu}\,
\lambda_i^a\,F_{\mu\nu,\,i}^a+h.c.
\eea

\medskip\noindent
where the last term was generated in (\ref{lgextra}) ($i$ labels the
gauge group).
Since the higgsinos
acquired a goldstino component ($\propto \psi_X/f$)
via mass mixing, the first line above induces additional $\cO(1/f)$
couplings of the higgsino to goldstino and to
$Z_\mu=(1/g)\,(g_2 V_2^3-g_1 \,V_1)_\mu$ with $g^2=g_1^2+g_2^2$.
After some calculations one finds the coupling $Z_\mu\,\chi_j^0\,\psi_X$:
\medskip\noindent
\bea\label{rrr}
\cL^{new}\!\!+\!\cL_0^{onshell}\!\!
&=&\!
\frac{1}{f\sqrt 2}\!
\sum_{j=1}^4\bigg[
\overline\psi_X\overline\sigma^\mu\,\chi_j^0\,Z_\mu\,
\big(\mu\,m_Z\,w_j\! -\! m_Z^2 v_j\big)\!-\!
\opsi_X (\overline\sigma^\mu\partial^\nu\!-\!
\overline\sigma^\nu\partial^\mu)
\chi_j^0 Z_{\mu\nu} v_j\bigg]\nonumber
\\
&+&\!h.c.
\eea

\medskip\noindent
where
\bea
w_j=\cos\beta\,\cX_{j4}^*-\sin\beta\,\cX_{j3}^*,\quad
v_j=-\sin\theta_w\,\cX_{j1}^*+\cos\theta_w\,\cX_{j2}^*,\quad
Z_{\mu\nu}=\partial_\mu Z_\nu-\partial_\nu Z_\mu
\eea

\medskip\noindent
If $m_{\chi_1^0}$ is lighter than $Z_\mu$ then a decay of the latter
into $\chi_1^0+\psi_X$ is possible.
The decay rate of this process is (with $j=1$):
\medskip
\bea
\Gamma_{Z\ra \psi_X\chi_j^0}&=&
\frac{m_Z^5}{32\pi f^2}
\Big[
\zeta_1\vert w_j\vert^2 +\zeta_2\,\vert v_j\vert^2
+\zeta_3\,(w_j\,v_j^*+w_j^*\,v_j)\Big]
\Big(1-\frac{m_{\chi_j}^2}{m_Z^2}\Big)^2
\label{r4}
\eea

\medskip\noindent
with $\zeta_1=2 (2+r^2)\,\mu^2/m_{Z}^2$,
$\zeta_2= 2(8+r^2)(1+2 r^2)$,
$\zeta_3=-2(4+5 r^2) \mu/m_Z$ where $r=m_{\chi_j}/m_Z$
(in (\ref{rrr}) and subsequent one can actually replace  $\mu$ by $m_{\chi_j}$ 
and $w_j\ra w_j^*$, with $\cX_{j4}\leftrightarrow \cX_{j3}$).

The decay rate  should be within
the LEP error for $\Gamma_Z$, which is
$2.3$ MeV \cite{pdg} (ignoring theoretical uncertainties
 which are small).
 From this, one finds a lower bound for $\sqrt f$, which can be as
 high as $\sqrt f\approx 700$ GeV for the parameter space considered
 previously in Figure~\ref{higgs1}, while  generic values are  $\sqrt
 f\sim \cO(400)$ GeV.
Therefore the results for the  increase of $m_h$, that needed
a value for $\sqrt f$ in the TeV region, escape this constraint.
This constraint does not apply if the lightest neutralino has a mass
larger than $m_Z$, when  the opposite decay ($\chi_j\ra Z\,\psi_X$)
takes  place (this can be arranged for example by a larger $m_{\lambda_1}$). 

There  also exists the interesting possibility of an invisible decay 
of $Z_\mu$ gauge boson into a pair of goldstino fields, that we review
here \cite{SK,Luty:1998np,Brignole:2003cm}. This is induced by the
following terms in the Lagrangian, after the Higgs field acquires a
VEV:
\medskip
\bea\label{coupl}\!\!\!\!\!\!
\cL^{new}+\cL^{onshell}_0\!\!\!\!
&\supset &\Big\{
\frac{1}{4\,f^2}\,\,
\overline\psi_X\overline\sigma^\mu\psi_X\,\,
(g_2 V_2^3-g_1\,V_1)_\mu\,(m_1^2 \,v_1^2/2-m_2^2 \,v_2^2/2)
\nonumber\\
&-&\!\!\!\!
\frac{1}{4}\,\overline \psi_{h_1^0}\overline\sigma^\mu\psi_{h_1^0}
(g_2 V_2^3-g_1\,V_1)_\mu
+
\frac{1}{4}\,\overline \psi_{h_2^0}\overline\sigma^\mu\psi_{h_2^0}
(g_2 V_2^3-g_1\,V_1)_\mu\,\Big\}+h.c.\quad
\eea

\bigskip\noindent
With (\ref{Gcom}) and (\ref{coupl})
one finds the coupling of $Z$ boson to
a pair of goldstinos:
\medskip\noindent
\bea
\cL^{new}+\cL^{onshell}_0
\supset
\frac{m_Z^2}{4\,f^2}\,\,
\overline\psi_X\,\overline\sigma^\mu\,\psi_X\,\,Z_\mu\,
\langle D_Z  \rangle+h.c.
\eea

\medskip\noindent
where
$\langle D_Z\rangle \equiv \cos\theta_W\,\,\langle
D_2^3\rangle-\sin \theta_W\, \langle D_1\rangle
=-(m_Z^2/{g})\cos 2\beta+\cO(1/f)
$. The decay rate is then
\bea
\Gamma_{Z\ra \psi_X\psi_X}
&=&
\frac{m_Z}{24\, \pi\,g^2}
\bigg[\frac{m_Z^4}{2\,f^2}\bigg]^2
\cos^2 2\beta
\eea

\medskip\noindent
in agreement with previous results obtained for $B'=0$
\cite{Luty:1998np,SK,Brignole:2003cm}.
The decay rate is independent of $m_A$ and should be within
the LEP error for $\Gamma_Z$
($2.3$ MeV \cite{pdg}).
One can then easily see that the increase of the Higgs
mass above the LEP bound (114.4 GeV) seen earlier in
Figure~\ref{higgs1} is consistent with the  current bounds for this
decay rate, which thus places only  mild constraints on $f$, below the
TeV scale ($\approx 200$ GeV) \cite{Luty:1998np,Brignole:2003cm}.

Similarly,
 $\cL^{new}$ can also induce Higgs decays into goldstino
pairs. The terms in $\cL^{new}$
that contribute to Higgs decays are $\cL_{F\,(2)}^{aux}$,
$\cL_D^{aux}$, $\cL_m^{extra}$ together with
the MSSM higgsino-Higgs-gaugino  coupling (last term in (\ref{tt})).
After using (\ref{Gcom}), expanding the Higgs
fields about their v.e.v., one finds:
\medskip
\bea\label{lcr}
\cL^{new}+\cL^{onshell}_0\supset
\frac{\mu\,v}{4\,f^2}\,m_A^2\,\cos
2\beta\,\,\opsi_X\opsi_X\,\Big[h_1^{0\,\prime}\sin\beta
-  h_2^{0\,\prime}\,\cos\beta\Big]
+h.c.+\cO(1/f^3)
\eea

\medskip\noindent
which, similarly to $Z$ couplings, is independent of gaugino masses.
Here $v=246$ GeV and
$h_i^0=1/\sqrt{2}\,\,(v_i+h_i^{0\, \prime}+i \sigma_i)$, $\langle
h_i^{0\,\prime}\rangle=0$,  $\langle \sigma_i\rangle=0$.
In the mass eigenstates basis one simply replaces
the square bracket in (\ref{lcr}) by $\big[
H^0\sin(\beta-\alpha)-h^0\,\cos(\beta-\alpha)\big]$.
One can also replace $m_A$ by
$m_A^2=m_h^2+m_H^2-m_Z^2+\cO(1/f^2)$,
where the Higgs masses can be taken to be  the MSSM values
(up to higher order corrections in $1/f$).
The decay rate of $h^0$ into a
pair of goldstinos is then
\bea
\Gamma_{h^0\ra
  \psi_X\psi_X}=\frac{m_h}{8\pi\,f^4}\,g_{h^0\psi_X\psi_X}^2
\eea

\medskip\noindent
where $g_{h^0\psi_X\psi_X}$ is the coupling of $h^0\psi_X\psi_X$ of
the above Lagrangian. For relevant values of $f$ above $\sim$1 TeV it turns
out that this decay rate is very small  relative to
other partial decay rates of the Higgs in the MSSM/SM. For example, for
a total decay rate near $10^{-3}$ GeV (valid near a Higgs
mass of order $\cO(100)$ GeV), the branching ratio of this decay mode is
well below  the usual ones and below that of SM Higgs going
 into $\gamma\gamma$, by a factor $\approx 10^{-3}-10^{-2}$.

\chapter{Summary of Results}

This part of the thesis consists of two different effective analyses in the context of MSSM.

In the first one, covered in chapters \ref{MSSM5}, \ref{PhenomenologyMSSM5} and \ref{MSSMHiggs56}, we considered an extension by the complete set of R-parity conserving, mass dimension 5 operators for the MSSM and by dimension 5 and 6 for its Higgs sector. This set included all supersymmetric and supersymmetry breaking terms, the latter being incorporated by the use of spurions. Some of these operators are not physical since they can be related to each other by field redefinitions. We performed the appropriate, spurion dependent, redefinitions that allowed us to write down the full irreducible set of dimension 5 and dimension 6 operators. We further restricted the parameter space by applying phenomenological constraints, in particular from flavor changing neutral currents. We then studied the phenomenological consequences of the model both in the production of new couplings and in the mass of the Higgs.

The new couplings include ``wrong" Higgs Yukawa terms which are also generated at one loop in pure MSSM. One significant effect of these terms is the $\tan\beta$ enhancement of the mass of the bottom quark. If the scale of the effective operators is at the multiTeV scale, the effective contribution is comparable or even bigger than the loop contribution. We also found couplings of type 2 quarks - 2 squarks and 2 quarks - 2 sleptons. These are also relevant for LHC since they contribute to processes of squark production. The corresponding pure MSSM channels become weaker for higher collision energy, contrary to the effective contribution which is simply suppressed by $1/M$. The two can become comparable for energy of the TeV scale as in LHC.

The effective analysis presented offers a solution to the little hierarchy problem of MSSM. This problem is related to the fact that the tree level calculation for $m_h$ in pure MSSM reveals an upper bound, equal to $m_Z=91.2$ GeV, which is in complete disagreement with the lower bound of 114 GeV from the LEPII experiment. The only way to overpass this discrepancy in pure MSSM is to suppose significant loop corrections implying very heavy stops or large stop mixing. In any case fine tuning is reintroduced and this is what we call the little hierarchy problem. However, we showed that effective operators can significantly raise the mass of the Higgs thus reducing the fine tuning. This result suggests an alternative interpretation of the little hierarchy. Instead of viewing it as a deficiency of MSSM, it can be viewed as an indication for new physics at the multiTeV range.

The second effective analysis, presented in chapter \ref{NonlinearMSSM}, is not related to some ``new physics" but to the SUSY breaking sector. Models of low energy SUSY breaking predict a very light gravitino. In the low energy regime, the dynamics of the gravitino can be accurately described by the dynamics of its goldstino component. So if the breaking scale is around TeV, apart from the pure MSSM spectrum we need to include the goldstino mode.

The effective description of the goldstino mode is done via nonlinear realization of supersymmetry. There are various ways to study such systems. We chose the language of ``constrained superfields" as the most general and easy to reproduce the couplings of goldstinos to MSSM fields. We wrote the full set of couplings and studied their phenomenological significance. One important effect is again related to the mass of the Higgs. It is shown that the presence of a goldstino can also increase $m_h$ providing us with yet another way to alleviate the little hierarchy, even without the hypothesis of new physics. Furthermore, we found that invisible decays of Higgs to goldstinos and other neutralinos can be of comparable size with the standard decay to two photons. Finally, assuming that the lightest neutralino is lighter than the Z gauge boson, we got a bound on the SUSY breaking scale of around 700 GeV from invisible Z boson decays.


\part{\textsc{Brane - Bulk Interactions in \\ N = 2 Global and Local Supersymmetry}}

\chapter{Preliminaries}\label{secintro}

\section{The Dirac Born Infeld Action as the Effective Action of a D-brane}

In 1934, a few years before the development of Quantum Electrodynamics, M. Born and L. Infeld proposed a generalization of Maxwell's electrodynamics that was free of the notorious divergence in the self-energy of the electron \cite{BornInfeld}. Their inspiration derived from how Special Relativity (SR) accommodated what they called ``the principle of finiteness'', that consistent theories should not allow physical quantities to become infinite.

In SR, the Newtonian kinetic energy of a particle is replaced by a function that imposes an upper limit in the velocity.

\be
{1\over 2}mv^2 \quad\rightarrow\quad mc^2\left(1-\sqrt{1-{v^2\over c^2}}\right).
\ee

The deeper reason behind this replacement is the \textit{principle of relativity}, that the kinetic action be invariant under Lorentz transformations. Born and Infeld suggested a similar replacement for electrodynamics
\be\label{BIold}
{1\over 2}(H^2-E^2)\quad\rightarrow\quad b^2\left(\sqrt{1+{1\over b^2}(H^2-E^2)}-1 \right),
\ee
where $b$ is a constant with the same dimension as the fields. They supported it by constructing a general expression for the Lorentz invariant action of a tensor field $A_{\mu\nu}$. In a few lines, this is what they did: Under a coordinate transformation, the measure $d^4x$ becomes $Jd^4x$ and the determinant $|A|$ becomes $J^{-2}|A|$, where $J$ is the Jacobian of the transformation. It is obvious then that $\sqrt{|A|}\,d^4x$ forms an invariant piece. As for any arbitrary tensor, we can split $A_{\mu\nu}$ into a sum of its symmetric and antisymmetric part. The symmetric part was identified with the metric $g_{\mu\nu}$ and the antisymmetric with the field strength $F_{\mu\nu}$. A general expression for an invariant Lagrangian is then:
\be
\cL=\sqrt{-|g+F|}+\alpha\sqrt{-|g|}+\beta\sqrt{-|F|} \ .
\ee
However, the last piece is a total derivative and can be ignored. Also, $\alpha=-1$ by the requirement that we reduce to Maxwell's electrodynamics in the limit of small fields. After restoring dimensions we find that in flat space the Lagrangian takes the form:
\be
\cL = b^2\left(1-\sqrt{-|\eta+{F\over b}|} \right)=b^2\left(1-\sqrt{1+{F_{\mu\nu}F^{\mu\nu}\over 2b^2}-{(F^{\mu\nu}\tilde{F}_{\mu\nu})^2\over 16b^4}} \right),
\ee
where $\tilde{F}_{\mu\nu}=\epsilon_{\mu\nu\rho\sigma}F^{\rho\sigma}/2$ is the dual field strength. We see that their derivation led to the suggested action (\ref{BIold}) up to the piece $F\tilde{F}$ that does not affect the resolution of the electron self energy problem. In fact, we see that the electric field $E$ has a maximum value $b$, in direct correspondence to the maximum velocity $c$ of a particle in SR. As a consequence, the electric potential at zero distance doesn't diverge as $1/r$ but rather takes a maximum value proportional to $\sqrt{b/e}$, with $e$ the electron charge.

The Born Infeld (BI) action offered an ingenious solution to the apparent divergence of the electric field at short distances. However, it was a classical solution to a problem that is purely quantum mechanical. The advent of Quantum Electrodynamics and renormalizable quantum field theories in the following years resolved, beyond many other things, the self energy problem.

Little attention was paid to the BI action until 50 years later. In a paper by E. Fradkin and A. Tseytlin in 1984, it was shown that the low energy effective action for open bosonic strings propagating in a background of constant field strength is given precisely by the BI action \cite{FradkinTseytlin}. The same action is obtained in the superstrings case, too \cite{PopeTownsend, MetsaevTseytlin}. In this framework, the maximal value `b' of the field strength is interpreted as the string tension $T=1/2\pi\alpha'$. At such extreme values, higher harmonics of the string can be excited and thus the energy of the field is transferred into these modes. In a way, the extended nature of strings {\it smears}  the singularity. This was a remarkable discovery as it provided a closed expression where $\alpha'$ corrections are summed up to all orders.

The connection with D-branes, which were discovered some years later, didn't take long to reveal. It was soon demonstrated that the effective action for the coupling of a D-brane with NSNS bulk fields is given by the Dirac-Born-Infeld (DBI) action \cite{Leigh}. The DBI action is merely a dimensional reduction of a generalization of BI action to include the coupling to the dilaton and the antisymmetric tensor. The effective action of a D-brane was extended after it was discovered that these non perturbative objects break half of the bulk supersymmetries and act as sources for the RR fields of the closed string spectrum \cite{Polchinski}. This introduced a second piece in the effective action given by Wess-Zumino terms \cite{Douglas}. All in all, the bosonic part of the world-volume effective action of a Dp-brane at the string tree level is given by:
\be
S_{Dp}=-T_p \int d^{p+1}x\, e^{-\phi}\left(\sqrt{-|g|}-\sqrt{-|g+2\pi\alpha'F+B|}\right)+\mu_p\int \sum_l e^{2\pi\alpha'F+B}\wedge C_l
\ee
at string frame. $T_p$ is the brane tension, $\mu_p$ is the brane's charge for the various RR fields denoted by $C_l$ (so $l$ is even in type IIB and odd in type IIA) while B is the NSNS 2-form.

In the previous paragraph we mentioned that D-branes are objects that break half of the bulk supersymmetries and that their low energy effective action is described by the DBI action. To be more precise, it has been shown that the broken half of the supersymmetry is realized nonlinearly on the worldvolume of the D-brane. These facts lead us to the following question: Is it possible to apply the tools of nonlinear realizations developed in the previous part of the thesis, in order to ``reproduce" the low energy effective action of a D-brane? In the following chapters we show that for the general case of $\cN=2$ bulk supersymmetry, it is. We do this by defining appropriate $\cN=2$ superfields and then upgrading the constrained superfields technique to $\cN=2$ superspace. The constraint breaks one supersymmetry leaving one linear and one nonlinear in the effective theory. The result comes out to be precisely the DBI action plus Wess - Zumino terms.

\section{Quaternion-K\"ahler and Hyper-K\"ahler Manifolds}

Supersymmetric Lagrangians of interacting matter typically contain complicated, field dependent terms in their kinetic part forming a nonlinear $\sigma$ model. An efficient way to study the structure of the allowed couplings is to view the fields as coordinates of a Riemannian manifold. Restrictions that supersymmetry imposes on the couplings are translated into restrictions on the corresponding manifold of the $\sigma$ model.

$\cN=1$ global supersymmetry requires that the manifold of hypermultiplet scalars is K\"ahler while for $\cN=1$ supergravity it is further restricted to be Hodge. Adding one more supersymmetry brings further conditions: The scalar manifold in global $\cN=2$ is restricted to be hyperK\"ahler while in local $\cN=2$ it is quaternion-K\"ahler. Since we will focus on $\cN=2$ supersymmetric models, we briefly present some basic facts about these two manifolds.

A quaternion-K\"ahler manifold is a 4n real dimensional K\"ahler manifold with holonomy contained in $Sp(2)\times Sp(2n)$. It has three complex structures
\be
J^iJ^k=-\delta^{ik}\mathbb{I}+\epsilon^{ikl}J^l
\ee
with $i,k,l=1,2,3$ and a hermitian metric such that, for each $i$
$$
g_{\alpha\beta}\,J^{i\,\alpha}_\kappa \,J^{i\,\beta}_\lambda=g_{\kappa\lambda}\ .
$$
It is also Einstein, which means that its Ricci tensor is proportional to the metric:
$$
R_{\alpha\beta}=2\rho(2+n)g_{\alpha\beta}\ .
$$
and is strictly non-vanishing. In addition, it has a self-dual Weyl curvature (Weyl tensor is the traceless component of Riemann tensor). In 4D (n=1) the holonomy is $Sp(2)\times Sp(2)\sim SO(4)$ so the holonomy condition is empty. In this case the proper condition is self-duality of the Weyl tensor. In $\cN=2$ nonlinear $\sigma$ models coupled to SUGRA the Einstein parameter is identified as $\rho=-k^2$ where $k^2=8\pi G_N$ ($G_N$ is Newton's constant). In the zero curvature limit ($k\rightarrow 0$) we obtain global supersymmetry and a manifold which is Ricci-flat (Ricci tensor is zero).

By properly taking the global supersymmetry limit in a SUGRA theory with matter couplings, we should reduce to some global matter coupling theory which, as we mentioned, is described by a hyper-K\"ahler manifold. Hyper-K\"ahler manifolds are defined as the 4n real dimensional, connected, Riemannian manifolds whose holonomy group is contained in $Sp(2n)$. All hyper-K\"ahler manifolds are also K\"ahler and Ricci-flat, that is $R_{\mu\nu}=0$. This matches with the zero curvature limit of the quaternion-K\"ahler. However, hyper-K\"ahler are not a subclass of quaternion-K\"ahler whose Ricci scalar and $Sp(2)$ connection are strictly non-zero.

\section{Superspace Conventions}\label{A1}

The notation used henceforth is somewhat different from the one of the previous part, being more suitable for the work done here. We present the notation as well as some ingredients that will be proven useful in the following chapters.

The $\cN=1$ supersymmetry variation of a superfield $V_1$ is
$\delta V_1 = ( \epsilon Q + \ov\epsilon\ov Q )V_1$, with supercharges verifying the algebra
\be
\label{conv3}
\{ÊQ_\alpha, \ov Q_\dalpha\} =  -2i (\sigma^\mu)_{\alpha\dalpha} \, \partial_\mu .
\ee
On $V_1$, the supersymmetry algebra is
\be
\label{conv4}
[ \delta_1 , \delta_2 ] V_1 = -2i \, ( \epsilon_1\sigma^\mu\ov\epsilon_2 
- \epsilon_2\sigma^\mu\ov\epsilon_1 ) \, \partial_\mu V_1.
\ee
The covariant derivatives
\be
\label{conv5}
D_\alpha = \frac{\partial}{\partial \theta^\alpha} - i(\sigma^\mu\ov\theta)_\alpha
\, \partial_\mu \, , 
\qquad\qquad
\ov D_\dalpha = \frac{\partial}{\partial \ov\theta^\dalpha} 
- i(\theta\sigma^\mu)_\dalpha \, \partial_\mu 
\ee
anticommute with supercharges and verify
\be
\label{conv6}
\{ÊD_\alpha, \ov D_\dalpha\} =  -2i (\sigma^\mu)_{\alpha\dalpha} \, \partial_\mu 
\ee
as well.

The second supersymmetry will transform $V_1$ into another superfield $V_2$ and these two will form an $\cN=2$ supermultiplet. It is known that the covariant derivatives themselves offer a good differential realization of the supersymmetry algebra; this is easily seen here by comparing (\ref{conv3}) and (\ref{conv6}). So we choose to realize the second supersymmetry algebra on the covariant derivatives by postulating the following transformations:
\be
\delta^* V_1=-{i\over \sqrt{2}}(\eta D+\ov{\eta D})V_2\ , \qquad \delta^*V_2=i\sqrt{2}(\eta D+\ov{\eta D})V_1 \ .
\ee
where $\eta_\alpha$ is the spinorial parameter of the second supersymmetry. What we have presented here is the realization of $\cN=2$ supersymmetry in terms of $\cN=1$ superfields. We will see later that for our purposes, we can also define $\cN=2$ chiral superfields, which will be very useful in simplifying various expressions.

The $\cN=1$ supersymmetry variations of the components $(z,\psi,f)$ of a chiral superfield 
$\Phi$, $\ov D_\dalpha\Phi=0$, are
\be
\label{conv12}
\begin{array}{rcl}
\delta z &=& \sqrt 2 \, \epsilon\psi \, , 
\crbig
\delta \psi_\alpha &=& -\sqrt2 \, [ f \epsilon_\alpha  
+ i(\sigma^\mu\ov\epsilon)_\alpha \partial_\mu z] \, ,
\crbig
\delta f &=& -\sqrt2 \, i \, \partial_\mu\psi\sigma^\mu\ov\epsilon.
\end{array}
\ee
The bosonic expansions of the chiral superfields that will appear later are:
\be
\label{convexp}
\begin{array}{rcl} 
W_\alpha(y,\theta) &=& \theta_\alpha d(y)
+{i\over 2}(\theta\sigma^\mu\ov{\sigma}^\nu)_\alpha F_{\mu\nu}(y) ,
\crbig
\chi_\alpha(y,\theta) &=& -{1\over 4}\theta_\alpha C(y)
+{1\over 4}(\theta\sigma^\mu\ov{\sigma}^\nu)_\alpha\, b_{\mu\nu}(y),
\crbig
\Phi(y,\theta) &=& \phi(y)-\theta\theta f_\phi(y),
\end{array}
\ee
and any other chiral superfield has an expansion similar to $\Phi$. In this notation $\ov{\chi}_{\dalpha}=(\chi_\alpha)^*$ but $\ov{W}_\dalpha=-(W_\alpha)^*$. Since 
$L = D^\alpha\chi_\alpha - \ov D_\dalpha\ov\chi^\dalpha$, the linear superfield has
bosonic expansion 
\be
\label{conv7}
\begin{array}{l}
L(x,\theta,\ov\theta) = C + \theta\sigma^\mu\ov\theta v_\mu + 
{1\over4}\theta\theta\ov{\theta\theta} \, \Box C, 
\crbig \hspace{2.9cm}
v_\mu = {1\over 2}\epsilon_{\mu\nu\rho\sigma}\partial^\nu b^{\rho\sigma}
= {1\over 2}\epsilon_{\mu\nu\rho\sigma}\partial^{[\nu}b^{\rho\sigma]}
= {1\over6} \epsilon_{\mu\nu\rho\sigma}H^{\nu\rho\sigma}.
\end{array}
\ee
With these expansions, 
$$
\Dint \left[- L^2 + {1\over2}(\Phi+\ov\Phi)^2 \right]
$$
is the Lagrangian of a free, canonically-normalized, single-tensor $\cN=2$ multiplet. Its bosonic
content is
$$
{1\over2} (\partial_\mu C)(\partial^\mu C) + {1\over12} H_{\mu\nu\rho}H^{\mu\nu\rho},
\qquad\qquad
H_{\mu\nu\rho} = 3\, \partial_{[\mu}b_{\nu\rho]}.
$$
For more details on the single tensor multiplet see section \ref{secST}.

The identities
\be
DD\,\theta\theta = \ov{DD}\,\ov{\theta\theta} = -4, 
\qquad\qquad
\Dint = -\frac{1}{4}\Fint \ov{DD} = -\frac{1}{4}\Fbarint DD,
\ee
only valid under a space-time integral $\int d^4x$, are commonly used. Also,
$$
\begin{array}{ll}
D_\alpha D_\beta = \frac{1}{2}\epsilon_{\alpha\beta} DD ,
\qquad\qquad
&\ov D_\dalpha \ov D_\dbeta = -\frac{1}{2}\epsilon_{\dalpha\dbeta}\ov{DD} ,
\crbig
[ D_\alpha , \ov{DD} ] = -4i(\sigma^\mu\ov D)_\alpha \partial_\mu ,
\qquad\qquad
&[ \ov D_\dalpha , DD ] = +4i(D\sigma^\mu)_\dalpha \partial_\mu ,
\crbig
DD\, W_\alpha = 4i(\sigma^\mu\partial_\mu\ov W)_\alpha ,
\qquad\qquad
&\ov{DD}\, \ov W_\dalpha = - 4i(\partial_\mu W\sigma^\mu)_\dalpha   .
\end{array}
$$

\chapter{The Linear N = 2 Maxwell-Dilaton System}\label{seclinear}

Our first objective is to describe, in the context of linear $\cN=2$ supersymmetry, 
the coupling of the single-tensor multiplet to $\cN=2$ super-Maxwell theory. Since these
two supermultiplets admit off-shell realizations, they can be described in superspace without reference 
to a particular Lagrangian.  Gauge transformations of the Maxwell multiplet use a single-tensor
multiplet, we then begin with the latter.

\section{The Single-Tensor Multiplet}\label{secST}

In global $\cN=1$ supersymmetry, a real antisymmetric tensor field $b_{\mu\nu}$ is described
by a chiral, spinorial superfield $\chi_\alpha$ with $8_B+8_F$ fields \cite{S}\footnote{The 
notation $m_B+n_F$
stands for `$m$ bosonic and $n$ fermionic fields'.}:
\be
\label{ST2}
\chi_\alpha = - {1\over4} \theta_\alpha (C+iC^\prime) 
+ {1\over4}(\theta\sigma^\mu\ov\sigma^\nu)_\alpha \, b_{\mu\nu} + \ldots
\qquad\qquad
(\,\ov D_\dalpha\chi_\alpha=0\,),
\ee
$C$ and $C^\prime$ being the real scalar partners of $b_{\mu\nu}$. 
The curl
$h_{\mu\nu\rho} = 3\, \partial_{[\mu} b_{\nu\rho]}$ is described by the real superfield
\be
\label{ST1}
L=D^\alpha \chi_\alpha -\ov D_\dalpha \ov\chi^\dalpha. 
\ee
Chirality of $\chi_\alpha$ implies linearity of $L$: $DDL=\ov{DD}L=0$. The linear superfield $L$
is invariant under the supersymmetric gauge transformation\footnote{$\Delta$ is an arbitrary real superfield.}
\be
\label{ST3}
\chi_\alpha \quad\longrightarrow\quad \chi_\alpha + {i\over4}\ov{DD}D_\alpha \Delta, \qquad\qquad
\ov\chi_\dalpha \quad\longrightarrow\quad \ov\chi_\dalpha + {i\over4}DD \ov D_\dalpha \Delta,
\ee
of $\chi_\alpha$: this is the supersymmetric extension of the invariance of $h_{\mu\nu\rho}$ under
$\delta b_{\mu\nu} = 2\,\partial_{[\mu}\Lambda_{\nu]}$. Considering bosons only, the gauge
transformation (\ref{ST3}) eliminates three of the six components of $b_{\mu\nu}$ and the scalar 
field $C^\prime$. Accordingly, $L$ only depends on the invariant curl $h_{\mu\nu\rho}$ and on the 
invariant real scalar $C$. The linear $L$ describes then $4_B+4_F$ fields.
Using either $\chi_\alpha$ or $L$, we will find two descriptions of the single-tensor multiplet of global
$\cN=2$ supersymmetry \cite{dWvH, LR, Ketal}.

In the gauge-invariant description using $L$, the $\cN=2$ multiplet is completed with a
chiral superfield $\Phi$ ($8_B+8_F$ fields in total). The second supersymmetry transformations (with parameter $\eta_\alpha$) are
\be
\label{ST4}
\begin{array}{rcl}
\delta^* L &=& -\frac{i}{\sqrt 2} (\eta D\Phi +\ov{\eta D}\ov \Phi) \,, 
\crbig
\delta^* \Phi &=&  i \sqrt2 \, \ov{\eta D}L \,,\qquad\qquad \delta^* \ov\Phi \,\,=\,\, i \sqrt2 \,\eta D L \,,
\end{array}
\ee
where $D_\alpha$ and $\ov D_\dalpha$ are the usual $\cN=1$ supersymmetry derivatives verifying
$\{ D_\alpha , \ov D_\dalpha \} = -2i (\sigma^\mu)_{\alpha\dalpha}\partial_\mu$. It is easily verified
that the $\cN=2$ supersymmetry algebra closes on $L$ and $\Phi$. 

We may try to replace $L$ by $\chi_\alpha$ with second supersymmetry transformation 
$\delta^*\chi_\alpha = -{i\over\sqrt2}\Phi\,\eta_\alpha$, as suggested when comparing eqs.~(\ref{ST1})
and (\ref{ST4}). However, with superfields $\chi_\alpha$ and $\Phi$ only, the $\cN=2$ algebra only closes
up to a gauge transformation (\ref{ST3}). This fact, and the unusual number $12_B+12_F$
of fields, indicate that $(\chi_\alpha,\Phi)$ is a gauge-fixed version of the off-shell 
$\cN=2$ multiplet. We actually need another chiral $\cN=1$ superfield $Y$ to close the 
supersymmetry algebra. The second supersymmetry variations are
\be
\label{ST5}
\begin{array}{rcl}
\delta^* Y &=& \sqrt2\, \eta\chi \, , 
\crbig
\delta^* \chi_\alpha &=& -{i\over\sqrt2} \Phi\,\eta_\alpha - {\sqrt2\over4} \eta_\alpha \, \ov{DD}\, \ov Y
-\sqrt2 i (\sigma^\mu\ov\eta)_\alpha \partial_\mu Y \, , 
\crbig
\delta^* \Phi &=&  2\sqrt2 i \left[\frac{1}{4}\,\ov{DD\eta\chi} 
+ i \partial_\mu\chi\sigma^\mu\ov\eta \right] .
\end{array}
\ee
One easily verifies that the $Y$--dependent terms in $\delta^*\chi_\alpha$ induce a gauge transformation
(\ref{ST3}). Hence, the linear $L$ and its variation $\delta^*L$ do not feel $Y$.  
The superfields $\chi_\alpha$, $\Phi$ and $Y$ have $16_B+16_F$ field components. Gauge 
transformation (\ref{ST3}) eliminates $4_B+4_F$ fields. To further eliminate $4_B+4_F$ fields,
a new gauge variation 
\be
\label{ST6}
Y \qquad\longrightarrow\qquad Y - {1\over2}\ov{DD} \Delta^\prime,
\ee
with $\Delta^\prime$ real, is then postulated. We will see below that this variation is actually dictated by 
$\cN=2$ supersymmetry. There exists then a gauge in which $Y=0$ but in this gauge the 
supersymmetry algebra closes on $\chi_\alpha$ only up to a transformation (\ref{ST3}).
This is analogous to the Wess-Zumino gauge of $\cN=1$ supersymmetry, but in our case, 
this particular gauge respects $\cN=1$ supersymmetry and gauge symmetry (\ref{ST3}).

Two remarks should be made at this point. Firstly, the superfield $Y$ will play an important role in the
construction of the Dirac-Born-Infeld interaction with nonlinear $\cN=2$ supersymmetry. As we will
see later on\footnote{See section \ref{secV1XY}.}, it includes a four-index antisymmetric tensor field 
in its highest component. Secondly, a constant ($\theta$--independent) background value 
$\langle\Phi\rangle$ breaks the second supersymmetry only, $\delta^*\chi_\alpha = 
-{i\over\sqrt2}\langle\Phi\rangle \eta_\alpha+\ldots\,\,$ It is a natural source of partial supersymmetry 
breaking in the single-tensor multiplet. Notice that the condition $\delta^*\langle\Phi\rangle =0$
is equivalent to $\ov D_\dalpha(D\chi-\ov{D\chi})=0$.

An invariant kinetic action for the gauge invariant single-tensor multiplet involves an arbitrary 
function solution of the 
three-dimensional Laplace equation (for the variables $L$, $\Phi$ and $\ov\Phi$) \cite{LR}:
\be
\label{ST7}
{\cal L}_{ST} = \Dint {\cal H } (L, \Phi, \ov\Phi) \,, \qquad\qquad
{\partial^2{\cal H }\over\partial L^2} + 2 {\partial^2{\cal H }\over\partial\Phi\partial\ov\Phi} =0.
\ee
In the dual hypermultiplet formulation the Laplace equation is replaced by a Monge-Amp\`ere equation.
We will often insist on theories with axionic shift symmetry $\delta\Phi = i c$ ($c$ real), dual to a double-tensor theory. In this case, ${\cal H}$ is a function of $L$ and $\Phi+\ov\Phi$ so that the general solution of Laplace equation is
\be
\label{ST8}
{\cal L}_{ST} = \Dint \, H({\cal V}) + {\rm h. c.} ,
\qquad\qquad
{\cal V} = L + {i\over\sqrt2} (\Phi+\ov\Phi ),
\ee
with an arbitrary analytic function $H({\cal V})$.

The single-tensor multiplet as well as its Poincar\'e duals will play a central role in what follows. For this reason in Appendix \ref{Equivalents} we give a detailed presentation of these multiplets and the duality transformations that switch from one to the other.

\section{The Maxwell Multiplet, Fayet-Iliopoulos Terms} \label{secMaxwell}

Take two real vector superfields $V_1$ and $V_2$. Variations
\be
\label{Max1}
\delta^* V_1 = -\frac{i}{\sqrt2}\Bigl[ \eta D + \ov{\eta D} \Bigr] V_2 \, , \qquad\qquad
\delta^* V_2 = \sqrt2 i \Bigr[ \eta D + \ov{\eta D} \Bigr] V_1
\ee
provide a representation of $\cN=2$ supersymmetry with $16_B+16_F$ fields. We may reduce
the supermultiplet by imposing on $V_1$ and $V_2$ constraints consistent with the second
supersymmetry variations: for instance, the single-tensor multiplet is obtained by requiring 
$V_1=L$ and $V_2=\Phi+\ov\Phi$. Another option is to impose a gauge invariance: we may impose that
the theory is invariant under\footnote{For clarity, we use the following convention for field variations:
$\delta^*$ refers to the second ($\cN=2$) supersymmetry variations of the superfields and component
fields; $\delta_{U(1)}$ indicates
the Maxwell gauge variations; $\delta$ appears for gauge variations of superfields or field 
components related (by supersymmetry) to $\delta b_{\mu\nu} = 2\,\partial_{[\mu} \Lambda_{\nu]}$.}
\be
\label{Max2}
\delta_{U(1)}\,V_1 = \Lambda_\ell \,, \qquad\qquad 
\delta_{U(1)}\,V_2 = \Lambda_c + \ov\Lambda_c \,,
\ee
where $\Lambda_\ell$ and $\Lambda_c$ form a single-tensor multiplet,
\be
\label{Max3}
\Lambda_\ell = \ov\Lambda_\ell \,, \qquad\qquad DD\Lambda_\ell=0,  \qquad\qquad
\ov D_\dalpha \Lambda_c=0,
\ee
with transformations (\ref{ST4}). Defining the gauge invariant superfields\footnote{Remember that
with this (standard) convention, $\ov W_\dalpha$ is {\it minus} the complex conjugate of $W_\alpha$.}
\be
\label{Max4}
\begin{array}{rclrcl}
W_\alpha &=& -\frac{1}{4}\,\ov{DD}D_\alpha \, V_2 \,, \qquad&\qquad
\ov W_\dalpha &=& -\frac{1}{4}\, DD \ov D_\dalpha \, V_2 \,,
\crbig
X &=& {1\over2}\, \ov{DD}\, V_1\,, & \ov X &=& {1\over2}\, DD\, V_1,
\end{array}
\ee
the variations (\ref{Max1}) imply\footnote{There is a phase choice in the definition of $X$: a
phase rotation of $X$ can be absorbed in a phase choice of $\eta$.}
\be
\label{Max5}
\begin{array}{l}
\delta^*X = \sqrt 2 \, i \, \eta^\alpha W_\alpha ,
\qquad\qquad\qquad\qquad
\delta^*\ov X = \sqrt 2 \, i \, \ov\eta_\dalpha \ov W^\dalpha ,
\crbig
\delta^* W_\alpha =  \sqrt 2 \, i \left[ \frac{1}{4}\eta_\alpha \ov{DD}\,\ov X 
+ i (\sigma^\mu\ov\eta)_\alpha \, \partial_\mu X \right] ,
\crbig
\delta^* \ov W_\dalpha = \sqrt 2 \, i \, \left[ \frac{1}{4} \ov\eta_\dalpha {DD}\, X 
- i (\eta\sigma^\mu)_\dalpha \, \partial_\mu \ov X \right].
\end{array}
\ee
While $(V_1,V_2)$ describes the $\cN=2$ supersymmetric extension of the gauge potential 
$A_\mu$, $(W_\alpha, X)$ is the multiplet of the gauge curvature $F_{\mu\nu} =
2\,\partial_{[\mu} A_{\nu]}$ \cite{Maxmult}.

The $\cN=2$ gauge invariant Lagrangian depends on the derivatives of a holomorphic prepotential 
${\cal F}(X)$:
\be
\label{Max6}
\begin{array}{rcl}
{\cal L}_{Max.} &=& {1\over4}\Fint \Bigl[ {\cal F}^{\prime\prime}(X)WW - {1\over2}{\cal F}^\prime(X)
\ov{DD}\,\ov X \Bigr] + {\rm c.c.}
\crbig
&=& {1\over4}\Fint {\cal F}^{\prime\prime}(X)WW + {\rm c.c.}
+ {1\over2} \Dint \Bigl[ {\cal F}^\prime(X)\ov X + \ov{\cal F}^\prime(\ov X) X \Bigr] 
+\partial_\mu(\ldots).
\end{array}
\ee

In the construction of the Maxwell multiplet in terms of $X$ and $W_\alpha$, one expects a triplet of 
Fayet-Iliopoulos terms,
\be
\label{Max7}
{\cal L}_{F.I.}  = - {1\over4} (\xi_1+ ia)\Fint X - {1\over4} (\xi_1-ia)\Fbarint \ov X + \xi_2 \Dint  V_2 ,
\ee
with real parameters $\xi_1$, $\xi_2$ and $a$. They may generate background values of the auxiliary
components $f_X$ and $d_2$ of $X$ and $V_2$ which in general break both supersymmetries:
\be
\label{Max8}
\delta^* X= \sqrt2i \, \eta\theta\, \langle d_2 \rangle + \ldots,
\qquad\qquad
\delta^*W_\alpha =\sqrt2i \, \eta_\alpha\, \langle \ov f_X \rangle + \ldots
\ee
In terms of $V_1$ and $V_2$ however, the relation $X= {1\over2}\ov{DD} V_1$ implies
that $\Im f_X$ is the curl of a three-index antisymmetric tensor (see section \ref{secV1XY}) and
that its expectation value is turned into an integration constant of the tensor field equation 
\cite{ANT, ATT}. As a consequence, 
$$
- {1\over4} (\xi_1+ia)\Fint X - {1\over4} (\xi_1-ia)\Fbarint \ov X = \xi_1 \Dint V_1 + {\rm derivative}
$$
and the Fayet-Iliopoulos Lagrangian becomes
\be
\label{Max9}
{\cal L}_{F.I.}  = \Dint [\xi_1 V_1 + \xi_2 V_2],
\ee
with two real parameters only. 

The Maxwell multiplet with superfields $(X,W_\alpha)$ and the single-tensor multiplet
$(Y,\chi_\alpha,\Phi)$ have a simple interpretation in terms of chiral superfields on $\cN=2$ 
superspace. We will use this formalism to construct their interacting Lagrangians in section
\ref{secchiralN=2}.

\section{The Chern-Simons Interaction} \label{secCS}

With a Maxwell field $F_{\mu\nu} = 2\,\partial_{[\mu} A_{\nu]}$ (in $W_\alpha$) and an antisymmetric 
tensor $b_{\mu\nu}$ (in $\chi_\alpha$ or $L$), one may expect the presence of a 
$b\wedge F$ interaction
$$
\epsilon^{\mu\nu\rho\sigma} b_{\mu\nu}F_{\rho\sigma} =
2\,\epsilon^{\mu\nu\rho\sigma} A_\mu\partial_{\nu}b_{\rho\sigma} + {\rm derivative}.
$$
This equality suggests that its $\cN=2$ supersymmetric extension also exists in two forms: 
either as an integral over chiral superspace of an expression depending on $\chi_\alpha$,
$W_\alpha$, $X$, $\Phi$ and $Y$, or as a real expression using $L$, $\Phi+\ov\Phi$, 
$V_1$ and $V_2$.

In the `real' formulation, the $\cN=2$ Chern-Simons term is\footnote{The dimensions in mass unit of our superfields are as follows: $V_1,V_2 : 0$~, $X, Y:1$~, $W_\alpha, \chi_\alpha: 3/2$~, $\Phi,L:2$. The coupling constant $g$ is then dimensionless. }
\be
\label{CS1}
{\cal L}_{CS} = -g \Dint \Bigl[ LV_2 + (\Phi+\ov\Phi) V_1 \Bigr],
\ee
with a real coupling constant $g$.
It is invariant (up to a derivative) under the gauge transformations (\ref{Max2}) of $V_1$ and $V_2$
with $L$ and $\Phi$ left inert. Notice that the introduction of Fayet-Iliopoulos terms for $V_1$ and 
$V_2$ corresponds respectively to the shifts $\Phi+\ov\Phi\rightarrow\Phi+\ov\Phi - \xi_1/g$ and
$L\rightarrow L - \xi_2/g$ in the Chern-Simons term.  

The `chiral' version uses the spinorial prepotential $\chi_\alpha$ instead of $L$.
Turning expression (\ref{CS1}) into a chiral integral and using $X={1\over2}\ov{DD}\,V_1$ leads 
to 
\be
\label{CS3}
{\cal L}_{CS,\,\chi} = g\Fint  \Bigl[ \chi^\alpha W_\alpha + {1\over2} \Phi X  \Bigr]
+ g\Fbarint \Bigl[ -\ov\chi_\dalpha\ov W^\dalpha + {1\over2} \ov\Phi \ov X  \Bigr] ,
\ee
which differs from ${\cal L}_{CS}$ by a derivative. The chiral version of the Chern-Simons 
term ${\cal L}_{CS,\chi}$ transforms as a derivative under the gauge variation (\ref{ST3}) of 
$\chi_\alpha$. Its invariance under constant shift symmetry of $\Im\Phi$ follows from 
$X={1\over2}\ov{DD}\,V_1$. It does not depend on $Y$.

The consistent Lagrangian for the Maxwell -- single-tensor system with Chern-Simons interaction
is then
\be
\label{CS4}
{\cal L}_{ST} + {\cal L}_{Max.} + {\cal L}_{CS}
\qquad\qquad{\rm or}\qquad\qquad
{\cal L}_{ST} + {\cal L}_{Max.} + {\cal L}_{CS,\,\chi}.
\ee
The first two contributions include the kinetic terms and self-interactions of the multiplets
while the third describes how they interact. Each of the three terms is separately $\cN=2$ supersymmetric.

Using a $\cN=1$ duality, a linear multiplet can be transformed 
into a chiral superfield with constant shift symmetry and the opposite transformation of course exists. Hence, performing both transformations, a single-tensor multiplet Lagrangian $(L,\Phi)$ with constant shift symmetry of the chiral $\Phi$ has a `double-dual' second version. Suppose that
we start with a Lagrangian where Maxwell gauge symmetry acts as a St\"uckelberg gauging
of the single-tensor multiplet:\footnote{Strictly speaking, the coupling constant $g$ in this theory
has dimension (energy)$^2$. There is an irrelevant energy scale involved in the duality transformation
of a dimension two $L$ into a dimension two chiral superfield. Hence, $g$ in eq.~(\ref{B4}) is again
dimensionless. }
\be
\label{B1}
{\cal L} = \Dint {\cal H}(L-gV_1, \Phi+\ov\Phi - gV_2) .
\ee
The shift symmetry of $\Im\Phi$ has been gauged and ${\cal L}$ 
is invariant under gauge transformations (\ref{Max2}) combined with
\be
\label{B3}
\delta_{U(1)} L = g\Lambda_\ell \,, \qquad\qquad
\delta_{U(1)} \Phi = g\Lambda_c \,,
\ee
and under $\cN=2$ supersymmetry if ${\cal H}$ verifies Laplace equation (\ref{ST7}). 
If we perform a double dualization $(L,\Phi+\ov\Phi)\rightarrow (\tilde\Phi + \ov{\tilde\Phi},
\tilde L)$, we obtain the dual theory 
\bea
\label{B4}
\tilde{\cal L}&=& \Dint \tilde{\cal H}(\tilde L, \tilde\Phi+\ov{\tilde\Phi}) 
+g \Fint \left[ \tilde\chi^\alpha W_\alpha + {1\over2}\tilde\Phi X \right] + {\rm c.c.}
\\ \nonumber
&=&\Dint \left[ \tilde{\cal H}(\tilde L, \tilde\Phi+\ov{\tilde\Phi}) - g\tilde LV_2 \right]
+{g\over2} \Fint \tilde\Phi X + {\rm c.c.}
\eea
where $\tilde{\cal H}$ is the result of the double Legendre transformation
\be
\label{B5}
\tilde{\cal H} (\tilde y, \tilde x) = {\cal H}(x,y) - \tilde xx - \tilde yy.
\ee
The dual theory is then the sum of the ungauged Lagrangian (\ref{ST7}) and of the Chern-Simons 
coupling (\ref{CS1}). This {\it single-tensor -- single-tensor} duality is actually $\cN=2$ covariant: if 
${\cal H}$ solves Laplace equation, so does $\tilde{\cal H}$, and every intermediate step of
the duality transformation can be formulated with explicit $\cN=2$ off-shell supersymmetry.

We have then found two classes of couplings of Maxwell theory to the single-tensor multiplet.
Firstly, using the supersymmetric extension of the $b\wedge F$ coupling, as in eqs.~(\ref{CS4}).
Secondly, using a St\"uckelberg gauging (\ref{B1}) of the single-tensor kinetic terms. The first
version only is directly appropriate to perform an electric-magnetic duality transformation. However, 
since the second version can always be turned into the first one by a 
single-tensor -- single-tensor duality, electric-magnetic duality of the second version requires this 
preliminary step: both theories have the same `magnetic' dual.

\section{The Significance of $V_1$, $X$ and $Y$} 
\label{secV1XY}

In the description of the $\cN=2$ Maxwell multiplet in terms of two $\cN=1$ real superfields, 
$V_2$ describes as usual the gauge potential $A_\mu$, a gaugino 
$\lambda_\alpha$ and a real auxiliary field $d_2$ (in Wess-Zumino gauge). We wish to clarify
the significance and the field content of the superfields $V_1$ and $X={1\over2} \ov{DD}V_1$, as 
well as the related content of the chiral superfield $Y$ used in the description in 
terms of the spinorial potential $\chi_\alpha$ of the single-tensor multiplet $(Y,\chi_\alpha,\Phi)$. 

The vector superfield $V_1$ has the $\cN=2$ Maxwell gauge variation 
$\delta_{U(1)} V_1 = \Lambda_\ell$, with
a real linear parameter superfield $\Lambda_\ell$. In analogy with the Wess-Zumino gauge 
commonly applied to $V_2$, there exists then a gauge where 
\be
\label{C1}
V_1(x,\theta,\ov\theta) = \theta\sigma^\mu\ov\theta\, v_{1\mu} -{1\over2} \theta\theta \, \ov x -{1\over2} \ov{\theta\theta} \, x
-{1\over\sqrt2} \theta\theta\ov{\theta\psi}_X -{1\over\sqrt2} \ov{\theta\theta}\theta\psi_X 
+{1\over2} \theta\theta\ov{\theta\theta} \, d_1.
\ee
This gauge leaves a residual invariance acting on the vector field $v_{1\mu}$ only:
\be
\label{C1a}
\delta_{U(1)} v_1^\mu=\frac{1}{2}\epsilon^{\mu\nu\rho\sigma}\partial_\nu \Lambda_{\rho\sigma} \, .
\ee
This indicates that the vector $v_1^\mu$ is actually a three-index antisymmetric tensor,
\be
\label{C2}
v_1^\mu = {1\over6}\epsilon^{\mu\nu\rho\sigma} A_{\nu\rho\sigma} ,
\ee
with Maxwell gauge invariance 
\be
\label{C3}
\delta_{U(1)} A_{\mu\nu\rho} = 3\,\partial_{[\mu}\Lambda_{\nu\rho]}.
\ee
By construction, $X = {1\over2}\ov{DD}V_1$ is gauge invariant. In chiral variables,
\be
\label{C4}
X(y,\theta) = x + \sqrt2 \, \theta\psi_X - \theta\theta (d_1+i\partial_\mu v_1^\mu).
\ee
Hence, while $\Re f_X = d_1$,
\be
\label{C5}
\Im f_X = \partial_\mu v_1^\mu = {1\over24}\epsilon^{\mu\nu\rho\sigma} F_{\mu\nu\rho\sigma},
\qquad\qquad
F_{\mu\nu\rho\sigma} = 4\, \partial_{[\mu}A_{\nu\rho\sigma]}
\ee
is the gauge-invariant curl of $A_{\mu\nu\rho}$. It follows that the field content (in Wess-Zumino gauge)
of $V_1$ is the second gaugino $\psi_X$, the complex scalar of the Maxwell multiplet $x$, a real
auxiliary field $d_1$ and the three-form field $A_{\mu\nu\rho}$, which corresponds to a single, 
non-propagating component field. The gauge-invariant chiral $X$ includes the four-form curvature 
$F_{\mu\nu\rho\sigma}$.

At the Lagrangian level, the implication of relations  (\ref{C5}) is as follows. Suppose that we compare
two theories with the same Lagrangian ${\cal L}(u)$ but either with $u=\phi$, a real scalar, or with
$u=\partial_\mu V^\mu$, as in eq.~(\ref{C5}). Since ${\cal L}(\phi)$ does not depend on $\partial_\mu\phi$, the scalar
$\phi$ is auxiliary. The field equations for both theories are
$$
{\partial\over\partial\phi}{\cal L}(\phi) =0 , \qquad\qquad
\partial_\nu \left. {\partial\over\partial u}{\cal L}(u) \right|_{u= \partial_\mu V^\mu} = 0
$$
The second case allows a supplementary integration constant $k$ related to the possible addition of
a `topological' term proportional to $\partial_\mu V^\mu$ to the Lagrangian \cite{ANT, ATT}:
$$
\left. {\partial\over\partial u}{\cal L}(u) \right|_{u= \partial_\mu V^\mu} = k.
$$
In the first case, the same integration constant appears if one considers the following modified 
theory and field equation:
$$
{\cal L}(\phi) - k\,\phi  \qquad\longrightarrow\qquad {\partial\over\partial\phi}{\cal L}(\phi) = k.
$$
Returning to our super-Maxwell case, the relation is $\phi=\Im f_X$ and the modification of the 
Lagrangian is then
\be
\label{C6}
-k \Im f_X = - {ik\over2} \Fint X + {\rm c.c.}
\ee
This is the third Fayet-Iliopoulos term, which becomes a `hidden parameter' \cite{ANT} when
using $V_1$ instead of $X$. 

Consider finally the single-tensor multiplet $(Y, \chi_\alpha, \Phi)$ and the supersymmetric extension
of the antisymmetric-tensor gauge symmetry, as given in Eqs.~(\ref{ST3}) and (\ref{ST6}):
$$
\delta Y = -{1\over2}\ov{DD} \Delta^\prime, \qquad\qquad
\delta \chi_\alpha =  {i\over4}\ov{DD} D_\alpha \Delta, \qquad\qquad
\delta \Phi =0.
$$
Using expansion (\ref{C4}), there is a gauge in which $Y$ reduces simply to
\be
\label{C7}
Y = -i \, \theta\theta\, \Im f_Y
\ee
and one should identify $\Im f_Y$ as a four-index antisymmetric tensor field, 
\be
\label{C8}
\Im f_Y = {1\over24} \, \epsilon^{\mu\nu\rho\sigma}C_{\mu\nu\rho\sigma},
\ee
with residual gauge invariance
\be
\label{C9}
\delta \, C_{\mu\nu\rho\sigma} = 4\, \partial_{[\mu}\Lambda_{\nu\rho\sigma]}.
\ee
The antisymmetric tensor $C_{\mu\nu\rho\sigma}$ describes a single field component 
which can be gauged away using
$\Lambda_{\nu\rho\sigma}$. Applying this extended Wess-Zumino 
gauge to the $\cN=2$ multiplet $(Y,\chi_\alpha,\Phi)$, the fields described by these $\cN=1$
superfields are as given in the following table. 

\begin{center}
\begin{tabular}{|c|c|c|c|}
\hline
$\cN=1$ superfield & Field & Gauge invariance & Number of fields \\
\hline 
$\chi_\alpha$ & $b_{\mu\nu}$ & $\delta b_{\mu\nu} = 2\,\partial_{[\mu}\Lambda_{\nu]}$
& $6_B-3_B=3_B$
\\ 
& $C$ && $1_B$
\\
& $\chi_\alpha$ && $4_F$
\\
$\Phi$ & $\Phi$ && $2_B$
\\
& $f_\Phi$ && $2_B$ (auxiliary)
\\
& $\psi_\Phi$ && $4_F$
\\
$Y$ & $C_{\mu\nu\rho\sigma}$ & 
$\delta \, C_{\mu\nu\rho\sigma} = 4\, \partial_{[\mu}\Lambda_{\nu\rho\sigma]}$
& $1_B-1_B=0_B$
\\
\hline
\end{tabular}
\end{center}
The propagating bosonic fields $b_{\mu\nu}$, $C$ and $\Phi$ (four bosonic degrees of freedom) 
have kinetic terms defined by Lagrangian ${\cal L}_{ST}$, eq.~(\ref{ST7}).

\section{Chiral N = 2 Superspace} \label{secchiralN=2}

Many results of the previous section can be reformulated in terms of chiral superfields on 
$\cN=2$ superspace. We now turn to a discussion of this framework, including an explicitly 
$\cN=2$ covariant formulation of electric-magnetic duality.

\subsection{Chiral N = 2 Superfields}\label{subchiralN=2}

A chiral superfield on $\cN=2$ superspace can be written as a function of $y^\mu,
\theta, \tilde\theta$:
\be
\label{chiral1}
\ov D_\dalpha \,{\cal Z} = \ov {\widetilde D}_\dalpha \,{\cal Z} = 0 \qquad\longrightarrow\qquad
{\cal Z} =  {\cal Z}(y,\theta,\tilde\theta)
\ee
with $y^\mu = x^\mu - i\theta\sigma^\mu\ov\theta - i\tilde\theta\sigma^\mu\ov{\tilde\theta}$ and 
$\ov D_\dalpha \,y^\mu = \ov {\widetilde D}_\dalpha \,y^\mu = 0$.
Its second supersymmetry variations are
\be
\label{chiral2}
\delta^*{\cal Z} = i(\eta\tilde Q + \ov \eta\ov{\tilde Q}){\cal Z},
\ee
with supercharge differential operators $\tilde Q_\alpha$ and $\ov{\tilde Q}_\dalpha$
which we do not need to explicitly write.
It includes four $\cN=1$ chiral superfields and $16_B+16_F$ component fields and we may use
the expansions
\be
\label{chiral2b}
\begin{array}{rcl}
{\cal Z}(y,\theta,\tilde\theta) &=& Z(y,\theta) + \sqrt2 \, \tilde\theta^\alpha\omega_\alpha(y,\theta)
 -\tilde\theta\tilde\theta F(y,\theta)
\crbig
&=& Z(y,\theta) + \sqrt2 \, \tilde\theta^\alpha\omega_\alpha(y,\theta)
 -\tilde\theta\tilde\theta \left[ {i\over2}\Phi_{\cal Z}(y,\theta) + {1\over4}\ov{DD}\, \ov Z(y,\theta) \right],
\end{array}
\ee
where $\tilde\theta$ and $\widetilde D_\alpha$ are the Grassmann coordinates and
the super-derivatives associated with the second supersymmetry. The second supersymmetry 
variations (\ref{chiral2}) are easily obtained by analogy with 
the $\cN=1$ chiral supermultiplet:
\be
\label{chiral2c}
\begin{array}{rcl}
\delta^* Z &=& \sqrt2 \, \eta\omega , 
\crbig
\delta^* \omega_\alpha &=& -\sqrt2 [ F \eta_\alpha
+ i (\sigma^\mu\ov\eta)_\alpha \, \partial_\mu Z ]
\,\,=\,\,
-{i\over\sqrt2} \Phi_{\cal Z}\,\eta_\alpha - {\sqrt2\over4} \eta_\alpha \, \ov{DD}\, \ov Z
-\sqrt2 i (\sigma^\mu\ov\eta)_\alpha \partial_\mu Z ,
\crbig
\delta^* F &=& -\sqrt2i\,\partial_\mu\omega\sigma^\mu\ov\eta,
\crbig
\delta^*\Phi_{\cal Z} &=& 2\sqrt2 i \left[\frac{1}{4}\,\ov{DD\eta\omega} 
+ i \partial_\mu\omega\sigma^\mu\ov\eta \right] .
\end{array}
\ee
We immediately observe that the second expansion (\ref{chiral2b}) leads to the second 
supersymmetry variations (\ref{ST5}) of a single-tensor multiplet $(Y=Z, \chi=\omega, 
\Phi=\Phi_{\cal Z})$. Similarly, the expansion
\be
\label{chiral4}
{\cal  W}(y, \theta, \tilde \theta) = X(y,\theta) + \sqrt2 i \,\tilde\theta W(y,\theta) 
- \tilde\theta\tilde\theta\, {1\over4}\ov{DD} \ov X (y,\theta),
\ee
which is obtained by imposing $\Phi_{\cal Z}=0$ in expansion (\ref{chiral2b}), leads to the 
Maxwell supermultiplet (\ref{Max5}) \cite{GSW}. 
The Bianchi identity $D^\alpha W_\alpha = \ov D_\dalpha \ov W^\dalpha$ is required by 
$\delta^*\Phi_{\cal Z}=0$.
The $\cN=2$ Maxwell Lagrangian (\ref{Max6}) rewrites then as an integral over chiral $\cN=2$ superspace,
\be
\label{chiral5}
{\cal L}_{Max.} = {1\over2}\Fint \int d^2\tilde\theta\, {\cal F}({\cal W})
+ {\rm c. c.},
\ee
and the Fayet-Iliopoulos terms (\ref{Max9}) can be written \cite{IZ}
\be
\label{DBI12}
{\cal L}_{F.I.} = \Dint [ \xi_1 V_1 + \xi_2V_2]
= -{1\over4}\Fint\int d^2\tilde\theta \left[ \tilde\theta\tilde\theta\,\xi_1 -\sqrt2 i \,\theta\tilde\theta \, \xi_2 
\right] {\cal W} + {\rm c.c.}
\ee 

Considering the unconstrained chiral superfield (\ref{chiral2b}) with $16_B+16_F$ fields, the reduction
to the $8_B+8_F$ components of the single-tensor multiplet is done by imposing gauge invariance
(\ref{ST3}) and (\ref{ST6}). In terms of $\cN=2$ chiral superfields, this gauge symmetry is simply
\be
\label{chiral6}
\delta {\cal Y} =  - \widehat {\cal W}, 
\ee
where $\widehat {\cal W}$ is a Maxwell $\cN=2$ superfield parameter (\ref{chiral4}). In terms of $\cN=1$ superfields,
this is 
\be
\label{chiral7}
\delta Y =  - \widehat X, \qquad\qquad 
\delta \chi_\alpha =  - i \widehat W_\alpha, \qquad\qquad 
\delta \Phi =  0,
\ee
as in eqs.~(\ref{ST3}) and (\ref{ST6}). Hence, a single-tensor superfield ${\cal Y}$ is a chiral
superfield ${\cal Z}$ with the second expansion (\ref{chiral2b}) and with gauge symmetry (\ref{chiral6}).

The chiral version of the Chern-Simons interaction (\ref{CS3}) can be easily written on 
$\cN=2$ superspace. Using ${\cal Y}$ with gauge invariance (\ref{chiral6}) and ${\cal W}$ to 
respectively describe the single-tensor and the Maxwell multiplets. Then 
\be
\label{chiral8}
{\cal L}_{CS,\chi} = ig \Fint \int d^2\tilde\theta\, {\cal Y}{\cal W} + {\rm c.c.}
\ee
It is gauge-invariant since for any pair of Maxwell superfields
\be
\label{chiral9}
i\Fint \int d^2\tilde\theta\, {\cal W}\widehat{\cal W} + {\rm c.c.}= {\rm derivative}.
\ee
Notice that the lowest component superfield $Y$ of ${\cal Y}$ does not contribute
to the field equations derived from ${\cal L}_{CS,\chi}$: it only contributes to this Lagrangian with a derivative.

Finally, a second method to obtain an interactive Lagrangian for the Maxwell--single-tensor system is then obvious. Firstly, a generic $\cN=2$ chiral superfield ${\cal Z}$ can always be written as
\be
\label{chiral10}
{\cal Z} = {\cal W} + 2g {\cal Y}.
\ee
It is invariant under the single-tensor gauge variation (\ref{chiral6}) if one also postulates that
\be
\label{chiral11}
\delta {\cal W} =  2g\,\widehat {\cal W},
\ee
which amounts to a $\cN=2$ St\"uckelberg gauging of the symmetry of the antisymmetric tensor.
With this decomposition, $F_{\mu\nu}$ and $b_{\mu\nu}$ only appear in the $\theta_\alpha
\tilde\theta_\beta$ component of ${\cal Z}$ through the gauge-invariant combination
$F_{\mu\nu}-gb_{\mu\nu}$.
The chiral integral
\be
\label{chiral12}
{\cal L} = {1\over2}\Fint \int d^2\tilde\theta\, {\cal F}({\cal W} + 2g {\cal Y})
+ {\rm c. c.} + {\cal L}_{ST}
\ee
provides a $\cN=2$ invariant Lagrangian describing $16_B+16_F$ (off-shell) interacting fields. 
There exists a gauge in which ${\cal W}=0$, in which case theory (\ref{chiral12}) describes a massive chiral $\cN=2$ superfield. 

Theory (\ref{chiral12}) is actually related to the Chern-Simons Lagrangian (\ref{CS4}) by electric-magnetic duality, as will be shown below.

\subsection{Electric-Magnetic Duality}
\label{subsectEMdual}

The description in chiral $\cN=2$ superspace of the Maxwell multiplet allows 
to derive a $\cN=2$ covariant version of electric-magnetic duality. The Maxwell Lagrangian (\ref{Max6}) supplemented by the Chern-Simons coupling (\ref{CS3}) can be written
\be
\label{Belec8}
{\cal L}_{electric} = \Fint\bigint d^2\tilde\theta\, \left[ {1\over2}{\cal F}({\cal W}) 
+ ig{\cal Y}{\cal W}\right] + {\rm c.c.},
\ee
adding eqs.~(\ref{chiral5}) and (\ref{chiral8}). Replace then ${\cal W}$ by an unconstrained chiral 
superfield $\hat{\cal Z}$ (with $\cN=1$ superfields $\hat Z$, $\hat\omega_\alpha$ and $\hat\Phi$) and
introduce a new Maxwell multiplet $\widetilde{\cal W}$ (with 
$\cN=1$ superfields $\widetilde X$ and $\widetilde W_\alpha$). Using 
$$
\widetilde X = {1\over2} \, \ov{DD}\, \widetilde V_1 \,, \qquad\qquad
\widetilde W_\alpha = -{1\over4} \, \ov{DD}D_\alpha \widetilde V_2 \,,
$$
we have
\be
\label{Belec9}
\begin{array}{rcl}
i\Fint\bigint d^2\tilde\theta\, \widetilde{\cal W} \hat{\cal Z} + {\rm c.c.}
&=& \Fint \left[ {1\over2}\hat\Phi\widetilde X + \hat\omega\widetilde W\right] + {\rm c.c.}
\crbig
&=& -\Dint\left[ \widetilde V_1(\hat\Phi+\ov{\hat\Phi}) 
+ \widetilde V_2 (D^\alpha\hat\omega_\alpha 
- \ov D_\dalpha\ov{\hat\omega}^\dalpha) \right].
\end{array}
\ee
Consider now the Lagrangian
\be
\label{Belec10}
{\cal L} =
\Fint\bigint d^2\tilde\theta\, \left[ {1\over2}{\cal F}(\hat{\cal Z}) 
+ {i\over2} \hat{\cal Z}( \widetilde{\cal W} + 2g{\cal Y})\right] + {\rm c.c.}
\ee
Invariance under the gauge transformation of the single-tensor superfield,
eq.~(\ref{chiral6}), requires a compensating gauge variation of 
$\widetilde{\cal W}$, as in eq.~(\ref{chiral11}).
Eliminating $\widetilde{\cal W}$ leads back to theory (\ref{Belec8}) with 
$\hat{\cal Z}={\cal W}$. This can be seen in two ways. Firstly, the condition
$$
i\Fint\bigint d^2\tilde\theta\, \widetilde{\cal W}\hat{\cal Z} + {\rm c.c.} = {\rm derivative}
$$
leads to $\hat{\cal Z} = {\cal W}$, a $\cN=2$ Maxwell superfield, up 
to a background value. Secondly, 
using eqs.~(\ref{Belec9}), we see that $\widetilde V_2$ imposes the Bianchi 
identity on $\hat\omega$ while $\widetilde V_1$ cancels 
$\hat\Phi$ up to an imaginary constant.\footnote{An unconstrained $\widetilde X$ 
would forbid this constant.} We will come back to the (important) role
of a nonzero background value in the next section. For the moment we 
disregard it.

On the other hand, we may prefer to eliminate $\hat{\cal Z}$, 
using its field equation
\be
\label{Belec11}
{\cal F}^\prime(\hat{\cal Z}) = - i{\cal V} \,,
\qquad\qquad 
{\cal V} \equiv  \widetilde{\cal W} + 2g{\cal Y} \,,
\ee
which corresponds to a Legendre transformation exchanging variables 
$\hat{\cal Z}$ and ${\cal V}$. Defining
\be
\label{Belec12}
\widetilde{\cal F}({\cal V}) = {\cal F}(\hat{\cal Z}) + i{\cal V}\hat{\cal Z},
\ee
we have
\be
\label{Belec13}
\widetilde{\cal F}^\prime({\cal V}) = i\hat{\cal Z} \,, \qquad\qquad 
{\cal F}^\prime(\hat{\cal Z}) = -i{\cal V} \,, \qquad\qquad
\widetilde{\cal F}^{\prime\prime}({\cal V}){\cal F}^{\prime\prime}(\hat{\cal Z}) 
= 1.
\ee
The dual (Legendre-transformed) theory is then
\be
\label{Belec14}
\widetilde {\cal L}_{magnetic} = {1\over2} \Fint\bigint d^2\tilde\theta\,
\widetilde{\cal F}(\widetilde{\cal W}+2g{\cal Y}) + {\rm c.c.}
\ee
or, expressed in $\cN=1$ superspace,\footnote{The free, 
canonically-normalized theory corresponds to ${\cal F}({\cal W}) = {1\over2}{\cal W}^2$
and $\widetilde {\cal F}({\cal V}) = {1\over2}{\cal V}^2$.}
\be
\label{Belec14b}
\begin{array}{rcl}
\widetilde {\cal L}_{magnetic}
&=& {1\over4} \Fint\Bigl[ \widetilde{\cal F}^{\prime\prime}(\widetilde X+2gY) 
\, (\widetilde W-2ig\chi)^\alpha(\widetilde W-2ig\chi)_\alpha
\crbig
&& \hspace{1.0cm}  -{1\over2}\widetilde{\cal F}^\prime(\widetilde X+2gY) \, \ov{DD}(\ov{\widetilde X}
+2g\ov Y) 
 - 2ig\,\widetilde{\cal F}^\prime(\widetilde X+2gY)\Phi \Bigr] + {\rm c.c.}
\end{array}
\ee
We then conclude that the presence of the Chern-Simons term in the 
electric theory induces a St\"uckelberg gauging in the dual magnetic 
theory.

As explained in ref.~\cite{IZ}, the situation changes when Fayet-Iliopoulos terms (\ref{DBI12}) 
are present in the electric theory. In the magnetic theory
coupled to the single-tensor multiplet, with Lagrangian (\ref{Belec14b}),
the gauging $\delta\widetilde{\cal W} = 2g\widehat{\cal W}$ forbids
Fayet-Iliopoulos terms for the magnetic Maxwell superfields $\widetilde 
V_1$ and $\widetilde V_2$. Spontaneous supersymmetry breaking by 
Fayet-Iliopoulos terms in the electric theory finds then a different origin in 
the magnetic dual.

For our needs, we only consider the Fayet-Iliopoulos term induced
by $V_1$, {\it i.e.}~we add 
\be
\label{Belec15}
{\cal L}_{FI} = \xi_1 \int d^4\theta\, V_1 = -{1\over4} \xi_1 
\Fint\int d^2\tilde\theta\, \tilde\theta\tilde\theta \,{\cal W}
+{\rm c.c.}
\ee
to ${\cal L}_{electric}$, eq.~(\ref{Belec8}). In turn, this amounts to
add 
$$
 -{1\over4} \xi_1 \Fint\int d^2\tilde\theta\, \tilde\theta\tilde\theta \, \hat{\cal Z}
+{\rm c.c.}
$$
to theory (\ref{Belec10}). But, in contrast to expression 
(\ref{Belec15}), this
modification is not invariant under the second supersymmetry: according to
the first eq.~(\ref{chiral2c}), its $\delta^*$ variation
$$
 -{\sqrt2\over4} \xi_1 \Fint \eta\omega
+{\rm c.c.}
$$
is not a derivative.\footnote{It would be a derivative if $\omega_\alpha$
would be replaced by the Maxwell superfield $W_\alpha$, as in 
eq.~(\ref{Belec15}).} To restore $\cN=2$ supersymmetry, we must 
deform the $\delta^*$ variation of $\widetilde W_\alpha - 2ig \chi_\alpha$
into
\be
\label{Belec16}
\delta^*_{deformed}(\widetilde W_\alpha - 2ig \chi_\alpha) =
{1\over\sqrt2}\xi_1 \eta_\alpha +
\delta^*(\widetilde W_\alpha - 2ig \chi_\alpha),
\ee
the second term being the usual, undeformed, variations 
(\ref{Max5}) and (\ref{ST5}). Hence, the magnetic theory has a goldstino
fermion and linear $\cN=2$ supersymmetry partially breaks to $\cN=1$, as a consequence of 
the electric Fayet-Iliopoulos term.
Concretely, the magnetic theory is now
\be
\label{Belec17}
\begin{array}{rcl}
\widetilde {\cal L}_{magnetic} &=& {1\over2} \Fint\bigint d^2\tilde\theta\,
\widetilde{\cal F}\Bigl(\widetilde{\cal W}+2g{\cal Y} + {i\over2}\xi_1
\tilde\theta\tilde\theta\Bigr) + {\rm c.c.}
\crbig
&=& {1\over2} \Fint\bigint d^2\tilde\theta\, \left[
\widetilde{\cal F}\Bigl(\widetilde{\cal W}+2g{\cal Y} \Bigr) 
+ {i\over2}\xi_1 \tilde\theta\tilde\theta\,
\widetilde{\cal F}^\prime\Bigl(\widetilde{\cal W}+2g{\cal Y} \Bigr) 
\right] + {\rm c.c.}
\crbig
&=& \left[ {1\over2} \Fint \bigint d^2\tilde\theta\, 
\widetilde{\cal F}\Bigl(\widetilde{\cal W}+2g{\cal Y} \Bigr) 
+ {i\over4}\xi_1 \Fint
\widetilde{\cal F}^\prime\Bigl(\widetilde X +2gY \Bigr) \right]
+ {\rm c.c.}
\end{array}
\ee
One easily checks that $\cN=2$ supersymmetry holds, using the deformed variations (\ref{Belec16}).

\chapter{Nonlinear N = 2 Supersymmetry and the DBI Action} \label{secDBI}

In the previous sections, we have developed various aspects of the coupling of a
Maxwell multiplet to a single-tensor multiplet in linear $\cN=2$ supersymmetry. With these tools,
we can now address our main subject: show how a Dirac-Born-Infeld Lagrangian (DBI) coupled to
the single-tensor multiplet arises from nonlinearization of the second supersymmetry.

It has been observed that the DBI Lagrangian with nonlinear second 
supersymmetry can be derived by solving a constraint invariant under $\cN=2$ supersymmetry 
imposed on the super-Maxwell theory \cite{BG, RT}. We start with a summary of this result, following 
mostly Ro\v cek and Tseytlin \cite{RT}, and we then generalize the method to incorporate the fields of the
single-tensor multiplet. 

\section{The N = 2 Super-Maxwell DBI Theory}

The constraint imposed on the $\cN=2$ Maxwell chiral superfield 
${\cal W}$ is \cite{RT}\footnote{See also Ref.~\cite{R} and very recently Ref.~\cite{SK} in the context of 
$\cN=1$ supersymmetry.}
\be
\label{DBI5}
{\cal W}^2- {1\over\kappa} \tilde\theta\tilde\theta \, {\cal W}
= \left( {\cal W}  - {1\over2\kappa} \tilde\theta\tilde\theta \right)^2 = 0.
\ee
It imposes a relation between the super-Maxwell Lagrangian superfield 
${\cal W}^2$ and the Fayet-Iliopoulos `superfield' $\tilde\theta\tilde\theta{\cal W}$, eq.~(\ref{Belec15}). 
The real scale parameter $\kappa$ has dimension (energy)$^{-2}$.
In terms of $\cN=1$ superfields, the constraint is equivalent to
\be
\label{DBI6}
X^2 =0 , \qquad\qquad XW_\alpha =0 , \qquad\qquad
WW - {1\over2} X\ov{DD}\ov X =  {1\over\kappa} X.
\ee
The third equality leads to
\be
\label{DBI7}
X = {2 \, WW \over  {2\over\kappa} + \ov{DD}\ov X}
\ee
which, since $W_\alpha W_\beta W_\gamma =0$, implies the first two conditions. Solving 
the third constraint amounts to express $X$ as a function of $WW$ \cite{BG}\footnote{See Appendix \ref{App2}.}.
The DBI theory is then obtained using as Lagrangian the Fayet-Iliopoulos
term (\ref{Belec15}) properly normalized:
\be
\label{DBI8}
{\cal L}_{DBI} = {1\over4\kappa} \Fint X + {\rm c.c}
= {1\over8\kappa^2} \left[ 1- \sqrt{- {\rm det}(\eta_{\mu\nu} + 2\sqrt2\kappa F_{\mu\nu} )} \right] 
+ \ldots
\ee
The constraints (\ref{DBI5}) and (\ref{DBI6}) are not invariant under 
the second linear supersymmetry, with variations $\delta^*$. However, 
one easily verifies that the three constraints (\ref{DBI6}) are invariant under 
the deformed, nonlinear variation 
\be
\label{DBI9}
\delta^*_{deformed}  W_\alpha =  \sqrt 2 \, i \left[ {1\over2\kappa}\eta_\alpha
+\frac{1}{4}\eta_\alpha \ov{DD}\,\ov X 
+ i (\sigma^\mu\ov\eta)_\alpha \, \partial_\mu X \right] ,
\ee
with $\delta^*X$ unchanged. 
The deformation preserves the $\cN=2$ supersymmetry algebra. It indicates 
that the gaugino spinor in $W_\alpha = -i\lambda_\alpha + \ldots$ 
transforms inhomogeneously,
$\delta^*\lambda_\alpha = -{1\over\sqrt2\kappa}\, \eta_\alpha+\ldots$, 
like a goldstino for the breaking of the second supersymmetry. 
In other words, at the level of the $\cN=2$ chiral superfield ${\cal W}$,
$$
\delta^*_{deformed} \, {\cal W} = -{1\over\kappa} \tilde\theta\eta + 
i\left(\eta\tilde Q + \ov\eta\ov{\tilde Q}\right){\cal W}
= i \left(\eta\tilde Q + \ov\eta\ov{\tilde Q}\right) \left({\cal W} - {1\over2\kappa} \tilde\theta
\tilde\theta\right).
$$
The deformed second supersymmetry variations $\delta^*_{deformed}$ 
act on ${\cal W}$ as the usual variations $\delta^*$ act on the shifted superfield
${\cal W}  - {1\over2\kappa} \tilde\theta\tilde\theta$. In fact, this superfield
transforms like a chiral $\cN=2$ superfield (\ref{chiral2b}) with $Z=X$, $\omega_\alpha = iW_\alpha$
verifying the Bianchi identity and with $\Phi_{\cal Z}=-i/\kappa$. The latter background value of 
$\Phi_{\cal Z}$ may be viewed as the source of the partial breaking of linear supersymmetry.

Hence, the scale parameter $\kappa$ introduced in the nonlinear constraint (\ref{DBI5})
appears as the scale parameter of the DBI Lagrangian and also as the order parameter of 
partial supersymmetry breaking. The Fayet-Iliopoulos term (\ref{DBI8}) has in principle an
arbitrary coefficient $-\xi_1/4$, as in eq.~(\ref{Max9}). We have chosen $\xi_1= -\kappa^{-1}$ to canonically
normalize gauge kinetic terms.

The DBI Lagrangian is invariant under electric-magnetic duality.\footnote{For instance, in the 
context of D3-branes of IIB superstrings, see Ref.~\cite{T}. Our procedure is inspired by 
Ref.~\cite{RT}.} In our $\cN=2$ case, the invariance
is easily established in the language of $\cN=2$ superspace. 
We first include the constraint as a field equation of the Lagrangian:
\be
\label{DBIEM1}
{\cal L}_{DBI} = \Fint\int d^2\tilde\theta \left[ {1\over4\kappa}\tilde\theta\tilde\theta\,{\cal W}
+ {1\over4}\Lambda\left( {\cal W} - {1\over2\kappa}\tilde\theta\tilde\theta \right)^2\,  \right] + {\rm c.c.}
\ee
The field equation of the $\cN=2$ superfield $\Lambda$ enforces (\ref{DBI5}). We then introduce
two unconstrained $\cN=2$ chiral superfields $U$ and $\Upsilon$ and the modified Lagrangian
$$
{\cal L}_{DBI} = \Fint\int d^2\tilde\theta \left[ {1\over4\kappa}\tilde\theta\tilde\theta\,{\cal W}
+ {1\over4}\Lambda U^2 
- {1\over2} \Upsilon\left(U - {\cal W} + {1\over2\kappa}\tilde\theta\tilde\theta\right)
 \right] + {\rm c.c}.
$$
Since the Lagrange multiplier $\Upsilon$ imposes $U={\cal W} - {1\over2\kappa}\tilde\theta\tilde\theta$,
the equivalence with (\ref{DBIEM1}) is manifest. But we may also eliminate ${\cal W}$ which only 
appears linearly in the last version of the theory. The result is 
$$
\Upsilon = -i\widetilde{\cal W} - {1\over2} \left( {1\over\kappa} - i\zeta \right)\tilde\theta\tilde\theta
$$
where $\widetilde{\cal W}$ is a Maxwell $\cN=2$ superfield dual to ${\cal W}$ and $\zeta$ an 
arbitrary real constant. 
As in subsection \ref{subsectEMdual}, $\cN=2$ supersymmetry of the theory with a Fayet-Iliopoulos 
term requires a nonlinear deformation of the $\delta^*$ variation of $\widetilde{\cal W}$:
$\widetilde{\cal W}  - {i\over2} \left( {1\over\kappa} - i\zeta \right)\tilde\theta\tilde\theta$ should be a 
`good' $\cN=2$ chiral superfield. Replacing $\Upsilon$ in the Lagrangian and taking $\zeta=0$ leads to
$$
{\cal L}_{DBI} = \Fint\int d^2\tilde\theta \left[  {1\over4}\Lambda U^2 
+ {i\over2} U \left[\widetilde{\cal W} - {i\over2\kappa} \tilde\theta\tilde\theta \right]
+ {i\over4\kappa}\widetilde{\cal W} \, \tilde\theta\tilde\theta 
 \right] + {\rm c.c}.
$$
Finally, eliminating $U$ gives the magnetic dual
\be
\label{DBIEM2}
{\cal L}_{DBI} = \Fint\int d^2\tilde\theta \left[  {1\over4\Lambda} 
\left(\widetilde{\cal W} - {i\over2\kappa} \tilde\theta\tilde\theta \right)^2
+ {i\over4\kappa}\widetilde{\cal W} \, \tilde\theta\tilde\theta 
 \right] + {\rm c.c}.
\ee
One easily verifies that the resulting theory has the same expression as the initial `electric'  
theory (\ref{DBI8}). The Lagrange multiplier $\Lambda^{-1}$ imposes constraint
(\ref{DBI5}) to $-i\widetilde{\cal W}$, which reduces to eq.~(\ref{DBI7}) applied to $-i\widetilde X$.
The Lagrangian is then given by the Fayet-Iliopoulos term for this superfield.

\section{Coupling the DBI Theory to a Single-Tensor Multiplet:\\ 
a Super-Higgs Mechanism without Gravity}\label{couplingDBIst}

The $\cN=2$ super-Maxwell DBI theory is given by a Fayet-Iliopoulos term
for a Maxwell superfield submitted to the quadratic constraint (\ref{DBI5}),
which also provides the source of partial supersymmetry breaking. 
The second supersymmetry is deformed by the constraint: it is 
${\cal W} - {1\over2\kappa}\tilde\theta\tilde\theta$ which
transforms as a regular $\cN=2$ chiral superfield. Instead of expression
(\ref{chiral8}), we are thus led to consider the following Chern-Simons 
interaction with the single-tensor multiplet:
\be
\begin{array}{rcl}
{\cal L}_{CS, def.} &=& ig\Fint\bigint d^2\tilde\theta \, {\cal Y}\left( {\cal W}
- {1\over2\kappa}\tilde\theta\tilde\theta\right)
+ {\rm c.c.}
\crbig
&=& g\Fint\left[ {1\over2}\Phi X + \chi^\alpha W_\alpha
- {i\over2\kappa} Y \right]
+ {\rm c.c.} + {\rm derivative.}
\end{array}
\ee
The new term induced by the deformation of $\delta^*W_\alpha$ 
is proportional to the four-form field described by the chiral superfield $Y$, 
as explained in section \ref{secV1XY} [see eq.~(\ref{C8})].
This modified Chern-Simons interaction, invariant under the deformed second supersymmetry 
variations, may be simply added to the Maxwell DBI theory (\ref{DBIEM1}). 
We then consider the Lagrangian
\be
\label{DBIa}
{\cal L}_{DBI} = \Fint\int d^2\tilde\theta \left[ ig{\cal Y}
\left( {\cal W} - {1\over2\kappa}\tilde\theta\tilde\theta\right)
-{1\over4} \xi_1\tilde\theta\tilde\theta\,{\cal W}
+ {1\over2} \Lambda\left( {\cal W} - {1\over2\kappa}\tilde\theta\tilde\theta \right)^2\,  \right] + {\rm c.c.},
\ee
for the constrained Maxwell and single-tensor multiplets, keeping the Fayet-Iliopoulos coefficient
$\xi_1$ arbitrary. For a coherent theory with a propagating single-tensor multiplet,
a kinetic Lagrangian ${\cal L}_{ST}$ [eq.~(\ref{ST7})] should also be added. 
Since
$$
\Fint\int d^2\tilde\theta \left[ ig{\cal Y}{\cal W}
-{1\over4}  \xi_1\tilde\theta\tilde\theta\,{\cal W}\right] + {\rm c.c.}
= \Fint \left[ g\,\chi W +{g\over2}\Phi X -{1\over4} \xi_1 X \right] 
+ {\rm c.c.} + {\rm deriv.},
$$
we see that the Fayet-Iliopoulos term is equivalent to a constant real 
shift of $\Phi$ which, according to variations (\ref{ST5}), partially breaks supersymmetry. 
We will choose to expand $\Phi$ around $\langle\Phi\rangle=0$ and keep $\xi_1\ne0$.

Again, the constraint (\ref{DBI5}) imposed by the Lagrange multiplier 
$\Lambda$ can be solved to express $X$ as a function of $WW$: $X=X(WW)$.
The result is \cite{BG}
\be
\label{XWWis}
X(WW) =
\kappa WW - \kappa^3 \ov{DD} \left[ { WW \ov{WW} \over 
1 + \kappa^2A + \sqrt{1 +2\kappa^2A + \kappa^4B^2}} \right],
\ee
where $A$ and $B$ are defined in Appendix \ref{App2}. The DBI Lagrangian coupled to the single-tensor
multiplet reads then
\be
\label{DBIb}
{\cal L}_{DBI} = 
\Fint\left[ {1\over4} \left(2g\Phi - \xi_1\right) X(WW) + g\chi^\alpha W_\alpha
- {ig\over2\kappa} Y \right]
+ {\rm c.c.} + {\cal L}_{ST}.
\ee
The bosonic Lagrangian depends on a single auxiliary field\footnote{
Since $X(WW)|_{\theta=0}$ is a function of fermion bilinears, 
the auxiliary $f_\Phi$ does not contribute to the 
bosonic Lagrangian and $\chi_\alpha$ does not include any auxiliary field.}, $d_2$ in $W_\alpha$ 
or $V_2$:
\be
\label{DBIc}
\begin{array}{rcl}
{\cal L}_{DBI,\, bos.} &=&  {1\over8\kappa}(2g\Re\Phi - \xi_1)
\left( 1-\sqrt{-8\kappa^2 d_2^2-\det (\eta_{\mu\nu}+2\sqrt 2\kappa\,F_{\mu\nu})}\right)
- {g\over2}Cd_2 
\crbig
&& + g\epsilon^{\mu\nu\rho\sigma}\left( {\kappa\over4}\Im\Phi F_{\mu\nu}F_{\rho\sigma}
- {1\over4}b_{\mu\nu}F_{\rho\sigma} 
+ {1\over24\kappa} C_{\mu\nu\rho\sigma} 
\right) + {\cal L}_{ST, \, bos.}.
\end{array}
\ee
The real scalar field $C$ is the lowest component of the linear superfield $L$. Contrary to
$\langle\Phi\rangle$, its background value is allowed by $\cN=2$ supersymmetry. However,
a non-zero $\langle C \rangle$ would induce a non-zero $\langle d_2\rangle$ which would  
spontaneously break the residual $\cN=1$ linear supersymmetry. This is visible in the bosonic 
action which, after elimination of
\be
\label{d2elec}
d_{2, \,bos.} = {gC\over2\kappa}  \sqrt{-\det (\eta_{\mu\nu}+2\sqrt 2\kappa\,F_{\mu\nu}) \over 
(2g\Re\Phi - \xi_1)^2 + 2g^2 C^2},
\ee
becomes
\be
\label{DBId}
\begin{array}{rcl}
{\cal L}_{DBI,\, bos.} &=&  {1\over8\kappa}(2g\Re\Phi - \xi_1)\left[ 1 -
\sqrt{1 + {2g^2 C^2 \over (2g\Re\Phi - \xi_1)^2}}
\sqrt{-\det (\eta_{\mu\nu}+2\sqrt 2\kappa\,F_{\mu\nu})}
\right]
\crbig
&& + g\epsilon^{\mu\nu\rho\sigma}\left( {\kappa\over4}\Im\Phi F_{\mu\nu}F_{\rho\sigma}
- {1\over4}b_{\mu\nu}F_{\rho\sigma} 
+ {1\over24\kappa} C_{\mu\nu\rho\sigma} 
\right) + {\cal L}_{ST, \, bos.}.
\end{array}
\ee
First of all, as expected, the theory includes a DBI Lagrangian for the Maxwell field strength 
$F_{\mu\nu}$, with scale $\sim \kappa$. With the Chern-Simons coupling to the single-tensor 
multiplet, the DBI term acquires a field-dependent coefficient,
\be
- {1\over8\kappa} \sqrt{(2g\Re\Phi - \xi_1)^2 + 2g^2 C^2Ê} \,
\sqrt{-\det (\eta_{\mu\nu}+2\sqrt 2\kappa\,F_{\mu\nu})}.
\ee
It also includes a $F\wedge F$ term which respects the axionic shift symmetry of $\Im\Phi$, a
$b\wedge F$ coupling induced by (linear) $\cN=2$ supersymmetry and a `topological' $C_4$ term
induced by the nonlinear deformation. These terms are strongly reminiscent of those found when
coupling a D-brane Lagrangian to IIB supergravity. The contribution of the four-form can be 
eliminated by a gauge choice of the single-tensor symmetry (\ref{C9}). We have however 
insisted on keeping off-shell (deformed) $\cN=2$ supersymmetry, hence the presence of this term. 

The theory also includes a semi-positive scalar potential\footnote{We only consider 
$2g\Re\Phi - \xi_1 > 0$, in order to have well-defined positive gauge kinetic terms.}
\be
\label{DBIpot}
V(C,\Re\Phi) = {2g\Re\Phi - \xi_1 \over8\kappa}
\left[ \sqrt{1 + {2g^2 C^2 \over (2g\Re\Phi - \xi_1)^2}} - 1 \right]
\ee
which vanishes only if $C$ is zero.\footnote{With respect to $\Re\Phi$, the potential is stationary, 
${\partial V\over\partial\Re\Phi}=0$, only if $C=0$. All local minima are then characterized by
$C=0$ and $\Re\Phi$ arbitrary and are then (supersymmetric) global minima.}
The scalar potential determines then $\langle C \rangle=0$ but leaves $\Re\Phi$ arbitrary. Since
$$
\langle d_2 \rangle = {g\langle C\rangle\over2\kappa} \,
\Bigl\langle (2g\Re\Phi - \xi_1)^2 + 2g^2 C^2 \Bigr\rangle^{-1/2},
$$
the vacuum line $\langle C \rangle=0$ is compatible with linear $\cN=1$ and deformed 
second supersymmetry. While $\Phi$ is clearly massless, $C$ has a mass term
$$
- {1\over2}M_C^2 \, C^2
= - {g^2\over4\kappa(2\Re\Phi - \xi_1)}  C^2.
$$
The same mass is acquired by the $U(1)$ gauge field coupled to the antisymmetric tensor 
$b_{\mu\nu}$, and by the goldstino (the $U(1)$ gaugino in $W_\alpha$) that forms a Dirac spinor 
with the fermion of the linear multiplet $\chi_\alpha$. In other words, the Chern-Simons coupling 
$\chi W$ pairs the Maxwell goldstino with the linear multiplet to form a massive vector, while the 
chiral multiplet $\Phi$ remains massless with no superpotential. 

At $\langle C \rangle=\langle \Re\Phi \rangle=0$, gauge kinetic terms are canonically normalized if
$\xi_1 = -\kappa^{-1}$. The Maxwell DBI theory (\ref{DBI8}) is of course recovered when the 
Chern-Simons interaction decouples with $g=0$.
Notice finally that the kinetic terms ${\cal L}_{ST}$ of the single-tensor multiplet are given by 
eq.~(\ref{ST7}), as with linear $\cN=2$ supersymmetry.
Since the nonlinear deformation of the second supersymmetry does not affect $\delta^*L$ or 
$\delta^*\Phi$ even if $\langle\Re\Phi\rangle\ne0$, the
function ${\cal H}$ remains completely arbitrary.

The phenomenon described above provides a first instance of a super-Higgs mechanism without 
gravity: the nonlinear goldstino multiplet is `absorbed' by the linear multiplet to form a massive 
vector $\cN=1$ superfield. One may wonder how this can happen without gravity; normally one 
expects that the goldstino can be absorbed only by the gravitino in local supersymmetry. The 
reason of this novel mechanism is that the goldstino sits in the same multiplet of the linear supersymmetry as a gauge field which has a Chern-Simons interaction with the tensor multiplet. 
This will become clearer in Section \ref{secQED}, where we will show by a change of variables  
that this coupling is equivalent to an ordinary gauge interaction with a charged hypermultiplet, 
providing non derivative gauge couplings to the goldstino. Actually, this particular super-Higgs
mechanism is an explicit realization of a phenomenon known in string theory where
the $U(1)$ field of the D-brane world-volume becomes 
in general massive due to a Chern-Simons interaction with the R--R antisymmetric tensor of a 
bulk hypermultiplet.\footnote{This can be avoided in the orientifold case: the $\cN=2$ bulk
supermultiplets are truncated by the orientifold projection.}

We have chosen a description in terms of the single-tensor multiplet because it admits an 
off-shell formulation well adapted to our problem.
Our DBI Lagrangian (\ref{DBIa}), supplemented with kinetic terms ${\cal L}_{ST}$, admits however 
several duality transformations. Firstly, since it only depends on ${\cal W}$, we may perform an
electric-magnetic duality transformation, as described in section \ref{secmagndual}. 
Then, for any choice
of ${\cal L}_{ST}$, we can transform the linear $\cN=1$ superfield $L$ into a chiral $\Phi^\prime$.
The resulting theory is a hypermultiplet formulation with superfields $(\Phi,\Phi^\prime)$ and 
$\cN=2$ supersymmetry realized only on-shell. As already
explained in section \ref{secCS}, the $b\wedge F$ interaction is replaced by a St\"uckelberg 
gauging of the axionic shift symmetry of the new chiral $\Phi^\prime$: the K\"ahler potential of 
the hypermultiplet formulation is a function of $\Phi^\prime+ \ov\Phi^\prime - gV_2$. 
Explicit formulae  are given in the next section and in 
section \ref{secQED} we will use this mechanism in the case of nonlinear $\cN=2$ QED. Finally, 
if kinetic terms ${\cal L}_{ST}$ also respect the shift symmetry of $\Im\Phi$, the chiral $\Phi$ can be
turned into a second linear superfield $L^\prime$, leading to two formulations
which are also briefly described below.

\section{Hypermultiplet, Double-Tensor and Single-Tensor Dual Formulations}  
\label{sechyperform}

As already mentioned, using the single-tensor multiplet is justified by the existence of an 
off-shell $\cN=2$ formulation. The hypermultiplet formulation, with two $\cN=1$ chiral superfields,
is however more familiar and the first purpose of this subsection is to translate our results into this
formalism.
In the DBI theory (\ref{DBIb}), the linear superfield $L$ only appears in
$$
\begin{array}{l}
{\cal L}_{ST} + g\Fint \chi^\alpha W_\alpha + {\rm c.c.}
= \Dint \left[{\cal H} (L, \Phi, \ov \Phi) + g L V_2\right] +{\rm derivative.}
\end{array}
$$
These contributions are not invariant under $\delta^*$ variations: the nonlinear deformation 
acts on $W_\alpha$ and on $V_2$. Nevertheless,
the linear superfield can be transformed into a new chiral superfield $\Phi^\prime$.
The resulting `hypermultiplet formulation' has Lagrangian
\be
\label{hyp1}
\begin{array}{rcl}
{\cal L}_{DBI,\, hyper.} &=& \Dint {\cal K}\Bigl(\Phi^\prime+\ov\Phi^\prime - gV_2 , \Phi, \ov\Phi \Bigr)
\crbig
&&+ \Fint\left[ {1\over4}\left(2g\Phi - \xi_1\right) X(WW) 
- {ig\over2\kappa} Y \right]
+ {\rm c.c.}
\end{array}
\ee
The K\"ahler potential is given by the Legendre transformation
\be
\label{hyp2}
{\cal K} (\Phi^\prime+\ov\Phi^\prime , \Phi, \ov\Phi )= 
{\cal H} (U,\Phi, \ov\Phi ) - U(\Phi^\prime + \ov \Phi^\prime),
\ee
where $U$ is the solution of
\be
\label{hyp3}
{\partial\over\partial U}{\cal H} (U,\Phi, \ov\Phi ) = \Phi^\prime + \ov \Phi^\prime.
\ee
In the single-tensor formulation, $\cN=2$ supersymmetry implies that ${\cal H}$ solves Laplace equation.
As a result of the Legendre transformation, the determinant of ${\cal K}$ is constant and the
metric is hyperk\"ahler \cite{LR}. It should be noted that the Legendre transformation defines the
new auxiliary scalar $f_{\Phi^\prime}$ of $\Phi^\prime$ according to
\be
\label{hyp4}
f_{\Phi^\prime} = \left({\partial^2  {\cal H} \over \partial U\partial \Phi}\right)_{\theta=0} \, f_\Phi.
\ee
Hence, the hypermultiplet formulation has the same number of independent auxiliary fields as the 
single-tensor version: $d_2$ and $f_\Phi$. 

The second supersymmetry variation $\delta^*$ of $\Phi^\prime$ is also defined by transformation (\ref{hyp3}): in the hypermultiplet formulation, $\cN=2$ is realized on-shell only, using the Lagrangian function. The nonlinear deformation of variations $\delta^*$ acts on $V_2$. Since 
$W_\alpha = -{1\over4}\ov{DD}D_\alpha V_2$, eq.~(\ref{DBI9}) indicates that
$$
\delta^* V_2 = {i\over\sqrt2\kappa}(\ov{\theta\theta}\theta\eta - \theta\theta\ov{\theta\eta})
+ \sqrt2i \, (\eta D + \ov{\eta D}) V_1.
$$ 
The $\kappa$-dependent term in the $\delta^*$ variation of the K\"ahler potential term in
${\cal L}_{DBI,\, hyper.}$ is then the same as the $\kappa$-dependent part in 
$g\,\delta^*\int d^2\theta\,\chi^\alpha W_\alpha + {\rm c.c}$, which is compensated by the 
variation of the four-form field. This can again be shown using 
eqs.~(\ref{hyp2}) and (\ref{hyp3}).
This hypermultiplet formulation will be used in Section \ref{secQED}, on the example of 
nonlinear DBI QED with a charged hypermultiplet.

For completeness, let us briefly mention two further formulations of the same DBI theory, using either 
a double-tensor, or a dual single-tensor $\cN=2$ multiplet. These possibilities appear if Lagrangian
(\ref{DBIb}) has a second shift symmetry of $\Im\Phi$. This is the case if the single-tensor kinetic
Lagrangian has this isometry:
$$
{\cal L}_{ST} = \Dint {\cal H} (L, \Phi+\ov \Phi) .
$$
We may then transform $\Phi$
into a linear superfield $L^\prime$ using an $\cN=1$ duality transformation. 
Keeping $L$ and turning $\Phi$ into $L^\prime$ leads to a double-tensor
formulation with superfields $(L,L^\prime)$. The Lagrangian has the form
\be
\label{DT}
{\cal L}_{DT} = \Dint {\cal G}\Bigl(L,L^\prime - g V_1(WW) \Bigr) 
- \Fint\left[ {1\over4} \xi_1 X(WW) - g\chi^\alpha W_\alpha
+ {ig\over2\kappa} Y \right]
+ {\rm c.c.} 
\ee
The function ${\cal G}$ is the Legendre transform of ${\cal H}$ with respect to its second variable 
$\Phi+\ov\Phi$ and the real superfield $V_1(WW)$ is defined by the equation
\be
\label{V1WWis}
X(WW) = {1\over2}\ov{DD}\, V_1(WW).
\ee
It includes the DBI gauge kinetic term in its $d_1$ component and the Lagrangian depends on the
new tensor $b_{\mu\nu}^\prime$ through the combination $3\,\partial_{[\mu} b^\prime_{\nu\rho]}
- g\,\omega_{\mu\nu\rho}$, where $\omega_{\mu\nu\rho}= 3\,A_{[\mu}F_{\nu\rho]}$ is the 
Maxwell Chern-Simons form. 

Finally, turning $\Phi$ and $L$ into $L^\prime$ and $\Phi^\prime$, leads to another single-tensor theory
with a St\"uckelberg gauging of both $\Phi^\prime$ and $L^\prime$, as in theory (\ref{B1}). In this case,
the Lagrangian is
\be
\label{STprime}
{\cal L}_{ST^\prime} = \Dint \widetilde{\cal H} \Bigl(\Phi^\prime+\ov\Phi^\prime - gV_2, 
L^\prime - g V_1(WW) \Bigr)
- \Fint\left[ {1\over4}\xi_1 X(WW) 
+ {ig\over2\kappa} Y \right]
+ {\rm c.c.} 
\ee
While in the double-tensor theory (\ref{DT}) the second nonlinear supersymmetry only holds
on shell, it is valid off shell in theory (\ref{STprime}).
The function $\widetilde{\cal H}$ verifies Laplace equation, as required by $\cN=2$ linear 
supersymmetry.\footnote{See eq.~(\ref{ST7}).} Using the supersymmetric Legendre transformation, 
one can show that the nonlinear deformation of $\delta^*V_2$, which affects $\delta^*\widetilde{\cal H}$,
is again balanced by the variation of the four-form superfield $Y$.

\section{The Magnetic Dual}  \label{secmagndual}

To perform electric-magnetic duality on theory (\ref{DBIa}), we first replace it with 
\be
\label{DBImb}
\begin{array}{rcl}
{\cal L}_{DBI} &=& \Fint\bigint d^2\tilde\theta \Bigl[ ig{\cal Y}
\left( {\cal W} - {1\over2\kappa}\tilde\theta\tilde\theta \right)
-{1\over4} \xi_1\tilde\theta\tilde\theta\,{\cal W}
\crbig
&& \hspace{2.2cm}  + {1\over4}\Lambda U^2
-{1\over2} \Upsilon\left(U - {\cal W}+ {1\over2\kappa}\tilde\theta\tilde\theta\right) \Bigr] 
+ {\rm c.c.} + {\cal L}_{ST}.
\end{array}
\ee
Both $U$ and $\Upsilon$ are unconstrained chiral $\cN=2$ superfields. The Lagrange multiplier
$\Upsilon$ imposes $U = {\cal W}-  {1\over2\kappa}\tilde\theta\tilde\theta$, which leads again to theory
(\ref{DBIa}). The first two terms, which have gauge and $\cN=2$ invariance properties related to 
the Maxwell character of ${\cal W}$ are left unchanged. The term quadratic in ${\cal W}$ has 
been turned into a linear one using the Lagrange multiplier. Hence,  the Maxwell superfield  
${\cal W}$, which contributes to Lagrangian (\ref{DBImb}) by
\be
\label{DBImc}
\Fint\bigint d^2\tilde\theta\, {\cal W}\left( ig{\cal Y} + {1\over2}\Upsilon 
-{1\over4} \xi_1\,\tilde\theta\tilde\theta \right) + {\rm c.c.},
\ee
can as well be eliminated: $\Upsilon$ should be such that this contribution is a derivative.
In terms of $\cN=1$ chiral superfields, ${\cal W}$ has components $X$ 
and $W_\alpha$ and since there exists two real superfields $V_1$ 
and $V_2$ such that $X={1\over2}\ov{DD}\, V_1$ and 
$W_\alpha = -{1\over4} \ov{DD}D_\alpha\, V_2$, we actually need to 
eliminate $V_1$ and $V_2$ with result
\be
\label{DBImd}
\Upsilon = -i \widetilde{\cal W} - 2 ig{\cal Y} + {1\over2} (\xi_1 + i \zeta)\,
\tilde\theta\tilde\theta.
\ee
In this expression, $\widetilde{\cal W}$ is a Maxwell $\cN=2$ superfield, the `magnetic dual' of
the eliminated ${\cal W}$.  There is a new 
arbitrary real deformation parameter $\zeta$, allowed by the field equation 
of $V_2$. Notice however that $\xi_1+i\zeta$ can be eliminated by a constant complex shift 
of $\Phi$. Invariance of $\Upsilon$ under the single-tensor gauge variation (\ref{chiral6})
implies that $\delta\widetilde{\cal W} = 2g \widehat{\cal W} =
-2g\delta{\cal Y}$ and
\be
\label{DBIme}
{\cal Z } \equiv  \widetilde{\cal W} + 2g{\cal Y}
\ee
is then a gauge-invariant chiral superfield. As already mentioned,
any unconstrained chiral $\cN=2$ superfield can be decomposed in this way and
our theory may as well be considered as a description of the chiral superfields ${\cal Z}$
and ${\cal Y}$ with Lagrangian
\be
\label{DBImf}
{\cal L}_{DBI} =
\Fint\bigint d^2\tilde\theta \Bigl[ {1\over4}\Lambda U^2 
+ i U \Bigl({1\over2}{\cal Z} + {i\over4} (\xi_1+i\zeta)\tilde\theta\tilde\theta \Bigr)
+ {i\over4\kappa}\tilde\theta\tilde\theta ({\cal Z} - 2g {\cal Y}) \Bigr] + {\rm c.c.} + {\cal L}_{ST}.
\ee
Invariance under the second supersymmetry implies that 
${\cal Z} + {i\over2}(\xi_1+i\zeta)\tilde\theta\tilde\theta$ transforms as a standard 
$\cN=2$ chiral superfield and then
\be
\label{DBImg}
\delta^*_{deformed} \, {\cal Z} = i(\xi_1+i\zeta)\tilde\theta\eta 
+ i(\eta\tilde Q + \ov\eta\ov{\tilde Q}){\cal Z}.
\ee
Eliminating $U$ leads finally to
\be
\label{DBImh}
\widetilde{\cal L}_{DBI} =
\Fint\bigint d^2\tilde\theta \Bigl[ {1\over4\Lambda}
\Bigl({\cal Z} + {i\over2}(\xi_1+i\zeta)\tilde\theta\tilde\theta \Bigr)^2
+ {i\over4\kappa}\tilde\theta\tilde\theta ({\cal Z} - 2g {\cal Y}) \Bigr] + {\rm c.c.} + {\cal L}_{ST},
\ee
which is the electric-magnetic dual of theory (\ref{DBIa}).\footnote{It reduces to eq.~(\ref{DBIEM2})
if $g=0$.}
The Lagrange multiplier superfield $\Lambda^{-1}$ implies now the constraint
\be
\label{DBImi}
0 = \left({\cal Z} + {i\over2}(\xi_1+i\zeta)\tilde\theta\tilde\theta \right)^2 
= {\cal Z}^2 + i(\xi_1+i\zeta)\tilde\theta\tilde\theta{\cal Z}.
\ee
Using the expansion (\ref{chiral2b}),
$$
{\cal Z}(y,\theta,\tilde\theta) = Z(y,\theta) + \sqrt2\,\tilde\theta\omega(y,\theta)
-\tilde\theta\tilde\theta\left[ {i\over2}\Phi_{\cal Z}(y,\theta) + {1\over4} \ov{DD} \ov Z(y,\theta)\right],
$$
with $Z=\widetilde X + 2g Y$, $\omega_\alpha =  i\widetilde W_\alpha + 2g \chi_\alpha$
and $\Phi_{\cal Z}=2g\Phi$, this constraint corresponds to
$$
Z^2 = 0,
\qquad\qquad
Z\omega_\alpha =0,
\qquad\qquad
{1\over2}Z\ov{DD}\ov Z +\omega\omega =
-iZ[\Phi_{\cal Z} - (\xi_1+i\zeta)].
$$
In this case, and in contrast to the electric case, the constraint leading to the DBI theory
is due to the scale $\langle\Phi_{\cal Z}\rangle=2g\langle\Phi\rangle$: we will actually choose $\zeta=0$,
absorb $\xi_1$ into $\Phi_{\cal Z}$ and consider the constraint ${\cal Z}^2=0$ with a non-zero 
background value $\langle\Phi_{\cal Z}\rangle$ breaking the second supersymmetry.
Our magnetic theory is then
\be
\label{DBImk}
\widetilde{\cal L}_{DBI} =
\Fint\bigint d^2\tilde\theta \Bigl[ {1\over4\Lambda}
{\cal Z}^2
+ {i\over4\kappa}\tilde\theta\tilde\theta ({\cal Z} - 2g {\cal Y}) \Bigr] + {\rm c.c.} + {\cal L}_{ST},
\ee
with constraints
\be
\label{DBImj}
Z^2 = 0,
\qquad\qquad
Z\omega_\alpha =0,
\qquad\qquad
{1\over2}Z\ov{DD}\ov Z +\omega\omega =
-iZ\Phi_{\cal Z},
\ee
the DBI scale arising from $\Phi_{\cal Z} = \phi_{\cal Z} + \langle\Phi_{\cal Z}\rangle$.
As in the Maxwell case, the third equation, which also reads
\be
\label{DBImk2}
Z = {i\omega\omega \over \Phi_{\cal Z}  - {i\over2}\ov{DD}\ov Z},
\ee
implies $Z\omega_\alpha = Z^2=0$ and allows to express $Z$ as a function of $\omega\omega$
and $\Phi$, $Z=Z(\omega\omega,\Phi)$, using $\Phi_{\cal Z} = 2g\Phi - \xi_1$. 
The magnetic theory (\ref{DBImk}) is then simply
\be
\label{DBIml}
\widetilde{\cal L}_{DBI} = -{1\over2\kappa} \Im
\Fint \Bigl[Z(\omega\omega,\Phi) - 2g Y \Bigr] + {\cal L}_{ST}.
\ee
It is the electric-magnetic dual of expression (\ref{DBIb}).
At this point, it is important to recall that $\omega$ and $\Phi$ are actually $\cN=1$ superfields
components of ${\cal Z} = \widetilde{\cal W} + 2g{\cal Y}$, {\it i.e.}
\be
\omega_\alpha = i\widetilde W_\alpha + 2g\chi_\alpha.
\ee
The kinetic terms for the single-tensor multiplet $(L,\Phi)$, $L=D\chi-\ov D\ov\chi$, are included in 
${\cal L}_{ST}$ while $Z(\omega\omega,\Phi)$ includes the DBI kinetic terms for the
Maxwell $\cN=1$ superfield $\widetilde W_\alpha$. As in the electric case, the magnetic theory has a 
contribution proportional to the four-form field included in $Y$. 

The third constraint (\ref{DBImj}) is certainly invariant under the variations (\ref{chiral2c}), using 
$Z\omega_\alpha=0$. But with a non-zero background value  $\Phi = \phi + \langle\Phi\rangle$,
the spinor $\omega_\alpha$ transforms nonlinearly, like a goldstino:\footnote{See eq.~(\ref{DBImg}).}
\be
\label{DBImm}
\delta^*\omega_\alpha = -{i\over\sqrt2} \langle\Phi\rangle\,\eta_\alpha
-{i\over\sqrt2} \phi\,\eta_\alpha - {\sqrt2\over4} \eta_\alpha \, \ov{DD}\, \ov Z
-\sqrt2 i (\sigma^\mu\ov\eta)_\alpha \partial_\mu Z.
\ee

\subsection{The Bosonic Lagrangian}

The bosonic Lagrangian included 
in the magnetic theory (\ref{DBIml}) is 
\be
\label{DBIbos1}
\begin{array}{rcl}
\widetilde{\cal L}_{DBI,bos.} &=& {\Re\Phi_{\cal Z}\over8\kappa} 
- {\Re\Phi_{\cal Z}\over8\kappa|\Phi_{\cal Z}|^2}
\Biggl\{ - |\Phi_{\cal Z}|^4 \,{\rm det} \left[\eta_{\mu\nu} - 2\sqrt2\,|\Phi_{\cal Z}|^{-1} 
(\widetilde F_{\mu\nu} - gb_{\mu\nu}) \right]
\crbig
&&
- 8 \tilde d{_2}^2 \, (|\Phi_{\cal Z}|^2 + 2g^2C^2) + 2 g^2 C^2 |\Phi_{\cal Z}|^2
\crbig
&& + 8gC\tilde d_2 \, \epsilon^{\mu\nu\rho\sigma}
(\widetilde F_{\mu\nu} - g \, b_{\mu\nu})(\widetilde F_{\rho\sigma} - g \, b_{\rho\sigma})
\Biggr\}^{1/2}
\crbig
&& - {\Im\Phi_{\cal Z}\over8\kappa|\Phi_{\cal Z}|^2} \left[ \epsilon^{\mu\nu\rho\sigma}
(\widetilde F_{\mu\nu} - g \, b_{\mu\nu})(\widetilde F_{\rho\sigma} - g \, b_{\rho\sigma})
- 4gC \widetilde d_2 \right]
\crbig
&&
+ {g\over24\kappa}\epsilon^{\mu\nu\rho\sigma} C_{\mu\nu\rho\sigma} + {\cal L}_{ST,bos.}.
\end{array}
\ee
It depends on a single auxiliary field, the Maxwell real scalar $\widetilde d_2$, with field equation
\be
\begin{array}{rcl}
\widetilde d_{2, \, bos.} &=& \displaystyle-\frac{g\,C}{2(|\Phi_{\cal Z}|^2+2g^2C^2)} \,
\epsilon^{\mu\nu\rho\sigma} 
(\widetilde F_{\mu\nu} - g \, b_{\mu\nu})(\widetilde F_{\rho\sigma} - g \, b_{\rho\sigma})
\crbig
&& \displaystyle
- \frac{g\,C\Im\Phi_{\cal Z}}{2|\Phi_{\cal Z}|^2}
\frac{\sqrt{-\det\Bigl(\eta_{\mu\nu}+\frac{2\sqrt2}{\sqrt{2g^2C^2+|\Phi_{\cal Z}|^2}}
(\widetilde F_{\mu\nu} - g \, b_{\mu\nu})\Bigr)}}
{\sqrt{(\Re\Phi_{\cal Z})^2+2g^2C^2}} \, .
\end{array}
\ee
Eliminating $\tilde d_2$ and using $\Phi_{\cal Z}=2g\Phi - \xi_1$ to reintroduce the superfield 
$\Phi$ of the single-tensor multiplet and the `original' Fayet-Iliopoulos term $\xi_1$, we 
finally obtain the magnetic, bosonic Lagrangian
\be
\label{DBIbos2}
\begin{array}{rcl}
\widetilde{\cal L}_{DBI,bos.} &=&  \displaystyle{2g\Re\Phi - \xi_1 \over 8\kappa} 
-{1\over8\kappa}\sqrt{(2g\Re\Phi -\xi_1)^2+ 2 g^2 C^2 }
\crbig
&& \hspace{2.3cm} \times \sqrt{-\det\Big(\eta_{\mu\nu}
-\frac{2\sqrt2}{\sqrt{2g^2 C^2 + |2g\Phi - \xi_1|^2}}(\widetilde F_{\mu\nu} - g b_{\mu\nu})\Big)}\Biggr)
\crbig
&& \displaystyle -{g\Im\Phi \over 4\kappa(2g^2C^2 
+ |2g\Phi - \xi_1|^2)}\epsilon^{\mu\nu\rho\sigma}(\widetilde F_{\mu\nu} - g b_{\mu\nu}) 
(\widetilde F_{\rho\sigma} - g b_{\rho\sigma})
\crbig
&& \displaystyle + {g\over24\kappa}\epsilon^{\mu\nu\rho\sigma} C_{\mu\nu\rho\sigma}
+ {\cal L}_{ST,bos.} \,.
\end{array}
\ee
As in the electric case, the DBI term has a field-dependent coefficient,
\be
- {1\over8\kappa} \sqrt{(2g\Re\Phi - \xi_1)^2 + 2g^2 C^2 } \,
\sqrt{-\det \Big(\eta_{\mu\nu}-\frac{1}{\sqrt{ 2g^2C^2 +|2g\Phi - \xi_1|^2}}
(\widetilde F_{\mu\nu}-gb_{\mu\nu})\Big)},
\ee
and, as expected, the scalar potentials of the magnetic and electric [eq.~(\ref{DBIpot})] theories
are identical.

Define the complex dimensionless field
\be
\label{DBIbos3}
S =  \kappa\sqrt{(2g\Re\Phi - \xi_1)^2 + 2g^2 C^2 } +2i\kappa g\Im\Phi,
\ee
for which $\kappa^{-2} |S|^2 = |2g\Phi-\xi_1|^2 + 2g^2C^2$. In terms of $S$, the magnetic theory
(\ref{DBIbos2}) rewrites as
\be
\label{S1}
\begin{array}{rcl}
\widetilde{\cal L}_{DBI,bos.} &=& \displaystyle{2g\Re\Phi - \xi_1 \over 8\kappa} 
-{1\over8\kappa^2} \Re{1\over S} \sqrt{-\det\Bigl( |S|\eta_{\mu\nu} 
- 2\sqrt2\kappa (\widetilde F_{\mu\nu} - g b_{\mu\nu})\Bigr)}
\crbig
&& \displaystyle + {1\over8} \Im {1\over S}\,
\epsilon^{\mu\nu\rho\sigma}(\widetilde F_{\mu\nu} - g b_{\mu\nu}) 
(\widetilde F_{\rho\sigma} - g b_{\rho\sigma})
+ {g\over24\kappa}\epsilon^{\mu\nu\rho\sigma} C_{\mu\nu\rho\sigma}
+ {\cal L}_{ST,bos.}
\crbig
&=& \displaystyle{2g\Re\Phi - \xi_1 \over 8\kappa} 
-{1\over8\kappa^2} \Re S \sqrt{-\det\Bigl( \eta_{\mu\nu} 
- 2\sqrt2\kappa |S|^{-1}(\widetilde F_{\mu\nu} - g b_{\mu\nu})\Bigr)}
\crbig
&& \displaystyle + {1\over8} \Im {1\over S}\,
\epsilon^{\mu\nu\rho\sigma}(\widetilde F_{\mu\nu} - g b_{\mu\nu}) 
(\widetilde F_{\rho\sigma} - g b_{\rho\sigma})
+ {g\over24\kappa}\epsilon^{\mu\nu\rho\sigma} C_{\mu\nu\rho\sigma}
+ {\cal L}_{ST,bos.} .
\end{array}
\ee
This is to be compared with the electric theory (\ref{DBId}):
\be
\label{S2}
\begin{array}{rcl}
{\cal L}_{DBI,\, bos.} &=&  \displaystyle {2g\Re\Phi - \xi_1 \over 8\kappa} - {1\over8\kappa^2} \Re S
\sqrt{-\det (\eta_{\mu\nu}-2\sqrt 2\kappa\,F_{\mu\nu})}
\crbig
&& \displaystyle + {1\over8}\Im S \, \epsilon^{\mu\nu\rho\sigma}F_{\mu\nu}F_{\rho\sigma}
- {g\over4}\epsilon^{\mu\nu\rho\sigma}b_{\mu\nu}F_{\rho\sigma} 
+ {g\over24\kappa} \epsilon^{\mu\nu\rho\sigma}C_{\mu\nu\rho\sigma} 
+ {\cal L}_{ST, \, bos.}.
\end{array}
\ee
Hence, the duality from the electric to the magnetic theory corresponds to the transformations
\be
\label{S3}
b_{\mu\nu} \,\rightarrow\, 0, \qquad
F_{\mu\nu} \,\rightarrow\, \widetilde F_{\mu\nu} - g b_{\mu\nu}, \qquad
S\,\rightarrow\, S^{-1}, \qquad
\eta_{\mu\nu}\,\rightarrow\, |S|\eta_{\mu\nu},
\ee
which can be also derived from electric-magnetic duality applied on the bosonic DBI theory
only.

\section{Double-Tensor Formulation and Connection with the String Fields}

In IIB superstrings compactified to four dimensions with eight 
residual supercharges, the dilaton belongs to a double-tensor supermultiplet. 
This representation of $\cN=2$ supersymmetry includes 
two Majorana spinors, two antisymmetric tensors $B_{\mu\nu}$ (NS--NS) and $C_{\mu\nu}$ (R--R)
with gauge symmetries
\be
\label{antisymsym}
\delta_{gauge}\,B_{\mu\nu} = 2\,\partial_{[\mu} \Lambda_{\nu]}, \qquad\qquad
\delta_{gauge}^{\,\prime} \,C_{\mu\nu} = 2\,\partial_{[\mu}\Lambda^\prime_{\nu]}
\ee
and two (real) scalar fields, the NS--NS dilaton and the R--R scalar, for a total of $4_B+4_F$ physical
states. In principle, both antisymmetric tensors can be dualized to pseudoscalar fields with 
axionic shift symmetry, in a version of the effective field theory where the dilaton belongs to a 
hypermultiplet with four scalars in a quaternion-K\"ahler
manifold possessing three perturbative shift isometries, since the R--R scalar 
has its own shift symmetry. It is easy to see that only two shift isometries, related to the two 
antisymmetric tensors, commute, 
while all three together form the Heisenberg algebra. Indeed, in the double-tensor basis, the R--R field 
strength is modified~\cite{IIBsugra} due to its anomalous Bianchi identity to 
$3\,\partial_{[\lambda}C_{\mu\nu]}-3\,C^{(0)}\partial_{[\lambda}B_{\mu\nu]}$. Thus, a shift of 
the R--R scalar $C^{(0)}$ by a constant $\lambda$ is accompanied by an appropriate 
transformation of $C_{\mu\nu}$ to leave its modified field-strength invariant:
\be
\label{Heisenberg}
\delta_{H}C^{(0)}=\lambda, \qquad \delta_H C_{\mu\nu}=\lambda B_{\mu\nu} .
\ee
It follows that $\delta_{gauge}$, $\delta_{gauge}'$ and $\delta_H$ verify the Heisenberg algebra, 
with a single non-vanishing commutator 
\be
\label{commutator}
\left[\delta_{gauge},\delta_H\right]=\delta_{gauge}^{\,\prime}\, .
\ee

More details about the Heisenberg algebra in local and global supersymmetry are given in chapter~\ref{UniversalHypermultipletLocalGlobalSupersymmetry} where we obtain the global supersymmetry limit of the universal hypermultiplet. Our aim is to use the Heisenberg algebra in order to establish the connection between the general formalism developed so far and string theory. This formalism would then describe the coupling of a D-brane with bulk fields in the limit of global supersymmetry.

To this end, we transform the $\cN=2$ double-tensor into a 
single-tensor representation by dualizing one of its two $\cN=1$ linear multiplet components $L'$, containing the R--R fields $C_{\mu\nu}$ and $C^{(0)}$, into a chiral basis $\Phi+\ov{\Phi}$. In this basis, the two R--R isometries correspond to constant complex shifts of the $\cN=1$ superfield $\Phi$. Imposing this symmetry to the kinetic function of eqs.~(\ref{ST7})--(\ref{ST8}), one obtains (up to total derivatives, after superspace integration):
\be
\label{kinetic}
{\cal H}(L,\Phi,\ov{\Phi})=\alpha\Bigl(-{1\over 3}L^3 + {1\over 2}L (\Phi+\ov{\Phi})^2\Bigr)
+\beta\Bigl(-L^2+{1\over2}(\Phi+\ov{\Phi})^2\Bigr)\, ,
\ee
where $\alpha$ and $\beta$ are constants. Note that the second term proportional to $\beta$ can be obtained from the first by shifting $L+\beta/\alpha$. For $\alpha=0$ however, it corresponds to the free case of quadratic kinetic terms for all fields of the single-tensor multiplet.
The coupling to the Maxwell goldstino multiplet is easily obtained 
using eqs.~(\ref{DBIc}), (\ref{V1WWis}) and (\ref{CS1}). Up to total derivatives, 
the action is:
\be
\label{DT1}
\begin{array}{rcl}
{\cal L} &= &\Dint\Big[\alpha\Big(-{1\over 3}L^3 + {1\over 2}L (\Phi+\ov{\Phi})^2\Big)
+\beta\Big(-L^2+{1\over2}(\Phi+\ov{\Phi})^2\Big)
\crbig
&&\hspace{.4cm}
-g(\Phi+\ov{\Phi})V_1(WW)\Big]
+g\Fint \Bigl[\chi^\alpha W_\alpha - {i\over 2 \kappa} Y -\frac{\xi_1}{4g}X(WW) \Bigr] + {\rm c.c.}
\end{array}
\ee
In general, the four-form field is not inert under the variation $\delta_H$ of eq.~(\ref{Heisenberg}) \cite{BT}.
In our single-tensor formalism, $\delta_H L=0$ and $\delta_H\Phi = c$ where $c$ is complex when combined with the axionic shift  $\delta'_{gauge}$ of ${\rm Im}\Phi$ dual to $C_{\mu\nu}$ of eq.~(\ref{antisymsym}); in addition
\be
\label{Yvar}
\delta_H Y = - ic\kappa X(WW).
\ee
With this variation, the Lagrangian, including the Chern-Simons interaction, is invariant under 
the Heisenberg symmetry.

We can now dualize back $\Phi+\ov{\Phi}$ to a second linear multiplet $L'$ by first
replacing it with a real superfield $U$:
\be
\label{DT2}
\begin{array}{rl}
{\cal L} = &\Dint\Big[\alpha\left(-{1\over 3}L^3 + {1\over 2}L U^2\right)
+\beta\left(-L^2+{1\over2}U^2\right)-U (mL'+ gV_1)\Big]
\crbig
&+g\Fint \Bigl[ \chi^\alpha W_\alpha - {i\over 2 \kappa} Y -\frac{\xi_1}{4g}X \Bigr] + {\rm c.c.},
\end{array}
\ee
where the constant $m$ corresponds to a rescaling of $L'$. Solving for $U$,
\be
\label{DT3}
U={mL'+gV_1\over \alpha L+\beta}\,,
\ee
delivers the double-tensor Lagrangian
\be
\label{DT4}
{\cal \widetilde L} =
\Dint\Big[ -{\alpha\over 3}L^3 -\beta L^2- {1\over2}{(mL'+gV_1)^2\over \alpha L
+\beta}\Big] 
+g\Fint \Bigl[ \chi^\alpha W_\alpha - {i\over 2 \kappa} Y - {\xi_1\over 4g}X \Bigr] + {\rm c.c.},
\ee
where as before $V_1= V_1(WW)$ and $X=X(WW)= {1\over2}\ov{DD}\,V_1(WW)$. 
It is invariant under variation (\ref{Yvar}) of the four-form superfield combined with
$\delta_H L^\prime = 2c(\alpha L+\beta)/m$.

After elimination of the Maxwell auxiliary field (choosing $m=\sqrt2$)
\be
d_{2, \,bos.} = {gC\over2\kappa}  \sqrt{-\det (\eta_{\mu\nu}+2\sqrt 2\kappa\,F_{\mu\nu}) \over 
\left({\sqrt2 g\,C^\prime\over \alpha C+\beta} - \xi_1\right)^2 + 2g^2 C^2} \, ,
\ee
the component expansion of the bosonic Lagrangian is 
\be
\label{DT8}
\begin{array}{rcl}
{\cal \widetilde L}_{bos.} &=& (\alpha C+\beta)\left[{1\over 2} (\partial_\mu C)^2+{1\over 2}
\partial_\mu\Big({C'\over \alpha C+\beta}\Big)^2 +{1\over 12} (3 \,\partial_{[\mu} b_{\nu\rho]})^2\right]  
\crbig
&&  +{1\over 12(\alpha C+\beta)}\left(3\,\partial_{[\mu} b'_{\nu\rho]}
+{g\kappa\over \sqrt{2}}\omega_{\mu\nu\rho}
-{C'\over \alpha C + \beta} 3\, \partial_{[\mu} b_{\nu\rho]}\right)^2 
\crbig
&& -{g\over 4 \kappa\sqrt2}({C'\over \alpha C+\beta}+{\xi_1\over\sqrt2 g})+ {g\over 4\kappa\sqrt{2}}
\sqrt{({C'\over \alpha C+\beta}+{\xi_1\over \sqrt2g})^2+C^2}
\sqrt{-\det(\eta_{\mu\nu}+2\sqrt{2}\kappa F_{\mu\nu})}
\crbig
&&- \frac{g}{4} \epsilon^{\mu\nu\rho\sigma} b_{\mu\nu} F_{\rho\sigma}
+{g\over 24\kappa}\epsilon^{\mu\nu\rho\sigma}C_{\mu\nu\rho\sigma} \,.
\end{array}
\ee
in terms of the Maxwell Chern-Simons form $\omega_{\nu\rho\sigma} = 3\, A_{[\nu} F_{\rho\sigma]}$.

This is the explicit expression of the interacting action (\ref{DT}) and the kinetic part for the double-tensor multiplet. It describes the global supersymmetry limit of the effective four dimensional action of a D-brane coupled to the universal dilaton hypermultiplet of the perturbative type II string. The precise identification of the fields will be done in section~\ref{seclimit} in the dual single-tensor basis but we can already see the similarities here: As mentioned previously, its general form in the local case depends also on two constant parameters, upon imposing the perturbative Heisenberg isometries, that correspond to the tree and one-loop contributions~\cite{oneloop}. We expect that these two parameters are related to $\alpha$ and $\beta$ of our action. Moreover, by identifying the two antisymmetric tensors $b_{\mu\nu}$ and $b^\prime_{\mu\nu}$ with the respective NS--NS $B_{\mu\nu}$ and R--R $C_{\mu\nu}$ and the combination $C'/(\alpha C+\beta)$ with the R--R 
scalar $C^{(0)}$, as the Heisenberg transformations indicate, one finds that the two actions match up to normalization factors depending on the NS--NS dilaton that should correspond to the scalar $C$.

\section{Nonlinear N = 2 QED}  \label{secQED}

We will now show that the effective theory presented above describing a super-Higgs 
phenomenon of partial (global) supersymmetry breaking can be identified with the Higgs 
phase of nonlinear $\cN=2$ QED, up to an appropriate choice of the single-tensor multiplet 
kinetic terms. We will then analyze its vacuum structure in the generally allowed parameter 
space.

In linear $\cN=2$ quantum electrodynamics (QED), the Lagrangian couples a hypermultiplet with 
two chiral superfields $(Q_1,Q_2)$ to the vector multiplet $(V_1,V_2)$ or $(X,W_\alpha)$. 
The $U(1)$ gauge transformations 
of the hypermultiplet are linear, and $Q_1$ and $Q_2$ have opposite $U(1)$ charges:
\be
\label{QED1}
{\cal L}_{QED} = \Dint \left[ \ov Q_1Q_1 e^{V_2} + \ov Q_2Q_2 e^{-V_2} \right]
+ \Fint {i\over\sqrt2} XQ_1Q_2 + {\rm c.c.} + {\cal L}_{Max.} + \Delta{\cal L},
\ee
where ${\cal L}_{Max.}$ includes (canonical) gauge kinetic terms and $\Delta{\cal L}$ 
contains three parameters:
\be
\label{QED1b}
\Delta{\cal L} = m\Fint Q_1Q_2 + {\rm c.c.} + \Dint [\xi_1V_1 + \xi_2V_2].
\ee
The hypermultiplet mass term with coefficient $m$ can be eliminated by a shift of $X$ and 
$\xi_{1,2}$ are the two Fayet-Iliopoulos coefficients. 
Since $\xi_1\int d^2\theta d^2\ov\theta\, V_1 = -{1\over4}\int d^2\theta\, \xi_1X + {\rm c.c.}$, the 
complete superpotential $w$ is
$$
w = \left({i\over\sqrt2} X + m\right) Q_1Q_2 - {1\over4}\xi_1X.
$$
There are six real auxiliary fields, $f_{Q_1}$, $f_{Q_2}$, $d_1$ and $d_2$ but only four are actually independent:\footnote{
We use the same notation for a chiral superfield $\Phi$, $Q_1$, $Q_2$, \dots and for its
lowest complex scalar component field.} $Q_1 \ov f_{Q_1} = Q_2 \ov f_{Q_2}$.
Since the metric is canonical, $\det K_{i\ov j} = 1$ and trivially hyperk\"ahler.
If $\xi_1=\xi_2=0$, the gauge symmetry is not broken and the hypermultiplet mass 
$m + i \langle X\rangle / \sqrt2 $ is arbitrary.
Any nonzero $\xi_1$ or $\xi_2$ induces $U(1)$ symmetry breaking with all fields having the same mass. In any case, $\cN=2$ supersymmetry remains unbroken at the global minimum.

In order to first bring the theory to a form allowing dualization to our single-tensor formulation,
we use the holomorphic field redefinition\footnote{This field redefinition  
has constant Jacobian.}
\be
\label{QED2}
\begin{array}{c}
Q_1 = a\, \sqrt\Phi \, e^{\Phi^\prime}, \qquad  \qquad
Q_2 = ia\, \sqrt\Phi \, e^{-\Phi^\prime} ,
\crbig
Q_1Q_2 = ia^2\Phi , \qquad  \qquad Q_1/Q_2 = -ie^{2\Phi^\prime} ,
\end{array}
\ee
with $a^2 = 1/\sqrt2$. The QED Lagrangian becomes
\be
\label{QED3}
\begin{array}{rcl}
{\cal L}_{QED} &=& {1\over\sqrt2} \Dint 
\sqrt{\Phi\ov\Phi}  \left[ e^{\Phi^\prime+\ov\Phi^\prime + V_2} 
+ e^{-\Phi^\prime-\ov\Phi^\prime - V_2}\right]
+ {\cal L}_{Max.} 
\crbig
&& 
+ \Fint \left[ -{1\over2} \Phi (X-\sqrt2im) - {1\over4}\xi_1 X \right] + {\rm c.c.} 
+ \xi_2\Dint V_2.
\end{array}
\ee
While the gauge transformation of $\Phi^\prime$ is $\delta_{U(1)}\Phi^\prime = \Lambda_c$, 
$\Phi$ is gauge invariant. Since the K\"ahler potential is now a function of 
$\Phi^\prime+\ov\Phi^\prime$, with a St\"uckelberg gauging of the axionic shift
of $\Phi^\prime$, the chiral $\Phi^\prime$ can be dualized to a linear $L$ using a $ N=1$ Legendre
transformation. The result is
\be
\label{QED5}
\begin{array}{rcl}
{\cal L}_{QED} &=& \Dint 
\left[ \sqrt{2\Phi\ov\Phi + L^2} - L \ln\left( \sqrt{2\Phi\ov\Phi + L^2} + L\right) 
\right] + {\cal L}_{Max.}
\crbig
&& -  \Fint \left[{1\over2}X\Phi + \chi^\alpha W_\alpha 
- {i\over\sqrt2}m\Phi + {1\over4}\xi_1X \right] + {\rm c.c.} + \xi_2\Dint V_2.
\end{array}
\ee
The dual single-tensor QED theory has off-shell $\cN=2$ invariance (the Laplace equation (\ref{ST7}) 
is verified) and the two multiplets are now coupled by a $\cN=2$ Chern-Simons interaction 
(\ref{CS3}). Notice that the free quadratic kinetic terms of the charged hypermultiplet lead to a 
highly non-trivial kinetic function in the single-tensor representation. Moreover, 
there are only four auxiliary fields, $f_\Phi$, $d_1$ and $d_2$.
The Legendre transformation defines the scalar field $C$ in $L$ as
\be
e^{2\Re\Phi^\prime} = {1\over\sqrt{2\Phi\ov\Phi}}\left( \sqrt{2\Phi\ov\Phi+C^2} + C \right),
\qquad
e^{-2\Re\Phi^\prime} = {1\over\sqrt{2\Phi\ov\Phi}}\left( \sqrt{2\Phi\ov\Phi+C^2} - C \right)
\ee
and eqs.~(\ref{QED2}) relate then $C$ and $\Phi$ with $Q_1$ and $Q_2$:
\be
\label{QED8}
C = |Q_1|^2 - |Q_2|^2, \qquad\qquad \Phi = -\sqrt2i\,Q_1Q_2.
\ee

According to eq.~(\ref{DBIb}), the nonlinear DBI version of $\cN=2$ QED is obtained by replacing
in Lagrangian (\ref{QED5}) $X$ by $X(WW)$, which includes DBI gauge kinetic terms, by
omitting ${\cal L}_{Max.}$ which is removed by the third constraint (\ref{DBI6}) and by adding the 
four-form term ${i\over2\kappa}\int d^2\theta\, Y + {\rm c.c.}$:
\be
\label{QED6}
\begin{array}{rcl}
{\cal L}_{QED, DBI} &=& \Dint 
\left[ \sqrt{2\Phi\ov\Phi + L^2} - L \ln\left( \sqrt{2\Phi\ov\Phi + L^2} + L\right)
+ \xi_2\, V_2 \right] 
\crbig
&& -\Fint\left[ \left({1\over2}\Phi + {1\over4}\xi_1 \right)  X(WW) - {i\over\sqrt2}m\Phi 
+ \chi^\alpha W_\alpha - {i\over2\kappa} Y \right]
+ {\rm c.c.} 
\end{array}
\ee
Notice that two additional terms appear compared to the action studied in Section~\ref{secDBI}: a 
Fayet-Iliopoulos term proportional to $\xi_2$ and a term linear in $\Phi$ which is also invariant 
under the second (nonlinear) supersymmetry (\ref{ST4}); they generate, together with $\xi_1$ 
the general parameter space of nonlinear QED coupled to a charged hypermultiplet.
Without loss of generality, we choose $m$ to be real, while
the choice $\xi_1 = -1/\kappa$ would canonically normalize gauge kinetic terms for a background 
where $\Phi$ vanishes.
We may return to chiral superfields $(\Phi,\Phi^\prime)$ or $(Q_1,Q_2)$ to write the DBI theory
as\footnote{See eq.~(\ref{hyp1}).}
\be
\label{QED4}
\begin{array}{rcl}
{\cal L}_{QED} &=& \Dint \left[ \ov Q_1Q_1 e^{V_2} + \ov Q_2Q_2 e^{-V_2} + \xi_2 V_2\right]
\crbig
&&
+\Fint\left[ \left( {i\over\sqrt2}Q_1Q_2 - {1\over4}\xi_1 \right)  X(WW) + mQ_1Q_2 
+ {i\over2\kappa} Y \right]
+ {\rm c.c.} 
\end{array}
\ee
Since $X(WW)|_{\theta=0}$ only depends on fermion fields, 
the auxiliary fields $f_1$ and $f_2$ only contribute to
the bosonic Lagrangian by a hypermultiplet mass term
$$
\Bigl( |f_1|^2 + |f_2|^2 \Bigr)_{bos.}= m^2 \left( |Q_1|^2 + |Q_2|^2 \right) 
$$
to be added to the scalar potential obtained from eq.~(\ref{DBIpot}) with the substitutions
$$
2 g\Re\Phi-\xi_1\,\longrightarrow\, 2\sqrt2\Im (Q_1Q_2)-\xi_1, \qquad\qquad
gC \,\longrightarrow\, C+\xi_2 = \xi_2 + |Q_1|^2 - |Q_2|^2
$$
(since we have chosen $g=1$). The complete potential is then\footnote{The auxiliary $d_2$ is given 
in eq.~(\ref{d2elec}).}
\be
\begin{array}{rcl}
V_{QED,DBI} &=& \displaystyle{1\over8\kappa}\left(2\sqrt2\Im (Q_1Q_2)-\xi_1 \right) \left[ 
\sqrt{1 + { 2[\xi_2 + |Q_1|^2 - |Q_2|^2]^2
\over [2\sqrt2\Im (Q_1Q_2)-\xi_1]^2}} - 1 \right]
\crbig
&& + m^2 \left( |Q_1|^2 + |Q_2|^2 \right). 
\end{array}
\ee
The analysis is then very simple. The first line vanishes only for 
\be
\label{vac1}
\langle \xi_2 + |Q_1|^2 - |Q_2|^2 \rangle = 0, \qquad\qquad
\langle 2\sqrt2\Im (Q_1Q_2)-\xi_1\rangle > 0.
\ee
The first condition is the usual $D$--term equation $\langle d_2\rangle =0$ for the Maxwell superfield.  
The second condition is necessary to have a well-defined DBI gauge kinetic term at the minimum. 
Hence, if $m=0$, conditions (\ref{vac1}), which can always be solved, define the vacuum of the theory. 
Choosing $\langle Q_1 \rangle = v$ and $\langle Q_2 \rangle=\sqrt{v^2+\xi_2}$, with $v$ real
(and arbitrary), we find 
a massive vector boson which, along with a real scalar and the two Majorana fermions
$$
{1\over \sqrt{2 v^2+\xi_2}} \, \left[ v \psi_{Q_1}- \sqrt{v^2+\xi_2}\, \psi_{Q_2} 
\right] \pm i\lambda ,
$$ 
makes a massive $\cN=1$ vector multiplet of mass $\sqrt{v^2+\xi_2/2}$. Hence the potentially 
massless gaugino $\lambda$, with its goldstino-like second supersymmetry variation 
$\delta^*\lambda_\alpha = - {1\over\sqrt2\kappa}\eta_\alpha + \ldots$, has been absorbed 
in the massive 
$U(1)$ gauge boson multiplet. This is possible only because the second supersymmetry 
transformation  of the four-form field compensates the gaugino nonlinear variation.
The fermion
$$
\sqrt{v^2 + \xi_2} \, \psi_{Q_1} + v \, \psi_{Q_2}
$$
is massless and corresponds to the fermion of the chiral superfield $\Phi$ in the single-tensor 
formalism, in agreement with our analysis in Section~\ref{couplingDBIst} [see below 
eq.~(\ref{DBIpot})]. With two real scalars, it belongs to a massless $\cN=1$ chiral multiplet.

If $m\ne0$, a supersymmetric vacuum has $\langle Q_1 \rangle = \langle Q_2 \rangle
=0$. It only exists if $\xi_2=0$ and $\xi_1\ne0$. The second condition is again to have DBI gauge 
kinetic terms on this vacuum. In this case, the $U(1)$ gauge symmetry is not broken, the goldstino 
vector multiplet remains massless and the hypermultiplet has mass $m$. If $m\ne0$, a nonzero 
Fayet-Iliopoulos coefficient $\xi_2$ breaks then $\cN=1$ linear supersymmetry. Note that the 
single-tensor formalism is appropriate for the description of the Higgs phase of nonlinear QED in a 
manifest $\cN=1$ superfield basis (with respect to the linear supersymmetry), while the charged 
hypermultiplet representation is obviously convenient for describing the Coulomb phase. 

One can finally expand the action (\ref{QED4}) in powers of $\kappa$ in order to find the lowest 
dimensional operators that couple the goldstino multiplet of partial supersymmetry breaking to the 
$\cN=2$ hypermultiplet. Besides the dimension-four operators corresponding to the gauge factors 
$e^{\pm V_2}$, one obtains a dimension-six superpotential interaction $\sim\kappa Q_1Q_2W^2$ 
coming from the solution of the nonlinear constraint $X=\kappa W^2+{\cal O}(\kappa^3)$; it amounts 
to a field-dependent correction to the $U(1)$ gauge coupling.

\chapter{The Universal Hypermultiplet in Local and Global Supersymmetry}
\label{UniversalHypermultipletLocalGlobalSupersymmetry}

\section{On the Heisenberg Algebra and Global Supersymmetry}\label{secglobal}

In the context of IIB superstrings, the Heisenberg algebra is generated by a combination of the gauge symmetries of the two antisymmetric tensors $B_{\mu\nu}$ (NS-NS) and $C_{\mu\nu}$ (R-R) and of the shift symmetry of the R-R scalar $C_0$:
\be
\label{Heis1}
\delta B_{\mu\nu} = 2 \, \partial_{[\mu}\Lambda_{\nu]} , \qquad\qquad
\delta C_{\mu\nu} = 2 \, \partial_{[\mu}\tilde \Lambda_{\nu]} + \lambda B_{\mu\nu} , \qquad\qquad
\delta C_0 = \lambda.
\ee
As a consequence, the theory depends on the invariant three-forms
\be
\label{Heis2}
H_{\mu\nu\rho} = 3\,\partial_{[\mu} B_{\nu\rho]}, \qquad\qquad
F_{\mu\nu\rho} = 3\,\partial_{[\mu} C_{\nu\rho]} - C_0 H_{\mu\nu\rho}
\ee
and on $\partial_\mu C_0$. The Heisenberg algebra follows from
\be
\label{Heis3}
[\delta_1 , \delta_2 ] \, C_{\mu\nu} = 2 \, \partial_{[\mu} \lambda_2\Lambda_{1\nu]}
- 2 \, \partial_{[\mu}  \lambda_1\Lambda_{2\nu]}.
\ee
After reduction to four dimensions, the gauge symmetries imply that each tensor can be dualized into a scalar field with axionic shift symmetry. The third global symmetry (with parameter $\lambda$) combines then with the axionic shifts to realize again the Heisenberg algebra on three scalar fields. 

Indeed, one obtains three scalar fields $\varphi$, $\tau$ and $\eta = C_0$, with Heisenberg variations
\be
\label{HeisScalar}
\delta\eta = c_X, \qquad \delta \varphi = c_Y, \qquad 
\delta \tau = c_Z - c_X\varphi\, .
\ee
The scalars $\varphi$ and $\tau$ are Poincar\'e dual to $C_{\mu\nu}$ and $B_{\mu\nu}$, respectively. The duality relations are, schematically, 
$$
\partial_\mu\varphi \quad \sim \quad \epsilon_{\mu\nu\lambda\rho} F^{\nu\lambda\rho},
\qquad\qquad 
\partial_\mu\tau + \eta\,\partial_\mu\varphi \quad \sim \quad 
\epsilon_{\mu\nu\lambda\rho} H^{\nu\lambda\rho}\, .
$$
The algebra is $[X,Y] \sim Z$, with $Y$ and $Z$ generating the axionic shifts (with parameters $c_Y$ and $c_Z$), while $X$ generates the shift of the R-R scalar (with parameter $c_X$). Notice that the central charge of the algebra is (depending on the representation) the gauge symmetry of the R-R tensor and the axionic symmetry of $\tau$, dual to the NS-NS tensor.

The Heisenberg algebra is extended by a fourth perturbative generator $M$ that rotates $X,Y$ and commutes also with the central charge$~Z$: 
\be
\label{Mis}
\delta_M \eta=c_M\varphi\, ,\qquad\delta_M \varphi= -c_M \eta\, ,\qquad
\delta_M \tau = {c_M\over2}(\eta^2-\varphi^2).
\ee
Equivalently, $M$ rotates the phase of the complex R-R scalar $\eta+i\varphi$.
As a result, the perturbative symmetry becomes the two-dimensional Euclidean group $E_2$ with central extension $Z$.

\subsection{Lagrangians}\label{secLagr}

Consider a $\cN=1$ globally supersymmetric theory with two superfields, a chiral $\Phi$ and a real linear $L$. 
It contains three real scalars, $\Re\phi=\Re\Phi|_{\theta=0}$, 
$\Im\phi=\Im\Phi|_{\theta=0}$, and $C=L |_{\theta=0}$, and
$L$ also depends on the curl of an antisymmetric tensor
$H_{\mu\nu\rho} = 3\,\partial_{[\mu} B_{\nu\rho]}$. The Lagrangian (up to two derivatives) is 
\be
{\cal L} = \Dint {\cal H}(L, \Phi, \ov\Phi) + \Fint W(\Phi) + \Fbarint \ov W(\ov\Phi)\, .
\ee
Besides the gauge invariance of $B_{\mu\nu}$ which does not act on the superfields, we also
impose a two-parameter global symmetry acting on $\Phi$ with variations
\be
\label{Hsym1}
\delta\Phi = \alpha-i\beta .
\ee
In this formulation, all three symmetries trivially commute. Nevertheless, in the version where 
$B_{\mu\nu}$ is dualized to a scalar, or in the version where $\Im\phi$ (for instance) is 
transformed into a second antisymmetric tensor, the three-parameter symmetry realizes a 
Heisenberg algebra acting either on three scalars according to eq.~(\ref{HeisScalar}), as in the hypermultiplet formulation of IIB 
strings compactified to four dimensions, or on two tensors and one scalar according to eqs.~(\ref{Heis1})
and (\ref{Heis3}). The Lagrangian compatible with the required symmetry (\ref{Hsym1}) has
\be
\label{Hsym2}
{\cal H}(L,\Phi,\ov\Phi) = {\cal F}(L) + [AL+B]\Phi\ov\Phi,
\qquad
W(\Phi)=k\Phi,
\ee
with an arbitrary function ${\cal F}(L)$ and real constants $A$ and $B$.\,\footnote{Of course, $B$ can be eliminated by a constant shift of $L$.} The constant $k$
generates a $C$--dependent potential $V= |k|^2/(AC+B)$ which does not admit a vacuum
if $A\ne0$. We take then $k=0$. 

The superfields $\Phi$ and $L$ provide an off-shell representation of the $\cN=2$ single-tensor multiplet. On the $\cN=1$ Lagrangian, the condition for a second supersymmetry is \cite{LR}
\be
\label{Hsym3}
{\partial^2{\cal H}\over\partial L^2} + 2 {\partial^2{\cal H}\over\partial\Phi\partial\ov\Phi}=0,
\ee
which in turn indicates that
\be
\label{Hsym4}
{\cal F}_{N=2}(L) = -{A\over3} \, L^3 - BL^2.
\ee
The same theory is given by
\be
\label{Hsym4b}
\widehat{\cal F}_{N=2}(L) = -{1\over3A^2} \, (AL+B)^3.
\ee
Hence, the $\cN=2$ theory compatible with complex shift symmetry of $\Phi$ is the sum
\be
\label{Hsym5}
{\cal L}_{N=2} = \Dint \left[ A \left(-{1\over3}L^3 + L\Phi\ov\Phi\right) + B (-L^2+\Phi\ov\Phi)\right]
\ee
of a trilinear interacting term and of a free term where the symmetry is trivial.
If canonical dimensions are assigned to $L$ and $\Phi$,  $A$ has dimension (mass)$^{-1}$ and $B$
is dimensionless.

Fur further use, we need the bosonic component expansion of this superfield theory.
Using (\ref{conv7}) and the expansion of $\Phi$
$$
\Phi(x,\theta,\ov\theta) = \phi(x) - i\theta\sigma^\mu\ov\theta\, \partial_\mu \phi 
- \theta\theta f - {1\over4}\theta\theta\ov{\theta\theta} \Box \phi ,
$$
we obtain\footnote{The auxiliary field $f$ vanishes.}
\be
\label{Hsymbos}
\begin{array}{rcl}
{\cal L}_{N=2, \, bos.} &=& (AC+B)\Bigl[ {1\over2}(\partial_\mu C)^2 
+ (\partial_\mu\phi)(\partial^\mu\ov\phi) + {1\over12} H^{\mu\nu\rho}H_{\mu\nu\rho} \Bigr]
\crbig
&& -{i\over12}A\, \epsilon^{\mu\nu\rho\sigma} (\ov\phi\,\partial_\mu\phi -\phi\,\partial_\mu\ov\phi)
H_{\nu\rho\sigma}.
\end{array}
\ee
Since, $\partial_{[\mu}H_{\nu\rho\sigma]}=0$, the variation (\ref{Hsym1}) of $\phi$ induces 
a total derivative. Kinetic terms are positive if $AC+B > 0$. If $A\ne0$, $B$ can be eliminated by shifting 
$C$. The (shifted) field $C$ will be assumed strictly positive and the two options are an interacting, 
cubic theory with $A>0$ and $B=0$, or the free theory $A=0$, $B>0$.

We may then perform two supersymmetric duality transformations \cite{S} on theory (\ref{Hsym2}), either
turning the linear $L$ into a chiral $S$ or turning the chiral $\Phi$ into a second linear multiplet $L^\prime$.
The first transformation leads to
\be
\label{Hsym7}
{\cal L} = \Dint \left[\widetilde{\cal F}({\cal Y}) + B\Phi\ov\Phi \right],
\ee
where $\widetilde{\cal F}({\cal Y})$ is the Legendre transform of ${\cal F}(L)$ and the variable 
is\footnote{Notice that $\int d^2\theta d^2\ov\theta\,\Phi\ov\Phi 
= {1\over A}\int d^2\theta d^2\ov\theta\, {\cal Y} + {\rm derivative}$.} 
${\cal Y}=S+\ov S + A\Phi\ov\Phi$. Invariance of ${\cal Y}$ under shift symmetries (\ref{Hsym1}) 
requires a compensating variation of $S$:
\be
\label{Hsym9}
\delta_H S = (\alpha\delta_X+\beta\delta_Y+\gamma\delta_Z)S 
= -A(\alpha+i\beta)\Phi + 2i \gamma, 
\ee
where the axionic shift symmetry of $\Im S$ is dual to the gauge symmetry of $B_{\mu\nu}$, and the subscripts $X,Y,Z$ make clear the correspondence with the transformations (\ref{HeisScalar}). Indeed, since
\be
\label{Hsym10}
[\delta_H^\prime,\delta_H] S \equiv -A(\alpha^\prime+i\beta^\prime) \delta_H\Phi 
+ A(\alpha+i\beta) \delta_H^\prime \Phi
= 2iA(\alpha^\prime\beta - \alpha\beta^\prime), 
\qquad
[\delta_H,\delta_H^\prime]\Phi = 0,
\ee
the chiral theory has Heisenberg symmetry. Moreover, the theory (\ref{Hsym7}) has
another symmetry $M$ rotating the chiral superfield $\Phi$, as already mentioned in the Introduction (see eq.~(\ref{Mis})).

For the $\cN=2$ single-tensor theory (\ref{Hsym5}), the dual hypermultiplet theory\footnote{With
positive K\"ahler metric.} is
\be
\label{Hsym11}
{\cal L}_{N=2} = \Dint {\cal K}({\cal Y})=
{2\over3A^2}\Dint\left( A{\cal Y} + B^2 \right)^{3/2}.
\ee
Eliminating some derivatives, the limiting case $A=0$ is a free theory. As required for a hyper-K\"ahler sigma-model, the determinant of the K\"ahler metric is constant (and positive).

A useful change of variable is
\be
\label{Hsym12b}
\hat S = S - {A\over2}\Phi^2, \qquad\qquad 
{\cal Y}= \hat S + \ov{\hat S} + {A\over2}(\Phi+\ov\Phi)^2.
\ee
and transformation (\ref{Hsym9}) becomes $\delta_H \hat S = -2A\alpha\Phi + 2i\gamma$.
With these variables, the transformations with parameters $\beta$ and $\gamma$ 
only act as shift symmetries of $\Im\Phi$ and $\Im\hat S$ respectively. In terms of variables ${\cal Y}$,
$\Im \hat S$, $\Re\Phi$ and $\Im\Phi$, one immediately deduces that the most general 
Heisenberg-invariant supersymmetric theory 
is of the form (\ref{Hsym7}).

Performing the second duality transformation of the chiral $\Phi$ into a linear $L^\prime$, 
always leads to the dual theory
\be
\label{Hsym14}
{\cal L} = \Dint \left[{\cal F}(L) - {1\over2} { {L^\prime}^2\over AL+B} \right],
\ee
with $\cal F$ given in eq.~(\ref{Hsym4}). Expression (\ref{Hsym14})
is actually the most general $\cN=1$ Lagrangian for $L$ and $L^\prime$ with symmetry 
\be
\label{Hsym13}
\delta L^\prime = \alpha(AL+B).
\ee
This transformation, which links the two antisymmetric tensors in $L$ and $L^\prime$ as in variation
(\ref{Heis1}), forms with their respective gauge symmetries a Heisenberg algebra realized as in type 
IIB strings.

Instead of $\Im\Phi$, we could have chosen to dualize $e^{ia}\Phi$
for any phase $a$, since
$$
\Dint (AL+B)\Phi\ov\Phi 
= {1\over2}\Dint (AL+B) (e^{ia}\Phi + e^{-ia}\ov\Phi)^2
+{\rm derivative}.
$$
The result would be again theory (\ref{Hsym14}). This is a consequence of symmetry $M$, which is however fixed by the choice of dualization and
does not act on $L^\prime$.

\subsection{Hyper-K\"ahler Metrics with Heisenberg Symmetry}\label{secHeis}

The K\"ahler coordinates defined by $\cN=1$ chiral superfields $S$ and $\Phi$ are not necessarily 
the most appropriate to describe a hyper-K\"ahler manifold. There is
a `standard' set of coordinates used to describe hyper-K\"ahler 
metrics with shift isometries in the literature. For comparison purposes, 
we define in this subsection these coordinates in terms of our superfield components.

For any hyper-K\"ahler manifold with a shift symmetry, one can find 
coordinates in which the metric has the Gibbons-Hawking form \cite{GH}
\be
\label{HK6}
ds^2 = f(\vec x) \, dx_i\, dx_i + f(\vec x)^{-1} (d\tau + \omega_i\, dx_i)^2 ,
\ee
with condition $\vec\nabla\times\vec\omega = \vec\nabla f$.
Imposing the requirement of a Heisenberg symmetry acting according to 
\be
\delta_H\, x_1 = \sqrt2\, \alpha, \qquad
\delta_H\, x_2 = -\sqrt2\, \beta, \qquad
\delta_H\, x_3 = 0, \qquad
\delta_H\, \tau = -\sqrt2\, \alpha \, x_2 + \gamma
\ee
also defines $d\tau + x_1\, dx_2$ as the invariant derivative of $\tau$ and indicates that 
$\vec\omega = (0,x_1,0)$. The value of $f(\vec x)$ follows then from 
$\vec\nabla\times\vec\omega = \vec\nabla f$.
This last condition is invariant under $\vec\omega\rightarrow\vec\omega 
+ \vec\nabla\lambda(\vec x)$, for any gauge function $\lambda(\vec x)$. 
In turn, invariance of 
the metric requires the compensating transformation $\tau\rightarrow\tau-\lambda(\vec x)$.

From the $\cN=2$ K\"ahler potential (\ref{Hsym11}), the K\"ahler metric
can be written\footnote{From here on, we do not distinguish
chiral superfields $S$ and $\Phi$ and their lowest complex scalar components.}
\be
\label{HK3}
\begin{array}{rcl}
ds^2 &=& {1\over2}(A{\cal Y}+B^2)^{-1/2}\left[ {1\over4} d{\cal Y}^2 
+ \Bigl(d\Im S +i\frac{A}{2}(\Phi\, d\ov\Phi - \ov\Phi \,d\Phi) \Bigr)^2\right]
\crbig
&& + (A{\cal Y}+B^2)^{1/2} \, d\Phi d\ov\Phi ,
\end{array}
\ee
using coordinates $({\cal Y}, \Im S, \Re\Phi, \Im\Phi)$. 
The supersymmetric duality transformation
from $L$ to $S$ exchanges a real scalar $C=L|_{\theta=0}$, invariant under Heisenberg variations, and
$\Re S$ with variation (\ref{Hsym9}).
The Legendre transformation 
defines the change of variable from ${\cal Y}$ to $C$:
\be
\label{HK4}
AC + B = \sqrt{A{\cal Y}+B^2}.
\ee
Then, in terms of coordinates $(C, \Im S, \Re\Phi, \Im\Phi)$, the metric becomes
\be
\label{HK5}
ds^2 = \displaystyle\frac{AC+ B}{2} \Bigl[ dC^2 + 2 \, d\Phi d\ov\Phi \Bigr]
\displaystyle + \frac{2}{(AC+ B)} \Bigl(d\tau + A\Re\Phi \,d\Im\Phi \Bigr)^2.
\ee
This is the Gibbons-Hawking metric (\ref{HK6}) with 
$\vec x = (\sqrt2\Re\Phi, \sqrt2\Im\Phi, C)$ and
$$
\tau = {1\over2}\left(\Im S - A\Re\Phi\Im\Phi\right) = {1\over2}\Im \hat S .
$$
The function
\be
f(\vec x) = {AC+B\over2}
\ee
solves the hyper-K\"ahler condition 
$\vec\nabla\times\vec\omega = \vec\nabla f$ with 
$\vec\omega=(0,{A\over2}x_1,0)$.
Choosing for instance $\lambda= -{A\over2}x_1x_2$ turns then
$\vec\omega$ into $(-{A\over2}x_2, 0,0)$ and $d\tau + {A\over2}x_1dx_2$
into $d\tau - {A\over2}x_2dx_1$. Similarly, 
a rotation of $\Phi$
$$
\delta_M \, x_1= mx_2, \qquad\qquad \delta_M \, x_2 = -mx_1,
$$
which is compatible with the shift symmetry (\ref{Hsym1}), corresponds to 
$\lambda(\vec x)= {Am\over4}(x_2^2-x_1^2)$. 
It is the isometry $M$ of metric (\ref{HK5}).

The conclusion is that the Gibbons-Hawking ansatz for the hyper-K\"ahler metric corresponds to coordinates where $\Re S$ is replaced by its Legendre dual $C$, which is also the lowest scalar component of the linear superfield dual to $S$.

\section{The Universal Hypermultiplet in N = 2 Supergravity}\label{seclocal}

Hypermultiplet scalars of $\cN=2$ supergravity live on $4n$--dimensional
quaternion-K\"ahler manifolds with holonomy included in $Sp(2n)\times Sp(2)$.
Supergravity requires that the curvature of these Einstein spaces is proportional to the gravitational 
coupling $\kappa^2$ \cite{BW}. Hence, the decoupling limit $\kappa\rightarrow0$ turns the hypermultiplet
manifold into a Ricci-flat hyper-K\"ahler space, as required by global $\cN=2$ supersymmetry \cite{AGF}. 
For a single hypermultiplet, or a four-dimensional quaternion-K\"ahler manifold, the defining condition 
on the holonomy is not pertinent since $Sp(2)\times Sp(2) \sim SO(4)$. The relevant condition is
then self-duality of the Weyl tensor.  

\subsection{The Calderbank-Pedersen Metric with Heisenberg Symmetry}\label{secCPH}

Calderbank and Pedersen \cite{CP} have classified all four-dimensional Einstein metrics with self-dual 
Weyl curvature and two commuting isometries. Using coordinates $(\rho, \eta, \varphi, \tau)$ with the
isometries acting as shifts of $\varphi$ and $\tau$, their metrics are written in terms of any single 
function $F(\rho,\eta)$ verifying
\be
\label{CPFis1}
{\partial^2 F\over\partial\rho^2} + {\partial^2 F\over\partial\eta^2} 
= {3F\over 4\rho^2}.
\ee
It is simple to see~\cite{oneloop} that metrics with Heisenberg symmetry are then obtained if $F$ does not depend 
on $\eta$, {\it i.e.} if\,\footnote{The metric does not make sense without the $\rho^{3/2}$ contribution 
to $F$ and the overall normalization of $F$ is a choice of coordinates. Our $\chi$ is
$\hat\chi$ in Ref.~\cite{oneloop}.} 
\be
\label{CPFis2}
\sqrt\rho \, F(\rho) = {1\over2}[\rho^2 - \chi],
\ee
with an arbitrary real parameter $\chi$. The Calderbank-Pedersen metric with Heisenberg symmetry 
(the CPH metric) reads then
\be
\label{CP1}
ds^2_{CPH}=\frac{\rho^2+\chi}{(\rho^2-\chi)^2}(d\rho^2+d\eta^2+d\varphi^2)
+\frac{4\rho^2}{(\rho^2-\chi)^2(\rho^2+\chi)}(d\tau+\eta\, d\varphi)^2\,.
\ee

The coordinate $\rho$ is positive, $\rho>0$, and positivity of the metric requires $\rho^2 + \chi>0$,
a stronger condition if $\chi$ is negative. 
It is an Einstein metric with negative curvature, and is K\"ahler only if $\chi=0$.
Notice that if $\chi\ne0$, the rescaling
$(\rho, \eta, \varphi, \tau)\rightarrow(|\chi|^{1/2}\rho, |\chi|^{1/2}\eta, |\chi|^{1/2}\varphi, |\chi|\tau)$
turns $\chi$ in metric (\ref{CP1}) into $\pm1$.
This is not true if we turn on string interactions, such as in the presence of D-branes where the dilaton, or equivalently the field $\rho$, couples to the Dirac-Born-Infeld (DBI) action in a non-trivial way (see section~\ref{seclimit}). For this reason, we keep explicitly $\chi$ throughout the paper.
We may use a new coordinate $V=\rho^2$ with metric
\be
\label{CP2}
ds^2_{CPH}=\frac{V+\chi}{(V-\chi)^2}\left({dV^2\over4V}+d\eta^2+d\varphi^2 \right)
+\frac{4V}{(V-\chi)^2(V+\chi)}\Bigl(d\tau+\eta\, d\varphi \Bigr)^2\,.
\ee
The particular case $\chi=0$ has extended symmetry: it is the $SU(2,1)/SU(2)\times U(1)$ metric
with K\"ahler potential
\be
\label{CP3}
K(\hat S,\ov{\hat S}, \Phi,\ov\Phi) = - \ln V, \qquad\qquad V = \hat S + \ov{\hat S} - (\Phi+\ov\Phi)^2,
\ee
and with $\Phi = {1\over\sqrt2}(\eta+i\varphi)$, $\tau = -{1\over2}\Im \hat S$.

The CPH metric is invariant under four isometry variations acting on coordinates 
$(\eta,\varphi,\tau)$:
\be
\label{CP4}
\begin{array}{rclrclrclrcl}
\delta_X\eta &=&  \sqrt2, \qquad
&\delta_Y\eta &=&  0, \qquad
&\delta_Z\eta &=&  0, \qquad
&\delta_M\eta &=& \varphi,
\crbig
\delta_X\varphi &=& 0, 
&\delta_Y\varphi &=& -\sqrt2,
&\delta_Z\varphi &=& 0, 
&\delta_M\varphi &=& -\eta,
\crbig
\delta_X\tau &=& -\sqrt2\,\varphi, 
&\delta_Y\tau &=& 0,
&\delta_Z\tau &=& 1, 
&\delta_M\tau &=& {1\over2}(\eta^2-\varphi^2 ).
\end{array}
\ee
The non-zero commutators are
\be
\label{CP5}
[X,Y] = 2Z, \qquad\qquad
[M,X] = Y, \qquad\qquad [M,Y]=-X.
\ee
Hence, $X$, $Y$ and $Z$ generate the Heisenberg algebra and $Z$ is a central extension of a
two-dimensional euclidean algebra generated by $M$ (which rotates $\varphi$ and $\eta$), 
$X$ and $Y$ (which translate $\varphi$ and $\eta$). With these conventions,
\be
\delta_H\,\Phi = (\alpha X + \beta Y + \gamma Z) \Phi = \alpha-i\beta,
\qquad\qquad
\delta_H\,\hat S =4\alpha\,\Phi - 2i\gamma
\ee
and $V$ is invariant.

The metric (\ref{CP2}) appears in the one-loop-corrected Lagrangian of the universal hypermultiplet of type II strings, reduced to four dimensions,
with the NS-NS and R-R tensors dualized to scalars with shift symmetry
\cite{oneloop}. At one-loop order, the four-dimensional dilaton field is related to 
coordinate $V$ and parameter $\chi$ by
\be
\label{dilaton}
e^{-2\phi_4} = V-\chi, \, \qquad\qquad \chi=-\chi_1, \qquad\qquad \chi_1 = {\chi_E\over12\pi},
\ee
where $\chi_E$ is the Euler number of the internal CY$_3$ manifold.
The real number $\chi_1$ encodes the one-loop correction \cite{oneloop,oneloopold}. Notice that this
relation also indicates that $V-\chi=V+\chi_1>0$, which is stronger than $V=\rho^2 > 0$ if the Euler 
number is negative $(\chi>0$). Since positivity of the CPH metric also requires $V+\chi>0$ if $\chi<0$, the
domain of $V$ is naturally restricted to $V>|\chi|$. 

The R-R scalar is
\be
\label{C0}
C_0 \equiv \eta\, ,
\ee
and is shifted by symmetry $X$. Finally,
Poincar\'e duality gives the following equivalences
$$
\begin{array}{rcl}
d\varphi \quad  &\sim& \quad F_3 = dC_2 - \eta \, dB_2 ,
\crbig
d\tau + \eta\,d\varphi  \quad &\sim& \quad 
H_3 = d B_2 .
\end{array}
$$
In the scalar version, the central charge is the shift $Z$ of $\tau$ (related to the NS-NS tensor $B_2$)
while in the two-tensor version, it is the gauge variation of the (R-R) tensor $C_2$.
Writing $\eta$ and $\varphi$ in a complex $\Phi$ is conventional: we 
always use
$$
\Phi = {1\over\sqrt2}(\eta + i \varphi).
$$

In the previous section, we found a unique four-dimensional hyper-K\"ahler manifold with Heisenberg 
symmetry. It also admits the fourth isometry $M$ rotating $\Phi$.
In the quaternion-K\"ahler case, the theorem of Calderbank-Pedersen \cite{CP} leads 
then to a very similar uniqueness conclusion. We will see how these two results are connected 
when taking an appropriate zero-curvature limit. But we first want to obtain the $\cN=2$ supergravity 
coupling of the universal hypermultiplet on the CPH manifold.

\subsection{Coupling to N = 2 Supergravity}\label{secsugra}

There are different methods to construct hypermultiplet couplings to $\cN=2$ supergravity. The
simplest procedure, which is however not the most general, is to use hypermultiplets coupled to 
local $\cN=2$ superconformal symmetry \cite{N=2conf} and to perform a quaternionic quotient 
\cite {G1, G2} using supplementary hypermultiplet(s) and non-propagating vector multiplet(s). 
In this section, we use this procedure to obtain the supergravity theory of the one-loop-corrected 
dilaton hypermultiplet.

Related constructions, using more general but also more complicated methods, can be found in 
ref.~\cite{ARV}, in the language of projective superspace or in 
ref.~\cite{CIV}, using harmonic superspace.

Conformal $\cN=2$ supergravity is the gauge theory of $SU(2,2|2)$, which has a 
$SU(2)_R\times U(1)_R$ $R$--symmetry with non-propagating gauge fields. Pure 
Poincar\' e $\cN=2$ supergravity is obtained from the superconformal coupling of one propagating 
vector multiplet\footnote{Its gauge field is the graviphoton.} (which may be charged under $U(1)_R$) 
and one hypermultiplet (charged under $SU(2)_R$) by gauge-fixing of the extraneous symmetries.
These two multiplets include in particular the compensating fields used in the gauge-fixing 
to the Poincar\'e theory.

For the superconformal construction of our particular hypermultiplet sigma-model, we also need a
physical hypermultiplet, with positive kinetic metric, to describe the dilaton multiplet. In addition,
for the quaternionic quotient, we need a non-propagating vector multiplet with gauge
field $W_\mu$, gauging a specific generator $T$ to be discussed below, and,
since the elimination of the algebraic vector multiplet involves three constraints and one gauge choice 
on scalar fields, we also need a third non-physical hypermultiplet. Its kinetic metric can have a positive
or negative sign, depending on the constraints induced  by the choice of $T$.
Hence, we need to consider the $\cN=2$ superconformal theory of two vector multiplets and three 
hypermultiplets. The superconformal hypermultiplet scalar sector has then an
`automatic' $Sp(2,4)$ global symmetry in which the gauge generator $T$ 
of the quaternionic quotient is chosen.

\subsection{$S\!p(2,4)$}\label{secsl24}

In the following, we consider three hypermultiplets coupled to (superconformal) $\cN=2$ supergravity.
One (compensating) hypermultiplet has negative signature, the physical hypermultiplet has positive 
signature, the third hypermultiplet, associated to the non-propagating vector multiplet, may have a positive 
or negative signature, depending on the constraints applied to the scalar fields. In any case, we are considering $Sp(2,4)$--invariant supergravity couplings 
of $\cN=2$ hypermultiplets. 

The hypermultiplet scalars are $A_i^\alpha$, with $SU(2)_R$ index $i=1,2$ and $Sp(2,4)$ index 
$\alpha=1,\ldots,6$. They transform in representation $({\bf6},{\bf2})$ of $Sp(2,4)\times SU(2)_R$. 
Their conjugates are\footnote{We follow the conventions of the second paper of ref.~\cite{N=2conf}.}
\be
\label{conf2}
A^i_\alpha = (A_i^\alpha)^* = \epsilon^{ij}\rho_{\alpha\beta} A^\beta_j
\ee
with $\rho^{\alpha\beta}\rho_{\beta\gamma} = - \delta^\alpha_\gamma$ and 
$\epsilon^{ij}\epsilon_{jk} = - \delta^i_k$. 
We choose the $Sp(2,4)$--invariant metric as 
\be
\label{conf3}
\rho = I_3 \otimes i\sigma_2 = \left( \begin{array}{cc} 0 & I_3  \\ -I_3  & 0 \end{array}\right)
\ee
and we use
\be
\label{conf4}
d = \left(\begin{array}{cc} \eta&0\\0&\eta\end{array}\right),
\qquad\qquad \eta = {\rm diag}( -1 , 1 , -1 ), \qquad\qquad
\rho \,d\, \rho = -d.
\ee
In our choice of $\eta$, direction 1 corresponds to the superconformal compensator, direction
2 to the physical hypermultiplet and our choice of quaternionic quotient will require a negative metric in direction 3; otherwise, our construction does not work. 
On scalar fields, $Sp(2,4)$ acts according to
\be
\label{conf5}
\delta A_i^\alpha = g \,{t^\alpha}_\beta A^\beta_i, \qquad\qquad
\delta A^i_\alpha = g \,{t_\alpha}^\beta A_\beta^i, \qquad\qquad
{t_\alpha}^\beta = -\rho_{\alpha\gamma} \,{t^\gamma}_\delta\, \rho^{\delta\beta}.
\ee
Since relation (\ref{conf2}) also implies ${t_\alpha}^\beta = ({t^\alpha}_\beta)^*$, the choice
(\ref{conf3}) and the invariance of $d^\alpha_\beta A_\alpha^i A^\beta_i$ lead to
\be
\label{conf7}
t = \left( \begin{array}{cc}  U & \eta Q \\  -\eta Q^* & U^* \end{array}  \right),
\qquad\qquad U^\dagger = -\eta U \eta, \qquad
Q = Q^\tau, \qquad t^\dagger = - d \, t \, d.
\ee
This is an element of $Sp(2,4)$: $U$ generates the $U(1,2)$ subgroup (9 generators) and
$Q$ (12 generators) generates $Sp(2,4)/ U(1,2)$.
The $(2\times2)$ matrix $A^\dagger \, d \, t \, A$, with matrix elements $A^i_\alpha d^\alpha_\beta {t^\beta}_\gamma A^\gamma_j$, is antihermitian, as required by gauge invariance of $A^\dagger d A$,
and traceless. 

\subsection{The Heisenberg Subalgebra of $S\!U(1,2)$ and $S\!p(2,4)$}\label{secHeislocal}

At string tree-level, the universal hypermultiplet of the dilaton in type II strings lives,
when formulated in terms of four real scalars, on the quaternion-K\"ahler and K\"ahler 
manifold $SU(1,2) / SU(2)\times U(1) = U(1,2)/U(2)\times U(1)$ \cite{Fer}. Since $U(1,2)=SU(1,2)\times
U(1)_0$ is maximal in $Sp(2,4)$, $Sp(2,4)$ has a unique generator commuting with $SU(1,2)$:
the generator of $U(1)_0$. At one-loop however, the isometry is reduced and includes the Heisenberg 
algebra which is known to be a subalgebra of $SU(1,2)$. We need to find the most general 
generator $T$ of $Sp(2,4)$ which commutes with a Heisenberg subalgebra. In the following 
subsections, we will perform the quaternionic quotient construction induced by the gauging 
of $T$. 

Since elements $U$ of the $U(1,2)$ algebra verify $U^\dagger = -\eta \, U \, \eta$ and we
have chosen $\eta = {\rm diag}( -1 , 1 , -1 )$, a generic $U$ is
\be
U = \left( \begin{array}{ccc} ia & A & B \\ \ov A & ib & C \\ -\ov B & \ov C & ic \end{array} \right),
\ee
with $a$, $b$, $c$ real, $A$, $B$, $C$ complex and elements of $SU(1,2)$ are traceless.
On a three-dimensional complex vector, $U(1,2)$ variations are $\delta A = UA$.

We may define the Heisenberg subalgebra as the $U(1,2)$ transformations leaving 
$A_1-A_2$ invariant:
$(\delta_H A)_1 - (\delta_H A)_1= (UA)_1 - (UA)_2 =0$. The transformations acting on $A_1$ and $A_2$ are generated by the following three elements 
\be
X = \left( \begin{array}{ccc} 0 & 0 & 1 \\ 0 & 0 & 1 \\ -1 & 1 & 0 \end{array} \right), \qquad
Y = \left( \begin{array}{ccc} 0 & 0 & i \\ 0 & 0 & i \\ i & -i & 0 \end{array} \right), \qquad
Z = \left( \begin{array}{ccc} i & -i & 0 \\ i & -i & 0 \\0 & 0 & 0 \end{array} \right)
\ee
which verify 
\be
0 = XZ = ZX = YZ = ZY=Z^2, \quad XY = - YX = Z, \quad
X^2 = Y^2 = iZ.
\ee
The Heisenberg algebra 
\be
\label{Heisdef}
[X,Y] = 2 Z , \qquad\qquad [X,Z] = [Y,Z] = 0
\ee
is then realized as a subalgebra of $SU(1,2)$, with variations
\be
\label{Heisvar}
\delta_H\, A = (\alpha X + \beta Y + \gamma Z) \, A
= \left(\begin{array}{ccc}  i\gamma & -i\gamma & \alpha + i\beta \\
i\gamma & -i\gamma & \alpha + i\beta \\ -\alpha + i\beta & \alpha - i\beta & 0
\end{array}\right)
\left(\begin{array}{c} A_1 \\ A_2 \\ A_3 \end{array}\right)
\ee
in the fundamental representation.
Since $Z$ is a central charge of the Heisenberg algebra, we are interested in the
elements of $U(1,2)$ which commute with $Z$. They form an algebra generated by
five elements, $U_0$, $M$, $X$, $Y$ and $Z$, with
\be
U_0 = i I_3, \qquad\qquad
M = i\left( \begin{array}{ccc} 1 & 0 & 0 \\ 0 & 1 & 0 \\ 0 & 0 & -2 \end{array} \right)
\ee
($U_0$ generates the abelian factor of $U(1,2)=SU(1,2)\times U(1)_0$).
Besides the Heisenberg algebra generated by $X,Y,Z$, we also have
\be
[ M, X ] = 3 Y, \qquad [ M, Y ] = -3 X
\ee
and $M$ generates a rotation of $(X,Y)$ leaving $X^2 + Y^2 = 2iZ$ invariant:
$[M,X^2+Y^2] = 2i[M,Z]=0$.

One then easily checks that the most general $U(1,2)$ generator which commutes with the 
Heisenberg algebra generated by $X,Y,Z$ is proportional to
\be
\label{Tis1}
\widehat T = U_0 + \chi\, Z =
i \left( \begin{array}{ccc} 1+\chi & -\chi & 0 \\ \chi & 1-\chi & 0 \\0 & 0 & 1 \end{array} \right), 
\qquad\qquad U_0 = iI_3,
\ee
where $\chi$ is an arbitrary real number. If $\chi=0$, $\widehat T=U_0$ commutes with the whole 
$U(1,2)$. If $\chi\ne0$, $\widehat T$ commutes with the Heisenberg algebra supplemented 
by $U_0$ and $M$.
The extension to $Sp(2,4)$ is straightforward. Requiring that 
\be
\label{Tis2}
T =  \left( \begin{array}{cc} \hat T & 0  \\  0 & \hat T^* \end{array}  \right)
\ee
in $Sp(2,4)$ commutes with an element of $Sp(2,4)/U(1,2)$ corresponds to find a (non\-zero)
symmetric matrix $Q$ in eq.~(\ref{conf7}) such that $\hat T^\dagger Q$ is also antisymmetric, 
which is impossible.\footnote{This would not be true for $\hat T=Z$, which commutes with a larger 
subalgebra of $Sp(2,4)$. The $U_0$ component is necessary.} Hence,
$T$ is also the most general generator in $Sp(2,4)$ which commutes with the Heisenberg algebra generated by $X$, $Y$ and $Z$ in $SU(1,2)$. It actually commutes with $X$, $Y$, $Z$, $M$ 
and $U_0$.

\subsection{N = 2 Supergravity Scalar Lagrangian}\label{secsugra2}

To construct the scalar kinetic metric, 
the relevant terms of the $\cN=2$ conformal supergravity Lagrangian are \cite{N=2conf, G1, G2}
\be
\label{conf9}
\begin{array}{rcl}
e^{-1}{\cal L} &=& d^\alpha_\beta (D_\mu A^\beta_i) (D^\mu A_\alpha^i)
+ (g \, d^\alpha_\beta \, A^i_\alpha {T^\beta}_\gamma A^\gamma_k \, Y^k_i + {\rm c.c.})
\crbig
&& +{1\over6} R ( - X_0\ov X_0 + d^\alpha_\beta A_\alpha^i A^\beta_i )
+ d (X_0 \ov X_0  + {1\over2} d^\alpha_\beta A_\alpha^i A^\beta_i).
\end{array}
\ee
The complex scalar $X_0$ is the partner of the graviphoton, $Y^i_j$,  $Y^i_i=0$, is the triplet of real
auxiliary scalars in the non-propagating vector multiplet with gauge field $W_\mu$ used in the
quaternionic quotient. The covariant derivatives are
\be
\label{conf10}
\begin{array}{rcl}
D_\mu A_i^\alpha &=& \partial_\mu A_i^\alpha - g^\prime W_\mu {T^\alpha}_\beta A^\beta_i
- g {V_{\mu i}}^j A_j^\alpha,
\crbig
D_\mu A^i_\alpha &=& \partial_\mu A^i_\alpha - g^\prime W_\mu {T_\alpha}^\beta A_\beta^i
- g {{V_\mu}^i}_j A^j_\alpha,
\end{array}
\ee
where $g$ and $g^\prime$ are $SU(2)_R$ and $U(1)_T$ coupling constant.
The (anti-hermitian) $SU(2)$ gauge fields ${V_{\mu\, i}}^j$, ${V_{\mu\, i}}^i=0$, and the 
real auxiliary scalar $d$ belong to the multiplet of superconformal gauge fields:
$$
{V_{\mu\, i}}^j = {i\over2}V_\mu^x {(\sigma^x)_i}^j, \qquad\qquad
{{V_\mu}^i}_j = \epsilon^{ik}\epsilon_{jl} {V_{\mu\, k}}^l  = ({V_{\mu\, i}}^j )^*.
$$
We will commonly use a matrix notation, with a $6\times2$ complex matrix $A$ and its $2\times6$ 
conjugate $A^\dagger$ replacing $A^\alpha_i$ and $A_\alpha^i$. 
Condition (\ref{conf2}) implies that $A$ contains six complex components only. It also implies,
in particular, that $A^\dagger dA = {1\over2}\Tr(A^\dagger dA) \, I_2$. Since 
$V_\mu=-V^\dagger_\mu$, the Lagrangian and the derivatives read
\be
\label{conf10b}
\begin{array}{rcl}
e^{-1}{\cal L} &=& \Tr (D_\mu A^\dagger)d(D^\mu A) +  g\Tr Y  A^\dagger d\,TA + {\rm c.c.}
\crbig
&& +{1\over6} R ( - X_0\ov X_0 + \Tr A^\dagger dA )
+ d( X_0\ov X_0 + {1\over2}\Tr A^\dagger dA );
\crbig
D_\mu A &=& \partial_\mu A - g^\prime W_\mu TA - g A V_\mu,
\crbig
D_\mu A^\dagger&=& \partial_\mu A^\dagger - g^\prime W_\mu A^\dagger T^\dagger
+ g V_\mu A^\dagger.
\end{array}
\ee
Constraints are obtained from the elimination of the auxiliary fields and from the gauge-fixing  
of dilatation symmetry in the Poincar\'e theory:
\begin{itemize}
\item 
Einstein frame gauge-fixing condition and $d$ auxiliary field equation:
\be
\label{conf11}
X_0\ov X_0 = {1\over\kappa^2}, \qquad\qquad
\Tr A^\dagger dA= - {2\over\kappa^2}.
\ee
The second condition is invariant under $SU(2)_R$ and $Sp(4,2)$. With an $SU(2)$ gauge choice, 
it allows to eliminate four scalar 
fields and would lead to the $Sp(4,2)/Sp(4)\times Sp(2)$ sigma-model.
\item
Auxiliary fields $Y^i_j$:
\be
\label{conf12}
A^\dagger d\,TA = 0. 
\ee
Since this $2\times2$ matrix is traceless and antihermitian, 
these conditions eliminate three scalars and the associated abelian gauge invariance 
removes a fourth field. 
\end{itemize}
The $SU(2)_R$ gauge fields ${V_{\mu i}}^j$ and the abelian $W_\mu$ have then algebraic 
field equations:
\begin{itemize}
\item
Gauge field $W_\mu$, associated with generator $T$:
\be
\label{conf13}
W_\mu = {\Tr(\partial_\mu A^\dagger d\,T A - A^\dagger d \, T \partial_\mu A) 
\over 2 g^\prime \Tr( A^\dagger T^\dagger d\,TA)}.
\ee
\item
$SU(2)_R$ gauge fields ${V_{\mu \, i}}^j$:
\be
\label{conf14}
V_\mu = -{ \partial_\mu A^\dagger d\,A - A^\dagger d\,\partial_\mu A \over 
g\Tr (A^\dagger d A)} .
\ee
According to the second eq.~(\ref{conf11}), the denominator is $- 2g/ \kappa^2$.
\end{itemize}
At this point, the scalar kinetic Lagrangian in theory (\ref{conf9}) reduces to
\be
\label{conf15}
\begin{array}{rcl}
e^{-1}{\cal L} &=& 
e^{-1}({\cal L}_{kin.} + {\cal L}_{T} + {\cal L}_{SU(2)} )
\crbig
&=&
\Tr (\partial_\mu A^\dagger) d (\partial^\mu A)
- {g^\prime}^2 \Tr(A^\dagger T^\dagger d\,TA) W^\mu W_\mu 
- {g^2\over\kappa^2} \Tr (V^\mu V_\mu).
\end{array}
\ee
The scalar fields are submitted to constraints (\ref{conf11}) and (\ref{conf12}) and the gauge fields
$W_\mu$ and ${V_{\mu\, i}}^j$ are defined by their field equations (\ref{conf13}) and (\ref{conf14}). 

To study the constraints (\ref{conf11}) and (\ref{conf12}) for our specific choice (\ref{Tis1})
and (\ref{Tis2}) of gauged generator $T$, we introduce two three-component complex vectors:
\be
\label{conf16}
A^\alpha_i = \left( \begin{array}{cc}  \vec A_+ & \vec A_- \\
-\vec A_-^* & \vec A_+^*  \end{array} \right),  \qquad\qquad
A_\alpha^i = \left( \begin{array}{cc}  \vec A_+^* & \vec A_-^* \\
-\vec A_- & \vec A_+  \end{array} \right),
\ee
verifying the reality condition (\ref{conf2}). On each doublet $A_{+a}$, $A_{-a}$, $a=1,2,3$, act two 
different $SU(2)$ groups. Firstly, the superconformal $SU(2)_R$ acts on $\pm$ indices. Secondly,
$Sp(2,4) \supset Sp(2)_1 \times Sp(2)_2 \times Sp(2)_3 \sim SU(2)_1 \times SU(2)_2 \times SU(2)_3$
and $(A_{+a}, -A_{-a}^*)$ is a doublet of $SU(2)_a$. One could define three quaternions
\be
\label{conf17}
Q_a = \left( \begin{array}{cc}  A_{+a} & A_{-a} \\ -A_{-a}^* & A_{+a}^*  \end{array} \right)
\qquad\qquad a=1,2,3
\ee
with a left action of $SU(2)_a$ and a right action of the superconformal $SU(2)_R$. 
They verify (for each $a$)
\be
\label{conf17a}
Q_a \, Q_a^\dagger  = Q^\dagger_a \, Q_a = \det Q_a \, I_2,
\qquad\qquad \det Q_a = |A_{+a}|^2 + |A_{-a}|^2.
\ee

The second condition (\ref{conf11}) from $\cN=2$ supergravity becomes:
\be
\label{conf19}
\vec A_+^* \cdot \vec A_+ + \vec A_-^* \cdot \vec A_- = -{1\over\kappa^2}, \qquad\qquad
\vec A^*\cdot\vec A = \vec A^\dagger \eta \vec A = -|A_1|^2 + |A_2|^2 - |A_3|^2 .
\ee
With eq.~(\ref{Tis2}), condition (\ref{conf12}) leads to three (real) equations:
\be
\label{conf22}
\begin{array}{rcl}
\vec A_+^\dagger \,i\eta \hat T\, \vec A_+ &=&  \vec A_-^\dagger \,i\eta \hat T\, \vec A_-   ,
\crbig
\vec A_-^\dagger \,i\eta \hat T \, \vec A_+  &=& 0  
\end{array}
\ee
($[i\eta \hat T]^\dagger = i\eta\hat T$). 
With the explicit form of $\hat T$, eq.~(\ref{Tis1}), and defining dimensionless fields 
$a_{\pm i} = \sqrt2\kappa A_{\pm i}$,
the four constraints (\ref{conf19}) and (\ref{conf22}) read finally
\be
\label{sol10}
\begin{array}{ll}
I:\qquad&
|a_{+1}|^2 + |a_{-1}|^2 - |a_{+2}|^2 - |a_{-2}|^2 + |a_{+3}|^2 + |a_{-3}|^2 = 2,
\crbig
II:\qquad&
-|a_{+1}|^2 + |a_{+2}|^2 - |a_{+3}|^2 - \chi |a_{+1}-a_{+2}|^2
\crbig &\hspace{2.1cm}
= -|a_{-1}|^2 + |a_{-2}|^2 - |a_{-3}|^2 - \chi |a_{-1}-a_{-2}|^2,
\crbig
III:\qquad&
0 = -a_{+1}\ov a_{-1} + a_{+2}\ov a_{-2} - a_{+3}\ov a_{-3}
- \chi (a_{+1}-a_{+2})(\ov a_{-1} - \ov a_{-2}) .
\end{array}
\ee
They are invariant under Heisenberg variations (\ref{Heisvar}) of $\vec a_+$ and $\vec a_-$.
The case $\chi=0$ has been considered by Galicki \cite{G1}. Since it leads to 
$SU(1,2)/SU(2)\times U(1)$, coordinates more appropriate for this larger isometry have been used.

\subsection{Solving the Constraints}\label{secconst}

To solve the constraints (\ref{sol10}), we insist on keeping in $\vec a_-$ a field $\Phi$ which 
transforms under the Heisenberg variations\footnote{See eq.~(\ref{Heisvar}).}
$\delta_H\, \vec a_- = (\alpha X+\beta Y+ \gamma Z)\,\vec a_-$ with a complex shift:
\be
\label{sol4}
\delta_H \, \Phi = \alpha - i \beta.
\ee
This is the case if $a_{-1}=a_{-2}$, and $a_{-3}$ is then invariant. We may define 
$\ov\Phi = a_{-1}/a_{-3}$ and constraint $III$ reduces to $a_{+3}= (a_{+2}-a_{+1})\Phi$.
Since 
$$
\delta_H \left({a_{+2} + a_{+1} \over a_{+2} - a_{+1} }\right) = - 2i\gamma 
+ 2(\alpha + i \beta){ a_{+3} \over a_{+2} - a_{+1}}
= -2i\gamma + 2 \Phi\, \delta_H\ov\Phi,
$$
we finally define 
\be
\label{sol5}
S =  {a_{+2} + a_{+1} \over a_{+2} - a_{+1}} + Y , \qquad\qquad
\delta_HS = -2i\gamma + 2(\alpha+i\beta)\Phi
\ee 
and the quantity
\be
\label{sol5b}
Y = S+\ov S - 2 \Phi\ov\Phi
\ee 
is invariant under Heisenberg variations. The algebra follows from
$[\delta_H^\prime,\delta_H] = (\alpha^\prime\beta-\alpha\beta^\prime)[X,Y]
= 2(\alpha^\prime\beta-\alpha\beta^\prime)Z$:
$$
[\delta_H^\prime, \delta_H]S = 2(\alpha^\prime+i\beta^\prime) \delta_H\Phi 
- 2(\alpha+i\beta) \delta_H^\prime\Phi
= -4i(\alpha^\prime\beta-\alpha\beta^\prime)
= 2(\alpha^\prime\beta-\alpha\beta^\prime)Z.
$$
These definitions are summarized in the choice
\be
\label{sol11b}
\vec a_- = {K\over\Delta} \left( \begin{array}{c} \ov\Phi \\ \ov\Phi \\ 1 \end{array}\right),
\qquad\qquad
\vec a_+ = {1\over\Delta}\left( \begin{array}{c} S-Y-1 \\ S-Y+1 \\  a \end{array} \right),
\ee
with complex fields $S$, $\Phi$ and $a$.
The four available gauge choices have been used to take $\Delta=|\Delta|$, $K=|K|$ 
and $a_{-1} = a_{-2}$.
Under Heisenberg variations, $\Delta$ and $K$ are invariant.
Hence, we are left with eight real scalar fields submitted to the four constraints (\ref{sol10}) 
which drastically simplify:
\be
\label{sol12}
\begin{array}{ll}
I:\qquad&
\Delta^2\Bigl(2 - |a_{+1}|^2 + |a_{+2}|^2 - |a_{+3}|^2 \Bigr)  = K^2,
\crbig
II:\qquad&
2(S+\ov S) - |a|^2 - 4 Y= 4\chi - K^2 ,
\crbig
III:\qquad&
a  = 2\Phi.
\end{array}
\ee
Hence, the solution is
\be
\label{sol16}
\vec a_- = \sqrt{Y + 2 \chi \over Y + \chi}
 \left( \begin{array}{c} \ov\Phi \\ \ov\Phi \\ 1 \end{array}\right),
\quad\qquad
\vec a_+ ={1\over\sqrt{2(Y + \chi)}}
\left( \begin{array}{c} S - Y - 1 \\ S- Y + 1 \\ 2\Phi \end{array} \right).
\ee
The solution implies $Y+\chi>0$ if $\chi>0$ or $Y+2\chi>0$ if $\chi<0$.
The scalar kinetic Lagrangian (\ref{conf15}) obtained from this solution is\footnote{
All fields and parameter $\chi$ are dimensionless.}
\be
\label{sol18a}
\begin{array}{rcl} 
\kappa^2 {\cal L} &=& \displaystyle
{ (Y+3\chi) \over 4(Y+2\chi)(Y+\chi)^2}(\partial_\mu Y)^2 - {2\over Y+\chi} \, \partial_\mu\Phi\,
\partial^\mu\ov\Phi
\crbig
&& \displaystyle
+ {1\over2(Y+\chi)(Y+3\chi)} \, \left[ \Im(\partial_\mu S - 2\ov\Phi\, \partial_\mu\Phi)\right]^2 
\crbig
&& \displaystyle
+ {1 \over 2(Y+\chi)^2} \, \left[ \Im(\partial_\mu S - 2\ov\Phi\, \partial_\mu\Phi)\right]^2
+ {4(Y+2\chi)\over (Y+\chi)^2} \, \partial_\mu\Phi\,\partial^\mu\ov\Phi. 
\end{array}
\ee
The first line comes from the basic scalar kinetic terms ${\cal L}_{kin.}$ in Lagrangian
(\ref{conf15}). The second line is the contribution ${\cal L}_T$ of the 
gauge field of $T$, the third line arises from the supergravity $SU(2)_R$ gauge fields. Each term is 
separately invariant under Heisenberg variations.
Collecting terms, the final form of the theory is
\be
\label{sol18b}
\begin{array}{rcl}
\kappa^2 {\cal L} &=& \displaystyle
{Y+3\chi \over (Y+\chi)^2 } \left[ {1\over4} {(\partial_\mu Y)^2 \over Y+2\chi}
+ 2 \partial_\mu\Phi \, \partial^\mu\ov\Phi\right]
\crbig
&& \displaystyle
+ {Y+2\chi \over (Y+3\chi)(Y+\chi)^2 } \Bigl( \partial_\mu \Im\hat S 
- 4 \Re\Phi\, \partial_\mu\Im\Phi \Bigr)^2,
\end{array}
\ee
where 
\be
\label{newS}
\hat S = S+\Phi^2,
\ee
for which $Y=\hat S + \ov{\hat S} - (\Phi+\ov\Phi)^2$ and 
$\Im(dS-2\ov\Phi \,d\Phi) = d\Im\hat S - 4\Re\Phi\, d\Im\Phi$. 
From the existence of solutions (\ref{sol16}) and positivity of the
Lagrangian, the range of $Y$ is 
$Y+\chi>0$ if $\chi>0$ and $Y+3\chi>0$ if $\chi<0$
Writing as usual
\be
\label{Lmetric}
{\cal L} = {1\over\kappa^2} \, g_{ab} (\partial_\mu q^a)(\partial^\mu q^b)
= G_{ab}(\partial_\mu q^a)(\partial^\mu q^b),
\ee
$q^a=(Y, \Re\Phi,\Im \Phi, \Im\hat S)$, and comparing $ds^2= g_{ab} \,dq^a dq^b$ with 
expression (\ref{CP2}), we see that the hypermultiplet kinetic metric $g_{ab}$ is the CPH metric with
\be
\label{sol19}
Y = V-2\chi = \rho^2-2\chi,
\ee
and with\footnote{This choice is not unique. We may for instance rotate $\Phi$ using isometry $M$.}
\be
\label{sol20}
\Phi = {1\over\sqrt2} (\eta+i\varphi), \qquad\qquad
\Im\hat S = - 2 \tau.
\ee
Positivity of kinetic terms is obtained if $V= \rho^2 >|\chi|$ which is, as explained at the end
of subsection \ref{secCPH}, the natural domain of $V$. 

As already observed, the case $\chi=0$ corresponds to 
the $SU(2,1)/SU(2)\times U(1)$ metric
\be
\label{sol8c}
ds^2 = {1\over Y^2}\left[ {1\over4}dY^2 + \Bigl(d\Im\hat S-4\Re\Phi \,d\Im\Phi\Bigr)^2 \right] 
+ {2\over Y}\, d\Phi d\ov\Phi .
\ee
With K\"ahler coordinates $\hat S$ and $\Phi$, the K\"ahler potential is $K=-\ln Y$, with 
$Y=V=\hat S+\ov{\hat S} - (\Phi+\ov\Phi)^2$.

This relatively simple construction of the one-loop-corrected dilaton hypermultiplet metric allows easily
to derive the full $\cN=2$ supergravity Lagrangian, using $\cN=2$ superconformal tensor calculus
\cite{N=2conf, G1, G2}.

\section{Zero-Curvature Hyper-K\"ahler Limit}\label{seclimit}

All quaternion-K\"ahler metrics are Einstein spaces with nonzero curvature. With one hypermultiplet, the scalar kinetic Lagrangian (\ref{Lmetric}) verifies \cite{BW}
\be
\label{Ricci1}
R_{ab} = -6\,g_{ab} = -6 \kappa^2\, G_{ab}.
\ee
The link with global $\cN=2$ supersymmetry is realized by defining a $\kappa\rightarrow0$ hyper-K\"ahler limit of the CPH metric (\ref{CP2}) or (\ref{sol18b}) in which, if feasible, the Heisenberg 
algebra does not contract to an abelian symmetry. 
As observed in Subsection \ref{secCPH}, the magnitude of $\chi$ can be eliminated by rescaling 
of the coordinates (in the absence of D-branes). We then have three $|\chi|$-independent cases to examine: firstly, positive 
$\chi$, with $V>0$; secondly, $\chi=0$ ($V>0$) which is $SU(1,2)/SU(2)\times U(1)$; thirdly, a 
negative $\chi$, with $V > |\chi|$. In each case, we should seek to find a parameter-free 
zero-curvature limit. The most interesting case turns out to be $\chi$ negative, which we first study. 

With $\chi$ negative, we are interested in the CPH metric in the region $V+\chi\sim0$.
We then apply to metric (\ref{CP2}) the following change of variables:
\be
\label{change}
\begin{array}{rclrcl}
V &=& 2|\chi| \, \kappa^{2/3} \mu^{-1/3}\,C- \chi \,, \qquad&\qquad
\varphi&=& \sqrt{|\chi|} \,\kappa^{2/3} \mu^{-1/3}\,\hat{\varphi}\,,
\crbig
\eta&=&\sqrt{|\chi|} \,\kappa^{2/3} \mu^{-1/3}\,\hat{\eta}\,,\qquad&\qquad
\tau&=&|\chi|\,\kappa^{4/3} \mu^{1/3}\,\hat{\tau}\,,
\end{array}
\ee
where $\mu$ is an arbitrary mass scale. Positivity of the metric, $V+\chi>0$ implies $C>0$. 
While the original fields 
are dimensionless, the new, hatted, fields $(C, \hat \phi,\hat\eta,\hat\tau)$
have canonical dimension. With this choice of dependence in $\kappa$, the resulting metric is
\be
\label{flat4}
\begin{array}{rcl} 
ds^2 \,\,=\,\, g_{ab} \, dq^adq^b&=& \displaystyle
{\kappa^2\over2} {\mu C \over \bigl[(\kappa\mu)^{2/3}C 
+ \mu \bigr]^2 } \left[  
{dC^2 \over 2\kappa^{2/3}\mu^{-1/3}C + 1} + d\hat\eta^2 + d\hat\varphi^2 \right]
\crbig
&& \displaystyle
+ {\kappa^2\mu^2 \over 2C} {2(\kappa\mu)^{2/3}C + \mu \over
[(\kappa\mu)^{2/3}C + \mu ]^2 }  
\left[ d\hat\tau + \frac{1}{\mu}\hat\eta d\hat\varphi \right]^2,
\end{array}
\ee
since $\chi = -|\chi|$.
Using this metric in Lagrangian (\ref{Lmetric}), the overall factor $\kappa^2$ cancels and we can 
take the limit $\kappa\rightarrow0$, with result
\be
\label{flat5}
{\cal L}_{\kappa\rightarrow0} = {C \over 2\mu}
\left[ (\partial_\mu C)^2  + (\partial_\mu \hat{\eta})^2 + (\partial_\mu \hat{\varphi})^2\right]
+ {\mu\over 2C}  \left[\partial_\mu\hat{\tau} + {1\over\mu}\hat{\eta}\,\partial_\mu\hat{\varphi}\right]^2\,.
\ee
This scalar Lagrangian has the hyper-K\"ahler metric with Heisenberg symmetry (\ref{HK5}) with $A=1/\mu$ and $B=0$ and with relations $\Phi = {1\over\sqrt2}(\hat\eta + i \hat \varphi)$, $\hat\tau=2\tau$.
As noticed earlier, parameter $B$ can always be absorbed in a shift of $C$, as long as 
$A\ne0$.

Notice that to obtain limit (\ref{flat5}), we only need the change of variables (\ref{change}) 
up to higher orders in $\kappa$. In particular, according to eq.~(\ref{dilaton}), we
may write the four-dimensional string dilaton as
\be
\label{gslimit}
\begin{array}{rcl}
e^{-2\phi_4} &=& 2|\chi|\kappa^{2/3} \mu^{-1/3}\,C - 2\chi ,
\crbig
\phi_4 &=& \langle\phi_4\rangle - \kappa^{2/3}\mu^{-1/3} \hat\phi_4 ,
\crbig
e^{-2\langle\phi_4\rangle} &=& -2\chi \,\,=\,\, 2|\chi|,
\qquad\qquad\qquad   C \,\, = \,\, 2\hat\phi_4 ,
\end{array}
\ee
in terms of the fluctuation $\hat\phi_4$ and of the background value $\langle\phi_4\rangle$.
Since $|\chi| = \chi_1 = \chi_E/(12\pi)$, we are considering the case of a positive Euler number
$\chi_E=2(h_{11}-h_{21})$, with $h_{11},h_{12}$ the corresponding Betti numbers of the CY$_3$ manifold. A typical example with a single hypermultiplet would be IIA strings on a CY$_3$ manifold with $h_{21}=0$.
Positivity-related questions with several hypermultiplets, as is in particular 
the case with a negative Euler number, should be reanalyzed. 

Comparing the scalings (\ref{change}) and the identification of the string
coupling in the last eq.~(\ref{gslimit}), we see that the R-R fields
$\eta$ and $\varphi$ carry as expected a supplementrary factor $g_{string}$.

We could also consider the single-tensor version of the theory. 
Dualizing $\hat\tau$ into $H_{\mu\nu\rho}$, we find
\be
\label{flat6}
\begin{array}{rcl}
{\cal L}_{\kappa\rightarrow0,ST} &=& \displaystyle
{C \over \mu }\left[ {1\over2}(\partial_\mu C)^2
+ {1\over12} H^{\mu\nu\rho} H_{\mu\nu\rho} 
+ (\partial_\mu\ov\Phi)(\partial^\mu\Phi) \right]
\crbig
&& \displaystyle
- {i\over 12\mu} \epsilon^{\mu\nu\rho\sigma} ( \ov\Phi\partial_\mu\Phi - \Phi\partial_\mu\ov\Phi )
 H_{\nu\rho\sigma} .
\end{array}
\ee
This is the bosonic sector (\ref{Hsymbos}) of the single-tensor theory 
(\ref{Hsym5}) with again $A=1/\mu$ and $B=0$. Then, for negative $\chi$,
the $\cN=2$ supergravity hypermultiplet with Heisenberg symmetry is 
described in the global supersymmetry limit by the unique nontrivial
theory with the same symmetry. 

For completeness, we may also consider the case of the CPH metric with 
positive $\chi$. The interesting limiting regions are $V\sim0$ and 
$V-\chi\sim 0$. If $V = \rho^2 \ll \chi$, 
\be
ds^2_{CPH} = {1\over\chi}( d\rho^2 + d\eta^2 + d\varphi^2) + {4\rho^2\over\chi^3}
(d\tau + \eta \,d\varphi)^2.
\ee
The appropriate rescalings are $(\rho,\eta,\varphi,\tau) =
(\sqrt\chi\kappa\hat\rho,\sqrt\chi\kappa\hat\eta,\sqrt\chi\kappa\hat\varphi,\chi\hat\tau)$ to obtain
\be
ds^2_{CPH} = \kappa^2 \left[ d\hat\rho^2 + d\hat\eta^2 + d\hat\varphi^2 
+ 4\hat\rho^2 (d\hat\tau + \kappa^2\,\hat\eta d\hat\varphi)^2 \right].
\ee
The Heisenberg symmetry acting on the rescaled fields has algebra $[X,Y]=2 \kappa^2 Z$.
In the limit $\kappa\rightarrow0$, it contracts to $[X,Y]=0$ and we find
\be
\lim_{\kappa\rightarrow0} \, {1\over\kappa^2} ds^2_{CPH} = d\hat\rho^2 + 4\hat\rho^2d\hat\tau^2 + d\hat\eta^2 + d\hat\varphi^2,
\ee
which is the trivial four-dimensional euclidean space.  
The second region of interest if $\chi>0$ is $V-\chi\sim0$. First, we change coordinates to
\be
V= 2\lambda C + \chi, \qquad \eta= \lambda \hat\eta/\sqrt\chi,
\qquad \varphi = \lambda\hat\varphi / \sqrt\chi, 
\qquad \tau =\lambda \hat\tau
\ee
and the metric for $\lambda\rightarrow0$ and $\chi$ finite reads 
\be
ds^2_{CPH}  = {1\over 2C^2} \left[ dC^2
+ d\hat\eta^2 + d\hat\varphi^2 + d\hat\tau^2 \right].
\ee
This limiting metric is $SO(1,4)/SO(4)$, again with $R_{ij}=-6g_{ij}$ and with radius 
$\sim\langle C \rangle$. In the large radius, zero-curvature limit, the metric is trivial.
Finally, in the $SU(1,2)/SU(2)\times U(1)$ case $\chi=0$, the zero-curvature limit is again trivial.

The conclusion is that in the zero-curvature limit, the CPH one-loop Lagrangian for the 
dilaton hypermultiplet is the hyper-K\"ahler $\cN=2$ sigma-model with Heisenberg symmetry
(\ref{Hsym5}). If the one-loop parameter $\chi$ is negative, then $A\ne0$ and the Heisenberg 
algebra has a non-trivial realization in this limit. If $\chi\ge0$ however, $A=0$ and the
limit of $\cN=2$ global supersymmetry is the free hypermultiplet. In the string context, the above non-trivial limit can be taken if the string coupling is tuned at a fixed value, according to the third line of eq.~(\ref{gslimit}), which applies with positive Euler number.

In chapter \ref{secDBI} we constructed the interaction of a hypermultiplet with the Dirac-Born-Infeld Maxwell Lagrangian. The hypermultiplet sector has a full linear $\cN=2$ supersymmetry while the second supersymmetry is nonlinearly realized on the Maxwell superfield $W_\alpha$. As an application of our results, we can easily use our identification of the string universal hypermultiplet. The bosonic DBI action, after
elimination of the Maxwell auxiliary field and using the single-tensor formulation, is\footnote{In chapter \ref{secDBI}, this is the electric version of the theory, induced by a $\cN=2$ Chern-Simons coupling $g B\wedge F$.}
\be
\label{DBI}
\begin{array}{rcl}
{\cal L}_{DBI} &=& \displaystyle
{1\over 8 {\cal F}}(2g \textrm{Re} \Phi- {1\over {\cal F}})\left[ 1-\sqrt{1+{2g^2C^2\over(2g\Re\Phi- {1\over{\cal F}} )^2}}\sqrt{-\det(\eta_{\mu\nu}+2\sqrt{2}{\cal F}F_{\mu\nu})} \right]
\crbig
&+&\displaystyle g \epsilon^{\mu\nu\rho\sigma}\left({{\cal F}\over 4} \textrm{Im} \Phi F_{\mu\nu}F_{\rho\sigma}-{1\over 4} B_{\mu\nu}F_{\rho\sigma}+{1\over 24{\cal F}}C_{\mu\nu\rho\sigma}\right).
\end{array}
\ee
In this expression, ${\cal F}$ is the breaking scale of the second,
nonlinearly realized supersymetry (with dimension (energy)$^{-2}$) and
$g$ is the Chern-Simons coupling\footnote{In contrast to chapter \ref{secDBI}, we have defined single-tensor fields with canonical dimension
so that $g$ has dimension (energy). We also chose the Fayet-Iliopoulos term to be $1/{\cal F}$ so that gauge kinetic terms are canonically normalized at
$\Re\Phi=0$.} (equal to the string coupling for a D3-brane).
The four-form field 
$C_{\mu\nu\rho\sigma}$ is a component of the single-tensor multiplet required by supersymmetry of the nonlinear theory [see section \ref{secV1XY}].

Since we have control of the kinetic Lagrangian of the universal string hypermultiplet in the global supersymmetry limit, we can then identify 
the single-tensor fields in terms of string fields. First, $C$ is the global
dilaton and  $B_{\mu\nu}$ is the NS-NS tensor. 
Then, the complex scalar $\Phi$ includes the R-R fields. The
supersymmetric minimum of the scalar potential included in theory (\ref{DBI})
implies $\langle C\rangle =0$ and $\Phi$ corresponds to 
flat directions of this vacuum. 

\chapter{Summary of Results}
\label{secfinal}

This part of the thesis constitutes a detailed study, in the context of global supersymmetry, of the D-brane effective action of $\cN=2$ compactifications in type II string theory, including both the gauge part as well as the couplings of the brane to bulk fields. From a field theoretic point of view, this is the interaction of the Maxwell goldstino multiplet of $\cN=2$ nonlinear supersymmetry to a hypermultiplet with at least one isometry. The hypermultiplet is described by its Poincar\'e dual single tensor multiplet where $\cN=2$ supersymmetry can be realized off shell. The nonlinear breaking of the second SUSY is realized with a supersymmetric constraint while the coupling of the single-tensor to the goldstino multiplet is realized with a supersymmetric generalization of the usual Chern-Simons term $B\wedge F$. This system has equivalent descriptions in terms of different chiral and tensor multiplets. We proved the equivalence of these descriptions by performing $\cN=1$ and $\cN=2$ Poincar\'e type dualities which led us to a net of theories summarized in the figure below.

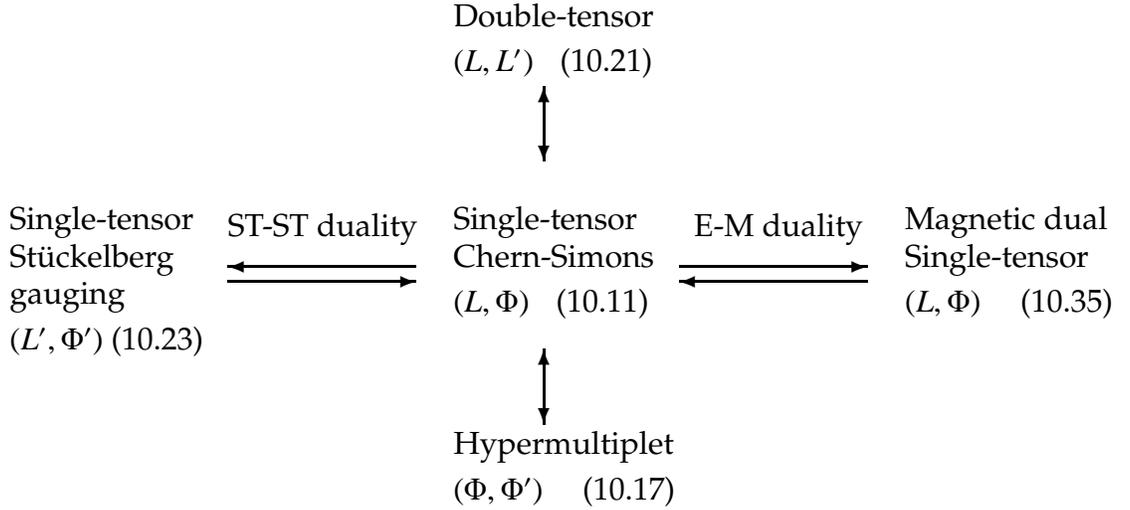
\begin{figure}[htbp]
\label{figurenet}
\begin{center}
\setlength{\unitlength}{1cm}%
\begin{picture}(15,7.5)(-0.5,0.5)
\thicklines

\put(0.1,4.1){Single-tensor}
\put(0.1,3.6){St\"uckelberg}
\put(0.1,3.1){gauging}
\put(0.1,2.5){$(L',\Phi')$ (\ref{STprime})}

\put(3.0,3.4){\vector(1,0){2.5}}
\put(5.5,3.6){\vector(-1,0){2.5}}
\put(3.0,4){ST-ST duality}

\put(6,4.1){Single-tensor}
\put(6,3.6){Chern-Simons}
\put(6,3){$(L,\Phi)$ \,\, (\ref{DBIb})}

\put(9.0,3.6){\vector(1,0){2.5}}
\put(11.5,3.4){\vector(-1,0){2.5}}
\put(9.2,4){E-M duality}

\put(12,4.1){Magnetic dual}
\put(12,3.6){Single-tensor}
\put(12,3){$(L,\Phi)$ \quad (\ref{DBIml})}

\put(7.2,5.2){\vector(0,1){0.8}}
\put(7.2,5.8){\vector(0,-1){0.8}}

\put(6,6.8){Double-tensor}
\put(6,6.2){$(L,L')$ \,\, (\ref{DT})}

\put(7.2,1.5){\vector(0,1){1}}
\put(7.2,2.5){\vector(0,-1){1}}

\put(6,1.1){Hypermultiplet}
\put(6,0.5){$(\Phi,\Phi')$ \quad (\ref{hyp1})}

\end{picture}%
\end{center}
\begin{quote} 
\caption{Web of dualities: double arrows indicate duality transformations preserving off-shell $\cN=2$  supersymmetry, simple arrows are $\cN=1$ off-shell dualities only, leading to theories with on-shell $\cN=2$ supersymmetry. The $\cN=1$ superfields and the related equations are indicated. }
\end{quote}
\end{figure}

Up to appropriate field redefinitions, this system is also equivalent to the Higgs phase of $\cN=2$ nonlinear QED coupled to a charged hypermultiplet. The system also explores a phase with all supersymmetries broken and a phase with the $U(1)$ gauge symmetry unbroken. In the Higgs phase an interesting phenomenon appears. The goldstino multiplet combines with the hypermultiplet to form a massive vector multiplet and a massless chiral multiplet. In the massive multiplet, the goldstino combines with a hypermultiplet fermion and becomes massive, thus realizing a new type of super-Higgs mechanism that doesn't involve a gravitino. This is possible because the hypermultiplet is charged under the $U(1)$ partner of the goldstino.

The next step is to find how the Lagrangian of our system eq.~(\ref{DBId}) relates with the global limit of the low energy effective D-brane action in $\cN=2$ compactifications. In other words, we have to relate the field basis used in our construction with the string basis of the universal hypermultiplet. To do that we need to specify the correct global limit of the universal hypermultiplet. At string tree level, the universal hypermultiplet is described by the symmetric coset $SU(2,1)/SU(2)\times U(1)$. At the quantum level this isometry structure reduces to the centrally extended Euclidean algebra $E_2$ which contains a Heisenberg subalgebra. Requiring that the same isometry structure survive in the global limit we found that apart from the trivial global limit of canonical kinetic terms (which destroys this isometry), there is also a limit leading to a hyperK\"ahler manifold. An independent derivation of the most general hyperK\"ahler manifold that satisfies the Heisenberg isometries had as a result precisely the same manifold that we obtained from this global limit. We could then identify the string basis of the system.


\appendix

\chapter{Coefficients for the Higgs Masses}
\label{CoefficientsHiggsMasses}

For completeness, we present the expressions of the coefficients in eq.~(\ref{dmh}):
\bea
\gamma_1^\pm&=&
\frac{  \pm v^2}{2 u^2 (1+u^2)^3\,w^{ 1/2}}
\nonumber\\
&\times &
\,\Big[(B_0m_0\mu_0)^2 \,(1+u^2)^4- 2 m_Z^2\,u^2 \,
\big[m_Z^2 (1-u^2)^2+(1+u^2)\,( 8 \mu_0^2\,u^2 \pm (u^2-1)\,w^{ 1/2}))
\big]\nonumber\\
&+&(B_0\,m_0\mu_0)\,u (1+u^2)^2\big[m_Z^2 \,(1+u^2)-(\pm\, w^{1/2} (1+u^2) +
16\mu_0^2\,u^2 )\big]
\Big]
\\
\gamma_2^\pm&=&
\frac{\pm v^2}{2 (1+u^2)^3 \,w^{ 1/2}}
\nonumber\\
&\times &
\Big[
(B_0m_0\mu_0)^2 (1+u^2)^4-2 m_Z^2 u^2 \big[8 \mu_0^2 (1+u^2)
+ m_Z^2 (1-u^2)^2 \pm w^{ 1/2}(1-u^4) \big]
\nonumber\\
&-&(B_0 m_0\mu_0)\,u\,(1+u^2)^2\big[
16 \mu_0^2 - m_Z^2 (1+u^2) \pm(1+u^2)\,w^{ 1/2}\big]\Big]
\\
\gamma_3^\pm\!&=&\!\!\gamma_4^\pm=\!
\frac{\pm v^2}{u\,(1+u^2)^2\,w^{1/2}}
\,\big\{\mu_0^2
\big[\!-\!B_0m_0\mu_0\,(1\!+\!u^2)^3\! +m_Z^2 u(1-\!6 u^2\! +u^4)\mp 
u(1\!+\!u^2)^2\,w^{1/2}\big]\nonumber\\
&+& B_0 m_0\mu_0 \,u^2\,(1+u^2)\,m_Z^2
+m_Z^2\,u^3\,(m_Z^2\mp w^{ 1/2})\big\}
\\[7pt]
\gamma_5^\pm&=&
\frac{\mp v^2}{8 u^3\,(1+u^2)^3\,w^{ 1/2}}
\Big[
 (B_0 m_0 \mu_0)^2 (1+u^2)^4\,(-1 +3 u^2)
- (B_0m_0\mu_0)\,u (1+u^2)^2\nonumber\\
&\times &\!\!\! \big[- 2 m_Z^2 (1+5 u^2)
+2\mu_0^2 \, (1+8 u^2 +25 u^4+2 u^6)
 \pm (1+u^2) (3 u^2-1)\,w^{ 1/2}\big]
\nonumber\\
&- &u^2\,m_Z^2 \big[
m_Z^2 (1-19 u^2 -u^4 +3 u^6)
- 2 \mu_0^2\,(1+u^2)(1 -16\,u^2-23 u^4+2 u^6)
\nonumber\\
&\pm &(1+u^2)^2(1+3 u^2)\,w^{ 1/2}\big]
+ 2 \mu_0^2\, u^2\,\big[\pm (1+u^2)^2\,(1-9 u^2 +2 u^4)
w^{ 1/2}\,\big]
\,\,\Big]
\\[7pt]
\gamma_6^\pm&=&
\frac{\pm v^2}{8 u^2\,(1+u^2)^3\,w^{ 1/2}}
\Big[
 (B_0 m_0 \mu_0)^2 \,u\,(1+u^2)^4\,(-3 + u^2)
- (B_0m_0\mu_0)\, (1+u^2)^2\nonumber\\
&\times &\!\!\! \big[ 2 m_Z^2 (5+ u^2)\,u^4
-2\mu_0^2 \, (2+25 u^2 +8 u^4+ u^6)
 \pm (1+u^2)\, (u^2-3)\,u^2\,w^{ 1/2}\big]
\nonumber\\
&+ &u\,m_Z^2 \big[
m_Z^2 (3- u^2 -19 u^4 + u^6)\,u^2
-2 \mu_0^2\,(1+u^2)(2 -23\,u^2-16 u^4+ u^6)
\nonumber\\
&\pm &u^2\,(1+u^2)^2(3+ u^2)\,w^{ 1/2}\big]
- 2 \mu_0^2\, u\,\big[\pm (1+u^2)^2\,(2-9 u^2 + u^4)
w^{ 1/2}\,\big]
\,\,\Big]\\[7pt]
\gamma_7^\pm&=&
\frac{\mp v^2 m_Z^2}{16 u^2 (1+u^2)^3\,w^{ 1/2}}
\Big[
-B_0 m_0\mu_0\,(1+u^2)(1+40 u^2 -114 u^4 +40 u^6+u^8)
\nonumber\\
&+& m_Z^2 \,(u+30 u^5 +u^9) \pm
u (1+u^2)^2 (1-10 u^2 +u^4) \,w^{ 1/2}
\Big]
\\[7pt]
\gamma_{x}^\pm&=&\!\!\!\frac{\pm 8\,(u^2-1)^2\,v^4}{u\,(1+u^2)^3\,w^{ 3/2}}
\, \,\big[ m_Z^2\,u-B_0m_0\mu_0\,(1+u^2)\big]
\big[ 2\,m_Z^2\,u-B_0m_0\mu_0\,(1+u^2)\big]\,m_0\,\mu_0
\\[7pt]
\gamma_{y}^\pm&=& \mp\frac{(-1+u^2)^2\,v^4}{(1+u^2)^4\,w^{ 3/2}}
\,\,\big[m_Z^2\,u-B_0\,m_0\,\mu_0\,(1+u^2)\big]^2\,(4 \,m_0^2)
\eea\bea
\gamma_{z}^\pm&=&\frac{ \mp v^4}{\mu_0^2\,u^2\,(1+u^2)^3\,w^{ 3/2}}
\\
&\times&\!\!\Big[
- 2\,(B_0m_0\mu_0)^3 \,u\,(1+u^2)^4
+m_Z^4\,u^2(1+u^2)
\big(4\,\mu_0^2(-1+u^2)^2-u^2 (2 m_Z^2\pm w^{1/2})\big)
\nonumber\\
&+&\!\! 2\, B_0 m_0\mu_0\,
m_Z^2\,u\,\big[-2 \mu_0^2(u^4-1)^2
+u^2 (m_Z^2(1-14 u^2+u^4)  \pm (u^4 - 6 u^2+1)\,w^{ 1/2}
)\big]\nonumber\\
&+&\!\!
(B_0 m_0\mu_0)^2\,(1+u^2)\big[\mu_0^2
\,(u^4-1)^2+u^2 (2m_Z^2\,(1-14 u^2+u^4) \mp(1+u^2)^2\,w^{1/2})\big]
\Big](4\mu_0^2)\nonumber
\eea

\chapter{The Solution of the Quadratic Constraint}\label{App2}

In sec.~\ref{secmagndual}, the quadratic constraint ${\cal Z}^2 =0$ must be solved to obtain the magnetic DBI theory coupled to a single-tensor multiplet. Using the expansion
$$
{\cal Z}(y,\theta,\tilde\theta) = Z(y,\theta) + \sqrt2\, \tilde\theta \omega(y,\theta)
- \tilde\theta\tilde\theta \left[ {i\over2}\Phi_{\cal Z} + {1\over4}\ov{DD} \ov Z(y,\theta) \right],
$$
in terms of the $\cN=1$ chiral superfields $Z$, $\omega_\alpha$ and $\Phi_{\cal Z}$, the constraint is
equivalent to the single equation
\be
\label{AppB1}
Z = - { \omega\omega \over i\Phi_{\cal Z} + {1\over2}\ov{DD}\ov Z} .
\ee
The electric constraint equation (\ref{DBI7}), which was solved by Bagger and Galperin
\cite{BG} using a method which applies to eq.~(\ref{AppB1}) as well, corresponds to the particular 
case $\omega_\alpha= iW_\alpha$, $\Phi_{\cal Z} = -i/\kappa$ and $Z=X$. Following then
Ref.~\cite{BG}, the solution of eq.~(\ref{AppB1}) is
\be
\label{AppB2}
Z(\omega\omega,\Phi_{\cal Z}) = 
{i\over \Phi_{\cal Z}}\left( \omega\omega + \ov{DD} \left[ { \omega\omega \ov{\omega\omega} \over 
|\Phi_{\cal Z}|^2 + A + \sqrt{|\Phi_{\cal Z}|^4 +2A|\Phi_{\cal Z}|^2 + B^2}} \right]\right),
\ee
where
$$
\begin{array}{rcl}
A &=& -{1\over2}(DD\,\omega\omega + \ov{DD}\,\ov{\omega\omega}) \,\, =\,\, A^*,
\crbig
B &=& -{1\over2}(DD\,\omega\omega - \ov{DD}\,\ov{\omega\omega})\,\, =\,\, -B^*.
\end{array}
$$
Another useful expression is
\be
\label{AppB3}
\begin{array}{l}
Z(\omega\omega,\Phi_{\cal Z}) = \displaystyle
{i\over \Phi_{\cal Z}} \Biggl( \omega\omega 
\crbig   \hspace{1.5cm}
\displaystyle + \ov{DD} \left[ 
{ \omega\omega \ov{\omega\omega} \over (DD\omega\omega) (\ov{DD}\ov{\omega\omega})}
\Bigl\{ |\Phi_{\cal Z}|^2 + A - \sqrt{|\Phi_{\cal Z}|^4 +2A|\Phi_{\cal Z}|^2 + B^2}\Bigr\}  \right] \Biggr).
\end{array}
\ee
In the text, we need the bosonic content of $Z(\omega\omega,\Phi_{\cal Z})$. We write:
\be
\label{AppB4}
\omega_\alpha (y,\theta)= \theta_\alpha\, \rho + {1\over2}(\theta\sigma^\mu\ov\sigma^\nu)_\alpha 
P_{\mu\nu} + \dots,
\ee
where $\rho$ is a complex scalar (2 bosons), $P_{\mu\nu}$ a real antisymmetric tensor (6 bosons)
and dots indicate omitted fermionic terms. Hence,
$$
\begin{array}{rcl}
\omega\omega &=& \theta\theta \left[ \rho^2 + {1\over2} P^{\mu\nu} P_{\mu\nu}  
+ {i\over4}\epsilon^{\mu\nu\rho\sigma}P_{\mu\nu} P_{\rho\sigma} \right] + \dots,
\crbig
A &=& 2(\rho^2 + \ov \rho^2) + 2 P^{\mu\nu} P_{\mu\nu} + \dots,
\crbig
B &=& 2(\rho^2 - \ov \rho^2) + i \epsilon^{\mu\nu\rho\sigma}P_{\mu\nu} P_{\rho\sigma} + \dots
\end{array}
$$
Since the bosonic expansion of $\omega_\alpha$ carries one $\theta_\alpha$, it follows from 
solution (\ref{AppB2}) that the 
bosonic $Z(\omega\omega,\Phi_{\cal Z})$ has a $\theta\theta$ component only, and that this 
component only depends on $\rho$, $P_{\mu\nu}$ and the lowest scalar component of 
$\Phi_{\cal Z}$ (which we also denote by $\Phi_{\cal Z}$). As a consequence, the bosonic 
$Z(\omega\omega,\Phi_{\cal Z})$ does not depend on the auxiliary scalar $f_{\Phi_{\cal Z}}$ of 
$\Phi_{\cal Z}$. We then find:
\be
\label{AppB5}
Z(\Phi_{\cal Z}, \omega\omega)_{bos.} = 
{i\ov\Phi_{\cal Z}\over |\Phi_{\cal Z}|^2}\omega\omega - {i\ov\Phi_{\cal Z}\over4|\Phi_{\cal Z}|^2} 
\theta\theta\left(|\Phi_{\cal Z}|^2 + A - \sqrt{|\Phi_{\cal Z}|^4 +2A|\Phi_{\cal Z}|^2 + B^2} 
\right)_{\theta=0}. 
\ee
The parenthesis is real. In terms of component fields:
\be
\label{AppB6}
\begin{array}{rcl}
Z &=& - {i \ov\Phi_{\cal Z} \over 4|\Phi_{\cal Z}|^2} \theta\theta \Big[ |\Phi_{\cal Z}|^2 
- i\epsilon^{\mu\nu\rho\sigma}P_{\mu\nu} P_{\rho\sigma} - 2 (\rho^2-\ov \rho^2)\Big]
\crbig
& & + {i \ov\Phi_{\cal Z} \over 4|\Phi_{\cal Z}|^2} \theta\theta\Bigl[ \Bigl(|\Phi_{\cal Z}|^2 
+ 2(\rho^2+\ov \rho^2)\Bigr)^2 - 16 \rho^2 \ov \rho^2 
+ 4 (\rho^2-\ov \rho^2)i\epsilon^{\mu\nu\rho\sigma}P_{\mu\nu} P_{\rho\sigma}
\crbig
& & \hspace{28mm} + 4 |\Phi_{\cal Z}|^2P^{\mu\nu} P_{\mu\nu} 
- \Bigl(\epsilon^{\mu\nu\rho\sigma}P_{\mu\nu} P_{\rho\sigma}\Bigr)^2 \Big]^{1/2} + \dots
\end{array}
\ee
The decomposition (\ref{DBIme}), ${\cal Z}= \widetilde{\cal W} + 2g {\cal Y}$, indicates that
\be
\label{AppB7}
\rho=-{g\over 2}C+i \widetilde d_2, 
\qquad P_{\mu\nu}= g b_{\mu\nu}- \widetilde F_{\mu\nu}\,,
\qquad\Phi_{\cal Z}=2g\Phi.
\ee
In Lagrangian (\ref{DBIml}), we need the imaginary part of the $\theta\theta$ component of
$Z(\omega\omega,\Phi_{\cal Z})$:
\be
\label{AppB8}
\begin{array}{rcl}
\Im Z(\omega\omega,\Phi_{\cal Z})|_{\theta\theta} &=& -{g\Re\Phi\over2} + {\Re\Phi\over8g|\Phi|^2}
\Biggl\{16g^4|\Phi|^4 + 8g^2|\Phi|^2(g^2C^2-4\tilde d_2^2) -16g^2C^2\tilde d_2^2
\crbig
&& + 16g^2 |\Phi|^2 (\widetilde F_{\mu\nu} - g \, b_{\mu\nu})(\widetilde F^{\mu\nu} - g \, b^{\mu\nu})
\crbig
&& + 8gC\tilde d_2 \, \epsilon^{\mu\nu\rho\sigma}
(\widetilde F_{\mu\nu} - g \, b_{\mu\nu})(\widetilde F_{\rho\sigma} - g \, b_{\rho\sigma})
\crbig
&&
- \Bigl[\epsilon^{\mu\nu\rho\sigma}
(\widetilde F_{\mu\nu} - g \, b_{\mu\nu})(\widetilde F_{\rho\sigma} - g \, b_{\rho\sigma}) \Bigr]^2
\Biggr\}^{1/2}
\crbig
&& + {\Im\Phi\over8g|\Phi|^2} \left[ \epsilon^{\mu\nu\rho\sigma}
(\widetilde F_{\mu\nu} - g \, b_{\mu\nu})(\widetilde F_{\rho\sigma} - g \, b_{\rho\sigma})
- 4gC \widetilde d_2 \right].
\end{array}
\ee
We now use 
\be
\label{AppB9}
\begin{array}{rcl}
-{\rm det} (|\Phi|\eta_{\mu\nu} + {\sqrt2\over g}\,P_{\mu\nu} ) 
&=& - |\Phi|^4 \,{\rm det} (\eta_{\mu\nu} + {\sqrt2\over g|\Phi|}\, P_{\mu\nu} )
\crbig
&=& |\Phi|^4 +{|\Phi|^2\over g^2}P^{\mu\nu}P_{\mu\nu} - 
{1\over16g^4}(\epsilon^{\mu\nu\rho\sigma}P_{\mu\nu}P_{\rho\sigma})^2 
\end{array}
\ee
to rewrite
\be
\label{AppB10}
\begin{array}{rcl}
\Im Z(\omega\omega,\Phi_{\cal Z})|_{\theta\theta} &=& -{g\Re\Phi\over2} + {\Re\Phi\over4g|\Phi|^2}
\Biggl\{ - 4g^4|\Phi|^4 \,{\rm det} \left[\eta_{\mu\nu} - {\sqrt2 \over g|\Phi|}(\widetilde F_{\mu\nu} 
-gb_{\mu\nu}) \right]
\crbig
&&
- 4 g^2 \tilde d_2^2 \Bigl(2|\Phi|^2 + C^2 \Bigr) + 2 g^4 C^2 |\Phi|^2
\crbig
&& + 2gC\tilde d_2 \, \epsilon^{\mu\nu\rho\sigma}
(\widetilde F_{\mu\nu} - g \, b_{\mu\nu})(\widetilde F_{\rho\sigma} - g \, b_{\rho\sigma})
\Biggr\}^{1/2}
\crbig
&& + {\Im\Phi\over8g|\Phi|^2} \left[ \epsilon^{\mu\nu\rho\sigma}
(\widetilde F_{\mu\nu} - g \, b_{\mu\nu})(\widetilde F_{\rho\sigma} - g \, b_{\rho\sigma})
- 4gC \widetilde d_2 \right].
\end{array}
\ee
As a check, choosing $\Phi = -1/(2g\kappa)$ and $g=0$ to decouple the single-tensor multiplet leads back to theory (\ref{DBI8}) since in that case $\tilde d_2=0$.

\chapter{Equivalent Descriptions of the Dilaton Multiplet}\label{Equivalents}

We present in detail three dual descriptions of the dilaton multiplet as well as the duality transformations that take us from one to another. We start by repeating the analysis of section \ref{secST} on the single-tensor multiplet, this time with more details, and then we go on to the hyper- and the two-tensor multiplets.

\section{The Single Tensor Formulation}

The single-tensor multiplet \cite{LR,tensor1,tensor2} is the $\cN=2$ extension of 
the antisymmetric tensor field $b_{\mu\nu}$
with gauge symmetry $\delta_{gauge}b_{\mu\nu} = 2\partial_{[\mu}\Lambda_{\nu]}$.
It admits two descriptions, either in terms of the gauge-invariant curl $\partial_{[\mu} b_{\nu\rho]}$ or in terms of the antisymmetric tensor field submitted to its gauge transformation. 

In the case of $\cN=1$ supersymmetry, a real linear superfield $L$, $DDL=0$, $L=\ov L$, 
describes the curl of the antisymmetric tensor. 
It can be expressed in terms of a chiral spinor potential including the antisymmetric tensor:
\be
\label{AST1}
L=D^\alpha \chi_\alpha -\ov D_\dalpha \ov\chi^\dalpha,
\ee
with $\ov D_\dalpha \chi_\alpha =0$. The gauge invariance of the two-form field acts on the
potential $\chi_\alpha$ according to
\be
\label{AST2}
\chi_\alpha \quad\longrightarrow\quad \chi_\alpha + i\ov{DD}D_\alpha \Delta, \qquad\qquad
\ov\chi_\dalpha \quad\longrightarrow\quad \ov\chi_\dalpha + iDD \ov D_\dalpha \Delta,
\ee
which, since $D^\alpha\ov{DD}D_\alpha = \ov D_\dalpha DD \ov D^\dalpha$,
leaves invariant the linear superfield $L$ for any real $\Delta$. The potential $\chi_\alpha$
includes the antisymmetric tensor in its $\theta$ component:
\be
\label{AST3}
\chi_\alpha = \ldots - {1\over4} \theta_\alpha C + {1\over2}(\theta\sigma^\mu\ov\sigma^\nu)_\alpha \,
b_{\mu\nu} + \ldots,
\ee
$C$ being the real scalar partner of $b_{\mu\nu}$. The two descriptions of the $\cN=2$ single-tensor 
multiplet use either $L$ or $\chi_\alpha$, completed with one or two chiral $\cN=1$ superfields.  

In the gauge-invariant description using $L$, the $\cN=2$ multiplet is completed with a
chiral superfield $\Phi$ ($8_B+8_F$ fields in total). The second supersymmetry transformations are
\be
\label{AST4}
\begin{array}{rcl}
\delta^* L &=& -\frac{i}{\sqrt 2} (\eta D\Phi +\ov{\eta D}\ov \Phi) \,, 
\crbig
\delta^* \Phi &=&  i \sqrt2 \, \ov{\eta D}L \,,\qquad\qquad \delta^* \ov\Phi \,\,=\,\, i \sqrt2 \,\eta D L \,,
\end{array}
\ee
The supersymmetry algebra closes (off-shell) on $L$ and $\Phi$.

Alternatively, in terms of $\chi_\alpha$ and $\Phi$, eqs. (\ref{AST3}) suggest the variations
\be
\label{AST5}
\begin{array}{rcl}
\delta^* \chi_\alpha &=& -\frac{i}{\sqrt2} \, \Phi \, \eta_\alpha \,,\qquad\qquad 
\delta^* \ov\chi_\dalpha \,\,=\,\, \frac{i}{\sqrt2} \,\ov\Phi \, \ov\eta_\dalpha \,,
\crbig
\delta^* \Phi &=&  2\sqrt2 i \left[\frac{1}{4}\,\ov{DD\eta\chi} 
+ i \partial_\mu\chi\sigma^\mu\ov\eta \right] ,
\crbig
\delta^* \ov\Phi &=& -2\sqrt2 i \left[\frac{1}{4}\, DD\eta\chi - i \eta\sigma^\mu\partial_\mu 
\ov\chi \right].
\end{array}
\ee
On $\chi_\alpha$ however, the supersymmetry algebra closes up to a gauge transformation 
(\ref{AST2}):
\be
\label{AST6}
\begin{array}{rcl}
[ \delta_1^* , \delta_2^* ] \chi_\alpha &=& -2i\,(\eta_2\sigma^\mu\ov\eta_1 - \eta_1\sigma^\mu\ov\eta_2)\,
\partial_\mu\chi_\alpha 
\crbig
&& + {i\over2}\,\ov{DD}D_\alpha \, \Bigl[ i \, \eta_1\theta \,\ov{\eta_2\chi}
- i \, \ov{\eta_1\theta} \,\eta_2\chi - i \, \eta_2\theta \,\ov{\eta_1\chi} + i \, \ov{\eta_2\theta} \,\eta_1\chi 
\Bigr].
\end{array}
\ee
This result suggests that the $\cN=1$ superfields $\Phi$ and $\chi_\alpha$ do not complete 
a true off-shell
supermultiplet of $\cN=2$ supersymmetry. Another hint is given by the degrees of freedom: $\Phi$ and
$\chi_\alpha$ contain $12_B+12_F$ fields and gauge invariance (\ref{AST2}), which
is only compatible with $\cN=1$, removes $4_B+4_F$ fields,
to give the expected $8_B+8_F$ degrees of freedom in $L$ and $\Phi$. We should then expect
that the $\cN=2$ supermultiplet of the potential $\chi_\alpha$ (including the antisymmetric tensor 
among its component fields) has $16_B+16_F$ fields, with an extended
gauge transformation using a Maxwell $\cN=2$ multiplet and removing $8_B+8_F$ components. 

From the structure of relation (\ref{AST6}), one may guess that the introduction of another chiral 
superfield $Y$ (with $4_B+4_F$ fields) with $\delta^* Y \sim \eta\chi$ would be appropriate if we also add 
to $\delta^*\chi_\alpha$ a gauge transformation proportional to
$$
i\,\ov{DD}D_\alpha \, [ i\eta\theta \ov Y - i \ov{\eta\theta} Y ]
= -\eta_\alpha \ov{DD}\,\ov Y - 4i(\sigma^\mu\ov\eta)_\alpha \, \partial_\mu Y.
$$
This modification, being a gauge transformation of $\chi_\alpha$, does not affect
$\delta^* L$. One then easily verifies that the second supersymmetry variations
\be
\label{AST7}
\begin{array}{rcl}
\delta^* Y &=& \sqrt2\, \eta\chi \, , 
\crbig
\delta^* \chi_\alpha &=& -{i\over\sqrt2} \Phi\,\eta_\alpha - {\sqrt2\over4} \eta_\alpha \, \ov{DD}\, \ov Y
-\sqrt2 i (\sigma^\mu\ov\eta)_\alpha \partial_\mu Y \, , 
\end{array}
\ee
with $\delta^*\Phi$ as in (\ref{AST5}), close the $\cN=2$ superalgebra. 

It is then natural to generalize gauge transformation (\ref{AST2}) to $\cN=2$, using a Maxwell
supermultiplet with $\cN=1$ superfields $\widehat W_\alpha$ and $\widehat X$:
\be
\label{AST8}
\delta_{gauge} \chi_\alpha = i \widehat W_\alpha, \qquad\qquad
\delta_{gauge} Y = \widehat X, \qquad\qquad
\delta_{gauge} \Phi = 0.
\ee
Since $L=D\chi - \ov D\ov\chi$, the Bianchi identity verified by $\widehat W$ implies the gauge 
invariance of $L$. 
The second variation, which is the same as transformation (\ref{AST2}), contains in particular 
$\delta_{gauge}\, b_{\mu\nu} = \widehat F_{\mu\nu}$. This $\cN=2$ gauge transformation removes
$8_B+8_F$ component fields, leaving as expected $8_B+8_F$ fields. 

It may be useful to remark that giving a constant background value to the chiral $\cN=1$ superfield 
$\Phi$ seems to break $\cN=2$ supersymmetry to $\cN=1$. According to the second variation 
(\ref{AST7}), $\chi_\alpha$ transforms like a Goldstino if $\Phi$ acquires a background value. 
The lowest component of $\chi_\alpha$ does however transform under gauge symmetry (\ref{AST2})
and a Goldstino is generated only if a gauge-invariant quantity is created in a theory where the 
single-tensor multiplet interacts with other fields. In a theory depending only on the gauge-invariant
$L$ and $\Phi$, a background value of $\Phi$ does not break the second supersymmetry: it is invariant
under transformations (\ref{AST4}).\footnote{A background value of the scalar $C$ in $\chi_\alpha$
[see expansion (\ref{AST3})] does not break supersymmetry. It corresponds to a constant background 
value of $L$.}

The chiral superfield $Y$ does not contain any physical state: neither $L$ nor $\phi$ do depend 
on $Y$. There is a gauge similar to the Wess-Zumino gauge of $\cN=1$ supersymmetry in which $Y=0$.
This gauge choice respects $\cN=1$ supersymmetry and gauge symmetry (\ref{AST2}). 

An invariant kinetic action for the single-tensor multiplet involves an arbitrary function solution of the 
three-dimensional Laplace equation (for the variables $L$, $\Phi$ and $\ov\Phi$) \cite{LR}:
\be
\label{AST9}
{\cal L}_{ST} = \Dint {\cal H } (L, \Phi, \ov\Phi) \,, \qquad\qquad
{\partial^2{\cal H }\over\partial L^2} + 2 {\partial^2{\cal H }\over\partial\Phi\partial\ov\Phi} =0.
\ee
It is in particular straightforward to show that
\be
\label{AST10}
{\cal L}_{ST} = \Dint \, H({\cal V}) + {\rm h. c.} ,
\ee
with
$$
{\cal V} = L + {i\over\sqrt2} (\Phi+\ov\Phi )
$$
transforms with a derivative under the second supersymmetry for any function $H({\cal V})$.  
It is also invariant under a constant shift of $\Im\Phi$, the symmetry which allows dualization
of $\Phi$ into the second linear superfield of the double-tensor multiplet.

\section{Hypermultiplet Formulation}

In terms of $\cN=1$ superfields, a hypermultiplet has two chiral superfields $\Phi$ and $T$.
The linear $L$ of the single-tensor multiplet has been dualized to a chiral $T$ with axionic 
shift symmetry. Since the duality involves a Legendre transformation using the Lagrangian 
function, the second supersymmetry transformations will not any longer hold off-shell when 
acting on $\Phi$ and $T$: the 
hypermultiplet does not admit an off-shell formulation.

We start with the single-tensor Lagrangian
\be
\label{Ahyp0}
{\cal L}_{ST} = \Dint {\cal H } (L, \Phi, \ov\Phi).
\ee
To dualize the theory, use a real vector superfield $U$ and rewrite
\be
\label{Ahyp1}
{\cal L}_{ST} = \Dint \left[{\cal H } (U, \Phi, \ov\Phi) - m (T+\ov T)U\right],
\ee
with an arbitrary real parameter $m$.
Eliminating $U$ with 
\be
\label{Ahyp2}
{\partial\over\partial U}\,{\cal H } (U, \Phi, \ov\Phi) = m (T+\ov T),
\ee
one obtains the dual hypermultiplet theory
\be
\label{Ahyp3}
\tilde{\cal L}_{ST} = \Dint K(T+\ov T, \Phi, \ov\Phi), \qquad
K(T+\ov T, \Phi, \ov\Phi) = {\cal H } \Bigl(u , \Phi, \ov\Phi \Bigr) - m (T+\ov T) u,
\ee
where $U=u(T+\ov T, \Phi, \ov\Phi)$ is the solution of the Legendre transformation (\ref{Ahyp2}).

One can then derive various relations between derivatives of the K\"ahler potential $K$
and derivatives of ${\cal H}$:
\be
\label{Ahyp4}
\begin{array}{rclrcl}
K_{T\ov T} &=&   - \displaystyle{m^2 \over {\cal H}_{UU}}, \qquad&\qquad
K_{\Phi\ov \Phi} &=& {\cal H}_{\Phi\ov\Phi} 
- \displaystyle{{\cal H}_{U\Phi}{\cal H}_{U\ov\Phi} \over {\cal H}_{UU}},
\crbig
K_{T\ov \Phi} &=&  m \,\displaystyle{{\cal H}_{U\ov\Phi} \over {\cal H}_{UU}}, \qquad&\qquad
K_{\Phi\ov T} &=& m \, \displaystyle{{\cal H}_{U\Phi} \over {\cal H}_{UU}},
\end{array}
\ee
using the notation
$$
{\cal H}_{UU} = {\partial^2{\cal H }\over\partial U^2} ,
\qquad\qquad
{\cal H}_{\Phi\ov\Phi} = {\partial^2{\cal H }\over\partial\Phi\,\partial\ov\Phi}, \qquad\ldots
$$
As a consequence, the determinant of the ($2\times2$) K\"ahler metric is
\be
\label{Ahyp4c}
K_{T\ov T} K_{\Phi\ov\Phi} - K_{T\ov \Phi} K_{\Phi\ov T} = 
- m^2 \, {{\cal H}_{\Phi\ov\Phi} \over {\cal H}_{UU}} \,.
\ee
In this $\cN=1$ Legendre transformation, the condition for $\cN=2$ supersymmetry has not been used.
Hence for a single-tensor multiplet, the second eq.~(\ref{AST9}) implies \cite{GP, AHS}
\be
\label{Ahyp5}
K_{T\ov T} K_{\Phi\ov\Phi} - K_{T\ov \Phi} K_{\Phi\ov T} =  {1\over2} m^2
\ee
(Monge-Amp\`ere equation).
This result implies Ricci-flatness which, for a two-dimen\-sio\-nal complex manifold, indicates
that the hypermultiplet scalar manifold is hyper-K\"ahler, as expected in general \cite{AGF}. 
Hypermultiplet scalar kinetic terms are\footnote{Positivity of kinetic terms requires 
that ${\cal H}_{UU}<0$.}
\be
\label{Ahyp6}
\begin{array}{l}
K_{T\ov T} \left[\partial_\mu T + {K_{\Phi\ov T}\over K_{T\ov T}} \partial_\mu\Phi \right]
\left[\partial^\mu\ov T + {K_{T\ov\Phi}\over K_{T\ov T}} \partial^\mu\ov\Phi \right]
+ {m^2\over2K_{T\ov T}} \, \partial_\mu\Phi \, \partial^\mu\ov\Phi 
\crbig \hspace{1.3cm}
= -{1\over{\cal H}_{UU}}\Bigl| m\,\partial_\mu T  - {\cal H}_{U\Phi}\,\partial_\mu\Phi 
\Bigr|^2 -{1\over2}{\cal H}_{UU}(\partial_\mu\Phi)(\partial^\mu\ov\Phi).
\end{array}
\ee
using the same notation $T$ and $\Phi$ for the chiral superfields and for their lowest 
scalar components. The chiral superfields $T$ and $\Phi$ are K\"ahler coordinates. 

One should remark that adding to ${\cal H}$ the quantity
\be
\label{Ahyp6a}
\Delta{\cal H} = L [ g(\Phi) + \ov g(\ov\Phi)Ê]
\ee
does not change the single-tensor theory:\footnote{It is a trivial solution of Laplace equation.}
its superspace integral is a derivative. Since 
$$
\Delta{\cal H}_U = g(\Phi) + \ov g(\ov\Phi), \qquad \qquad
\Delta{\cal H}_{U\Phi} = g_\Phi (\Phi),
$$
the Legendre transformation (\ref{Ahyp2}) and the kinetic terms (\ref{Ahyp6}) are affected
by a modification of $T$:
\be
\label{Ahyp6b}
T \qquad\longrightarrow\qquad T - {g(\Phi)\over m}. 
\ee
Hence, for a given single-tensor theory defined by the function ${\cal H}$, we have a family
of hypermultiplet theories generated by the arbitrary function $g(\Phi)$. In other words, 
the chiral superfield dual to $L$ can be defined as $T - {g(\Phi)\over m}$, for any function $g$. 

The hyper-K\"ahler scalar metric is commonly expressed in ``mixed" coordinates where $u$, the solution 
of the Legendre transformation (\ref{Ahyp2}), is used instead of $\Re T$.  Defining then coordinates
\be
\label{Ahyp7}
q^a = (\tau, x^i) = (\Im T, \sqrt2\Re\phi, \sqrt2\Im\phi, u), \qquad
a=0,i, \quad i=1,2,3,
\ee
the line-element can be written
\be
\label{Ahyp8}
\begin{array}{rcl}
ds^2 &=& g_{ab} \, dq^a \, dq^b
\crbig
&=& - {{\cal H}_{UU}\over4}  \, du^2 + {\cal H}_{\Phi\ov\Phi} \, d\Phi\,d\ov\Phi
- {m^2\over{\cal H}_{UU}} \left[ d\Im T + {i\over2m} ({\cal H}_{U\Phi}\, d\Phi - 
{\cal H}_{U\ov\Phi}\, d\ov\Phi )  \right]^2 .
\end{array}
\ee
With the condition for $\cN=2$ supersymmetry, ${\cal H}_{\Phi\ov\Phi}  =
- {1\over2}{\cal H}_{UU}$, this is
\be
\label{Ahyp9}
\begin{array}{rcl}
ds^2 &=& - {{\cal H}_{UU}\over4} \, [ du^2 +2 \, d\Phi\,d\ov\Phi]
- {m^2\over{\cal H}_{UU}} \left[ d\Im t + {i\over2m} ({\cal H}_{U\Phi}\, d\Phi - 
{\cal H}_{U\ov\Phi}\, d\ov\Phi ) \right]^2
\crbig
&=& {m\over2}\Bigl( V \, dx^i\,dx^i + V^{-1}[ d\tau - \omega^i\, dx^i]^2  \Bigr), 
\end{array}
\ee
with functions $V(x^i)$ and $\omega^i(x^j)$ given by
\be
\label{Ahyp10}
V = -{{\cal H}_{UU}\over2m},\qquad  \omega^1 = {\Im {\cal H}_{U\Phi}\over\sqrt2 m} ,\qquad  
\omega^2 =  {\Re {\cal H}_{U\Phi}\over\sqrt2 m}, \qquad  \omega^3 = 0  .
\ee
Using again the condition for $\cN=2$ supersymmetry, which implies that the metric is
hyper-K\"ahler, one finds that
\be
\label{Ahyp11}
\vec\nabla\,V = \vec\nabla\wedge\vec \omega .
\ee
This indicates that $V$ solves Laplace equation
\be
\label{Ahyp12}
\partial^i \partial^i \, V = (\partial_u^2 + 2\,\partial_\Phi\partial_{\ov\Phi}) V =0,
\ee
in agreement with its definition (\ref{Ahyp10}). A (four-dimensional) hyper-K\"ahler metric with shift symmetry of $\tau=\Im T$ is then defined by $V$ and $\omega^i$ related by equations (\ref{Ahyp11})
\cite{GH}. Given a metric of this form, 
the single-tensor formulation of the $\cN=2$ supersymmetric theory is then obtained by integrating
eqs. (\ref{Ahyp10}) to find ${\cal H}$. Notice that eq.~(\ref{Ahyp11}) remains valid if
$$
\vec\omega \quad\longrightarrow\quad \vec\omega + \vec\nabla\, {\cal F},
$$
for an arbitrary real function ${\cal F}$. The metric is unchanged if coordinate $\tau$ is changed according 
to
$$
\tau \quad\longrightarrow\quad \tau +  {\cal F}.
$$
Comparing with eqs. (\ref{Ahyp6a}) and (\ref{Ahyp6b}), one sees that 
${\cal F} = {1\over\sqrt2m}\,\Im g(\Phi)$.

The K\"ahler formulation with complex coordinates $T$ and $\Phi$ is defined by relations
\be
\label{Ahyp13}
K_{T\ov T} = {m\over2 V}, \qquad\qquad
K_{\Phi \ov T} = - {m\over\sqrt2\, V}(\omega^2 + i \omega^1)
\ee
($\omega^3=0$) and by the Legendre transformation $K_T = -m u$ [see eqs. (\ref{Ahyp4})
and (\ref{Ahyp3})].

Notice that if the theory is also invariant under the shift of $\Im\Phi$, then is ${\cal H}$ a real function of
$L$ (or $U$) and $\Phi+\ov\Phi$ and $\omega^1=0$. Relation (\ref{Ahyp11}) implies then that $V$ 
does not depend on $x^2$: obviously, $V$ does not depend on $\Im \Phi$. 

As an example, the Taub-NUT metric is considered in Appendix \ref{secTaubNUT}.

\section{Two-Tensor Formulation}

Similarly, we can turn $\Phi+\ov\Phi$ into a second linear superfield $L^\prime$ to obtain the
two-tensor formulation of the kinetic Lagrangian (\ref{Ahyp0}). Rewriting
it as
\be
\label{Adt1}
{\cal L}_{ST} = \Dint \left[{\cal H } (L,V) - m L^\prime\,V\right],
\ee
with an unconstrained real superfield $V$ to impose $V=\Phi+\ov\Phi$ and an arbitrary parameter $m$.
If we instead eliminate $V$ by its field equation
\be
\label{Adt2}
{\cal H}_V = mL^\prime, \qquad\qquad {\cal H}_V = {\partial\over\partial V} {\cal H}(L,V),
\ee
the resulting two-tensor theory is
\be
\label{Adt3}
{\cal L}_{2T} = \Dint {\cal G} (L,L^\prime) , \qquad\qquad
{\cal G} (L,L^\prime) = {\cal H} (L,V) - m L^\prime\, V,
\ee
with $V$ replaced by the solution $V(L,L^\prime)$ of eq.~(\ref{Adt2}).
Again the Legendre transformation generates relations between derivatives of ${\cal G}$ and
${\cal H}$:
\be
\label{Adt4}
{\cal G}_{LL} = {\cal H}_{LL} - {{\cal H}_{LV}^2 \over {\cal H}_{VV}},
\qquad
{\cal G}_{LL^\prime} = m {{\cal H}_{LV} \over {\cal H}_{VV} } ,
\qquad
{\cal G}_{L^\prime L^\prime} = - {m^2\over {\cal H}_{VV}}.
\ee
As in the hypermultiplet formulation, we have a determinant relation
\be
\label{Adt5}
{\cal G}_{LL}\,{\cal G}_{L^\prime L^\prime} - {\cal G}_{LL^\prime}^2 = -m^2
{ {\cal H}_{LL} \over {\cal H}_{VV} }.
\ee

The bosonic kinetic terms of the two-tensor formulation can then be written
\be
\label{Adt6}
\begin{array}{rcl}
{\cal L}_{2T, kin.} &=& -{1\over4} {\cal G}_{LL} \Bigl[ (\partial_\mu C)(\partial_\mu C)
+ {1\over12} H_{\mu\nu\rho} H^{\mu\nu\rho} \Bigr]
\crbig
&& -{1\over4} {\cal G}_{L^\prime L^\prime} \Bigl[ (\partial_\mu C^\prime)(\partial_\mu C^\prime)
+ {1\over12} H_{\mu\nu\rho}^\prime H^{\prime\,\mu\nu\rho} \Bigr]
\crbig
&& -{1\over2} {\cal G}_{LL^\prime} \Bigl[ (\partial_\mu C)(\partial_\mu C^\prime)
+ {1\over12} H_{\mu\nu\rho} H^{\prime\,\mu\nu\rho} \Bigr]
\crbig
&=& - {1\over4} {\cal H}_{LL} \Bigl[ (\partial_\mu C)(\partial_\mu C)
+ {1\over12} H_{\mu\nu\rho} H^{\mu\nu\rho} \Bigr]
\crbig
&& + {m^2\over4{\cal H}_{VV}} \Bigl[ (\partial_\mu C^\prime -{1\over m} {\cal H}_{LV}\,\partial_\mu C)
(\partial^\mu C^\prime -{1\over m} {\cal H}_{LV}\,\partial^\mu C)
\crbig
&& \hspace{1.5cm} + {1\over12} (H_{\mu\nu\rho}^\prime -{1\over m} {\cal H}_{LV}\,H_{\mu\nu\rho})
(H^{\prime\,\mu\nu\rho} -{1\over m} {\cal H}_{LV}\,H^{\mu\nu\rho}) \Bigr] ,
\end{array}
\ee
with $H_{\mu\nu\rho} = 3\,\partial_{[\mu} B_{\nu\rho]}$ and $H_{\mu\nu\rho}^\prime 
= 3\,\partial_{[\mu} B_{\nu\rho]}^\prime$ and, as before, $V$ should be replaced by the 
solution $V(L,L^\prime)$.

The condition imposed by the second supersymmetry has not been imposed yet. In the
single-tensor formulation, $\cN=2$ supersymmetry is obtained if ${\cal H}_{LL} = -2{\cal H}_{VV}$.
The two-tensor version (\ref{Adt3}) has then $\cN=2$ supersymmetry if
\be
\label{Adt7}
{\cal G}_{LL}\,{\cal G}_{L^\prime L^\prime} - {\cal G}_{LL^\prime}^2 = 2 m^2,
\ee
{\it i.e.} if the determinant is a positive constant.
Bosonic kinetic terms of the $\cN=2$ theory are then
\be
\label{Adt8}
\begin{array}{rcl}
{\cal L}_{2T, kin.} &=& - {m^2\over2{\cal G}_{L^\prime L^\prime}} \Bigl[ (\partial_\mu C)(\partial_\mu C)
+ {1\over12} H_{\mu\nu\rho} H^{\mu\nu\rho} \Bigr]
\crbig
&& - {1\over4}{\cal G}_{L^\prime L^\prime} \Bigl[ (\partial_\mu C^\prime 
+ {{\cal G}_{LL^\prime}\over {\cal G}_{L^\prime L^\prime}} \,\partial_\mu C)
(\partial^\mu C^\prime + {{\cal G}_{LL^\prime}\over {\cal G}_{L^\prime L^\prime}}\,\partial^\mu C)
\crbig
&& \hspace{1.5cm} + {1\over12} (H_{\mu\nu\rho}^\prime 
+ {{\cal G}_{LL^\prime}\over {\cal G}_{L^\prime L^\prime}}\,H_{\mu\nu\rho})
(H^{\prime\,\mu\nu\rho} + {{\cal G}_{LL^\prime}\over {\cal G}_{L^\prime L^\prime}}\,H^{\mu\nu\rho}) \Bigr] ,
\crbig
&=& - {1\over4} {\cal H}_{LL} \Bigl[ (\partial_\mu C)(\partial_\mu C)
+ {1\over12} H_{\mu\nu\rho} H^{\mu\nu\rho} \Bigr]
\crbig
&& - {m^2\over2{\cal H}_{LL}} \Bigl[ (\partial_\mu C^\prime -{1\over m} {\cal H}_{LV}\,\partial_\mu C)
(\partial^\mu C^\prime -{1\over m} {\cal H}_{LV}\,\partial^\mu C)
\crbig
&& \hspace{1.5cm} + {1\over12} (H_{\mu\nu\rho}^\prime -{1\over m} {\cal H}_{LV}\,H_{\mu\nu\rho})
(H^{\prime\,\mu\nu\rho} -{1\over m} {\cal H}_{LV}\,H^{\mu\nu\rho}) \Bigr] .
\end{array}
\ee
While the first supersymmetry imposes a relation between scalar and tensor kinetic terms,
the second imposes a specific relation between the kinetic terms of the two linear superfields. 

In comparing with the reduction of a IIB supergravity Lagrangian, one should then choose
a gravity frame in which the relation between scalar and tensor kinetic terms is verified. The first
supersymmetry and kinetic terms (\ref{Adt6}) are then sufficient for this choice. 

\chapter{Obtaining the Taub-NUT Metric from Conformal Supergravity}

\section{$S\!U(2,1)/S\!U(2)\times U(1)$ and its Global Hy\-per-K\"ah\-ler Limit}

The superconformal construction of the $\cN=2$ $SU(2,1)/SU(2)\times U(1)$ sigma-model coupled to $\cN=2$ supergravity starts with one vector multiplet (for the graviphoton) and three hypermultiplets. However, with these states only, eliminating auxiliary fields and imposing Poincar\'e gauge conditions would lead to the $Sp(4,2)\,/\, Sp(4)\times Sp(2)$ theory. We need an additional non-propagating vector multiplet with gauge field $W_\mu$ to eliminate four more scalars and to reduce the theory 
to $SU(2,1)/SU(2)\times U(1)$. The vector field will be used to gauge a $U(1)$ or $SO(1,1)$ subgroup of $Sp(4,2)$ with generator $T$. This is very much similar to what we do in section \ref{seclocal} where we obtain the universal hypermultiplet from conformal $\cN=2$ supergravity.

The basic difference here is that in order to reduce to a Taub-NUT metric, we need to start with a different signature for $\eta$:
\be
\eta={\rm diag}(-1,1,1)
\ee
The first steps of writing down the supergravity scalar Lagrangian and imposing the proper constraints is exactly the same as in subsec.~\ref{secsugra2} until eq.~(\ref{conf19}) where the the different choice of signature appears explicitly:
\be
\label{appconf19}
\vec A_+^* \cdot \vec A_+ + \vec A_-^* \cdot \vec A_- = -{1\over\kappa^2}, \qquad\qquad
\vec A^*\cdot\vec A = \vec A^\dagger \eta \vec A = -|A_1|^2 + |A_2|^2 + |A_3|^2 .
\ee
From that point on, in order to obtain the Taub-NUT metric we proceed as follows. We first define
\be
\label{conf18}
q_a =  {1\over\kappa}\, Q_a Q_1^{-1}, 
\ee
and $q_1= {1\over\kappa}\,  I_2$ will not be used herebelow.
Defining the new coordinates $q_a$ left invariant by the superconformal $SU(2)$ is equivalent to identify the superconformal $SU(2)$ with $SU(2)_1$ and choose a gauge for $Q_1$. 
Explicitly,

$$
q_a = \left( \begin{array}{cc}  q_{+a} & q_{-a} \\ -q_{-a}^* & q_{+a}^*  \end{array} \right)
= {1\over\kappa\det Q_1} \left( \begin{array}{cc}  
A_{+a}A_{+1}^* + A_{-a}A_{-1}^* &\quad -A_{+a}A_{-1} + A_{-a}A_{+1}  \crbig
-A_{-a}^*A_{+1}^* + A_{+a}^*A_{-1}^* &\quad A_{-a}^* A_{-1} + A_{+a}^*A_{+1} 
\end{array} \right).
$$

Similarly,

$$
Q_a = \kappa \left( \begin{array}{cc}  
q_{+a}A_{+1} - q_{-a}A_{-1}^* &\quad q_{+a}A_{-1} + q_{-a}A_{+1}^*  \crbig
-q_{-a}^*A_{+1} - q_{+a}^*A_{-1}^* &\quad -q_{-a}^* A_{-1} + q_{+a}^*A_{+1}^*
\end{array} \right).
$$

The second condition (\ref{appconf19}) is now written as
\be
\label{appconf20}
- \det Q_1 + \det Q_2 + \det Q_3 = -{1\over\kappa^2}, \qquad
\det Q_1 = {1 \over \kappa^2(1 - \kappa^2\det q_2 - \kappa^2\det q_3)}.
\ee
Both $Q_a$ and $q_a$ have dimension (mass)$^1$ and they verify $\det Q_a \le \kappa^{-2}$,
$\det q_a \le \kappa^{-2}$.
We will use the $SU(2)$ symmetry to choose 
\be
\label{appconf20b}
A_{+1} = \sqrt{\det Q_1} = A_{+1}^*, \qquad A_{-1}=0, \qquad\quad
q_a = {1\over\kappa\sqrt{\det Q_1}} Q_a \quad (a=2,3).
\ee
Notice that with this choice $A_{+1}$ and $q_a$ are respectively of order $\kappa^{-1}$ and $\kappa^0$. Actually, in the global supersymmetry limit 
$\kappa\rightarrow0$, the constraint reduces to $A_{+1}=\kappa^{-1}$. 
The $SU(2)$ gauge fields and their contributions to the Lagrangian are of
order $\kappa^2$. 

With the above choices, the sigma-model Lagrangian for the scalar fields becomes
\be
\label{appconf20c}
\begin{array}{rcl}
{\cal L}_{scalar} &=&
2\kappa^2A_{+1}^2\Bigl[ 
(\partial_\mu q_{+2}) (\partial_\mu q_{+2}^*) 
+ (\partial_\mu q_{-2}) (\partial_\mu q_{-2}^*)
\crbig
&& \hspace{1.4cm} + (\partial_\mu q_{+3}) (\partial_\mu q_{+3}^*) 
+ (\partial_\mu q_{-3}) (\partial_\mu q_{-3}^*) \Bigr]
\crbig
&& + {1\over2} \kappa^6 A_{+1}^4 [\partial_\mu(\det q_2 + \det q_3)]^2
\crbig
&&- {g^2\over\kappa^2} \, {V_{\mu\, i}}^j {V_{\mu\, j}}^i
\crbig
&& + {g^\prime}^2 \,W_\mu W^\mu \, d^\alpha_\beta\, 
{T^\gamma}_\alpha {T^\beta}_\delta A_\gamma^iA^\delta_i 
\crbig
&=& {\cal L}_{0} + {\cal L}_{A_{+1}} + {\cal L}_{SU(2)} + {\cal L}_{O(1,1)},
\end{array} 
\ee
with $A_{+1}$ as in the first eq.~(\ref{appconf20b}). Notice that the term in the 
third line is 
$$
{\cal L}_{A_{+1}} = {2\over\kappa^2}\partial_\mu\ln(\kappa A_{+1})\,\partial^\mu\ln(\kappa A_{+1}).
$$
It vanishes in the limit $\kappa\rightarrow0$. The $SU(2)$ 
gauge fields do not depend on derivatives of $A_{+1}$: 
\be
\label{conf20d}
\begin{array}{rcl}
{\cal L}_{SU(2)} &=& - {g^2\over\kappa^2} \, {V_{\mu\, i}}^j {V_{\mu\, j}}^i
\,\,=\,\,
-{1\over4}\kappa^6 A_{+1}^4\,\Tr\, \Bigl[
q_2^\dagger\stackrel{\leftrightarrow}{\partial_\mu}q_2 + 
q_3^\dagger\stackrel{\leftrightarrow}{\partial_\mu}q_3 \Bigr]^2
\crbig
&=& -{1\over2}\kappa^6 A_{+1}^4 \Bigl[
( q_{+a}^*\stackrel{\leftrightarrow}{\partial_\mu}q_{+a} - 
q_{-a}^*\stackrel{\leftrightarrow}{\partial_\mu}q_{-a})^2
+ 4 (q_{+a}^*\stackrel{\leftrightarrow}{\partial_\mu}q_{-a}) 
(q_{-a}^*\stackrel{\leftrightarrow}{\partial_\mu}q_{+a})
\Bigr], 
\end{array}
\ee
where $a$ is summed over values $a=2,3$ only. This contribution also
cancels in the limit $\kappa\rightarrow0$ where ${\cal L}_{scalar, \,
\kappa\rightarrow0}={\cal L}_{0} + {\cal L}_{SO(1,1)}$, with
$\kappa A_{+1}=1$.

With $g^\prime=0$ and without the constraint (\ref{conf12}), one obtains 
the sigma-model $HP^2 = Sp(4,2)/ Sp(4)\times Sp(2)$. 
Expressed in terms of the 
quaternion $(2\times2)$ matrices $q_2$ and $q_3$, it reads:
\be
\label{conf20e}
\begin{array}{rcl}
{\cal L}_{HP^2} &=& 
\kappa^2 A_{+1}^2\, \Tr [ (\partial_\mu q_2)^\dagger(\partial^\mu q_2) + (\partial_\mu q_3)^\dagger(\partial^\mu q_3) ]
\crbig
&& + \kappa^6A_{+1}^4 \,\Tr[ (q_2^\dagger\partial_\mu q_2
+ q_3^\dagger\partial_\mu q_3)(\partial^\mu q_2^\dagger \, q_2
+ \partial^\mu q_3^\dagger \,q_3 ) ].
\end{array}
\ee
In the limit $\kappa\rightarrow0$, $\kappa A_{+1}\rightarrow1$ and the
sigma-model metric is trivial.

If we choose the $U(1)$ generator $T$ as in eq. (\ref{Tis2}):
\be
\label{conf21}
T = \left( \begin{array}{cc}  \hat T & 0 \\
0 & \hat T^*  \end{array} \right), \qquad\qquad \hat T^\dagger = -\eta \hat T \eta,
\ee
then constraint (\ref{appconf19}) leads to three (real) equations:
\be
\label{appconf22}
\begin{array}{rcl}
\vec A_+^\dagger \,i\eta \hat T\, \vec A_+ &=&  \vec A_-^\dagger \,i\eta \hat T\, \vec A_-   ,
\crbig
\vec A_-^\dagger \,i\eta \hat T \, \vec A_+  &=& 0  
\end{array}
\ee
($[i\eta \hat T]^\dagger = i\eta\hat T$). With the $SO(1,1)$ generator
\be
\label{conf23}
\hat T = \left( \begin{array}{ccc}
0 & \lambda & 0 \\ \lambda & 0 & 0 \\ 0 & 0 & i
\end{array}\right)
\ee
($\lambda$ real) the three constraints are:
\be
\label{conf24}
\begin{array}{rcl}
\lambda( A_{+2}^* A_{+1} - A_{+1}^* A_{+2} ) + i A_{+3}^* A_{+3}
&=&  \lambda( A_{-2}^* A_{-1} - A_{-1}^* A_{-2} ) + i A_{-3}^* A_{-3},
\crbig
\lambda( A_{-2}^* A_{+1} - A_{-1}^* A_{+2} ) + i A_{-3}^* A_{+3} &=& 0.
\end{array}
\ee
These conditions survive in the global supersymmetry limit $\kappa\rightarrow0$, where also $\det Q_1\rightarrow\kappa^{-2}$, if $\lambda A_{+1}$ has a finite limit. Since $\kappa A_{+1}\rightarrow1$, we then assume that\footnote{$\ell$ has dimension (mass)$^1$.} $\lambda = \ell\kappa$. In terms of the coordinates $q_a$, the conditions are:
$$
\begin{array}{l}
\ell \Bigl[ (|A_{+1}|^2 - |A_{-1}|^2) (q_{+2}^* - q_{+2}) - 2A_{+1}A_{-1}q_{-2}^* 
+ 2A_{+1}^*A_{-1}^*q_{-2}  \Bigr]
\crbig
\hspace{.7cm} =
i \Bigl[ ( |A_{-1}|^2 - |A_{+1}|^2 ) ( |q_{+3}|^2 -  |q_{-3}|^2 )) 
+ 2A_{+1}A_{-1} q_{+3}q_{-3}^* + 2A_{+1}^*A_{-1}^* q_{-3}q_{+3}^* \, \Bigr] ,
\crbig
\ell \Bigl[ A_{+1}A_{-1}^* (q_{+2}^* - q_{+2}) + A_{+1}A_{+1} q_{-2}^* 
+ A_{-1}^*A_{-1}^* q_{-2} \Bigr]
\crbig
\hspace{.7cm} =
i \Bigl[ A_{+1}A_{-1}^*( |q_{-3}|^2 - |q_{+3}|^2) - A_{+1}A_{+1}q_{+3}q_{-3}^*
+  A_{-1}^*A_{-1}^*q_{+3}^*q_{-3} \Bigr].
\end{array}
$$
Using $SU(2)$ symmetry to choose as earlier $A_{-1}=0$, we obtain
\be
\label{conf25}
\begin{array}{rcl}
i \ell\, (q_{+2}^* - q_{+2}) &=&  |q_{+3}|^2 - |q_{-3}|^2 \, ,
\crbig
\ell \,q_{-2} &=& i \, q_{+3}^* q_{-3},
\end{array}
\ee
independent of $\kappa$. In the limiting case $\ell=0$, $q_3=0$ and the resulting
constraint (\ref{appconf20}) leads to the four-dimensional $Sp(2,2)/Sp(2)\times Sp(2)$.
As a $SO(1,1)$ gauge choice, we may 
take $\Re q_{+2}=0$, which leads to
\be
\label{appconf26}
\begin{array}{rcl}
q_{+2} &=&  {i\over2\ell} \,(|q_{+3}|^2 - |q_{-3}|^2) \, ,
\crbig
q_{-2} &=& {i\over\ell} \, q_{+3}^* q_{-3} \,,
\crbig
\det q_2 &=& |q_{+2}|^2 + |q_{-2}|^2 \,\,=\,\, {1\over4\ell^2}\, (\det q_3)^2.
\end{array}
\ee
With $A_{-1}=0$ and $A_{+1}$ real, the unconstrained fields are $q_{\pm3}$, with
$q_{\pm2}$ given by eqs. (\ref{appconf26}) and with relations
\be
\label{conf27}
\begin{array}{rcl}
q_3 &=& \displaystyle{1\over\kappa\sqrt{\det Q_1}} \left( \begin{array}{cc}
A_{+3} & A_{-3} \\ -A_{-3}^* & A_{+3}^* \end{array} \right) ,
\crbig
A_{+1} &=& \sqrt{\det Q_1} \,\,=\,\, {1\over\kappa} \left[ 1 - \kappa^2 \det q_3 - {\kappa^2\over4\ell^2}
(\det q_3)^2 \right]^{-1/2}.
\end{array}
\ee

In terms of quaternion matrices, conditions (\ref{appconf26}) correspond to
\be
\label{conf27b}
q_2 = {i\over2\ell}\, q_3^\dagger \,J\, q_3, \qquad\qquad
J =  \left(\begin{array}{cc} 1&0 \\ 0&-1 \end{array}\right).
\ee

With the gauge choices $A_{-1}= \Re q_{+2}=0$ and $A_{+1}$ real, the
$SO(1,1)$ gauge field reads
\be
\label{conf40}
W_\mu = {i\over2g} \, { q_{+3}^*\stackrel{\leftrightarrow}{\partial_\mu}q_{+3} 
+ q_{-3}^*\stackrel{\leftrightarrow}{\partial_\mu}q_{-3}  \over 
\ell^2 + \det q_3 -{\kappa^2\over4}(\det q_3)^2 }
\ee
in terms of $q_{\pm3}$. Its contribution to the scalar Lagrangian is
\be
\label{conf41}
{\cal L}_{SO(1,1)} =
{1\over2}\kappa^2 A_{+1}^2 \, 
{ (q_{+3}^*\stackrel{\leftrightarrow}{\partial_\mu}q_{+3} 
+ q_{-3}^*\stackrel{\leftrightarrow}{\partial_\mu}q_{-3})^2  \over 
\ell^2 + \det q_3 -{\kappa^2\over4}(\det q_3)^2 } . 
\ee

To calculate the various contributions to the scalar Lagrangian 
(\ref{appconf20c}), we introduce new (real) coordinates $(r,\theta,\phi,\tau)$:
\be
\label{conf42}
q_{+3} = r \cos{\theta\over2} \, e^{i(\phi+\tau)/2}, \qquad\qquad
q_{-3} = r \sin{\theta\over2} \, e^{-i(\phi-\tau)/2}.
\ee
With these variables,
$$
\begin{array}{l}
\det q_3 = r^2,
\crbig
|dq_{+3}|^2 + |dq_{-3}|^2 = dr^2 + {r^2\over4}( d\theta^2 + \sin^2\theta \,
d\phi^2 ) + {r^2\over4} (d\tau + \cos\theta\,d\phi)^2,
\crbig
q_{+2} = {ir^2\over2\ell} \cos\theta,
\qquad\qquad
q_{-2} = {ir^2\over2\ell} \sin\theta \, e^{-i\phi},
\crbig
|dq_{+2}|^2 + |dq_{-2}|^2 = {r^2\over\ell^2} \Bigl[dr^2 + {r^2\over4}(d\theta^2
+ \sin^2\theta\,d\phi^2) \Bigr] ,
\crbig
\kappa A_{+1} = [1 - \kappa^2 r^2 - {\kappa^2\over4\ell^2}r^4 ]^{-1/2}.
\end{array}
$$
The basic scalar kinetic terms become
\be
\label{conf43}
\begin{array}{rcl}
{\cal L}_0 &=& 2\kappa^2A_{+1}^2 \Bigl[
|\partial_\mu q_{+2}|^2 + |\partial_\mu q_{-2}|^2+|\partial_\mu q_{+3}|^2 
+ |\partial_\mu q_{-3}|^2 \Bigr]
\crbig
&=& 2\kappa^2A_{+1}^2\biggl[ \left( 1 + {r^2\over\ell^2}\right) 
\left[ (\partial_\mu r)^2 + {r^2\over4}\{ (\partial_\mu \theta)^2 
+ \sin^2\theta \,(\partial_\mu \phi)^2 \}\right] 
\crbig
&& + {r^2\over4} (\partial_\mu \tau + \cos\theta\,\partial_\mu \phi)^2 \biggr].
\end{array}
\ee
The contribution of the $SO(1,1)$ gauge field is
\be
\label{conf44}
{\cal L}_{SO(1,1)} = 
-{r^4\over2} \kappa^2 A_{+1}^2
{( \partial_\mu\tau + \cos\theta \, \partial_\mu\phi )^2 
\over \ell^2 + r^2 - {\kappa^2\over4} r^4}.
\ee
The constribution of the $SU(2)$ gauge fields is
\be
\label{conf45}
{\cal L}_{SU(2)} = {1\over2} \kappa^6 A_{+1}^4 \, r^4 \biggl[
(\partial_\mu\tau + \cos\theta\,\partial_\mu\phi)^2
+ \biggl( 1 + {r^2\over2\ell^2}\biggr)^2
\Bigl\{ (\partial_\mu\theta)^2 + \sin^2\theta(\partial_\mu\phi)^2 \Bigr\}
\biggr].
\ee
Finally
\be
\label{conf46}
{\cal L}_{A_{+1}} =  2\kappa^6 A_{+1}^4 \left( 1 + {r^2\over2\ell^2}\right)^2
\, r^2 (\partial_\mu r)(\partial^\mu r).
\ee
Both ${\cal L}_{SU(2)}$ and ${\cal L}_{A_{+1}}$ vanish (like $\kappa^2$) 
in the limit $\kappa\rightarrow0$.
Then, summing the four contributions leads to the scalar Lagrangian
\be
\label{conf47}
\begin{array}{rcl}
{\cal L} &=&
{1\over2\ell^2}\kappa^4A_{+1}^4
\left( 1 + {\ell^2\over r^2} - {\kappa^2 r^2\over4} \right)
\left[ 4r^2(\partial_\mu r)^2 + r^4 \{ (\partial_\mu \theta)^2 
+ \sin^2\theta \,(\partial_\mu \phi)^2 \}\right] 
\crbig
&& 
+ {1\over2\ell^2}\kappa^4A_{+1}^4 \,
{ (\ell^2 + {\kappa^2 r^4\over4})^2
\over 1 + {\ell^2\over r^2} - {\kappa^2 r^2\over4} }
\, (\partial_\mu \tau + \cos\theta\,\partial_\mu \phi)^2.
\end{array}
\ee
If we define a new variable $R=r^2/\ell$, choosing a positive $\ell$, the theory becomes
\be
\label{conf49}
\begin{array}{rcl}
{\cal L} &=&
{1\over2}\kappa^4A_{+1}^4
\left( 1 + {\ell\over R} - {\kappa^2 \ell R\over4} \right)
 \left[ (\partial_\mu R)^2 + R^2 \{ (\partial_\mu \theta)^2 
+ \sin^2\theta \,(\partial_\mu \phi)^2 \}\right] 
\crbig
&& 
+ {1\over2}\kappa^4A_{+1}^4
{ (1 + {\kappa^2 R^2\over4})^2
\over 1 + {\ell\over R} - {\kappa^2 \ell R\over4} } \, \ell^2
\, (\partial_\mu \tau + \cos\theta\,\partial_\mu \phi)^2 ,
\end{array}
\ee
where\footnote{Positivity implies $R\le {2\over\kappa}(\sqrt{1 + \kappa^2\ell^2} - \kappa\ell)$.}
\be
\label{conf48}
\kappa^4 A_{+1}^4 = \left[1 - \kappa^2\ell R - {\kappa^2R^2\over4}
\right]^{-2}.
\ee
The parameter $\ell$ defines the energy scale of the field $R$ while 
the length $\kappa$ defines the curvature of the quaternionic manifold. 
The metric defined by these kinetic terms is Einstein with
\be
\label{conf48a}
R_{ab} = -6\kappa^2 \, g_{ab} \,,
\ee
as expected for a single hypermultiplet quaternionic space \cite{BW}.

The limit $\kappa\rightarrow0$ leads to
\be
\label{conf50}
\begin{array}{rcl}
{\cal L}_{\kappa\rightarrow0}
&=& 
{1\over2} \biggl[ \left( 1 + {\ell\over R}\right) 
\left[ (\partial_\mu R)^2 + R^2 \{ (\partial_\mu \theta)^2 
+ \sin^2\theta \,(\partial_\mu \phi)^2 \}\right] 
\crbig
&&  +  {\ell^2\over 1+ {\ell\over R}}
 (\partial_\mu \tau + \cos\theta\,\partial_\mu \phi)^2 \biggr].
\end{array}
\ee
We will see later [eq.~(\ref{TNUT4})] that the metric of this scalar Lagrangian is the Taub-NUT metric with $2M=\ell$. 

There are four isometries acting on $\theta$, $\phi$ and $\tau$. Three are
the spherical symmetries of $(\partial_\mu \theta)^2 
+ \sin^2\theta \,(\partial_\mu \phi)^2$, the fourth isometry is the shift of $\tau$.
Explicitly, the metric is invariant under 
\be
\begin{array}{rcl}
\delta\theta &=& \sin\phi\, c_2 + \cos\phi\, c_3,
\crbig
\delta\phi &=& c_1 +\rm{cotg}\,\theta(\cos\phi\,c_2 - \sin\phi\,c_3),
\crbig
\delta\tau &=& c_4 - {1\over\sin\theta}( \cos\phi\, c_2 - \sin\phi\, c_3  ).
\end{array}
\ee
where $C_I$, $I=1,2,3,4$ are the real parameters of the isometries. 
The $SU(2)$ algebra is verified by transformations
with parameters $c_1$, $c_2$ and $c_3$.

We introduce cartesian coordinates $x_i$, $i=1,2,3$ instead of the polar
coordinates $R,\theta,\phi$:
$$
x_1 = R \sin\theta\cos\phi, \qquad
x_2 = R \sin\theta\sin\phi, \qquad
x_3 = R \cos\theta.
$$
Using
$$
{x_1 dx_2 - x_2 dx_1 \over x_1^2 + x_2^2} = d\phi,
\qquad\qquad
{x_3\over R} = \cos\theta, \qquad
R = \sqrt{x_1^2 + x_2^2 + x_3^2},
$$
We can rewrite our Lagrangian in the following form:
\be
\label{conf51}
{\cal L} = F(R) \, (\partial_\mu x_i)(\partial^\mu x_i)
+ G(R) (\partial_\mu\tau + \omega_i \partial_\mu x_i)^2.
\ee
We find
\be
\label{conf52}
\begin{array}{rcl}
F(R) &=& \displaystyle{ {1\over2}\kappa^4A_{+1}^4
\left[ 1 + {\ell\over R} - {\kappa^2\ell R\over4} \right], }
\crbig
G(R) &=& \displaystyle{ {1\over2}\kappa^4A_{+1}^4
{ (1 + {\kappa^2 R^2\over4})^2
\over 1 + {\ell\over R} - {\kappa^2 \ell R\over4} } \, \ell^2,}
\crbig
\omega_1 &=& \displaystyle{  - {x_2x_3 \over R(x_1^2+x_2^2)},}
\qquad
\omega_2 \,\,=\,\, \displaystyle{ {x_1x_3 \over R(x_1^2+x_2^2)},}
\qquad
\omega_3\,\,=\,\,0.
\end{array}
\ee
In the limit $\kappa\rightarrow0$, $F(R)G(R) = \ell^2/4$.

Notice that
$$
{d\over dR}\left[ 1 + {\ell\over R} - {\kappa^2\ell R\over4} \right] = -{\ell\over R^2}
\left[ 1 + {\kappa^2 R^2 \over4} \right]
$$

In a set of KÓahler coordinates $z^i=(T,\Phi)$, one can in general write
$$
ds^2 = K_{T\ov T} \left(dT + {K_{\Phi\ov T}\over K_{T\ov T}} d\Phi \right)
\left(d\ov T + {K_{T\ov\Phi}\over K_{T\ov T}} d\ov\Phi\right)
+ { \det K_{i\ov j} \over K_{T\ov T}} \, d\Phi d\ov\Phi. 
$$
For an Einstein space with $R_{i\ov j} = \partial_i\partial_{\ov j}\ln\det K_{k\ov l} = \Delta K_{i\ov j}$,
$$
ds^2 = K_{T\ov T} \left(dT + {K_{\Phi\ov T}\over K_{T\ov T}} d\Phi \right)
\left(d\ov T + {K_{T\ov\Phi}\over K_{T\ov T}} d\ov\Phi\right)
+ {Ae^{\Delta K} \over K_{T\ov T}} \, d\Phi d\ov\Phi,
$$
where $A$ is an arbitrary positive constant.\footnote{$A$ could be in principle a harmonic 
function $f(T,\Phi) + \ov f(\ov T, \ov\Phi)$ but this case is irrelevant for us.}
Defining $K=-n\ln Y$, the line element is 
$$
ds^2 = K_{T\ov T} \left(dT + {K_{\Phi\ov T}\over K_{T\ov T}} d\Phi \right)
\left(d\ov T + {K_{T\ov\Phi}\over K_{T\ov T}} d\ov\Phi\right)
+ {AY^{-n\Delta} \over K_{T\ov T}} \, d\Phi d\ov\Phi. 
$$
If we further assume that the K\"ahler potential $K$ is a function of $T+\ov T$, $\Phi$ and $\ov\Phi$,
since $T$ is dual to a linear superfield, 
$$
dK_T = K_{T\ov T} \, d (T+\ov T) + K_{\Phi\ov T}d\Phi + K_{T\ov\Phi} d\ov\Phi ,
$$
and the line element becomes
$$
\begin{array}{rcl}
ds^2 &=& K_{T\ov T} \left(d\Re T + {K_{\Phi\ov T}\over 2K_{T\ov T}} d\Phi 
+ {K_{T\ov\Phi}\over 2K_{T\ov T}} d\ov\Phi \right)^2
\crbig
&&
+ K_{T\ov T} \left(d\Im T + {K_{\Phi\ov T}\over 2iK_{T\ov T}} d\Phi
- {K_{T\ov\Phi}\over 2iK_{T\ov T}} d\ov\Phi \right)^2
+ {AY^{-n\Delta} \over K_{T\ov T}} \, d\Phi d\ov\Phi
\crbig
&=& {1\over 4K_{T\ov T}} (dK_T)^2  
+ K_{T\ov T} \left(d\Im T + {K_{\Phi\ov T}\over 2iK_{T\ov T}} d\Phi
- {K_{T\ov\Phi}\over 2iK_{T\ov T}} d\ov\Phi \right)^2
+ {AY^{-n\Delta} \over K_{T\ov T}} \, d\Phi d\ov\Phi .
\end{array}
$$

\section{Taub-NUT}  \label{secTaubNUT}

The Taub-NUT (Taub-Newman-Unti-Tamburino) metric \cite{TaubNUT} describes 
a four-dimen\-sio\-nal euclidean space with self-dual curvature. It is then Ricci-flat and a solution of the
vacuum Einstein equations. Hence, it is also hyper-K\"ahler and appropriate to describe
the scalar sector of a globally $\cN=2$ hypermultiplet theory. 

The Taub-NUT metric is commonly expressed in coordinates where
\be
\label{TNUT1}
ds^2_{TN} = { r+M \over r-M} \, dr^2 + (r^2-M^2)\,(\sigma_1^2+\sigma_2^2)
+ 4M^2\, { r-M \over r+M} \, \sigma_3^2.
\ee
The one-forms
\be
\label{TNUT2}
\begin{array}{rcl}
\sigma_1 &=& \cos\tau \, d\theta + \sin\tau \sin\theta \, d\varphi, 
\crbig
\sigma_2 &=& -\sin\tau \, d\theta + \cos\tau \sin\theta \, d\varphi, 
\crbig
\sigma_3 &=& \cos\theta \, d\varphi +  d\tau
\end{array}
\ee
verify
\be
\label{TNUT3}
d\sigma_x = -\epsilon_{xyz} \,\sigma_y\wedge\sigma_z \qquad\qquad
(x,y,z=1,2,3).
\ee
The coordinates $\tau$, $\theta$ and $\varphi$ are angular variables ($0\le\theta\le\pi$,
$0\le\tau\le4\pi$, $0\le\varphi\le2\pi$), $r>M$ and $M$ is a (real) parameter. 
A more convenient form is obtained by shifting the singularity from $r=M$ to
$R=0$ with the redefinition $R = r-M$. The metric becomes \cite{GH}
\be
\label{TNUT4}
ds^2_{TN} = V [ dR^2 + R^2\, d\Omega] + {4M^2\over V}[ d\tau + \cos\theta\, d\varphi ]^2,
\qquad\qquad
V = 1 + {2M\over R},
\ee
where
\be
\label{TNUT5}
d\Omega = \sigma_1^2 + \sigma_2^2 = d\theta^2 + \sin^2\theta\, d\varphi^2.
\ee
This form is reminiscent of a (euclidean) Schwarzschild metric. Since
\be
\label{TNUT4a}
{1 \over 4M^2}\, ds^2_{TN} = {\rho+1\over\rho} [ d\rho^2 + \rho^2\, d\Omega] 
+ {\rho\over \rho+1}[ d\tau + \cos\theta\, d\varphi ]^2,
\qquad\quad
\rho = {R \over 2M},
\ee
the constant $2M$ sets the scale of the radial coordinate $R$.
Notice that the determinant of the metric is
\be
\label{TNUT6}
\det g_{ab} = 4M^2V^2R^4\sin^2\theta = 4M^2R^2 \sin^2\theta (R+2M)^2.
\ee
It would be constant in K\"ahler coordinates.

In cartesian coordinates $q^a=(\tau, x^i)$, with $dx^idx^i= dR^2 + R^2 d\Omega$,
the metric is
\be
\label{TNUT7}
ds^2_{TN} = 2M\left( {V\over2M} \, dx^i\,dx^i + {2M\over V}[ d\tau - \omega^i\, dx^i]^2 \right), 
\ee
with
\be
\label{TNUT8}
\omega^1 = - { x^2x^3\over R({x^1}^2 + {x^2}^2)},  \qquad
\omega^2 = { x^1x^3\over R({x^1}^2 + {x^2}^2)},  \qquad
\omega^3 = 0.
\ee
The relation \cite{GH}
\be
\label{TNUT9}
\vec\nabla {V\over2M} = \vec \nabla \wedge \vec \omega ,
\ee
which is required for four-dimensional hyperk\"ahler manifold, is verified. 

The Taub-NUT metric (\ref{TNUT4}) is invariant under $SU(2)\times U(1)$ isometries
\be
\label{TNUT9a}
\begin{array}{rcl}
\delta\theta &=& \sin\phi\, c_2 + \cos\phi\, c_3,
\crbig
\delta\phi &=& c_1 +\rm{cotg}\,\theta(\cos\phi\,c_2 - \sin\phi\,c_3),
\crbig
\delta\tau &=& c_4 - {1\over\sin\theta}( \cos\phi\, c_2 - \sin\phi\, c_3  ),
\end{array}
\ee
where $c_I$, $I=1,2,3,4$ are constant real parameters. 
The $SU(2)$ algebra is generated by transformations
with parameters $c_1$, $c_2$ and $c_3$. On cartesian coordinates $x^i$, the action
of the $SU(2)$ isometries is
\be
\label{TNUT9b}
\delta x^1 = - c_1x^2 + c_3 x^3 ,
\qquad
\delta x^2 = c_1 x^1  + c_2x^3 ,
\qquad
\delta x^3 = - c_3x^1  - c_2 x^2 .
\ee
On the K\"ahler coordinate $\Phi = (x^1+ix^2)/\sqrt2$, 
\be
\label{TNUT9c}
\delta \Phi = ic_1\,\Phi + {1\over\sqrt2}( c_3+ic_2 ) U \,,
\qquad
\delta U = -{1\over\sqrt2}( c_3- ic_2 ) \Phi - {1\over\sqrt2}( c_3+ic_2 ) \ov\Phi \,, 
\ee
where $U=x^3$, leaving $U^2 + 2\Phi\ov\Phi$ invariant.\footnote{The phase rotation of $\Phi$
has parameter $c_1$.} 

The single-tensor $\cN=2$ theory leading to the Taub-NUT scalar manifold in the 
hypermultiplet formulation is defined by the function
\be
\label{TNUT10}
{\cal H}(L,\Phi,\ov\Phi) = -{1\over2}\Bigl[L^2- \Phi\ov\Phi\Bigr]
+2M\left[ \sqrt{L^2 +2\Phi\ov\Phi} - L\ln\left( L +  \sqrt{L^2 +2\Phi\ov\Phi} \right) \right],
\ee
obtained by integrating eqs. (\ref{Ahyp10}). The real superfield $L^2+2\Phi\ov\Phi$ is
$R^2$. Since the action of isometries does not respect in general the chiral or linear 
nature of a superfield, we do not expect ${\cal H}$ to be invariant, but the line element (the kinetic terms) should be invariant.

In the hypermultiplet formulation, the line element
reads
\be
\label{TNUT11}
ds^2_{TN} = {1\over2} {\cal V} \, d\Phi\, d\ov\Phi + {\cal V}^{-1} \left| m \, dT + M \, {\cal U}\, {d\Phi\over\Phi}  \right|^2 ,
\ee
where
\be
\label{TNUT12}
{\cal V} = 1 + {2M\over\sqrt{U^2+2\Phi\ov\Phi}}, \qquad\qquad
{\cal U} = 1 - {U \over \sqrt{U^2+2\Phi\ov\Phi}}
\ee
and $U$ is defined (as a function of $T+\ov T$ and $\Phi\ov\Phi$) by the Legendre 
transformation (\ref{Ahyp2}):
\be
\label{TNUT13}
U + 2M\,\ln\left( U +  \sqrt{U^2+2\Phi\ov\Phi} \right)
= -m (T+\ov T) .
\ee
This equation cannot be analytically inverted. The determinant of the K\"ahler metric is constant,
as in eq.~(\ref{Ahyp5}), and the second eq.~(\ref{TNUT9c}) indicates that the $SU(2)$ isometries
act on $T+\ov T$ according to
\be
\label{TNUT13b}
\delta (T+\ov T) =   {1\over \sqrt2m} 
\left( 1 + {2M\over U+R} \right)
\left[ ( c_3- ic_2 ) \Phi + ( c_3+ic_2 ) \ov\Phi \right].
\ee

To compare eqs. (\ref{TNUT11}) and (\ref{TNUT7}), we need to rewrite $ds^2_{TN}$ in coordinates
$(\tau, x^i)$ with $x^i = (\sqrt2\Re\Phi,\sqrt2\Im\Phi, U)$ and $R^2= x^ix^i = U^2+2\Phi\ov\Phi$.
Hence,
$$
{\cal V} = 1 + {2M\over R} = V(R), \qquad\qquad
{\cal U} = 1 - {U \over R}
$$
and, according to eq.~(\ref{TNUT13}),
$$
2m \, d\Re T = -dU -2M \, {dU+dR \over U + R}
= - V\, dU - {2M\over R(U+R)}\, d(\Phi\ov\Phi) .
$$
We first obtain
$$
\begin{array}{rcl}
ds^2_{TN} &=& 
V^{-1} \left[ m\,d\Re T
+ {M{\cal U}\over 2\Phi\ov\Phi} \, d(\Phi\ov\Phi) \right]^2 + {V\over2}\, d\Phi d \ov\Phi
\crbig
&& + V^{-1} \left[ m\,d\Im T + {M{\cal U}\over \Phi\ov\Phi} (\Re\Phi\,d\,\Im\Phi - \Im\Phi\,d\,\Re\Phi )
\right]^2 .
\end{array}
$$
Since ${M{\cal U} \over 2\Phi\ov\Phi } = { M \over R(U+R) }$, we have
\be
\label{TNUT14}
\begin{array}{rcl}
ds^2_{TN} &=& {1\over4} \, V \, \Bigl[ (dU)^2 + 2 \, d\Phi d\ov\Phi \Bigr]
\crbig
&& + V^{-1} \left[ m\,d\Im T + {2M\over R(U+R)} (\Re\Phi\,d\,\Im\Phi - \Im\Phi\,d\,\Re\Phi )
\right]^2 
\crbig
&=& {1\over4} \left( V \, dx^i dx^i + 4m^2 V^{-1} 
\Bigl[ d\Im T +  {M\over m R(U+R)}( x^1 dx^2 - x^2 dx^1 ) \Bigr]^2  \right).
\end{array}
\ee
Finally, we set $m=M$ and use
$$
\begin{array}{rcl}
{1\over R(U+R)}( x^1 dx^2 - x^2 dx^1 ) &=& -{U\over R(R^2-U^2)}( x^1 dx^2 - x^2 dx^1 )
+ {i\over2} \, d \, \ln (\ov\Phi/\Phi)
\crbig
&=& - \omega_i \, dx^i + {i\over2} \, d \, \ln (\ov\Phi/\Phi),
\end{array}
$$
with $\omega_i$ as in eq.~(\ref{TNUT8}). Finally,
\be
\label{TNUT15}
ds^2_{TN} = {1\over4} \left( V \, dx^i dx^i + 4M^2 V^{-1} 
\Bigl[ d\Im T + {i\over2} \, d \, \ln (\ov\Phi/\Phi) - \omega_i \, dx^i \Bigr]^2  \right).
\ee
Comparison with expression (\ref{TNUT7}) indicates that the fourth coordinate is
\be
\label{TNUT16}
\tau = \Im T + {i\over2} \, \ln (\ov\Phi/\Phi).
\ee
The action of $SU(2)\times U(1)$ isometries on $\tau$ is
\be
\label{TNUT17}
\delta \tau = c_4 -{R\over2\sqrt2}{c_2-ic_3 \over \Phi} -{R\over2\sqrt2}{c_2+ic_3 \over \ov\Phi}.
\ee
Hence,
\be
\begin{array}{rcl}
\delta \Im T &=& c_4 - c_1 -{1\over\sqrt2}\, {1\over R-U} [ (c_2+ic_3)\Phi
+ (c_2-ic_3)\ov\Phi]
\crbig
\delta T &=& i(c_4 - c_1) + {1\over\sqrt2} \left[ {R\over\Phi\ov\Phi} + {1\over2M} \right](c_3-ic_2)\Phi
- {1\over\sqrt2} \left[ {U\over\Phi\ov\Phi} - {1\over2M} \right](c_3+ic_2)\ov\Phi
\end{array}
\ee

To summarize, K\"ahler coordinates $T$ and $\Phi$ of the Taub-NUT metric are related to standard 
variables by $(\tau, x^i) = (\tau, \sqrt2\Re\Phi, \sqrt2\Im\Phi, U)$. Eq.~(\ref{TNUT16}) defines 
$\Im T$ while the Legendre  transformation (\ref{TNUT13}) gives implicitly $\Re T$.

\end{document}